\documentclass[11pt, a4paper, copyright, goog]{google}

\usepackage[authoryear, sort&compress, round]{natbib}
\usepackage{booktabs}
\usepackage{graphicx}
\usepackage{xcolor}

\usepackage{amsmath,amsfonts,bm}

\def\eqref#1{equation~\ref{#1}}

\def\1{\bm{1}}

\DeclareMathAlphabet{\mathsfit}{\encodingdefault}{\sfdefault}{m}{sl}
\SetMathAlphabet{\mathsfit}{bold}{\encodingdefault}{\sfdefault}{bx}{n}

\usepackage{amsmath}
\usepackage{amssymb}
\usepackage{float}
\usepackage{makecell}
\usepackage{multirow}
\usepackage{xltabular}
\usepackage{adjustbox}
\usepackage{wrapfig}

\newsavebox{\tabnotebox}

\newcommand{\sysname}{CLASHarness}

\keywords{multi-agent systems, conflict resolution, computer use, benchmarks}

\uselogo{}

\title{CLASHarness: Benchmarking Multi-Agent Conflict in Computer Use via a Controllable and Verifiable Generation Pipeline}

\correspondingauthor{ylai76@emory.edu, ruixifan@google.com}
\reportnumber{0001}
\renewcommand{\today}{2026-09-26}

\author[1,2]{Yuxiang Lai}
\author[3]{Jike Zhong}
\author[1,6]{Chengzhi Liu}
\author[4]{Ming Li}
\author[1,7]{Haonan Wang}
\author[2,5]{Xiaofeng Yang}
\author[2]{Liang Zhao}
\author[1]{Youbao Tang}
\author[1]{Ruixi Fan}

\affil[1]{\thepa{}{}}
\affil[2]{Emory University}
\affil[3]{University of Southern California}
\affil[4]{University of Florida}
\affil[5]{University of Chicago}
\affil[6]{University of California, Santa Barbara}
\affil[7]{Columbia University}
\begin{abstract}
Vision-language models (VLMs) are increasingly deployed in multi-agent systems to automate complex computer-use workflows, prompting numerous benchmarks to evaluate and advance agent capabilities. However, existing benchmarks mostly assume consistent, cooperative environments. In contrast, real-world collaboration rarely runs on consistent information: peer or sub-agents frequently report outdated states, issue contradictory instructions, or compete over shared permissions and files, forcing an agent to check others' outputs and verify ground truth in the operating environment. Evaluating such conflict resolution remains challenging: manual task authoring cannot scale across complex systems, while generating tasks via unconstrained LLM generation produces narrative drift, answer leakage, and unsatisfiable grading rubrics that leave tasks unsolvable. We introduce \sysname{} (\textbf{C}onflict-\textbf{L}oaded \textbf{A}gent \textbf{S}cenario \textbf{H}arness), a controllable framework for generating verifiable multi-agent conflicts in computer use, accompanied by a benchmark of 600 multi-round tasks. By combining hierarchical scenario planning with an eight-step verified micro-pipeline, \sysname{} pre-commits decisive visual facts before rendering and compiles hybrid code and semantic graders directly from executed reference trajectories, guaranteeing that every task is solvable by construction. Evaluating 15 frontier models on \sysname{} reveals three core bottlenecks. First, when outputs of sub-agents conflict with the environment, weaker models blindly follow misleading information, whereas stronger models will inspect visual evidence to verify ground truth. Second, resolving conflicts is much harder than finding them: while agents can spot inconsistencies, actively fixing the system state drops task success significantly. Third, coordination protocols remain a universal bottleneck: even frontier models fail over 58\% of closing actions, frequently forgetting to release shared locks or hand off tasks to sub-agents. \sysname{} provides a scalable, verifiable foundation for diagnosing coordination failures and building reliable multi-agent systems.
\par\vspace{6pt}\noindent\raisebox{-0.20\height}{\includegraphics[height=1.10em]{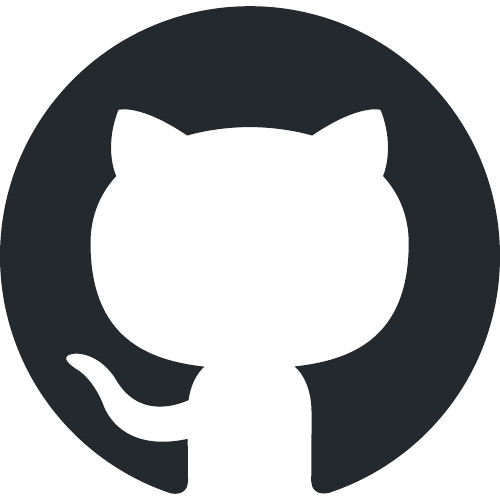}}\hspace{5pt}\textbf{Code \& Dataset:}~\href{https://github.com/google-research/CLASHarness}{\texttt{https://github.com/google-research/CLASHarness}}
\end{abstract}

\begin{document}

\maketitle

\section{Introduction}
\label{sec:intro}

As vision-language models (VLMs) transition to tool-augmented computer-use scenarios, multi-agent systems are increasingly deployed to automate complex collaborative workflows \citep{li2023camel, wu2023autogen, du2023improving}, prompting numerous benchmarks to evaluate and advance agent capabilities \citep{zhou2024webarena, jimenez2024swe, xie2024osworld, xu2026theagentcompany}. However, existing computer-use benchmarks mostly assume consistent, cooperative single-agent environments \citep{xie2024osworld, zhou2024webarena}, while multi-agent evaluations remain largely confined to pure-text dialogue \citep{cemri2026multi, pappu2026multi}. In contrast, real-world collaboration rarely runs on consistent information: peer or sub-agents frequently report outdated states, issue contradictory instructions, or compete over shared permissions and files, triggering severe coordination failures \citep{xie2024adaptive, xu2024knowledge, su2024conflictbank, khatua2026cooperbench, zhang2026silo}. For example, when a sub-agent relies on a stale cache to report a deployment as successful and urges routing production traffic, while live telemetry shows failing health checks, an agent resolving the conflict cannot blindly trust the peer report. Instead, it must check others' outputs, inspect visual dashboards to verify ground truth in the operating environment, and execute a configuration rollback.

Evaluating such conflict resolution faces a fundamental challenge in benchmark construction: manual task curation is labor-intensive and cannot scale across complex operating environments \citep{mialon2024gaia, deng2023mind2web}, while generating tasks via unconstrained LLM prompting produces uncontrollable scenarios suffering from narrative drift, answer leakage, ungrounded visual assets, and unsatisfiable grading rubrics that leave tasks unsolvable \citep{zheng2023judging, lu2025agentrewardbench, peng2026can}.

To address these challenges, we introduce \sysname{} (\textbf{C}onflict-\textbf{L}oaded \textbf{A}gent \textbf{S}cenario \textbf{H}arness), a controllable generation framework and benchmark for synthesizing verifiable multi-agent conflict tasks in computer use. Instead of relying on unconstrained prompting, \sysname{} follows three core design principles to guarantee scenario validity by design: (1)~\textbf{Hierarchical Phased Synthesis:} scenario generation is decomposed into macro scenario planning and verified micro-execution, preventing context overflow and drift across rounds; (2)~\textbf{Pre-Committed Multimodal Grounding:} decisive ground-truth values are committed prior to environment rendering, ensuring visual evidence is truly necessary rather than decorative; and (3)~\textbf{Execution-Compiled Verification:} hybrid evaluation checkers, combining exact code checks and scoped semantic criteria, are compiled directly from executed reference trajectories, guaranteeing that every task is solvable.

We summarize our principal contributions as follows:
\begin{itemize}[leftmargin=*,itemsep=1.5pt]
  \item \textbf{Controllable Generation Framework (\sysname{}):} A hierarchical synthesis pipeline combining declarative outline planning with verified micro-execution, compiling hybrid checkers from executed reference trajectories to guarantee every task is solvable without manual scripting (\S\ref{sec:harness}).
  \item \textbf{600-Round Multimodal Conflict Benchmark:} An extensive benchmark spanning 20 operational domains, 600 executable conflict rounds, 7,970 pre-committed visual assets, and 3,760 hybrid checkers enforcing grounded visual reasoning (\S\ref{sec:benchmark}).
  \item \textbf{Comprehensive Diagnostic Findings:} Across 15 evaluated models, we uncover three systemic bottlenecks: (1)~\emph{Visual Grounding vs.\ Blind Trust:} weaker models blindly follow misleading information, whereas stronger models inspect visual evidence to verify ground truth (\S\ref{sec:taxonomy}); (2)~\emph{Resolution vs.\ Detection:} while agents can spot inconsistencies, actively fixing system state drops task success significantly (\S\ref{sec:conflict-gate}); and (3)~\emph{Coordination Breakdown:} even frontier models fail over 58\% of closing actions, often forgetting to release shared locks or hand off tasks to sub-agents (\S\ref{sec:taxonomy}).
\end{itemize}

\section{Related Work}
\label{sec:related}

\noindent\smallskip\textbf{Agentic Computer-Use and Multi-Agent Benchmarks.} Recent benchmarks evaluate tool-augmented agents on software engineering, OS execution, and web navigation \citep{mialon2024gaia, zhou2024webarena, jimenez2024swe, xie2024osworld, liu2024agentbench, bonatti2024windows, xu2026theagentcompany, yuan2026osworld2, li2026windowsworld, jang2026odysseys, li2026weavebench}, while multi-agent frameworks explore collaborative problem solving \citep{li2023camel, wu2023autogen, du2023improving} alongside emerging coordination failure studies \citep{cemri2026multi, pappu2026multi, khatua2026cooperbench, zhang2026silo, tessera2026benchmarking}. Whereas existing benchmarks assume internally consistent, cooperative environments, \sysname{} evaluates agents under explicit \emph{factual and temporal} contradictions across tools, files, visual evidence, and peer dialogues.

\noindent\smallskip\textbf{Knowledge Conflicts and Multimodal Grounding.} Prior work on knowledge conflicts studies contradictions between model memory and retrieved text in pure-text QA \citep{xu2024knowledge, xie2024adaptive, su2024conflictbank, wang2023resolving, wang2025retrieval, zhao2026exploring, peng2026navigating}, while multimodal benchmarks emphasize graphical navigation without conflicting signals \citep{koh2024visualwebarena, deng2023mind2web}. \sysname{} advances beyond static unimodal evaluations by pre-committing visual ground truths across dashboards, terminal outputs, and system diagrams, and incorporating temporal dynamics where stale evidence and expiring authorizations require grounded action.

\noindent\smallskip\textbf{Test-Time Compute and Reasoning Dynamics.} While scaling test-time reasoning compute consistently boosts performance on mathematical and formal domains \citep{snell2025scaling, guo2025deepseek, muennighoff2025s1}, open-ended agentic environments can trigger overthinking and repetitive action loops \citep{chen2025not, cuadron2025danger}. \sysname{} provides an empirical testbed evaluating how reasoning effort impacts multi-turn execution under operational conflict.

\section{The Generation Harness}
\label{sec:harness}

\begin{figure}[!t]
  \centering
  \includegraphics[width=\linewidth]{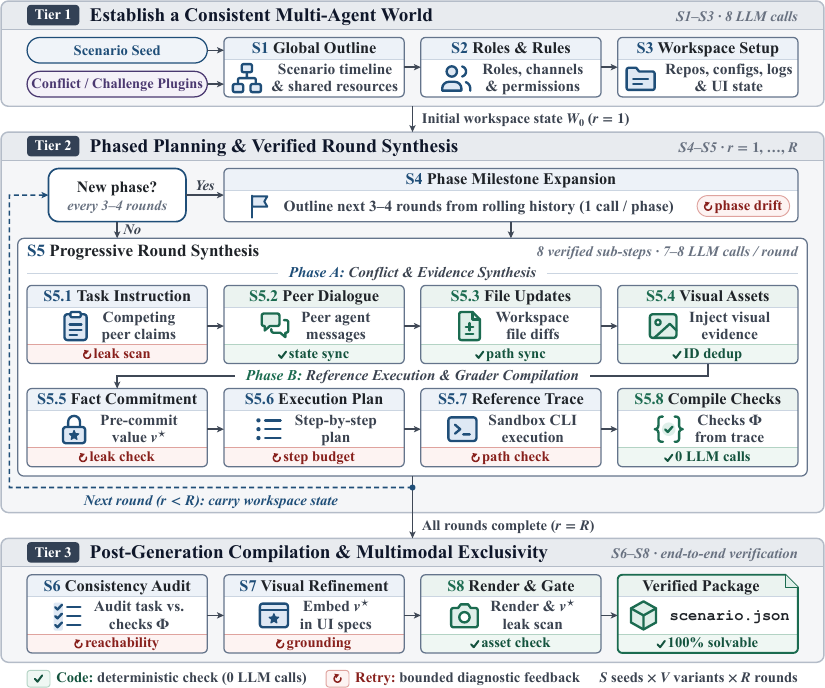}
  \caption{\textbf{Hierarchical generation pipeline in \sysname{}.} Tier~1 initializes a shared multi-agent world from seeds and plugins. Tier~2 refreshes phase milestones every 3--4 rounds and generates each round through eight explicit sub-steps. Ground truth $v^\star$ is committed before rendering; executed reference effects ground deterministic checks and scoped semantic criteria. The dashed return path carries workspace state to the next round ($r<R$); after $R=10$ rounds, generation proceeds to Tier~3 for consistency auditing, visual refinement, and rendering. Visual evidence and checkers share the committed $v^\star$, with text-exclusivity gates enforced when visuals are required. Green tags mark deterministic processing; red tags mark bounded correction with diagnostic feedback. Detailed safeguards appear in Appendix Table~\ref{tab:harness-safeguards}.}
  \label{fig:arch}
\end{figure}

\subsection{Problem Formalization}
\label{sec:formalization}

We formalize a multi-agent conflict task as a \emph{conflict round} defined by the tuple $\mathcal{R} = (\mathcal{W}_0, I, \mathcal{T}, v^{\star}, \Phi)$, where $\mathcal{W}_0$ denotes the initial computer-use workspace containing operating system state, file trees, asynchronous multi-agent communication transcripts, and rendered visual evidence. The instruction $I$ specifies the operational objective and explicitly frames competing claims from peer agents without revealing which assertion is grounded in fact. $\mathcal{T}$ represents the typed action space; executing $c \in \mathcal{T}$ transitions environment state and appends structured records to an audit effect log $\mathcal{E}_{\text{log}}$. The \emph{decisive value} $v^{\star}$ is the ground-truth fact resolving the contradiction. Finally, $\Phi = \{ \phi_1, \dots, \phi_m \}$ is a hybrid evaluation suite combining exact code checks and targeted semantic criteria evaluated over the final workspace state and $\mathcal{E}_{\text{log}}$.

This formulation frames conflict resolution as active investigation and decision-making rather than passive text classification. At inference time, the agent is provided with $(\mathcal{W}_0, I, \mathcal{T})$ and alerted to the contradiction, but must autonomously investigate multimodal evidence across $\mathcal{W}$, identify $v^\star$, and execute operational state changes (e.g., protocol handovers, configuration patches, rollbacks). Checks $\Phi$ are compiled directly from executed reference trajectories (\S\ref{sec:grading}), guaranteeing that every task is solvable. Rounds are grouped into 10-round \emph{packages} sharing an evolving workspace, generated from a structured scenario \emph{seed} (\S\ref{sec:benchmark}).

\subsection{Hierarchical Phased Generation Pipeline}
\label{sec:pipeline}

Synthesizing multi-round, multi-agent conflict scenarios poses a fundamental systems challenge. Under single-pass prompting, generator models must simultaneously author complex operational contexts, enforce conflicting agent identities, manipulate multi-file workspaces, and invent evaluation checkers. Single-pass generation frequently causes broken schemas, inconsistent storylines across rounds, leaked answers, and impossible grading checks. \sysname{} resolves these challenges through \emph{hierarchical phased outline planning}, decomposing synthesis into an eight-stage pipeline across three coordinated tiers with in-loop self-correction (Figure~\ref{fig:arch}).

\noindent\smallskip\textbf{Tier 1: Establishing a Consistent Multi-Agent World (Stages 1--3).} Realistic conflict scenarios require a coherent operational environment before specific conflicts can be introduced. Rather than synthesizing individual rounds in isolation, Tier~1 establishes the global multi-agent foundation across the entire package from structured scenario seeds and declared plugin effects. First, in Stage~1 (\emph{Global Scenario Outline}), the harness constructs the overall scenario progression, milestone timeline, and shared infrastructure allocations across all subsequent rounds. Building upon this global outline, Stage~2 (\emph{Subagent Roles \& Rules}) establishes agent roles, communication channels, and permission boundaries to prevent role drift and unauthorized tool access across rounds. Finally, Stage~3 (\emph{Workspace Setup}) sets up the initial operating workspace, generating code repositories, configuration files, system logs, and UI environments establishing cross-round layout consistency.

\noindent\smallskip\textbf{Tier 2: Phased Planning \& Verified Round Synthesis (Stages 4--5).} Each scenario package consists of $R=10$ sequential rounds. Accumulating the complete workspace and dialogue history across all rounds into a single prompt quickly exhausts context limits and causes inconsistencies. \sysname{} resolves this with a two-level planning hierarchy:
\textbf{High-Level Phase Planning (Stage~4):} The $R$ rounds are partitioned into sequential phases of $P_s \in [3, 4]$ consecutive rounds, such as Rounds 1--3, 4--6, and 7--10. At the start of each phase, Stage~4 (\emph{Phase Milestone Expansion}) outlines high-level milestones for the upcoming 3--4 rounds with a single model call using a rolling window of recent history, maintaining scenario coherence without overloading the prompt.

\textbf{Fine-Grained Round Synthesis (Stage~5):} Within each phase, the harness synthesizes each round sequentially through an eight-step verified micro-pipeline in Stage~5 (\emph{Progressive Round Synthesis}): authoring task instructions that frame competing claims while strictly isolating the factual resolution, synthesizing peer agent dialogue history, applying workspace file updates, injecting visual assets, pre-committing the decisive ground-truth value $v^\star$ under automated leakage scanning, drafting step-by-step execution plans, executing reference command-line trajectories in a live sandbox, and compiling exact code checks $\Phi$ directly from recorded state changes without model calls. Workspace modifications from round $r$ persist and carry over to round $r+1$. Decomposing round synthesis into isolated sub-steps confines potential failures to narrow execution scopes, enabling bounded in-loop retries without discarding valid package state.

\noindent\smallskip\textbf{Tier 3: Post-Generation Compilation \& Multimodal Exclusivity (Stages 6--8).} To guarantee that compiled benchmarks are verifiably solvable and genuinely multimodal, Tier~3 performs comprehensive end-to-end verification. A cross-stage consistency audit verifies natural-language task instructions against reference trajectories and compiled grading checks, eliminating mismatches between task goals and grading criteria in Stage~6 (\emph{Consistency Audit}). Next, Stage~7 (\emph{Multimodal Refinement}) synthesizes declarative HTML/Canvas specifications for visual evidence assets, adapting each asset to its platform's custom UI layout while embedding the decisive value $v^\star$ alongside realistic operational data. Stage~8 (\emph{Render \& Exclusivity Gate}) then renders high-resolution browser screenshots in a headless environment. To prevent agents from bypassing visual inspection via text shortcuts, automated text-leakage scans verify that the decisive value $v^\star$ appears exclusively within rendered visual assets for visually grounded tasks (\S\ref{sec:grading}), compiling the final verified $\mathtt{scenario.json}$.

\noindent\smallskip\textbf{Deterministic Normalization and Bounded Self-Correction.} Generating multi-round scenarios often introduces minor formatting inconsistencies or logical conflicts. Discarding and regenerating an entire 10-round package upon encountering an error would be slow and expensive. Instead, \sysname{} resolves generation defects through a two-tier self-correction mechanism (Figure~\ref{fig:arch}):
First, \textbf{deterministic code normalization} corrects formatting, syntax, and schema errors---such as missing fields or invalid file paths---directly in code with zero model invocations (green tags in Figure~\ref{fig:arch}).
Second, \textbf{bounded semantic retries} resolve logical mismatches: when inconsistencies arise between task instructions and grading checks, the harness confines retries strictly to the affected sub-step, providing diagnostic error feedback (red tags in Figure~\ref{fig:arch}) rather than restarting the package.
Because retries are restricted to isolated sub-steps, total API calls scale linearly with round count: $\text{Total} = 8 + \lceil R / P_s \rceil + K \cdot R + A$, where $K \approx 7\text{--}8$ calls per round and $A$ is the visual asset count, averaging 95 calls per 10-round package (Appendix~\S\ref{sec:app-pipeline}). This makes benchmark synthesis scalable and cost-effective while ensuring all generated tasks are solvable.

\subsection{Declarative Controllability and Modular Plugins}
\label{sec:extensions}
\label{sec:difficulty}

\noindent\smallskip\textbf{Structured Seed Specifications.} Scenario structure is defined via declarative seed schemas governing conflict types, interaction protocols, tool families, and evidence modalities, allowing users to customize parameters and synthesize new domains on demand (Appendix~\S\ref{sec:app-difficulty}).

\noindent\smallskip\textbf{Modular Plugins and State Effects.} To support modular extensions without modifying evaluation engines, \sysname{} separates domain mechanics into 23 modular plugins (16 conflict mechanics, 7 challenge plugins; Table~\ref{tab:coverage}). Each plugin $\mathcal{P}$ and tool command $c \in \mathcal{T}$ declares operational effects $\mathcal{E}(\mathcal{P})$ and $\mathcal{E}(c)$, such as file edits or permission changes. Valid actions and grading criteria match declared effects: $\mathcal{T}_{\text{valid}} = \{ c \in \mathcal{T} \mid \mathcal{E}(c) \cap \mathcal{E}(\mathcal{P}) \neq \emptyset \}$. Users can synthesize new conflicts or custom APIs by authoring a seed schema and registering state effects, without procedural simulation.

\subsection{Verifiable Grading and Multimodal Grounding}
\label{sec:normalize}
\label{sec:grading}

\noindent\smallskip\textbf{Executable Reference Grader Compilation.} In unconstrained generation, natural-language rubrics often specify impossible requirements. \sysname{} compiles checkers directly from executed reference trajectories: exact-match checks verify specific identifiers, while semantic criteria evaluate scoped file modifications (Appendix~\S\ref{sec:app-grading}), eliminating unreachable checks across the benchmark.

\noindent\smallskip\textbf{Multimodal Pre-Commitment \& Exclusivity Gates.} Mirroring cross-boundary ticket attachments, point-in-time incident snapshots, and physical sensor captures where underlying DOM or source access is unavailable, visual assets are provided as standalone image files. Committing $v^{\star}$ \emph{prior} to rendering aligns visual evidence with compiled checkers. Across 600 rounds (\S\ref{sec:benchmark}), 346 require visual evidence (\texttt{role="required"}), with automated leakage scans ensuring $v^{\star}$ appears exclusively in rendered images rather than text.

\section{The \sysname{} Benchmark: Scale, Diversity, and Validity}
\label{sec:benchmark}
\label{sec:scale}
\label{sec:validity}

\noindent\smallskip\textbf{Benchmark Scale and Structural Diversity.} Using \sysname{}, we construct a primary evaluation benchmark comprising 20 structured scenario seeds across diverse operational domains, 60 package variants (3 variants per seed), and 600 executable computer-use conflict rounds ($R=10$). Across these 600 rounds, the benchmark spans 3,760 evaluation checks (3,284 required) and 7,970 rendered visual evidence assets. Reference trajectories average 10.28 steps against a communicated step budget $b=20$ (Appendix~\S\ref{sec:app-corpus-coverage}). Task difficulty is empirically measured by reference trajectory length (7--18 steps; $b / \text{steps} \in [1.11, 2.86]$). Synthesizing the 20 seeds with an alternate model family yields an aligned parallel benchmark with low trajectory correlation (Pearson $r = 0.186$, avg.\ 12.48 steps; \S\ref{sec:generator-ablation}), confirming that the generator shapes execution complexity while seeds govern task semantics. Round diversity is strictly enforced: candidate $v^\star$ values cannot reuse past answers, and only 1.6\% of semantic criteria repeat within a package (2,086 of 2,137 criteria are globally distinct; Appendix~\S\ref{sec:app-diversity}).

\noindent\smallskip\textbf{Hybrid Rubric Composition and Evaluator Stability.} Of the 3,284 scored checks, 34.9\% ($1{,}147$) are deterministic code checks and 65.1\% ($2{,}137$) are targeted semantic criteria evaluated by an LLM judge (Table~\ref{tab:validity}). Every semantic check is scored via 3-sample self-consistency majority voting, achieving 97.5\% unanimous agreement across $19{,}419$ recorded judgments. Re-grading all high-effort rounds using independent frontier judges (Claude Sonnet 5, Claude Opus 5) shows high consistency across judges: relative model rankings remain nearly identical ($\rho \ge 0.98$, Cohen's $\kappa \ge 0.90$; Appendix Table~\ref{tab:judge-app}). In addition, primary diagnostic findings (visual grounding F3, evidence anchoring F4, terminal actions F5, temporal deadlines F6) are evaluated strictly via deterministic code checks, ensuring conclusions hold independently of model judging.

\noindent\smallskip\textbf{Cross-Metric Consistency and Capability Decoupling.} Model rankings under code and semantic rubrics closely agree ($\rho = +0.90$). Importantly, individual check evaluations reveal that visual evidence localization and semantic conflict resolution have almost zero correlation ($\rho = +0.02$). This confirms that visual grounding and conflict resolution are independent capabilities requiring separate evaluation rather than composite scores.

\begin{table}[t]
  \centering
  \footnotesize
  \setlength{\tabcolsep}{4pt}
  \renewcommand{\arraystretch}{0.9}
  \caption{\textbf{Main evaluation results across 15 models on \sysname{} (600 rounds, reported as mean $\pm$ SE).} \emph{Pass Rate}: percentage of tasks satisfying all required checks. \emph{Mean Score}: average percentage of individual checks satisfied per round. Task difficulty is categorized by ground-truth reference trajectory length (\emph{Easy}: $\le 9$ steps, \emph{Mid}: 10--11 steps, \emph{Hard}: $\ge 12$ steps). Best results are in \textbf{bold}, second-best \underline{underlined}. $^\dagger$Text-only models evaluated using OCR transcripts of visual assets. Appendix Tables~\ref{tab:main-high}--\ref{tab:main-low} provide full 95\% CIs and bootstrap details.}
  \label{tab:main}
  \begin{adjustbox}{max width=\linewidth}
  \begin{tabular}{l c | c c c c | c c c c | c c}
    \toprule
    \multirow{2}{*}{\textbf{Model}} & \textbf{Think} & \multicolumn{4}{c|}{\textbf{Pass Rate (\%)}}
      & \multicolumn{4}{c|}{\textbf{Mean Score (\%)}}
      & \multicolumn{2}{c}{\textbf{Avg. Cost}} \\
    \cmidrule(lr){3-6} \cmidrule(lr){7-10} \cmidrule(lr){11-12}
    & \textbf{Level} & \textbf{All} & \textbf{Easy} & \textbf{Mid} & \multicolumn{1}{c|}{\textbf{Hard}}
          & \textbf{All} & \textbf{Easy} & \textbf{Mid} & \multicolumn{1}{c|}{\textbf{Hard}}
          & \textbf{\$ } & \textbf{Tok (M)} \\
    \midrule
    \multirow{3}{*}{Claude Opus 5}                 & High & 25.2{\scriptsize$\pm$1.8} & 32.2{\scriptsize$\pm$3.0} & \underline{25.0}{\scriptsize$\pm$2.9} & \underline{13.6}{\scriptsize$\pm$2.9} & \underline{64.2}{\scriptsize$\pm$1.2} & 67.4{\scriptsize$\pm$1.9} & \underline{68.5}{\scriptsize$\pm$1.8} & \underline{51.8}{\scriptsize$\pm$2.6} & 2{,}631 & 511 \\
                                                        & Med  & \underline{25.3}{\scriptsize$\pm$1.8} & \underline{33.9}{\scriptsize$\pm$3.1} & 24.6{\scriptsize$\pm$2.9} & 12.1{\scriptsize$\pm$2.7} & 63.3{\scriptsize$\pm$1.2} & \underline{68.3}{\scriptsize$\pm$1.9} & 66.5{\scriptsize$\pm$1.9} & 49.7{\scriptsize$\pm$2.5} & 2{,}245 & 439 \\
                                                        & Low  & 23.5{\scriptsize$\pm$1.7} & 30.5{\scriptsize$\pm$3.0} & 23.7{\scriptsize$\pm$2.9} & 11.4{\scriptsize$\pm$2.7} & 60.9{\scriptsize$\pm$1.3} & 66.2{\scriptsize$\pm$1.9} & 64.3{\scriptsize$\pm$1.9} & 46.5{\scriptsize$\pm$2.6} & 1{,}848 & 364 \\
    \addlinespace[2.5pt]
    \multirow{3}{*}{Claude Opus 4.8}               & High & 22.8{\scriptsize$\pm$1.7} & 30.1{\scriptsize$\pm$3.0} & 23.7{\scriptsize$\pm$2.8} & 9.3{\scriptsize$\pm$2.4} & 62.1{\scriptsize$\pm$1.2} & 65.6{\scriptsize$\pm$2.0} & 67.4{\scriptsize$\pm$1.8} & 47.7{\scriptsize$\pm$2.5} & 2{,}933 & 571 \\
                                                        & Med  & 23.8{\scriptsize$\pm$1.7} & 29.7{\scriptsize$\pm$3.0} & 24.6{\scriptsize$\pm$2.9} & 12.9{\scriptsize$\pm$2.8} & 61.8{\scriptsize$\pm$1.2} & 65.1{\scriptsize$\pm$2.0} & 66.6{\scriptsize$\pm$1.9} & 48.5{\scriptsize$\pm$2.5} & 2{,}646 & 516 \\
                                                        & Low  & 21.3{\scriptsize$\pm$1.7} & 28.0{\scriptsize$\pm$2.9} & 21.4{\scriptsize$\pm$2.7} & 10.0{\scriptsize$\pm$2.6} & 60.8{\scriptsize$\pm$1.2} & 64.0{\scriptsize$\pm$2.0} & 65.8{\scriptsize$\pm$1.9} & 47.3{\scriptsize$\pm$2.5} & 2{,}296 & 449 \\
    \addlinespace[2.5pt]
    \multirow{3}{*}{Claude Sonnet 5}               & High & 20.8{\scriptsize$\pm$1.7} & 28.0{\scriptsize$\pm$2.9} & 21.4{\scriptsize$\pm$2.7} & 7.9{\scriptsize$\pm$2.3} & 59.6{\scriptsize$\pm$1.3} & 63.9{\scriptsize$\pm$2.0} & 64.1{\scriptsize$\pm$1.9} & 45.3{\scriptsize$\pm$2.5} & 1{,}031 & 497 \\
                                                        & Med  & 17.0{\scriptsize$\pm$1.5} & 25.4{\scriptsize$\pm$2.8} & 17.0{\scriptsize$\pm$2.5} & 2.9{\scriptsize$\pm$1.4} & 57.0{\scriptsize$\pm$1.3} & 61.5{\scriptsize$\pm$2.0} & 61.8{\scriptsize$\pm$2.0} & 41.7{\scriptsize$\pm$2.4} & 836 & 405 \\
                                                        & Low  & 13.8{\scriptsize$\pm$1.4} & 20.8{\scriptsize$\pm$2.6} & 12.9{\scriptsize$\pm$2.3} & 3.6{\scriptsize$\pm$1.6} & 53.7{\scriptsize$\pm$1.2} & 59.1{\scriptsize$\pm$1.9} & 55.8{\scriptsize$\pm$1.9} & 41.3{\scriptsize$\pm$2.3} & 708 & 345 \\
    \midrule
    GPT-6 Astra                                    & High & \textbf{28.7}{\scriptsize$\pm$1.9} & \textbf{35.2}{\scriptsize$\pm$3.1} & \textbf{28.1}{\scriptsize$\pm$3.1} & \textbf{18.6}{\scriptsize$\pm$3.3} & \textbf{66.8}{\scriptsize$\pm$1.2} & \textbf{68.9}{\scriptsize$\pm$2.0} & \textbf{70.1}{\scriptsize$\pm$1.8} & \textbf{58.2}{\scriptsize$\pm$2.6} & 3{,}003 & 663 \\
    \addlinespace[2.5pt]
    \multirow{3}{*}{GPT-5.6 Sol}                   & High & 24.5{\scriptsize$\pm$1.7} & 31.4{\scriptsize$\pm$3.0} & 24.1{\scriptsize$\pm$2.9} & \underline{13.6}{\scriptsize$\pm$2.9} & 63.3{\scriptsize$\pm$1.2} & 66.7{\scriptsize$\pm$1.9} & 66.8{\scriptsize$\pm$1.8} & 51.7{\scriptsize$\pm$2.6} & 2{,}165 & 454 \\
                                                        & Med  & 23.0{\scriptsize$\pm$1.7} & 29.2{\scriptsize$\pm$3.0} & 24.6{\scriptsize$\pm$2.9} & 10.0{\scriptsize$\pm$2.5} & 62.8{\scriptsize$\pm$1.2} & 66.2{\scriptsize$\pm$1.9} & 66.4{\scriptsize$\pm$1.9} & 51.1{\scriptsize$\pm$2.4} & 1{,}939 & 406 \\
                                                        & Low  & 20.8{\scriptsize$\pm$1.7} & 26.3{\scriptsize$\pm$2.9} & 22.8{\scriptsize$\pm$2.8} & 8.6{\scriptsize$\pm$2.4} & 61.0{\scriptsize$\pm$1.2} & 64.4{\scriptsize$\pm$1.9} & 65.6{\scriptsize$\pm$1.9} & 48.1{\scriptsize$\pm$2.4} & 1{,}582 & 330 \\
    \addlinespace[2.5pt]
    \multirow{3}{*}{GPT-5.6 Terra}                 & High & 20.0{\scriptsize$\pm$1.6} & 27.1{\scriptsize$\pm$2.9} & 20.1{\scriptsize$\pm$2.7} & 7.9{\scriptsize$\pm$2.3} & 60.2{\scriptsize$\pm$1.2} & 63.9{\scriptsize$\pm$2.0} & 64.2{\scriptsize$\pm$1.9} & 47.7{\scriptsize$\pm$2.4} & 731 & 382 \\
                                                        & Med  & 20.0{\scriptsize$\pm$1.6} & 25.8{\scriptsize$\pm$2.9} & 20.1{\scriptsize$\pm$2.7} & 10.0{\scriptsize$\pm$2.5} & 58.9{\scriptsize$\pm$1.3} & 62.8{\scriptsize$\pm$2.0} & 61.7{\scriptsize$\pm$2.0} & 47.9{\scriptsize$\pm$2.5} & 636 & 332 \\
                                                        & Low  & 17.7{\scriptsize$\pm$1.6} & 23.7{\scriptsize$\pm$2.8} & 17.9{\scriptsize$\pm$2.6} & 7.1{\scriptsize$\pm$2.2} & 57.2{\scriptsize$\pm$1.2} & 61.9{\scriptsize$\pm$1.9} & 60.0{\scriptsize$\pm$2.0} & 44.6{\scriptsize$\pm$2.5} & 563 & 293 \\
    \addlinespace[2.5pt]
    \multirow{3}{*}{GPT-5.6 Luna}                  & High & 19.8{\scriptsize$\pm$1.6} & 23.7{\scriptsize$\pm$2.7} & 22.3{\scriptsize$\pm$2.8} & 9.3{\scriptsize$\pm$2.5} & 60.4{\scriptsize$\pm$1.2} & 62.0{\scriptsize$\pm$2.0} & 65.0{\scriptsize$\pm$1.9} & 50.1{\scriptsize$\pm$2.5} & 86 & 450 \\
                                                        & Med  & 17.7{\scriptsize$\pm$1.6} & 22.0{\scriptsize$\pm$2.7} & 19.2{\scriptsize$\pm$2.7} & 7.9{\scriptsize$\pm$2.3} & 59.1{\scriptsize$\pm$1.2} & 62.1{\scriptsize$\pm$1.9} & 62.2{\scriptsize$\pm$1.9} & 49.1{\scriptsize$\pm$2.5} & 68 & 357 \\
                                                        & Low  & 19.5{\scriptsize$\pm$1.6} & 25.0{\scriptsize$\pm$2.8} & 18.3{\scriptsize$\pm$2.6} & 12.1{\scriptsize$\pm$2.7} & 59.0{\scriptsize$\pm$1.2} & 62.1{\scriptsize$\pm$1.9} & 62.3{\scriptsize$\pm$1.9} & 48.4{\scriptsize$\pm$2.6} & 59 & 305 \\
    \midrule
    \multirow{3}{*}{Gemini 3.7 Flash}              & High & 20.3{\scriptsize$\pm$1.6} & 26.7{\scriptsize$\pm$2.9} & 21.9{\scriptsize$\pm$2.8} & 7.1{\scriptsize$\pm$2.2} & 58.1{\scriptsize$\pm$1.3} & 60.8{\scriptsize$\pm$2.1} & 63.0{\scriptsize$\pm$2.0} & 45.9{\scriptsize$\pm$2.3} & 307 & 395 \\
                                                        & Med  & 21.5{\scriptsize$\pm$1.7} & 27.1{\scriptsize$\pm$2.9} & 24.1{\scriptsize$\pm$2.9} & 7.9{\scriptsize$\pm$2.3} & 61.8{\scriptsize$\pm$1.2} & 65.1{\scriptsize$\pm$1.9} & 65.5{\scriptsize$\pm$1.9} & 50.2{\scriptsize$\pm$2.4} & 250 & 324 \\
                                                        & Low  & 20.5{\scriptsize$\pm$1.6} & 27.1{\scriptsize$\pm$2.9} & 20.5{\scriptsize$\pm$2.7} & 9.3{\scriptsize$\pm$2.4} & 60.0{\scriptsize$\pm$1.2} & 64.5{\scriptsize$\pm$2.0} & 63.4{\scriptsize$\pm$1.9} & 47.0{\scriptsize$\pm$2.5} & 203 & 265 \\
    \addlinespace[2.5pt]
    \multirow{3}{*}{Gemini 3.6 Flash}              & High & 18.5{\scriptsize$\pm$1.6} & 24.6{\scriptsize$\pm$2.8} & 21.4{\scriptsize$\pm$2.7} & 3.6{\scriptsize$\pm$1.6} & 58.6{\scriptsize$\pm$1.3} & 61.7{\scriptsize$\pm$2.0} & 63.5{\scriptsize$\pm$2.0} & 45.6{\scriptsize$\pm$2.2} & 565 & 363 \\
                                                        & Med  & 20.2{\scriptsize$\pm$1.6} & 27.1{\scriptsize$\pm$2.9} & 21.9{\scriptsize$\pm$2.8} & 5.7{\scriptsize$\pm$2.0} & 60.2{\scriptsize$\pm$1.2} & 63.1{\scriptsize$\pm$2.0} & 64.7{\scriptsize$\pm$1.9} & 48.2{\scriptsize$\pm$2.4} & 508 & 327 \\
                                                        & Low  & 18.5{\scriptsize$\pm$1.6} & 23.3{\scriptsize$\pm$2.7} & 19.2{\scriptsize$\pm$2.6} & 9.3{\scriptsize$\pm$2.5} & 59.5{\scriptsize$\pm$1.2} & 62.0{\scriptsize$\pm$1.9} & 64.2{\scriptsize$\pm$1.9} & 48.0{\scriptsize$\pm$2.5} & 323 & 211 \\
    \addlinespace[2.5pt]
    \multirow{3}{*}{Gemini 3.1 Pro}                & High & 12.0{\scriptsize$\pm$1.3} & 15.7{\scriptsize$\pm$2.4} & 12.1{\scriptsize$\pm$2.2} & 5.7{\scriptsize$\pm$2.0} & 48.7{\scriptsize$\pm$1.3} & 49.3{\scriptsize$\pm$2.1} & 53.6{\scriptsize$\pm$2.0} & 39.9{\scriptsize$\pm$2.3} & 578 & 280 \\
                                                        & Med  & 9.8{\scriptsize$\pm$1.2} & 11.4{\scriptsize$\pm$2.1} & 10.3{\scriptsize$\pm$2.1} & 6.4{\scriptsize$\pm$2.1} & 47.9{\scriptsize$\pm$1.2} & 47.9{\scriptsize$\pm$2.0} & 53.5{\scriptsize$\pm$2.0} & 39.0{\scriptsize$\pm$2.4} & 490 & 239 \\
                                                        & Low  & 8.3{\scriptsize$\pm$1.1} & 10.2{\scriptsize$\pm$2.0} & 8.5{\scriptsize$\pm$1.8} & 5.0{\scriptsize$\pm$1.9} & 46.3{\scriptsize$\pm$1.2} & 48.0{\scriptsize$\pm$1.9} & 50.3{\scriptsize$\pm$1.9} & 36.9{\scriptsize$\pm$2.3} & 436 & 213 \\
    \addlinespace[2.5pt]
    \multirow{3}{*}{Gemini 3.5 Flash}              & High & 3.0{\scriptsize$\pm$0.7} & 3.8{\scriptsize$\pm$1.2} & 2.7{\scriptsize$\pm$1.1} & 2.1{\scriptsize$\pm$1.2} & 25.9{\scriptsize$\pm$1.1} & 21.6{\scriptsize$\pm$1.7} & 29.8{\scriptsize$\pm$1.7} & 27.2{\scriptsize$\pm$2.1} & 712 & 445 \\
                                                        & Med  & 3.2{\scriptsize$\pm$0.7} & 3.8{\scriptsize$\pm$1.2} & 3.6{\scriptsize$\pm$1.2} & 1.4{\scriptsize$\pm$1.0} & 29.9{\scriptsize$\pm$1.1} & 27.3{\scriptsize$\pm$1.8} & 32.3{\scriptsize$\pm$1.8} & 30.5{\scriptsize$\pm$2.1} & 703 & 440 \\
                                                        & Low  & 13.5{\scriptsize$\pm$1.4} & 17.4{\scriptsize$\pm$2.4} & 14.3{\scriptsize$\pm$2.4} & 5.7{\scriptsize$\pm$2.0} & 47.3{\scriptsize$\pm$1.3} & 45.9{\scriptsize$\pm$2.3} & 52.6{\scriptsize$\pm$2.1} & 40.8{\scriptsize$\pm$2.3} & 547 & 349 \\
    \midrule
    Kimi K3                                    & High & 21.8{\scriptsize$\pm$1.7} & 26.3{\scriptsize$\pm$2.8} & 23.2{\scriptsize$\pm$2.9} & 12.1{\scriptsize$\pm$2.7} & 55.9{\scriptsize$\pm$1.4} & 55.8{\scriptsize$\pm$2.3} & 61.6{\scriptsize$\pm$2.2} & 47.0{\scriptsize$\pm$2.7} & 2{,}173 & 706 \\
    DeepSeek V4 Pro$^\dagger$                  & High & 19.8{\scriptsize$\pm$1.6} & 24.2{\scriptsize$\pm$2.8} & 20.5{\scriptsize$\pm$2.7} & 11.4{\scriptsize$\pm$2.7} & 58.1{\scriptsize$\pm$1.3} & 58.8{\scriptsize$\pm$2.1} & 64.4{\scriptsize$\pm$1.9} & 46.9{\scriptsize$\pm$2.6} & 32 & 70 \\
    GLM-5.2 FP8$^\dagger$                      & High & 14.7{\scriptsize$\pm$1.4} & 18.2{\scriptsize$\pm$2.5} & 16.5{\scriptsize$\pm$2.5} & 5.7{\scriptsize$\pm$2.0} & 48.7{\scriptsize$\pm$1.3} & 48.2{\scriptsize$\pm$2.3} & 55.0{\scriptsize$\pm$2.1} & 39.4{\scriptsize$\pm$2.4} & 120 & 82 \\
    DeepSeek V4 Flash$^\dagger$                & High & 8.5{\scriptsize$\pm$1.1} & 8.5{\scriptsize$\pm$1.8} & 11.2{\scriptsize$\pm$2.1} & 4.3{\scriptsize$\pm$1.7} & 37.5{\scriptsize$\pm$1.3} & 34.5{\scriptsize$\pm$2.2} & 42.7{\scriptsize$\pm$2.2} & 34.4{\scriptsize$\pm$2.3} & N/A & 95 \\
    \bottomrule
  \end{tabular}
  \end{adjustbox}
\end{table}

\section{Behavioral Analysis}

\subsection{Main Evaluation Results: End-to-End Conflict Resolution Performance}
\label{sec:main-results}

\noindent\smallskip\textbf{Why Conflict Scenarios Break Frontier Models: The Competence-Resolution Gap.} Table~\ref{tab:main} shows why multi-agent conflicts are much harder than standard single-agent tasks. In our scenarios, peer or sub-agents give misleading or outdated text reports (e.g., claiming a broken deployment is healthy), shared resources are locked by competing agents, and system states change asynchronously. To pass a task, an agent must distrust false peer claims, verify ground truth from visual dashboards, override the faulty state, and complete strict closing handovers. Because models must satisfy every step in this chain, low pass rates do not mean complete failure: frontier models pass most individual checks ($55\%$--$67\%$), led by GPT-6 Astra at $66.8\%$. With deeper exploration (averaging 26.1 steps) and strong visual grounding, GPT-6 Astra handles complex conflict trajectories better than baseline models ($18.6\%$ on hard tasks vs.\ $13.6\%$ for Claude Opus 5). Yet even GPT-6 Astra solves only $28.7\%$ of tasks overall. Rarely hitting its execution limit ($1.0\%$), it frequently terminates prematurely on its own without finishing the required multi-agent cleanup---such as releasing shared locks or handing off tasks to sub-agents.

\begin{table}[!t]
  \centering
  \small
  \caption{\textbf{Failure concentration and error correlation dynamics (High Effort).} Measures whether execution errors cascade into task failure. \emph{Co-Failure Lift} $> 1.0$ indicates correlated mistakes: on easy tasks, early errors typically trigger complete task failure, whereas hard tasks exhibit gradual, independent error accumulation ($\text{lift} \approx 1.0$). \emph{Indep.} gives expected pass rates under independent check failures; asterisks ($^*$) denote statistically significant clustering.}
  \label{tab:concentration}
  \begin{adjustbox}{max width=\linewidth}
  \begin{tabular}{l | c c c | c c c c | c c}
    \toprule
    \multicolumn{1}{c}{} & \multicolumn{3}{c|}{\textbf{Entire Benchmark (\%)}} & \multicolumn{4}{c|}{\textbf{Co-Failure Lift}}
      & \multicolumn{2}{c}{\textbf{Lift $-$ Peer Mean}} \\
    \cmidrule(lr){2-4} \cmidrule(lr){5-8} \cmidrule(lr){9-10}
    \textbf{Model} & \textbf{Check} & \textbf{Pass} & \multicolumn{1}{c|}{\textbf{Indep.}}
          & \textbf{All (Ceil.)} & \textbf{Easy} & \textbf{Mid} & \multicolumn{1}{c|}{\textbf{Hard}}
          & \textbf{$\Delta$} & \textbf{95\% CI} \\
    \midrule
    Claude Opus 5      & \underline{63.0} & \underline{25.2} & \underline{10.8} & 1.49 (2.70) & 2.10\rlap{$^{*}$} & 1.37\rlap{$^{*}$} & 1.04 & $+0.188$ & $[+0.110, +0.268]$ \\
    Claude Opus 4.8    & 60.9 & 22.8 & 9.3 & 1.41 (2.56) & 1.95\rlap{$^{*}$} & 1.04 & 1.02 & $+0.102$ & $[+0.054, +0.153]$ \\
    Claude Sonnet 5    & 58.2 & 20.8 & 7.7 & 1.39 (2.39) & 1.71\rlap{$^{*}$} & 1.16\rlap{$^{*}$} & 1.04\rlap{$^{*}$} & $+0.081$ & $[+0.034, +0.131]$ \\
    \midrule
    GPT-6 Astra        & \textbf{66.7} & \textbf{28.7} & \textbf{13.8} & 1.47 (3.00) & 1.42 & 0.96 & 1.01 & $+0.149$ & $[+0.070, +0.229]$ \\
    GPT-5.6 Sol        & 62.2 & 24.5 & 10.2 & 1.42 (2.65) & 1.09 & 1.00 & 1.03 & $+0.116$ & $[+0.059, +0.176]$ \\
    GPT-5.6 Terra      & 59.3 & 20.0 & 8.3 & 1.37 (2.46) & 1.15 & 0.99 & 1.02 & $+0.062$ & $[+0.019, +0.106]$ \\
    GPT-5.6 Luna       & 59.8 & 19.8 & 8.6 & 1.36 (2.49) & 1.53\rlap{$^{*}$} & 0.96 & 1.02 & $+0.047$ & $[+0.002, +0.094]$ \\
    \midrule
    Gemini 3.7 Flash   & 57.5 & 20.3 & 7.3 & 1.30 (2.35) & 1.53\rlap{$^{*}$} & 1.11\rlap{$^{*}$} & 1.00 & $-0.024$ & $[-0.081, +0.030]$ \\
    Gemini 3.6 Flash   & 58.0 & 18.5 & 7.6 & 1.28 (2.38) & 1.34\rlap{$^{*}$} & 1.06 & 0.99 & $-0.036$ & $[-0.092, +0.017]$ \\
    Gemini 3.1 Pro     & 48.8 & 12.0 & 3.8 & 1.19 (1.95) & 1.38\rlap{$^{*}$} & 0.99 & 1.01 & $-0.136$ & $[-0.175, -0.097]$ \\
    Gemini 3.5 Flash   & 28.1 & 3.0 & 0.5 & 1.05 (1.39) & 1.08\rlap{$^{*}$} & 1.01 & 0.99\rlap{$^{*}$} & $-0.288$ & $[-0.338, -0.242]$ \\
    \midrule
    Kimi K3            & 55.8 & 21.8 & 6.5 & 1.40 (2.26) & 2.26\rlap{$^{*}$} & 1.17\rlap{$^{*}$} & 1.06\rlap{$^{*}$} & $+0.087$ & $[+0.037, +0.141]$ \\
    DeepSeek V4 Pro    & 57.7 & 19.8 & 7.4 & 1.35 (2.36) & 1.29 & 1.09\rlap{$^{*}$} & 1.02 & $+0.031$ & $[-0.013, +0.074]$ \\
    GLM-5.2 FP8        & 48.2 & 14.7 & 3.6 & 1.27 (1.93) & 1.81\rlap{$^{*}$} & 1.16\rlap{$^{*}$} & 1.00 & $-0.053$ & $[-0.099, -0.008]$ \\
    DeepSeek V4 Flash  & 38.8 & 8.5 & 1.6 & 1.16 (1.63) & 1.49\rlap{$^{*}$} & 1.06\rlap{$^{*}$} & 1.00 & $-0.177$ & $[-0.222, -0.133]$ \\
    \bottomrule
  \end{tabular}
  \end{adjustbox}
\end{table}

\noindent\smallskip\textbf{Diminishing and Inconsistent Returns from Test-Time Compute.} While test-time reasoning compute boosts formal math benchmarks, Table~\ref{tab:main} shows that in multi-agent conflict environments, scaling effort from Low to Medium to High yields diminishing returns or hurts performance. For top models (Claude Opus 5, GPT-5.6 Luna), increasing effort from Low to High brings minimal gains ($+1.7$\,pp and $+0.3$\,pp), leveling off at Medium effort. For flash models (Gemini 3.6 Flash, Gemini 3.7 Flash), Medium effort achieves peak accuracy ($20.2\%$, $21.5\%$) while High slightly drops. Worst of all, on Gemini 3.5 Flash, higher effort causes pass rates to drop sharply from $13.5\%$ to $3.0\%$, as internal thinking loops crowd out tool actions within the step budget.

\subsection{Failure Taxonomy: Content Accuracy vs.\ Protocol Execution}
\label{sec:taxonomy}

To diagnose why agents fail under conflict, we map all $3{,}284$ evaluation checks per model into three distinct competencies spanning six diagnostic failure modes (Appendix~\S\ref{sec:app-taxonomy}): \textbf{Visual Grounding}, inspecting visual conflict evidence (F3); \textbf{Content Accuracy}, covering messaging (F1), file modifications (F2), and evidence citations (F4); and \textbf{Structural Protocol Compliance}, executing terminal actions (F5) and synchronizing time windows (F6).

\noindent\smallskip\textbf{Resolving Text-Visual Contradictions vs.\ Multi-Agent Synchronization.}
First, \textbf{visual grounding vs.\ blind trust separates strong and weak models}: when sub-agent outputs conflict with the environment, weaker models blindly follow misleading text, whereas stronger models inspect visual evidence to verify ground truth and cite exact conflict evidence (outperforming weaker baselines by 31--65\,pp, with evidence citation F4 showing a 15.2\,pp spread even among top models; Appendix~\S\ref{sec:app-pairwise}).
Second, \textbf{coordination protocols remain a universal bottleneck}: when conflicts involve competing locks, permission boundaries, or expiring holds, models struggle to execute closing actions (F5, $58\%$--$72\%$ failure rate) or synchronize time windows (F6, $83\%$--$97\%$ failure rate). Even GPT-6 Astra and Claude Opus 5 fail over $58\%$ of closing actions, frequently forgetting to release shared locks or hand off tasks to sub-agents.
Finally, statistical tests across conflict categories show no domain specialization (Bonferroni-corrected $p \ge 0.094$; Appendix~\S\ref{sec:capability-axes}), confirming that these failures reflect core multi-agent coordination bottlenecks.

\subsection{Failure Concentration and Error Dynamics}
\label{sec:concentration}

While models achieve similar overall pass rates, they fail in fundamentally different ways: do errors trigger a chain reaction, or do models fail separate checks independently? To answer this, we evaluate \emph{co-failure lift}---the likelihood that a second check fails given that one has already failed, relative to independent chance (Table~\ref{tab:concentration}; Appendix~\S\ref{sec:app-concentration}). Comparing tasks by ground-truth step length reveals two distinct patterns (Table~\ref{tab:main}): on short tasks (9 or fewer steps), errors trigger a chain reaction, with 11 of 15 models exhibiting co-failure lift significantly above $1.0$, indicating that an early mistake causes the entire task to fail. In contrast, on long tasks (12 or more steps), lift drops near $1.0$ ($0.99$--$1.06$), showing failures on long tasks stem from gradual mistakes rather than single fatal errors.

\subsection{The Conflict-Resolution Gate: Isolating Resolution from Investigation}
\label{sec:conflict-gate}

To separate contradiction detection from operational execution, \sysname{} evaluates two distinct task demands: rounds where agents must resolve conflicts through state-changing actions (\emph{Resolve-gated}), and rounds with identical conflicts requiring only investigation (\emph{Investigate-only}). Control tasks confirm that performance drops stem from operational execution rather than context growth.

\begin{wraptable}[15]{r}{0.46\textwidth}
  \vspace{-\intextsep}
  \centering
  \footnotesize
  \caption{\textbf{The conflict-resolution gate.} Pass rates when tasks require
  actively \emph{resolving} a conflict compared to merely \emph{investigating} it.
  Treated rounds require state-changing operational protocols, whereas control
  rounds require only passive diagnosis over identical round positions. High Effort,
  15 evaluated models, $9{,}000$ rounds (details in Table~\ref{tab:conflict-gate-full},
  Appendix~\S\ref{sec:app-conflict-gate}).}
  \label{tab:conflict-gate}
  \begin{adjustbox}{max width=\linewidth}
  \begin{tabular}{l r r r}
    \toprule
    & \multicolumn{2}{c}{\textbf{Pass rate (\%)}} & \\
    \cmidrule(lr){2-3}
    \textbf{Seed arm} & \textbf{Resolve} & \textbf{Investigate} & \textbf{$\Delta$} \\
    \midrule
    Treated (12 seeds)       & 11.0 & 28.1 & $-17.1$ \\
    Control (8 seeds)        & 20.4 & 19.3 & $+1.1$ \\
    \midrule
    \textbf{Difference-in-differences} & \multicolumn{3}{r}{$\mathbf{-18.2}$} \\
    \quad 95\% CI (seed bootstrap) & \multicolumn{3}{r}{$[-29.1, -7.2]$} \\
    \quad Exact test $p$ & \multicolumn{3}{r}{.052} \\
    \bottomrule
  \end{tabular}
  \end{adjustbox}
\end{wraptable}

As shown in Table~\ref{tab:conflict-gate}, requiring active conflict resolution drops pass rates from $28.1\%$ to $11.0\%$ ($-17.1$\,pp), while control performance is unchanged ($+1.1$\,pp), yielding a net difference-in-differences penalty of $\mathbf{-18.2}$\,pp ($95\%$ CI $[-29.1, -7.2]$, exact $p = .052$). Pass rates drop from $28.1\%$ under investigation alone, to $12.3\%$ under partial execution, falling to $4.4\%$ upon full protocol completion.

\noindent\smallskip\textbf{Resolution vs.\ Detection.} On mean check score, the gap between resolving and investigating is only $-7.9$\,pp ($p = .252$): while agents easily spot inconsistencies, task success drops by $-18.2$\,pp when they must actively fix the system state by reverting changes, revoking permissions, or handing off locks (Appendix~\S\ref{sec:app-conflict-gate}).
\subsection{Compute Efficiency, Reasoning Effort, and Economic Costs}
\label{sec:efficiency}

Examining test-time compute reveals an \emph{effort inversion}: reasoning token share spans from $8.8\%$ for Claude Opus 5 to $90.5\%$ for Gemini 3.5 Flash (Table~\ref{tab:main}), yet correlates negatively with task success across the 20-step API models ($\rho = -0.697$). Rather than improving planning, extra thinking traps weaker models in repetitive loops: while top models rarely run out of steps (hitting the 20-step cap on only $0.5\%$ of rounds for Claude Opus 5 and $1.2\%$ for GPT-5.6 Sol), Gemini 3.5 Flash runs out of steps on $89.2\%$ of rounds (Appendix Table~\ref{tab:cost}). This planning failure directly drives high economic cost: across all models at high effort (\$11,737 across 87,493 calls), cost per solved task spans a 55-fold range, from \$0.72 for GPT-5.6 Luna to \$39.54 for Gemini 3.5 Flash (Appendix~\S\ref{sec:app-cost}).

\section{Ablations}
\label{sec:ablations}

\begin{figure}[!t]
  \centering
  \includegraphics[width=\linewidth]{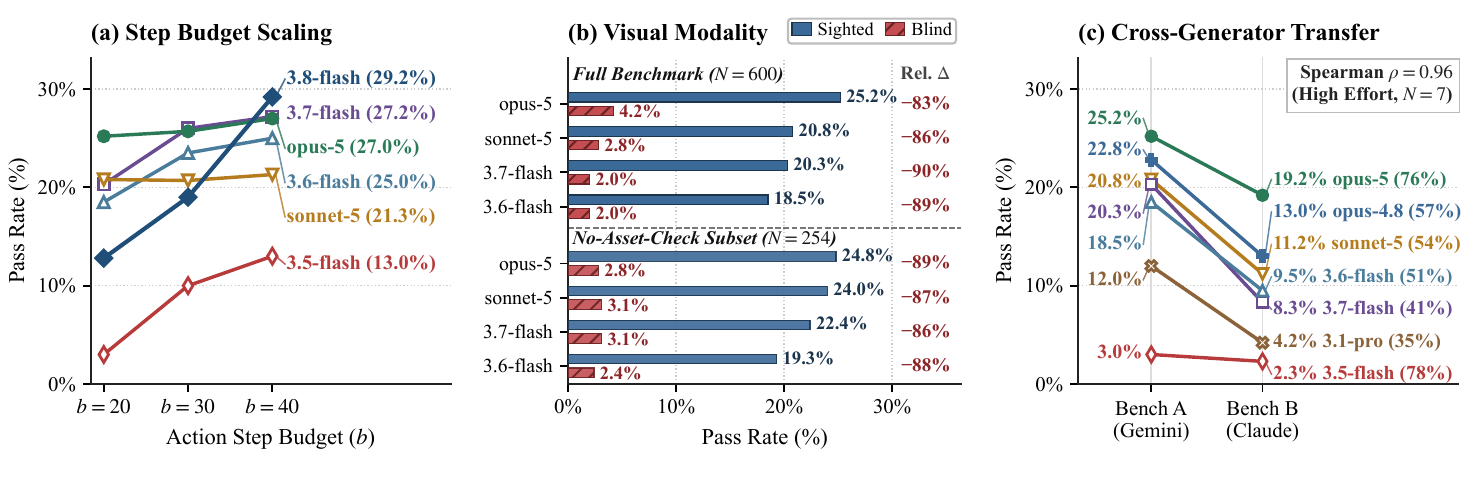}
  \caption{\textbf{Harness parameter ablations.} \emph{(a) Step budget:} Expanding action horizons unlocks frontier models with high exploration demands while separating efficient planners from thinking loops. \emph{(b) Visual modality:} Agent performance collapses without image inputs, confirming visual evidence is essential for conflict resolution. \emph{(c) Generator model:} Relative rankings remain highly consistent across parallel benchmarks synthesized by different generator models.}
  \label{fig:ablation}
\end{figure}

\noindent\smallskip\textbf{Cross-Generator Stability and Task Difficulty.}
\label{sec:generator-ablation}
To test whether findings depend on the scenario generator, we synthesized an alternate suite using Claude Sonnet 5 alongside default Gemini 3.6 Flash (Figure~\ref{fig:ablation}c). Although Sonnet yields longer reference trajectories (12.5 vs.\ 10.3 steps) and lower pass rates, model rankings stay consistent ($\rho = 0.96$) without self-preference (Appendix~\S\ref{sec:app-generator-ablation}).

\noindent\smallskip\textbf{Action Budget Limits and Planning Horizons.}
\label{sec:budget-ablation}
Varying step budgets from 20 to 40 steps (Figure~\ref{fig:ablation}a) reveals three planning patterns:
First, \textbf{step-efficient models} (Claude Opus 5, Sonnet 5) solve tasks concisely (mean 10.3 steps), rarely hitting the 20-step cap ($<3.3\%$) with minimal gain from extra steps ($<2$\,pp).
Second, \textbf{loop-prone models} (Gemini 3.5 Flash) hit the cap on $89.2\%$ of rounds due to repetitive tool calls, inflating cost to \$8.69 per solved task.
Third, \textbf{horizon-sensitive models} benefit from extended exploration: with deeper exploration, GPT-6 Astra reaches $\mathbf{28.7\%}$, while Gemini 3.8 Flash jumps from $12.8\%$ (20 steps) to $\mathbf{29.2\%}$ (40 steps) and $\mathbf{30.7\%}$ at 60 steps ($p = .0098$), cutting cost per solved task by $21\%$ to \$3.50 (Appendix~\S\ref{sec:app-3-8-flash}).

\noindent\smallskip\textbf{Visual Necessity: Sighted vs.\ Blind Evaluation.}
\label{sec:modality-ablation}
Withholding visual evidence causes agent performance to collapse across all models (Figure~\ref{fig:ablation}b). Even on tasks where image inspection is not explicitly enforced by the rubric, blind agents retain only 11.1\%--13.1\% of sighted pass rates ($p < 10^{-11}$), confirming visual assets provide decisive rather than decorative evidence (Appendix~\S\ref{sec:app-modality-ablation}).

\section{Limitations}
\label{sec:limits}

We note three scope limitations:
(1)~\textbf{Semantic Judging:} Semantic checks rely on an LLM judge without human studies, mitigated via 3-sample voting and multi-judge validation ($\rho \ge 0.98$, $\kappa \ge 0.90$; \S\ref{sec:validity}).
(2)~\textbf{Asynchronous Traces:} Conflict resolution is evaluated through logs and workspace traces rather than live chatting, ensuring reproducibility while mirroring ticket and review workflows.
(3)~\textbf{Diagnostic Focus:} Scenarios focus on conflict rounds without conflict-free baselines, isolating conflict handling via ablation gates (\S\ref{sec:conflict-gate}).
\section{Conclusion}
\label{sec:conclusion}

In this work, we introduced \sysname{}, a controllable generation framework and benchmark for synthesizing verifiable multi-agent conflict scenarios in computer use. By combining hierarchical outline planning with verified micro-execution, pre-committed visual ground truths, and execution-compiled hybrid checkers, \sysname{} guarantees task solvability by construction across 600 multi-round tasks and 20 operational domains. Our evaluation of 15 frontier models uncovers three fundamental bottlenecks: weaker models blindly trust misleading peer reports whereas stronger models actively inspect visual evidence to verify ground truth; actively fixing the system state is much harder than spotting inconsistencies; and closing coordination protocols---such as releasing shared locks and completing sub-agent handoffs---remain a universal failure mode across all model families. Beyond static evaluation, \sysname{}'s modular plugin architecture and automated verification pipeline provide both a rigorous diagnostic testbed and a scalable data engine for post-training next-generation VLMs to build reliable multi-agent systems.

{\setlength{\bibsep}{2.5pt}
\bibliography{main}}

\section*{AI Use Statement}
In this work, we used generative AI tools as generation engines in the \sysname{} hierarchical pipeline to expand structured scenario seeds for benchmark data generation under deterministic validation and reference-trajectory compilation (\S\ref{sec:harness}, \S\ref{sec:validity}), and for language polishing during manuscript preparation. We have reviewed all AI-assisted work, verified and tested all generated tasks and code for correctness, and take responsibility for the final content of this work, including text, claims, and artifacts produced with the aid of generative AI.

\section*{Reproducibility Statement}
The generation harness, conflict plugins, asset generators, and evaluation checkers are fully specified with deterministic seeds and reproducible reference trajectories (Appendix~\ref{sec:app-harness}), with tasks and grading surfaces detailed in Appendix~\ref{sec:app-coverage}. Every reported quantity is generated by released analysis scripts and verified mechanically: all table bodies are byte-diffed against script outputs, and every statistic in the prose of \S\ref{sec:validity}--\S\ref{sec:ablations} is programmatically asserted to prevent inconsistencies. A single validation command recomputes all counts from raw source files and flags any numerical drift.


\clearpage
\appendix
\raggedbottom
\numberwithin{table}{section}
\numberwithin{figure}{section}
\renewcommand{\thetable}{\thesection\arabic{table}}
\renewcommand{\thefigure}{\thesection\arabic{figure}}
\section*{\Large\centering{Appendix}}
\section{Overview and Table of Contents of Supplementary Material}
\label{sec:app-index}
\label{sec:app}

This supplementary document provides comprehensive algorithmic specifications, formal derivations, dataset statistics, extended benchmark leaderboards, and ablation studies supporting the findings presented in the main paper. In addition to this PDF document, the accompanying Supplementary Material archive includes standalone interactive HTML scenario visualizers (\texttt{index.html}) for exploring multi-round conflict dependency graphs, workspace state diffs, sub-agent communication transcripts, native $3{,}840 \times 2{,}160$ visual assets, and compiled evaluation rubrics across all 20 benchmark seed families.

\smallskip
\noindent\textbf{Table of Contents of Supplementary Material:}
\begin{itemize}[leftmargin=*,itemsep=3.5pt]
  \item \textbf{Appendix~\ref{sec:app-harness}: The \sysname{} Generation Pipeline: Algorithmic and Implementation Details}
  \begin{itemize}[leftmargin=1.5em,itemsep=1.5pt]
    \item \S\ref{sec:app-formalization}: Problem Formalization and Grounded Locality Axioms
    \item \S\ref{sec:app-pipeline}: Hierarchical Phased Generation Pipeline ($O(R)$ API Complexity)
    \item \S\ref{sec:app-safeguards}: Harness Generation Safeguards ($\mathtt{M1}$--$\mathtt{M14}$; Table~\ref{tab:harness-safeguards})
    \item \S\ref{sec:app-difficulty}: Declarative Controllability and 3-Tier Constraint Enforcement
    \item \S\ref{sec:app-plugins}: Modular Plugins and State Effects for Extensibility
    \item \S\ref{sec:app-assets}: Multimodal Asset Generation and Pre-Commitment (Figure~\ref{fig:asset-gallery})
    \item \S\ref{sec:app-grading}: Verifiable Grader Compilation and Exclusivity Gates
    \item \S\ref{sec:app-worked-round}: A Worked Round, end to end (Figures~\ref{fig:round-anatomy}, \ref{fig:round-cached-total}; Table~\ref{tab:round-checks})
    \item \S\ref{sec:app-diversity}: Task Diversity and State Continuity
  \end{itemize}

  \item \textbf{Appendix~\ref{sec:app-coverage}: Benchmark Composition, Coverage, and Validity}
  \begin{itemize}[leftmargin=1.5em,itemsep=1.5pt]
    \item \S\ref{sec:app-corpus-coverage}: Benchmark Coverage and Structural Diversity (Table~\ref{tab:coverage})
    \item \S\ref{sec:app-slack}: Continuous Measured Difficulty via Ground-Truth Trajectory Length
    \item \S\ref{sec:app-validity}: Hybrid Grading Validity, Judge Integrity, and Competency Decoupling (Table~\ref{tab:validity})
  \end{itemize}

  \item \textbf{Appendix~\ref{sec:app-main}: Comprehensive Experimental Results and Baseline Diagnostics}
  \begin{itemize}[leftmargin=1.5em,itemsep=1.5pt]
    \item \S\ref{sec:app-leaderboards}: Multi-Effort Evaluation Leaderboards (Tables~\ref{tab:main-high}, \ref{tab:main-medium}, \ref{tab:main-low})
    \item \S\ref{sec:app-pairwise}: Pairwise Statistical Significance and Frontier Compression
    \item \S\ref{sec:app-concentration}: Failure Concentration Dynamics
    \item \S\ref{sec:capability-axes}: Capability Breakdown by Conflict Category (Table~\ref{tab:capability})
    \item \S\ref{sec:app-conflict-gate}: The Conflict-Resolution Gate: Robustness, Replication, and Compliance (Tables~\ref{tab:conflict-gate-full}, \ref{tab:conflict-gate-app})
    \item \S\ref{sec:app-taxonomy}: Fine-Grained Diagnostic Failure Taxonomy (Figure~\ref{fig:taxonomy})
  \end{itemize}

  \item \textbf{Appendix~\ref{sec:app-efficiency}: Compute Efficiency, Token Economics, and Planning Horizons}
  \begin{itemize}[leftmargin=1.5em,itemsep=1.5pt]
    \item \S\ref{sec:app-cost}: Economic Cost and API Breakdown (Table~\ref{tab:cost})
    \item \S\ref{sec:effort}: Reasoning Effort Inversion Dynamics ($\rho = -0.697$)
    \item \S\ref{sec:app-step-budget}: Execution Step Budget Saturation
  \end{itemize}

  \item \textbf{Appendix~\ref{sec:app-corpora}: Extended Ablation Studies and Robustness Analyses}
  \begin{itemize}[leftmargin=1.5em,itemsep=1.5pt]
    \item \S\ref{sec:app-generator-ablation}: Cross-Generator Invariance and Parallel Benchmark Design (Table~\ref{tab:corpora-design})
    \item \S\ref{sec:app-budget-ablation}: Action Budget Limits and Planning Horizon Dynamics ($b=20 \to 30 \to 40$)
    \item \S\ref{sec:app-3-8-flash}: Frontier Horizon Sensitivity: The Gemini 3.8 Flash Case Study (Table~\ref{tab:38-flash-ablation})
    \item \S\ref{sec:app-modality-ablation}: Visual Necessity and Multimodal Grounding Ablation
    \item \S\ref{sec:app-judge-ablation}: Judge Selection and Cross-Judge Stability (Table~\ref{tab:judge-app})
  \end{itemize}

  \item \textbf{Appendix~\ref{app:report}: Every Exhibit from the Generated Behavioral Report}
  \begin{itemize}[leftmargin=1.5em,itemsep=1.5pt]
    \item All 23 Tables, Vector Charts, and CSS Panels, in the Report's Own Order
  \end{itemize}
\end{itemize}

\section{The Generation Harness: Algorithmic and Implementation Details}
\label{sec:app-harness}

\subsection{Problem Formalization and Grounded Locality Axioms}
\label{sec:app-formalization}

Our benchmark evaluates how agents resolve conflicts through active investigation in operating environments, rather than through passive text classification. The evaluating agent receives competing claims from peer agents without knowing in advance which claim matches the true system state. Formally, a conflict round is defined by the tuple $\mathcal{R} = (\mathcal{W}_0, I, \mathcal{T}, v^{\star}, \Phi)$ (\S\ref{sec:formalization}), governed by three structural principles:
\begin{enumerate}[leftmargin=*,itemsep=2pt]
  \item \textbf{Grounded Locality:} The decisive ground-truth entity $v^{\star}$ is strictly localized within the workspace $\mathcal{W}_0$. In 346 of the 600 tasks, $v^{\star}$ resides exclusively within a rendered visual asset without textual duplicates in the workspace (\S\ref{sec:exclusivity}).
  \item \textbf{Action-Based Resolution:} Task completion requires executing concrete action commands $c \in \mathcal{T}$, such as configuration edits, access control updates, or deployment rollbacks recorded in an append-only effect log $\mathcal{E}_{\text{log}}$, rather than simply outputting text answers.
  \item \textbf{Satisfiable Verification:} Satisfaction predicates $\Phi = \{ \phi_1, \dots, \phi_m \}$ are compiled directly from executed reference trajectories (\S\ref{sec:grading}), guaranteeing that all tasks are solvable by construction.
\end{enumerate}

\subsection{Hierarchical Phased Generation Pipeline}
\label{sec:app-pipeline}

Directly generating long-horizon, multi-agent scenarios with unconstrained single-pass prompting causes common failure modes: format collapse, narrative drift, premature answer leaks, and unsatisfiable grading predicates. \sysname{} resolves this by decomposing scenario synthesis into phased macro-planning and verified micro-execution across eight coordinated stages in three tiers (Figure~\ref{fig:arch}):
\begin{itemize}[leftmargin=*,itemsep=2pt]
  \item \textbf{Tier 1: Package Initialization (Stages 1--3, 8 calls):} Stage~1 (\emph{Global Outline}, 3 calls) establishes the scenario background, agent roles, timeline events, and resource allocations. Stage~2 (\emph{Subagents \& Interaction Rules}, 2 calls) compiles role boundaries ($\mathtt{M8}$) and dependency matrices ($\mathtt{M5}$, $\mathtt{M7}$). Stage~3 (\emph{Background Workspace Setup}, 3 calls) initializes codebases, system logs, session transcripts, and platform UI environments ($\mathtt{M3}$, $\mathtt{M12/13}$).
  \item \textbf{Tier 2: Round Generative Loop (Stages 4--5):} For each round $r = 1 \dots R$, Stage~4 (\emph{Phase Expansion}, 1 call/phase) dynamically projects milestone timelines over rolling attention windows ($P_s \approx 3\text{--}4$ rounds) based on accumulated execution history, avoiding context saturation. Stage~5 (\emph{Progressive Round Micro-Pipeline}, $K \in [7, 8]$ sub-calls) decomposes round synthesis into eight isolated sub-steps: (5.1) task instructions with boundary enforcement ($\mathtt{M1/M8}$), (5.2) peer agent messages and dialogue history ($\mathtt{M4}$), (5.3) environment file updates ($\mathtt{M3/M9}$), (5.4) visual asset injection ($\mathtt{M14}$), (5.5) decisive fact commitment ($v^\star$) under automated leakage scanning, (5.6) step-by-step execution planning ($\mathtt{M2}$), (5.7) reference CLI trajectory execution in a sandbox ($\mathtt{M9}$, $\mathtt{M11}$), and (5.8) deterministic predicate compilation ($\mathtt{M10}$, 0 calls). Narrow sub-calls isolate failure modes and enable bounded in-loop retries without discarding valid package state.
  \item \textbf{Tier 3: Post-Generation Compilation \& Audit (Stages 6--8):} Stage~6 (\emph{Consistency Audit}, $\le M=3$ retry iterations) performs a secondary cross-stage check reconciling instructions with evaluation predicates ($\mathtt{M6}$, $\mathtt{M9}$, $\mathtt{M10}$). Stage~7 (\emph{Multimodal Refinement}, 1 call per asset $a \in \mathcal{A}$, where $A = |\mathcal{A}|$) synthesizes declarative HTML/Canvas specifications for visual evidence assets. Stage~8 (\emph{Render \& VLM Audit}, 0 calls) executes headless browser capture, verifies visual exclusivity, and compiles identifiers ($\mathtt{M14}$) into the final $\mathtt{scenario.json}$.
\end{itemize}

\noindent\smallskip\textbf{Linear Invocation Complexity.} Structuring synthesis into phased windows ensures that total API calls scale strictly linearly with round count $R$:
\begin{equation}
  \text{Total API Calls} = 8 + \lceil R / P_s \rceil + K \cdot R + A, \qquad 7 \le K \le 8
\end{equation}
where $P_s$ is the phase span and $A$ is the asset count, averaging 95 API calls per 10-round package across the 600-round benchmark.

\clearpage
\subsection{Harness Generation Safeguards (M1--M14)}
\label{sec:app-safeguards}

To ensure all 600 tasks are solvable and structurally consistent without expensive regeneration, \sysname{} uses automated safeguards with a two-part defense: \emph{fix formatting defects directly in code ($0$ LLM cost), and retry semantic errors with in-loop diagnostic feedback}. Table~\ref{tab:harness-safeguards} details the 14 generation failure modes ($\mathtt{M1}$--$\mathtt{M14}$). These safeguards ensure \emph{generation and compiler integrity}, distinct from the six behavioral failure modes (\S\ref{sec:app-taxonomy}, \S\ref{sec:taxonomy}) observed in evaluated models.

\begin{table}[H]
  \centering
  \small
  \caption{\textbf{Generation failure modes ($\mathtt{M1}$--$\mathtt{M14}$) and dual-defense safeguards.} Mechanical defects are repaired deterministically in code ($0$ LLM cost); semantic drift triggers bounded in-loop retries with diagnostic error feedback.}
  \label{tab:harness-safeguards}
  \adjustbox{max width=\linewidth}{%
  \begin{tabular}{@{}lllll@{}}
    \toprule
    \textbf{Mode} & \textbf{Defect Name} & \textbf{Failure Mechanism \& Symptoms} & \textbf{Stage} & \textbf{Safeguard \& Fix Mechanism} \\
    \midrule
    \multicolumn{5}{@{}l}{\textit{Category I: Prompt Integrity and Planning Depth}} \\
    \midrule
    $\mathtt{M1}$ & Prompt Asset Leak & Decisive token or asset ID leaked in task prompt & 5.1 & Instruction sanitization \& bounded retry \\
    $\mathtt{M2}$ & Trajectory Collapse & Reference playbook collapses to 2--3 trivial steps & 5.6 & 5-phase shape \& min-step budget retry \\
    \midrule
    \multicolumn{5}{@{}l}{\textit{Category II: Schema, Formatting, and State Integrity (Deterministic Code Repairs, 0 Calls)}} \\
    \midrule
    $\mathtt{M3}$ & Empty File Crash & Missing \texttt{content} on empty workspace files & S3, 5.3 & Deterministic coercion \texttt{None} $\to$ \texttt{""} \\
    $\mathtt{M4}$ & Mock State Drift & Nested \texttt{state\_effects} schema prevents effect execution & 5.2 & Automatic dictionary-to-list state flattening \\
    $\mathtt{M5}$ & Step Dict Crash & Emitting dictionary instead of string in dependencies & S2 & Programmatic step schema coercion \\
    $\mathtt{M7}$ & Missing Role Identifiers & Nested schema missing \texttt{agent\_role} or mock fields & S2 & In-stage schema validator \& retry \\
    \midrule
    \multicolumn{5}{@{}l}{\textit{Category III: Persona, Tool, and Filesystem Boundaries}} \\
    \midrule
    $\mathtt{M6}$ & Persona Name Drift & Subagent character names drift across rolling stages & S1, S2, 5.1 & Canonical entity registry \& prompt bounds \\
    $\mathtt{M8}$ & Tool Access Overreach & Demanding shell tools ungranted in seed spec & S2, 5.1 & Negative-capability prompt enforcement \\
    $\mathtt{M9}$ & File-Path Desync & Reference commands target non-existent file paths & 5.3, 5.7 & Real-time workspace path sync \& retry \\
    $\mathtt{M11}$ & Hardcoded Sandbox Paths & Absolute script paths fail in container sandbox & 5.7 & Deterministic path relativization \\
    \midrule
    \multicolumn{5}{@{}l}{\textit{Category IV: Verifiable Evaluation and Multimodal Grounding}} \\
    \midrule
    $\mathtt{M10}$ & Unsatisfiable Graders & Reference trajectory fails its own generated checks & 5.5, 5.8 & Fact pre-commitment ($v^\star$) \& code-derived graders \\
    $\mathtt{M12/13}$ & Asset ID Collisions & Path-like asset IDs create doubled phantom IDs & S3 & ID sanitization \& platform prefix normalization \\
    $\mathtt{M14}$ & Snapshot ID Desync & Orphan asset checks or un-rendered snapshots & 5.4, S8 & Multi-round lock-step reconciliation \\
    \bottomrule
  \end{tabular}}
\end{table}

\subsection{Declarative Controllability and 3-Tier Constraint Enforcement}
\label{sec:app-difficulty}

Scenario structure and difficulty are declared through extensible seed schemas governing conflict types, interaction protocols, tool families, and evidence modalities. The harness enforces constraints via three complementary tiers: (1)~\textbf{Structural Code Rejection \& In-Loop Retry} for hard invariants, such as minimum trajectory length, entity persistence, and prompt leakage prevention; (2)~\textbf{Deterministic Schema Coercion} in code for mechanical format repairs without model calls; and (3)~\textbf{Prompt Conditioning} for semantic role framing and scenario context.

\subsection{Modular Plugins and State Effects for Extensibility}
\label{sec:app-plugins}

To support straightforward extensions without modifying evaluation code, \sysname{} separates domain mechanics into 23 plugins across 7 plugin categories: 16 conflict mechanics, spanning the 6 conflict categories, and 7 challenge plugins (Table~\ref{tab:coverage}). Each plugin $\mathcal{P}$ and tool command $c \in \mathcal{T}$ declares its operational effects $\mathcal{E}(\mathcal{P})$ and $\mathcal{E}(c)$. Valid actions and grading criteria are determined automatically by matching declared effects:
\begin{equation}
  \mathcal{T}_{\text{valid}} = \{ c \in \mathcal{T} \mid \mathcal{E}(c) \cap \mathcal{E}(\mathcal{P}) \neq \emptyset \}
\end{equation}
Adding novel conflict mechanics requires only registering a new plugin module with its declared effects, supported by 19 skill implementations covering 22 registered specifications across the codebase.

\subsection{Multimodal Asset Generation and Pre-Commitment}
\label{sec:app-assets}
\label{sec:assets}
\label{sec:consistency}

\begin{figure}[!t]
  \centering
  \begin{minipage}[t]{0.48\linewidth}\centering
    \includegraphics[width=\linewidth]{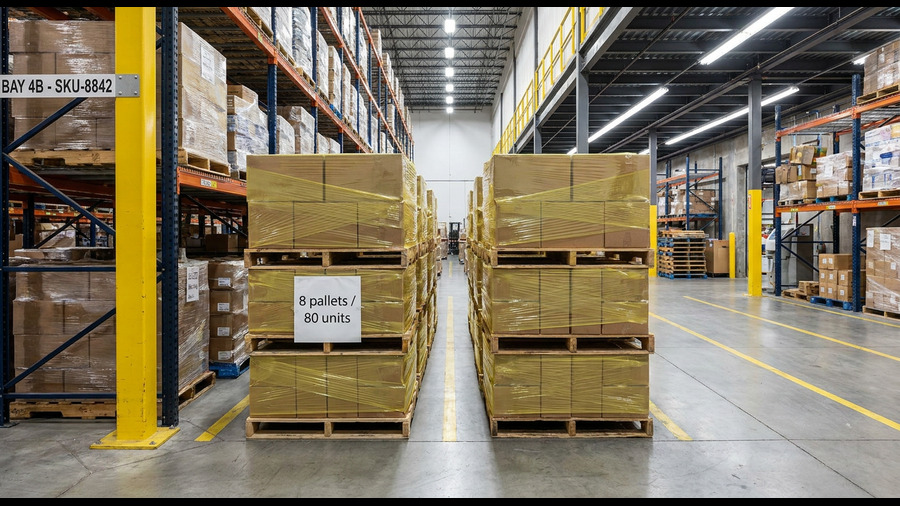}\\[1.5pt]
    {\footnotesize warehouse bay\\ reads \texttt{BAY 4B - SKU-8842}\\ {\scriptsize\texttt{distributed\_\allowbreak{}lock\_\allowbreak{}shared\_\allowbreak{}resource}}\par}
  \end{minipage}\hfill
  \begin{minipage}[t]{0.48\linewidth}\centering
    \includegraphics[width=\linewidth]{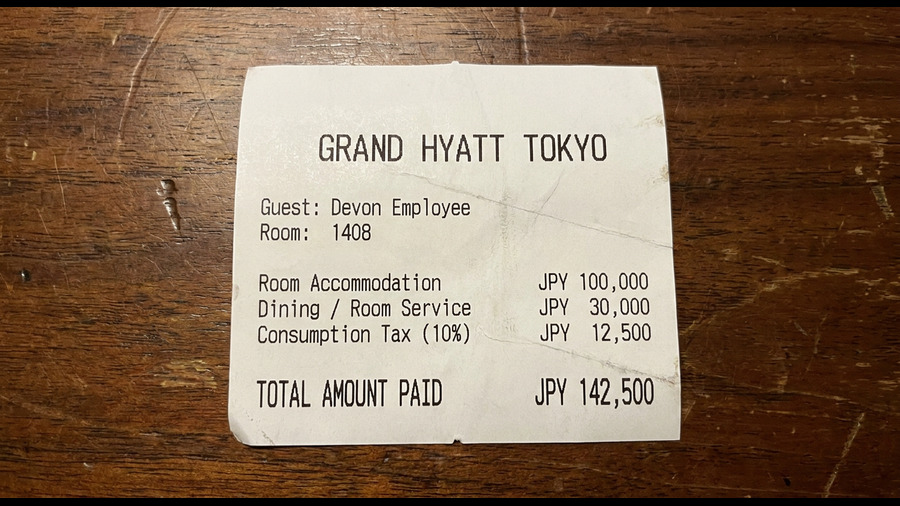}\\[1.5pt]
    {\footnotesize hotel receipt\\ reads \texttt{JPY 142,500}\\ {\scriptsize\texttt{expense\_\allowbreak{}reimbursement\_\allowbreak{}conflict}}\par}
  \end{minipage}\\[7pt]
  \begin{minipage}[t]{0.48\linewidth}\centering
    \includegraphics[width=\linewidth]{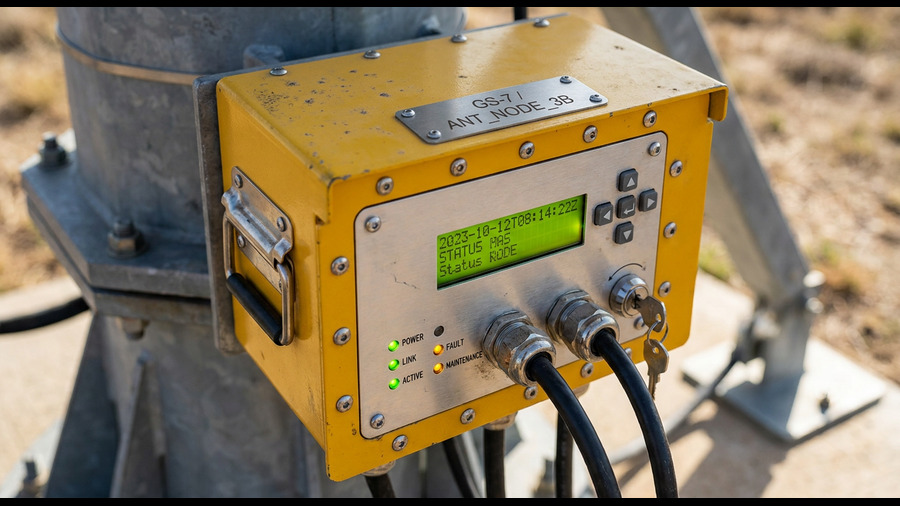}\\[1.5pt]
    {\footnotesize antenna diagnostic panel\\ reads \texttt{GS-7 ANT\_NODE\_38}\\ {\scriptsize\texttt{satellite\_\allowbreak{}ground\_\allowbreak{}station\_\allowbreak{}handover}}\par}
  \end{minipage}\hfill
  \begin{minipage}[t]{0.48\linewidth}\centering
    \includegraphics[width=\linewidth]{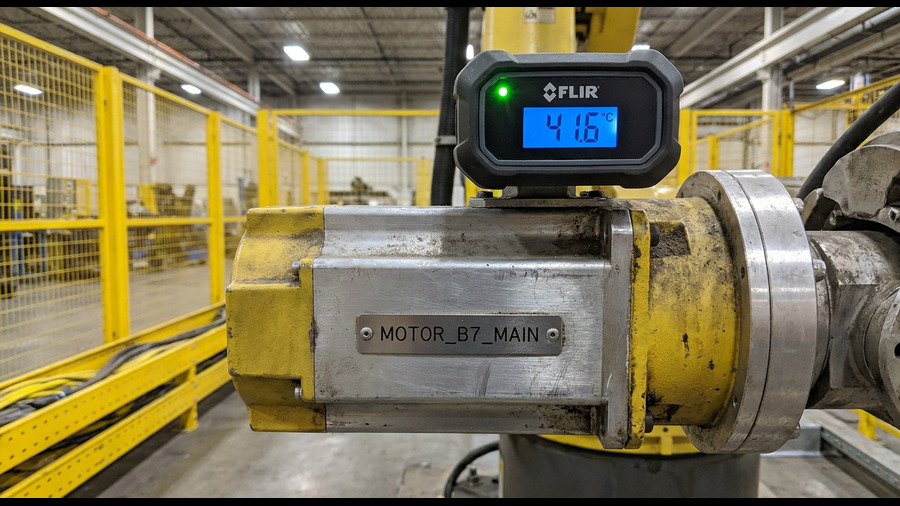}\\[1.5pt]
    {\footnotesize industrial IR thermometer\\ reads \texttt{41.6}\\ {\scriptsize\texttt{warehouse\_\allowbreak{}robotics\_\allowbreak{}safety\_\allowbreak{}halt}}\par}
  \end{minipage}
  \caption{\textbf{Photographic asset scenes in \sysname{}.} Shows the four real-world scenes used across all 19 photographic assets in the benchmark. Each photograph contains a decisive ground-truth string (shown below each image) required to resolve the corresponding conflict.}
  \label{fig:asset-gallery}
\end{figure}

\begin{figure}[!t]
  \centering
  \begin{minipage}[t]{0.315\linewidth}\centering
    \includegraphics[width=\linewidth]{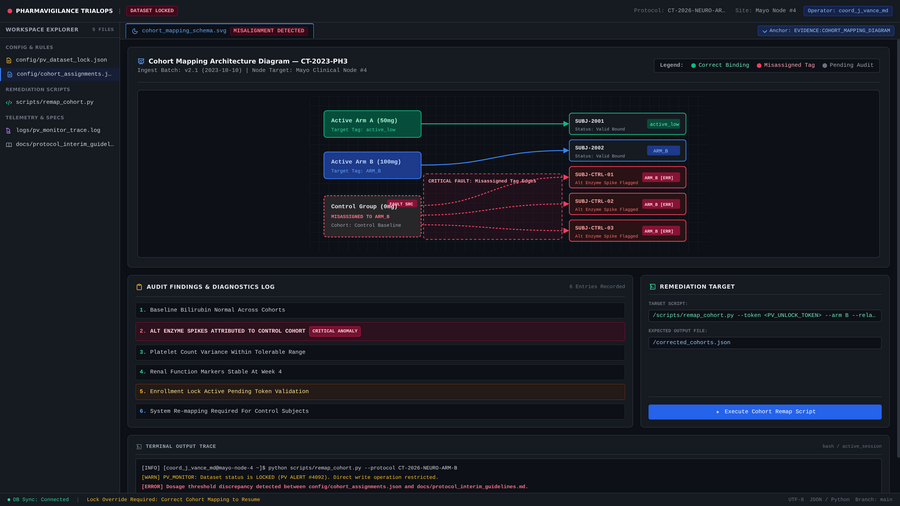}\\[1.5pt]
    {\scriptsize \texttt{linear\_\allowbreak{}dark} (321)\par}
  \end{minipage}\hfill
  \begin{minipage}[t]{0.315\linewidth}\centering
    \includegraphics[width=\linewidth]{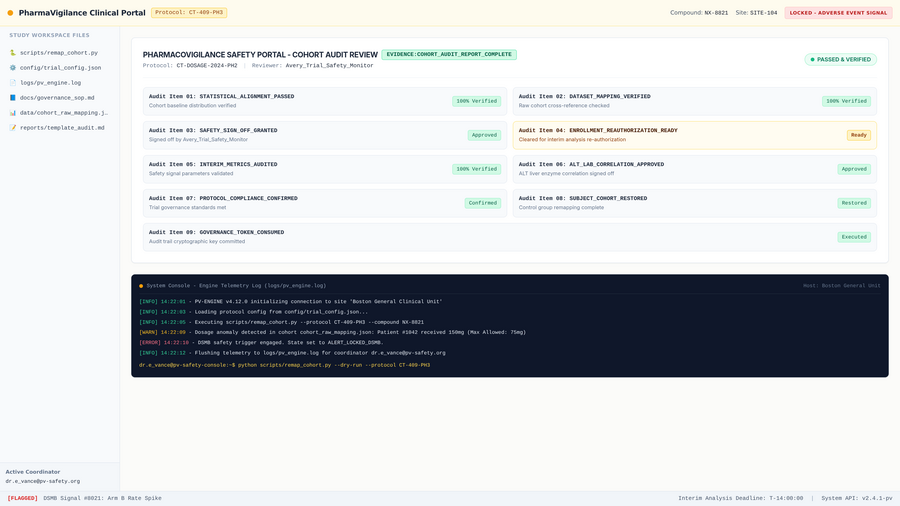}\\[1.5pt]
    {\scriptsize \texttt{modern\_\allowbreak{}clean} (250)\par}
  \end{minipage}\hfill
  \begin{minipage}[t]{0.315\linewidth}\centering
    \includegraphics[width=\linewidth]{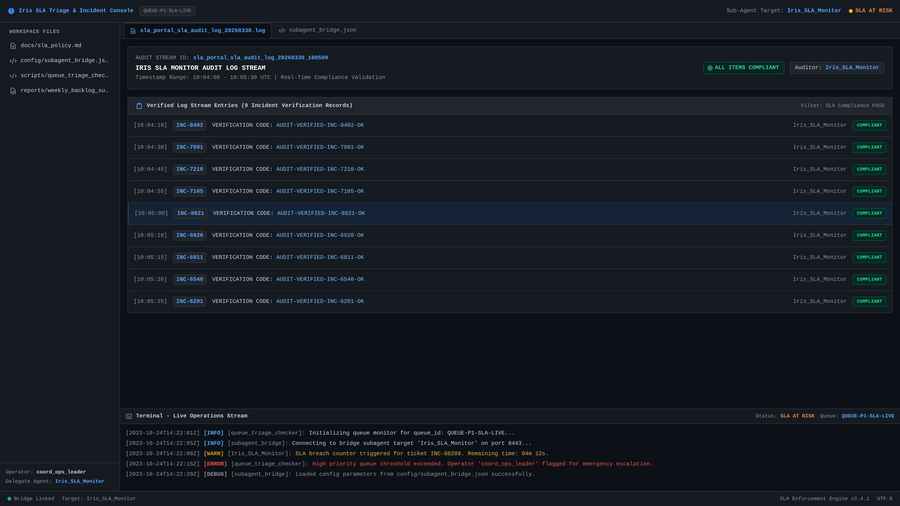}\\[1.5pt]
    {\scriptsize \texttt{github\_\allowbreak{}dark} (168)\par}
  \end{minipage}\\[7pt]
  \begin{minipage}[t]{0.315\linewidth}\centering
    \includegraphics[width=\linewidth]{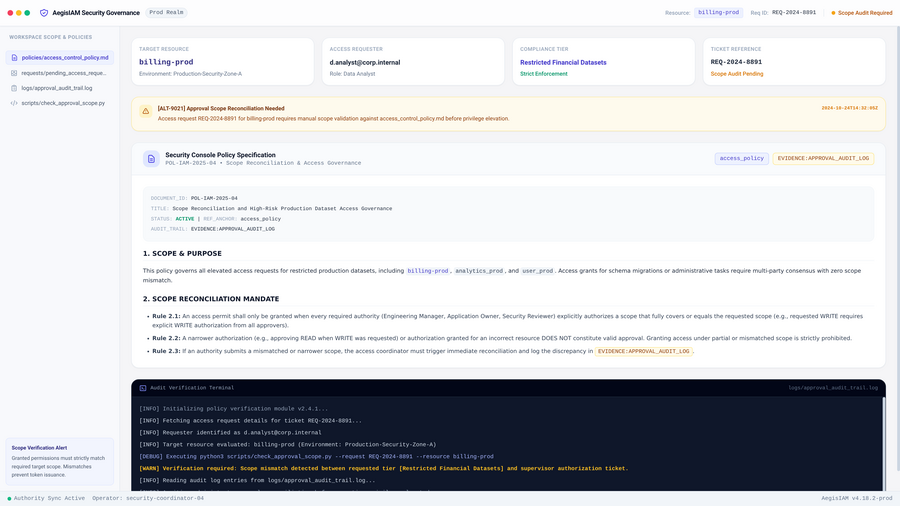}\\[1.5pt]
    {\scriptsize \texttt{apple\_\allowbreak{}developer} (132)\par}
  \end{minipage}\hfill
  \begin{minipage}[t]{0.315\linewidth}\centering
    \includegraphics[width=\linewidth]{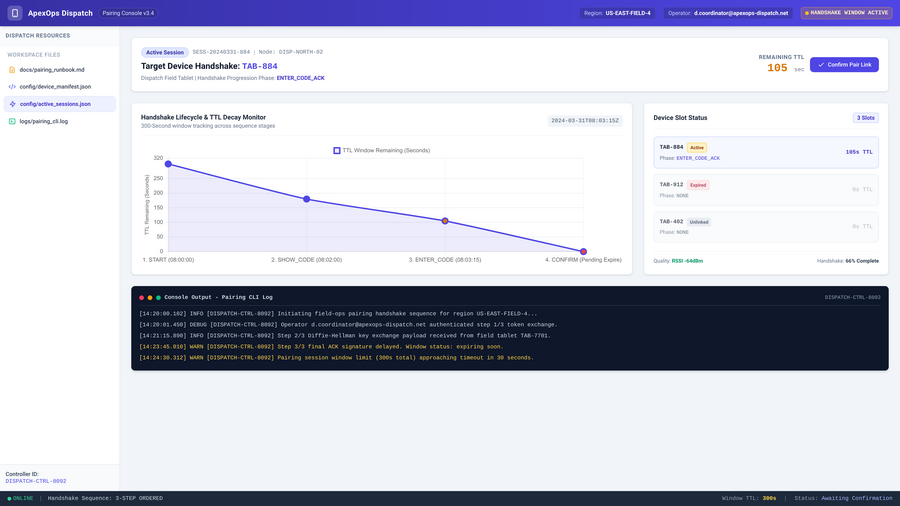}\\[1.5pt]
    {\scriptsize \texttt{material\_\allowbreak{}design} (124)\par}
  \end{minipage}\hfill
  \begin{minipage}[t]{0.315\linewidth}\centering
    \includegraphics[width=\linewidth]{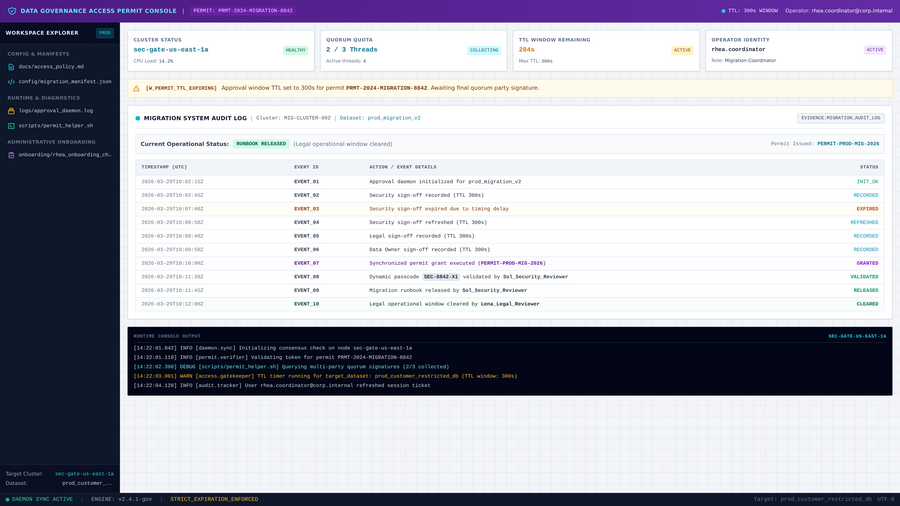}\\[1.5pt]
    {\scriptsize \texttt{datadog\_\allowbreak{}dashboard} (99)\par}
  \end{minipage}\\[7pt]
  \begin{minipage}[t]{0.315\linewidth}\centering
    \includegraphics[width=\linewidth]{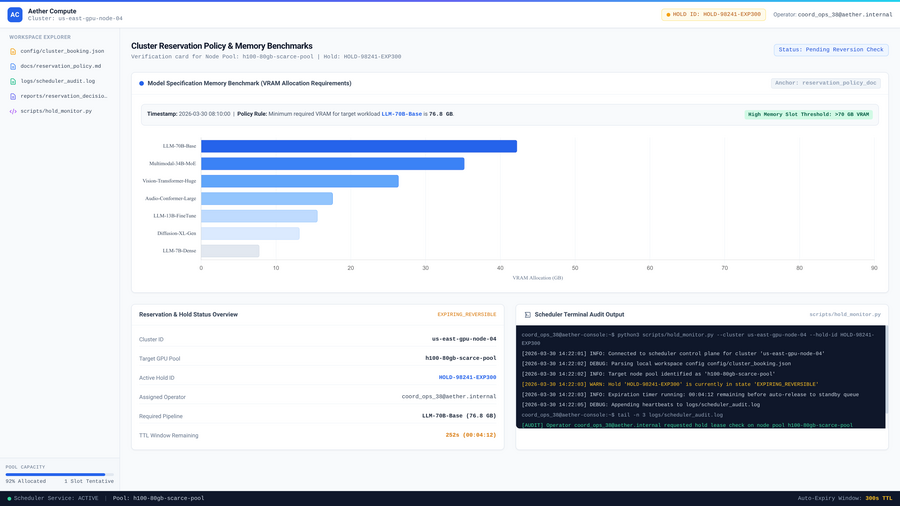}\\[1.5pt]
    {\scriptsize \texttt{stripe\_\allowbreak{}dashboard} (93)\par}
  \end{minipage}\hfill
  \begin{minipage}[t]{0.315\linewidth}\centering
    \includegraphics[width=\linewidth]{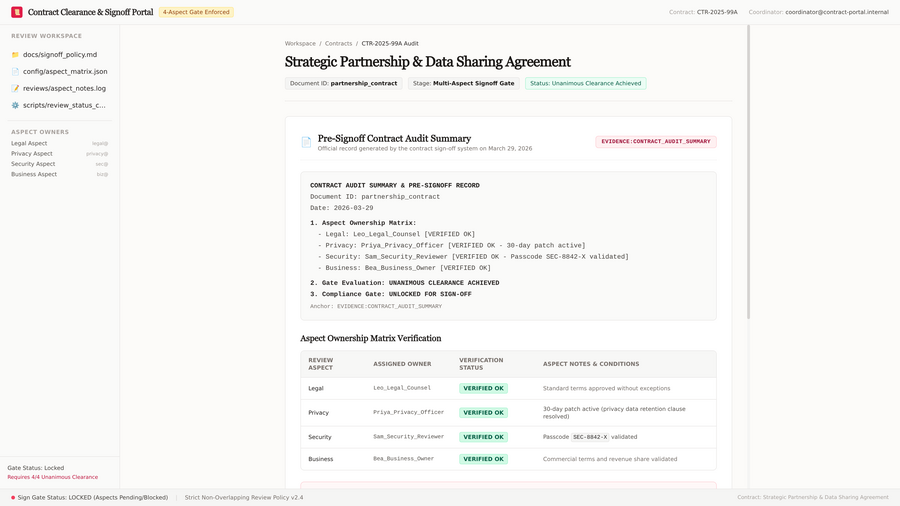}\\[1.5pt]
    {\scriptsize \texttt{notion\_\allowbreak{}docs} (74)\par}
  \end{minipage}\hfill
  \begin{minipage}[t]{0.315\linewidth}\centering
    \includegraphics[width=\linewidth]{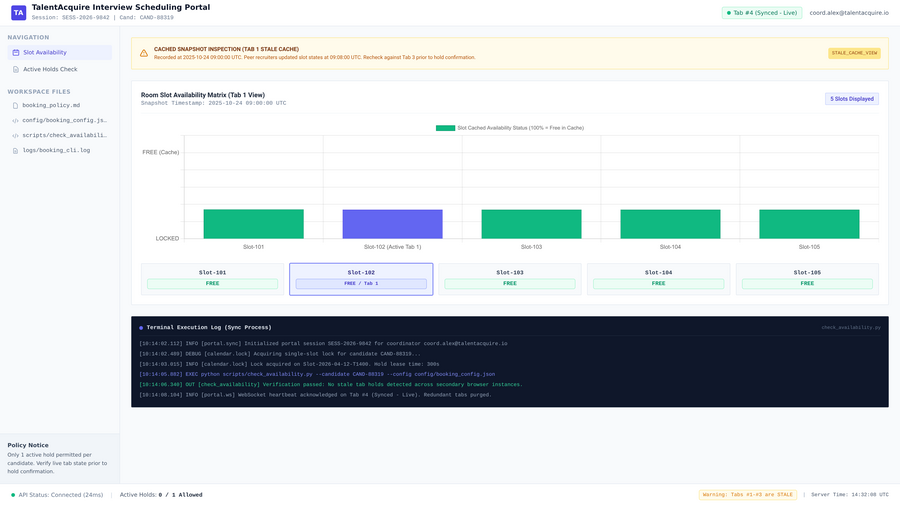}\\[1.5pt]
    {\scriptsize \texttt{flat\_\allowbreak{}minimalist} (68)\par}
  \end{minipage}\\[7pt]
  \begin{minipage}[t]{0.315\linewidth}\centering
    \includegraphics[width=\linewidth]{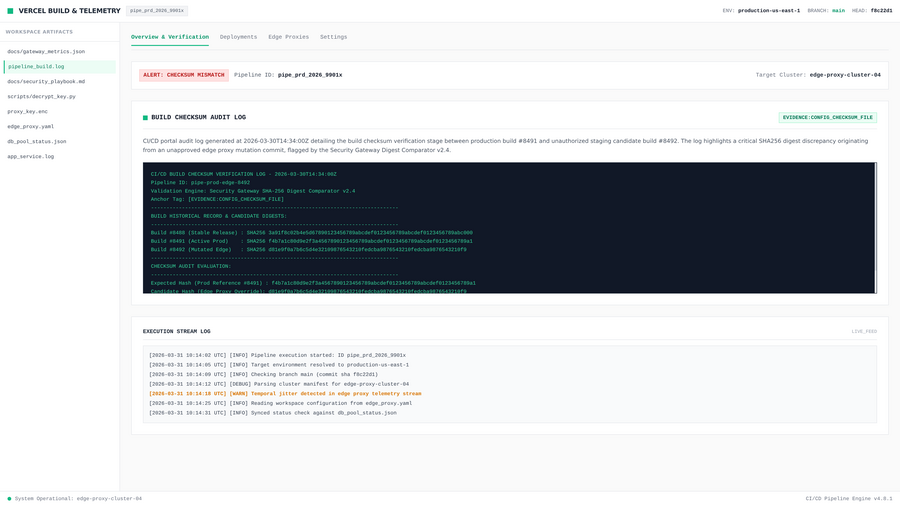}\\[1.5pt]
    {\scriptsize \texttt{vercel\_\allowbreak{}minimalist} (56)\par}
  \end{minipage}\hfill
  \begin{minipage}[t]{0.315\linewidth}\centering
    \includegraphics[width=\linewidth]{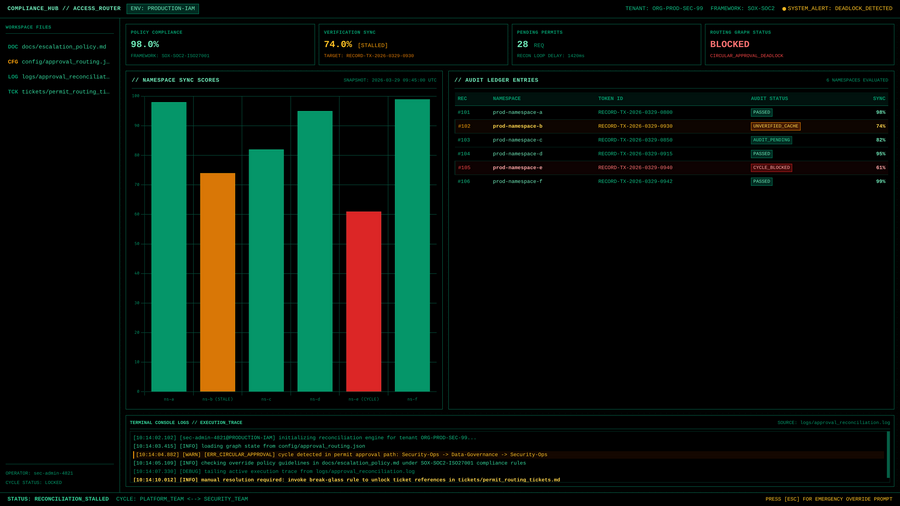}\\[1.5pt]
    {\scriptsize \texttt{bloomberg\_\allowbreak{}terminal} (55)\par}
  \end{minipage}\hfill
  \begin{minipage}[t]{0.315\linewidth}\centering
    \includegraphics[width=\linewidth]{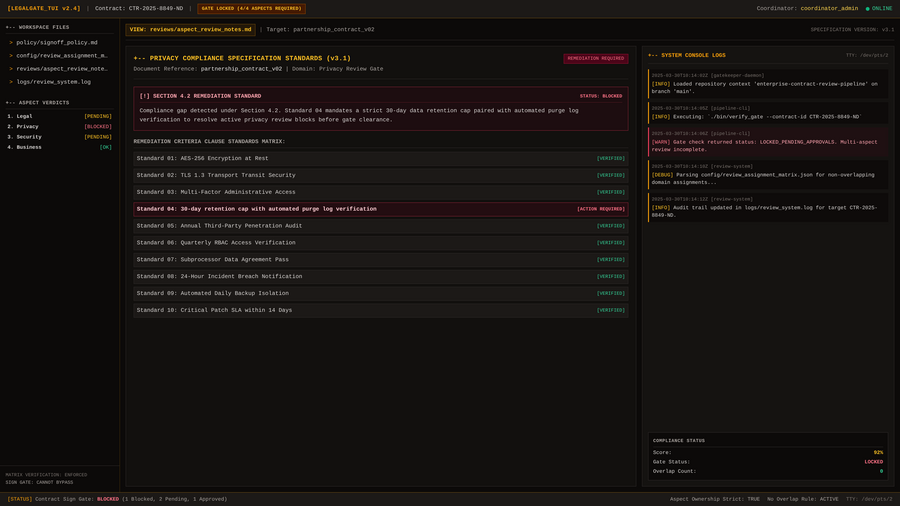}\\[1.5pt]
    {\scriptsize \texttt{terminal\_\allowbreak{}tui} (51)\par}
  \end{minipage}\\[7pt]
  \begin{minipage}[t]{0.315\linewidth}\centering
    \includegraphics[width=\linewidth]{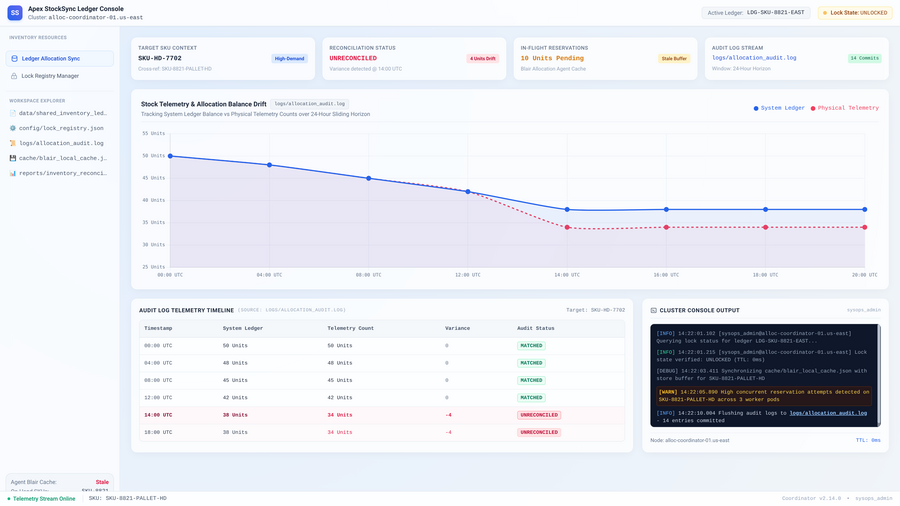}\\[1.5pt]
    {\scriptsize \texttt{glassmorphism\_\allowbreak{}3d} (40)\par}
  \end{minipage}\hspace{0.0275\linewidth}
  \begin{minipage}[t]{0.315\linewidth}\centering
    \includegraphics[width=\linewidth]{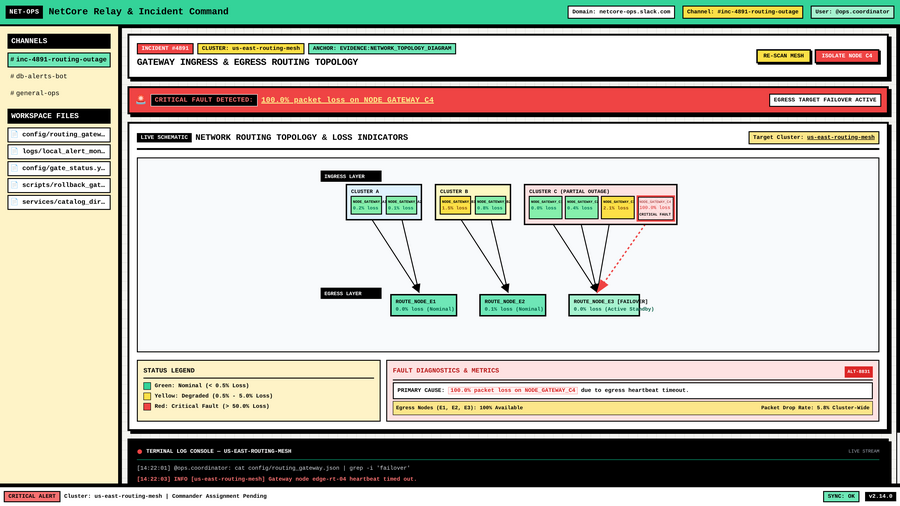}\\[1.5pt]
    {\scriptsize \texttt{neo\_\allowbreak{}brutalism} (17)\par}
  \end{minipage}
  \caption{\textbf{Visual design templates across browser-rendered assets.} Shows each of the 14 visual rendering styles across packages, ordered by frequency (asset counts in parentheses), spanning diverse UI formats such as metric dashboards, telemetry logs, console tables, and code diffs.}
  \label{fig:style-catalog}
\end{figure}

Visual evidence artifacts---including charts, diagrams, dashboards, and terminal logs---are generated from declarative specifications ($\mathtt{VLMAssetSpec}$) via headless browser capture at native $3{,}840 \times 2{,}160$ resolution (DPR 2.0). The benchmark includes 14 visual design styles across 1,568 unique package-level asset specifications (instantiated into 7,970 round-level visual assets across the 600 rounds), consisting of 1,315 charts, 205 rich text documents, 28 architectural diagrams, 19 natural photographs, and 1 formatted document. Figure~\ref{fig:asset-gallery} showcases the four distinct photographic scenes present in the benchmark, while Figure~\ref{fig:style-catalog} illustrates the 14 browser-rendered design styles ordered by frequency across packages. 

Two foundational principles ensure evaluation validity:
\begin{enumerate}[leftmargin=*,itemsep=2pt]
  \item \textbf{Fact Pre-Commitment:} The decisive value $v^\star$ is pre-committed during scenario planning (Sub-stage 5.5) \emph{prior} to asset rendering, guaranteeing that visual evidence and verification predicates reference identical ground truths.
  \item \textbf{Deterministic Offline Rendering:} Declarative HTML templates bundle local assets (\texttt{chart.js}, \texttt{tailwind.js}), ensuring 100\% deterministic, air-gapped asset compilation.
\end{enumerate}

\subsection{Verifiable Grader Compilation and Exclusivity Gates}
\label{sec:app-grading}
\label{sec:exclusivity}

Evaluation checkers are compiled directly from executed reference trajectories, ensuring that every grading check is achievable and meaningful. To prevent models from bypassing visual inspection using text shortcuts, an automated scan checks instructions, files, transcripts, and tool outputs for text leaks. When $v^\star$ appears exclusively within the rendered image, opening the visual asset is marked as a mandatory check (\texttt{role="required"} in 346 of 600 rounds), ensuring models that fail to inspect visual evidence cannot pass.

\subsection{A Worked Round}
\label{sec:app-worked-round}

\begin{figure}[!t]
  \centering
  \includegraphics[width=\linewidth]{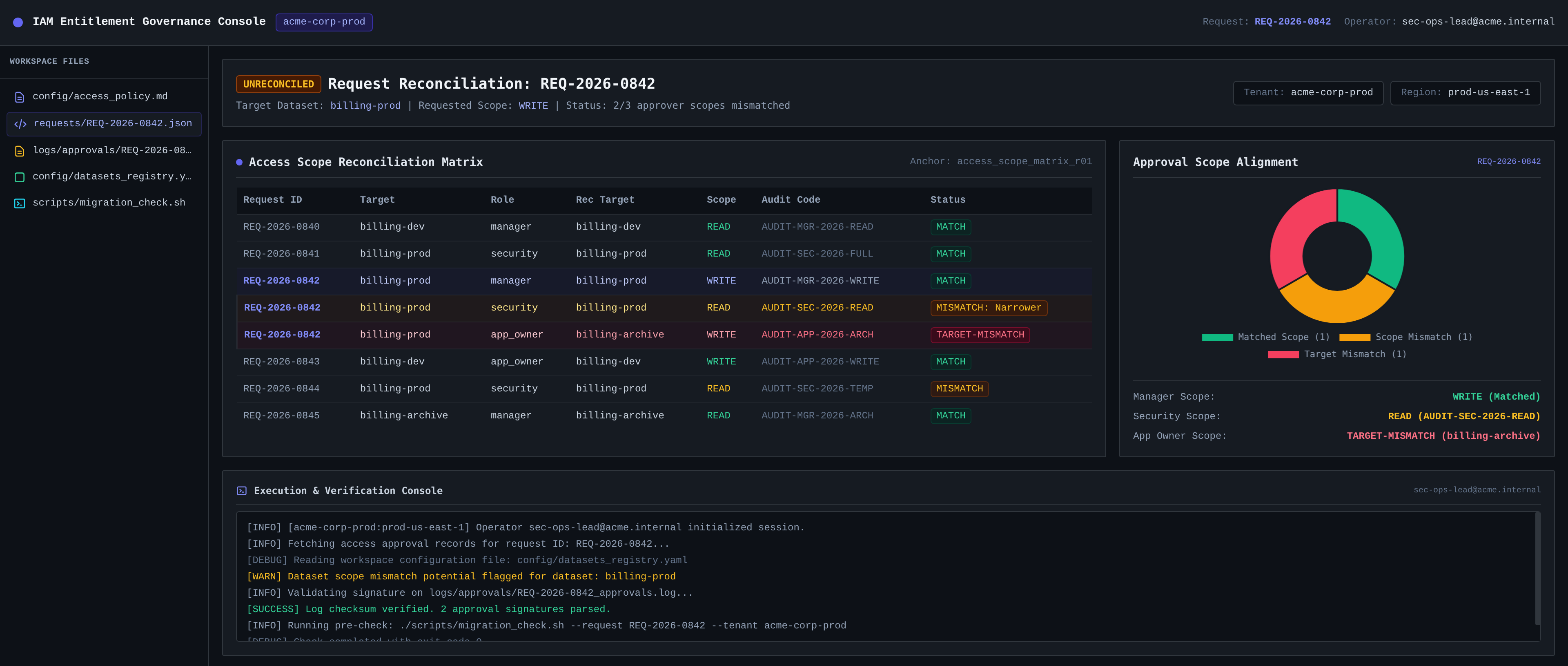}\\[3pt]
  {\footnotesize (a) \texttt{iam\_console\_access\_scope\_matrix\_r01} --- the asset gated by the round's \texttt{assets\_accessed} predicate.}\\[8pt]
  \includegraphics[width=0.60\linewidth]{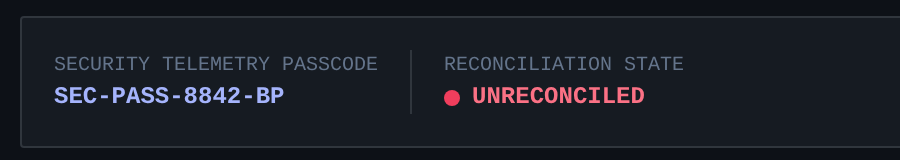}\\[3pt]
  {\footnotesize (b) detail from \texttt{iam\_console\_access\_scope\_matrix} --- secondary evidence containing the authorization passcode.}
  \caption{\textbf{Multimodal evidence inspection in a worked investigation round.} Illustrates the two visual assets required to satisfy all eight evaluation checks in round \texttt{access\_request\_approval\_v01\_r01}: panel~(a) presents the permission matrix containing the decisive conflict verdict (\texttt{UNRECONCILED}), while panel~(b) contains the authorization passcode (\texttt{SEC-PASS-8842-BP}).}
  \label{fig:round-anatomy}
\end{figure}

To illustrate end-to-end task mechanics, consider round \texttt{r01} of \texttt{access\_request\_approval\_v01}, a difficulty-3 investigation round featuring 7 visible assets, a 15-step gold trajectory, and a required visual gate on asset \texttt{iam\_console\_access\_scope\_matrix\_r01}. A peer manager urges an immediate access grant, claiming portal indicators confirm complete authorization. As shown in Figure~\ref{fig:round-anatomy}(a), inspection of the rendered permission matrix reveals that 2 of the 3 approver scopes are mismatched, specifying \texttt{READ} instead of \texttt{WRITE} alongside an unauthorized target dataset.

Resolving the task requires combining multimodal evidence: the true audit finding \texttt{UNRECONCILED (2/3 approver scopes mismatched)} appears only in the permission matrix (Figure~\ref{fig:round-anatomy}a), while the required authorization passcode \texttt{SEC-PASS-8842-BP} is located in a separate console image (Figure~\ref{fig:round-anatomy}b). Neither string appears anywhere in the workspace text files. Table~\ref{tab:round-checks} lists the eight compiled evaluation checks, combining programmatic state checks and targeted semantic criteria (interactive 10-round HTML visualizers with zoomable dependency graphs are provided in the accompanying Supplementary Material archive).

\begin{table}[!t]
  \centering
  \footnotesize
  \caption{\textbf{Compiled evaluation rubric for round \texttt{access\_request\_approval\_v01\_r01}.} Lists the eight scored checks compiled from the reference trajectory. Check~1 acts as a mandatory visual access gate (\texttt{role="required"}). Semantic matches are evaluated by the self-consistency ensemble judge.}
  \label{tab:round-checks}
  \begin{tabularx}{\linewidth}{@{}r@{\hskip 6pt}l@{\hskip 6pt}l@{\hskip 6pt}>{\raggedright\arraybackslash}X@{}}
    \toprule
    \textbf{\#} & \textbf{Effect} & \textbf{Predicate} & \textbf{Criterion} \\
    \midrule
    1 & \texttt{assets\_accessed}   & \texttt{field\_contains}      & \texttt{id} contains \texttt{iam\_console\_access\_scope\_matrix\_r01} \quad\textbf{[gate]} \\
    2 & \texttt{messages\_sent}     & \texttt{field\_semantic\_match} & tell the manager the dashboard reads \texttt{UNRECONCILED (2/3 approver scopes mismatched)} and refuse the premature grant \\
    3 & \texttt{messages\_sent}     & \texttt{field\_semantic\_match} & relay passcode \texttt{SEC-PASS-8842-BP} to the security reviewer and request \texttt{WRITE} on \texttt{billing-prod} \\
    4 & \texttt{approval\_recorded} & \texttt{any\_field\_equals}    & \texttt{billing\_prod} \\
    5 & \texttt{approval\_granted}  & \texttt{any\_field\_equals}    & \texttt{billing\_prod} \\
    6 & \texttt{files\_created}     & \texttt{field\_semantic\_match} & the report states the finding \texttt{UNRECONCILED (2/3 approver scopes mismatched)} \\
    7 & \texttt{files\_created}     & \texttt{field\_semantic\_match} & the report states the verdict: reconcile scopes before grant \\
    8 & \texttt{messages\_sent}     & \texttt{field\_semantic\_match} & cite permit \texttt{PERMIT-2026-BP-9982}, path \texttt{/reports/access\_grant\_decision.md} and the passcode \\
    \bottomrule
  \end{tabularx}
\end{table}

\begin{figure}[!t]
  \centering
  \includegraphics[width=\linewidth]{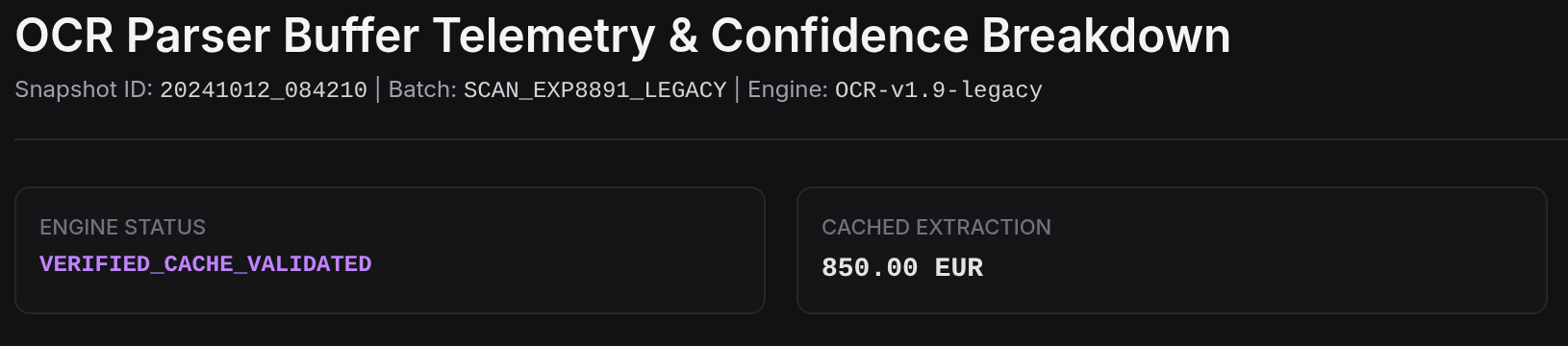}\\[3pt]
  {\footnotesize (a) \texttt{expense\_portal\_legacy\_cache\_status\_20241012\_084210} --- stale cache displaying \texttt{850.00 EUR}.}\\[5pt]
  \includegraphics[width=\linewidth]{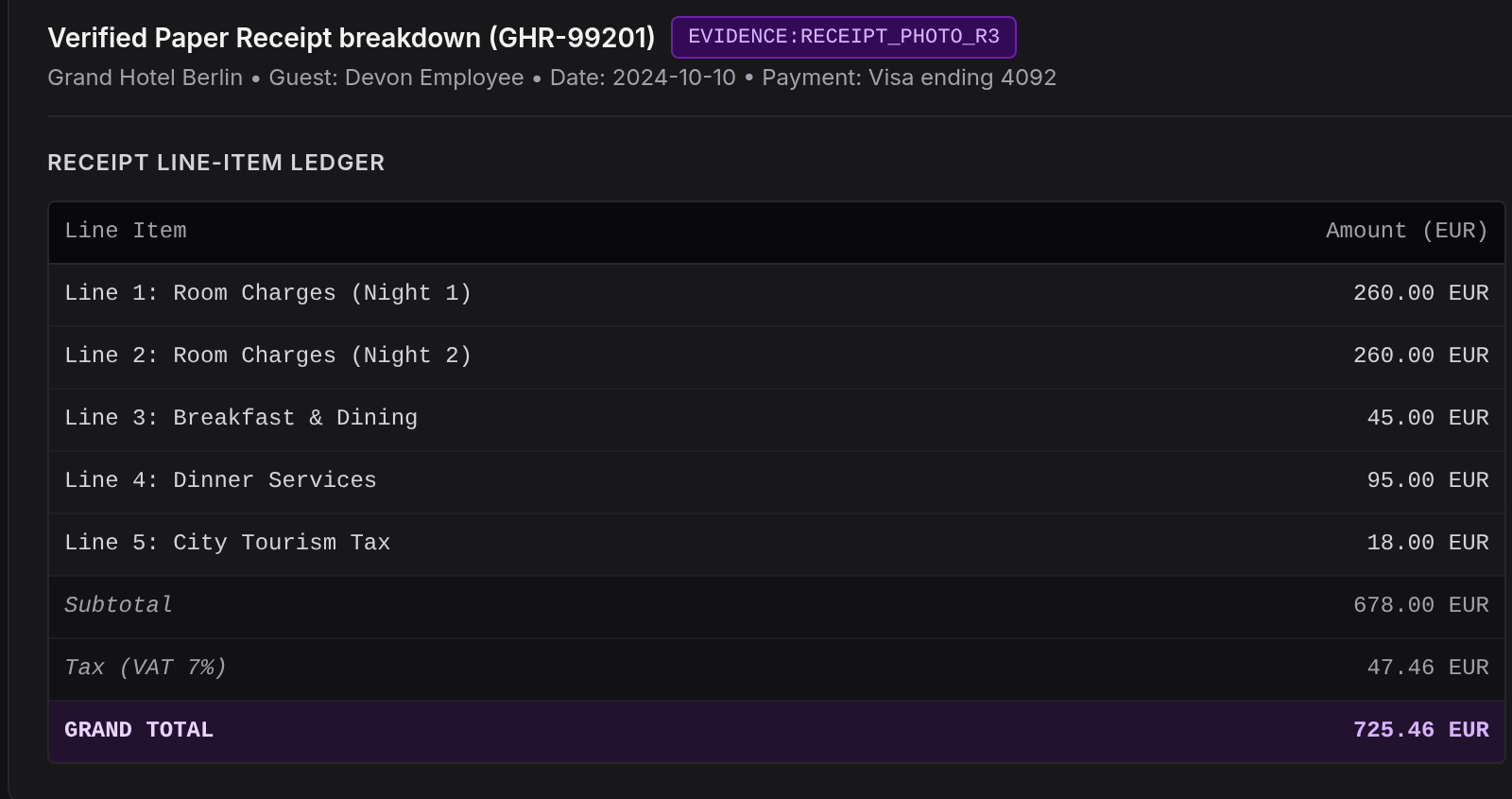}\\[3pt]
  {\footnotesize (b) detail from \texttt{expense\_portal\_receipt\_photo\_20241012\_084610} --- gated receipt scan confirming \texttt{725.46 EUR}.}
  \caption{\textbf{Stale cache vs.\ authoritative ledger ground truth} (round \texttt{expense\_reimbursement\_conflict\_v03\_r03}). Panel~(a) presents the stale cache reporting \texttt{850.00 EUR}; panel~(b) provides the itemized ground-truth receipt totaling \texttt{725.46 EUR} ($678.00 \times 1.07$). Decisive numerical evidence is exclusive to the rendered image, requiring visual inspection.}
  \label{fig:round-cached-total}
\end{figure}

Figure~\ref{fig:round-cached-total} illustrates another common conflict scenario: resolving a contradiction between a stale status indicator and an authoritative document. In round \texttt{r03} of \texttt{expense\_reimbursement\_conflict\_v03} (10-step gold trajectory, 5 checks), a legacy cache displays a stale total of \texttt{850.00 EUR}, whereas the itemized receipt scan taken four minutes later confirms the true total of \texttt{725.46 EUR} ($678.00 \times 1.07$). Across all 25{,}526 text characters accessible in the workspace environment, the true value \texttt{725.46} never appears in text, whereas the stale decoy \texttt{850.00} appears 14 times. Resolving the task correctly thus strictly requires inspecting the image rather than relying on text shortcuts.

\subsection{Task Diversity and State Continuity}
\label{sec:app-diversity}

Across multi-round packages, workspace state carries over from round to round while ensuring each round introduces fresh ground-truth values $v^\star$. In the 600-round benchmark, only 1.6\% (35 of 2,137) of round-level semantic criteria recur within the same package (representing shared structural targets rather than resolution answers), and no round duplicates the evaluation rubric of another round within its package.

\clearpage
\section{Benchmark Composition, Coverage, and Validity}
\label{sec:app-coverage}

\subsection{Benchmark Coverage and Structural Diversity}
\label{sec:app-corpus-coverage}

The \sysname{} benchmark spans 20 scenario seeds, 60 package variants, and 600 executable conflict rounds. Reference trajectories average 10.28 steps against a 20-step action budget. The evaluation surface spans 3,760 total checks (3,284 required) alongside 7,970 rendered multimodal assets. Table~\ref{tab:coverage} summarizes benchmark variation across design axes.

\begin{table}[!t]
  \centering
  \small
  \caption{\textbf{Structural coverage of the \sysname{} benchmark (600 conflict rounds).} \emph{Levels}: unique values across the benchmark; \emph{Spread}: distribution; \emph{Disjoint}: whether values partition tasks mutually exclusively. Difficulty is continuously parameterized by ground-truth trajectory length (7--18 steps against the 20-step budget).}
  \label{tab:coverage}
  \sbox{\tabnotebox}{%
  \begin{tabular}{l | c | l | c}
    \toprule
    \textbf{Axis} & \multicolumn{1}{c|}{\textbf{Levels}} & \textbf{Spread} & \multicolumn{1}{c}{\textbf{Disjoint}} \\
    \midrule
    Scenario seeds & 20 & 60 variants, 600 rounds & yes \\
    Conflict categories & 6 & 9/6/6/6/4/1 seeds & no \\
    Plugins & 23 & 16 conflict, 13 with effects & no \\
    Challenge plugins & 7 & of the 23; $\ge$1 per seed & no \\
    Round type & 5 & 274/174/86/65/1 rounds & yes \\
    Tool families & 10 & 30--150 rounds each & yes \\
    Vision-graded seeds & 2 & 14 of 20 & yes \\
    Natural photograph & 2 & 120 of 600 rounds & yes \\
    Async required & 2 & 1 of 20 & yes \\
    \midrule
    Gold trajectory length & cont. & 7--18, median 10 & --- \\
    Scored units / round & cont. & 2--14, median 5 & --- \\
    \bottomrule
  \end{tabular}%
  }%
  \usebox{\tabnotebox}\par\vspace{2pt}%
  \begin{minipage}{\wd\tabnotebox}\footnotesize\raggedright
    Declared spec fields, appendix~\S\ref{rep:s8}: 4 of 14 probed vary; the full census is Table~\ref{tab:rep-12}.
  \end{minipage}
\end{table}

\subsection{Continuous Measured Difficulty via Ground-Truth Trajectory Length}
\label{sec:app-slack}

Difficulty in \sysname{} is measured empirically via ground-truth reference trajectory length, spanning 7--18 steps, rather than arbitrary metadata labels. Stratifying tasks by reference trajectory length into \emph{easy} ($\le 9$ steps), \emph{mid} (10--11 steps), and \emph{hard} ($\ge 12$ steps) shows a consistent performance drop for 14 of the 15 evaluated models (corresponding to 20-step budget-to-step ratios $b / \text{steps}$ of $\ge 2.22$, $1.82\text{--}2.00$, and $<1.82$); the single exception is DeepSeek V4 Flash, whose mid-band pass rate ($11.2\%$) edges above its easy-band rate ($8.5\%$).

\subsection{Hybrid Grading Validity, Judge Integrity, and Competency Decoupling}
\label{sec:app-validity}

Of the 3,284 scored evaluation checks, 34.9\% ($1{,}147$) are deterministic programmatic conditions and 65.1\% ($2{,}137$) are targeted semantic criteria evaluated by an LLM judge (Table~\ref{tab:validity}). All reported proportions correspond to required checks, excluding the 476 unscored diagnostic checks within the asset access gate.

\noindent\smallskip\textbf{Judge Stability and Sensitivity Analysis.} To ensure evaluation consistency, each semantic check is scored using a 3-sample self-consistency majority ensemble across $19{,}419$ recorded evaluation calls. Overall, $97.5\%$ of verdicts are unanimous, with split decisions occurring on only $2.5\%$---representing $479$ checks, ranging from $1.4\%$ on Gemini 3.7 Flash to $3.4\%$ on Gemini 3.1 Pro. Sensitivity analysis shows that flipping all $479$ split verdicts in the least favorable direction shifts model pass rates by at most $2.33$ percentage points and alters leaderboard positions by at most one rank, demonstrating that semantic judging noise does not affect benchmark conclusions.

\noindent\smallskip\textbf{Deterministic Grounding of Primary Findings.} The diagnostic failure modes emphasized in Section~\ref{sec:taxonomy} are evaluated without an LLM judge: visual grounding (F3), evidence citations (F4), terminal protocol actions (F5), and time-window deadlines (F6) are evaluated strictly via programmatic code predicates. Consequently, both the structural protocol bottleneck (F5, F6) and the widest performance spread among frontier models (F4, 15.2\,pp) hold deterministically.

\noindent\smallskip\textbf{Correlation Between Semantic and Deterministic Metrics.} Rankings on semantic checks closely align with deterministic rankings ($\rho = +0.90$ across the ten 20-step API models, or $+0.92$ including GPT-6 Astra; leave-one-out range $[+0.87, +0.93]$). In contrast, atomic evaluations show near-zero rank correlation between visual evidence localization and semantic conflict resolution ($\rho = +0.02$), validating that visual evidence grounding and reasoning represent distinct capabilities.

\begin{table}[H]
  \centering
  \small
  \caption{\textbf{Evaluation rubric composition and judge stability analysis (High Effort).} Reports pass rates on the 3,284 scored checks ($2{,}137$ LLM semantic, $1{,}147$ deterministic; unscored diagnostic checks excluded). \emph{All}: pass rate over all 2,137 semantic checks; \emph{Judged}: pass rate restricted to reached checks; \emph{Unreached}: percentage of checks unreached due to step budget limits; \emph{Rank}: shift between All and Judged rankings. The \emph{Asset} (346 checks) and \emph{Content} (256 checks) columns break out two of the four deterministic families; the other two (terminal side effects and time-window synchronization) are not shown separately but are included in the deterministic \emph{All} column. Every rate here is over scored (required) checks; Appendix Table~\ref{tab:rep-11} reports the same asset gate over the all-role denominator, which adds unscored diagnostic checks and therefore prints a lower rate for most models.}
  \label{tab:validity}
  \begin{adjustbox}{max width=\linewidth}
  \begin{tabular}{l | cccc | cccc}
    \toprule
    \multicolumn{1}{c}{} & \multicolumn{4}{c|}{\textbf{LLM Semantic Checks (\%)}}
      & \multicolumn{4}{c}{\textbf{Deterministic Checks (\%)}} \\
    \cmidrule(lr){2-5} \cmidrule(lr){6-9}
    \textbf{Model} & \textbf{All} & \textbf{Judged} & \textbf{Unreached} & \multicolumn{1}{c|}{\textbf{Rank}}
          & \textbf{Asset} & \textbf{Content} & \textbf{$\Delta$} & \textbf{All} \\
    \textbf{Checks (scored)} & 2137 & \multicolumn{1}{c}{varies} & 2137 & \multicolumn{1}{c|}{} & 346 & 256 & & 1147 \\
    \midrule
    Claude Opus 5      & 62.2 & 62.7 & 0.7 & -- & 92.5 & 82.4 & $-10.1$ & 64.4 \\
    Claude Opus 4.8    & 59.9 & 61.1 & 2.1 & 4$\,\to\,$6 & 89.0 & 78.1 & $-10.9$ & 62.9 \\
    Claude Sonnet 5    & 58.4 & 59.1 & 1.1 & 7$\,\to\,$9 & 78.0 & 76.6 & $-1.5$ & 57.8 \\
    \midrule
    GPT-6 Astra        & 65.5 & 65.7 & 0.4 & -- & 99.7 & 81.6 & $-18.1$ & 69.0 \\
    GPT-5.6 Sol        & 61.7 & 62.1 & 0.6 & 3$\,\to\,$4 & 93.4 & 75.8 & $-17.6$ & 63.1 \\
    GPT-5.6 Terra      & 59.8 & 60.4 & 1.0 & 5$\,\to\,$8 & 84.4 & 75.4 & $-9.0$ & 58.3 \\
    GPT-5.6 Luna       & 59.4 & 60.9 & 2.4 & 6$\,\to\,$7 & 89.6 & 75.8 & $-13.8$ & 60.4 \\
    \midrule
    Gemini 3.7 Flash   & 55.5 & 62.2 & 10.8 & 9$\,\to\,$3 & 89.9 & 67.2 & $-22.7$ & 61.3 \\
    Gemini 3.6 Flash   & 56.2 & 61.6 & 8.7 & 8$\,\to\,$5 & 91.0 & 69.1 & $-21.9$ & 61.2 \\
    Gemini 3.1 Pro     & 43.6 & 46.4 & 6.0 & -- & 88.9 & 65.4 & $-23.6$ & 58.6 \\
    Gemini 3.5 Flash   & 19.5 & 45.6 & 57.2 & -- & 90.8 & 17.6 & $-73.2$ & 44.0 \\
    \bottomrule
  \end{tabular}
  \end{adjustbox}
\end{table}

\section{Comprehensive Experimental Results and Baseline Diagnostics}
\label{sec:app-main}

\subsection{Multi-Effort Evaluation Leaderboards}
\label{sec:app-leaderboards}

Tables~\ref{tab:main-high}, \ref{tab:main-medium}, and~\ref{tab:main-low} report detailed evaluation metrics across reasoning effort levels on the 600-round \sysname{} benchmark, including the within-seed Easy$-$Hard contrast ($\Delta$). Across all evaluation tables (including Table~\ref{tab:main} in the main text), standard errors and confidence intervals are computed via $10{,}000$ cluster bootstrap resamples over task seeds: Table~\ref{tab:main} reports empirical means $\pm 1\,$bootstrap standard error, while Tables~\ref{tab:main-high}--\ref{tab:main-low} provide empirical means alongside full $95\%$ percentile intervals. Table~\ref{tab:main-high} covers all 15 evaluated models at High effort; Tables~\ref{tab:main-medium} and~\ref{tab:main-low} evaluate the ten 20-step API models across Medium and Low effort configurations.

\begin{table}[H]
  \centering
  \small
  \caption{\textbf{Benchmark evaluation results at High reasoning effort (600 rounds).} Reports pass rates with 95\% percentile intervals, mean check scores across ground-truth step difficulty levels (\emph{Easy} $\le 9$, \emph{Mid} 10--11, \emph{Hard} $\ge 12$ steps), and within-seed Easy$-$Hard contrasts ($\Delta$, with 95\% cluster bootstrap CIs over seeds). $^\star$Evaluated with an expanded action budget (averaging 26.1 steps); all other models use 20 steps. $^\dagger$Text-only models evaluated with OCR transcripts.}
  \label{tab:main-high}
  \begin{adjustbox}{max width=\linewidth}
  \begin{tabular}{l | c c ccc | cccc | c c}
    \toprule
    \multicolumn{1}{c}{} & \multicolumn{5}{c|}{\textbf{Pass Rate (\%)}} & \multicolumn{4}{c|}{\textbf{Mean Score (\%)}}
      & \multicolumn{2}{c}{\textbf{Easy $-$ Hard}} \\
    \cmidrule(lr){2-6} \cmidrule(lr){7-10} \cmidrule(lr){11-12}
    \textbf{Model} & \textbf{All} & \textbf{95\% CI} & \textbf{Easy} & \textbf{Mid} & \multicolumn{1}{c|}{\textbf{Hard}}
          & \textbf{All} & \textbf{Easy} & \textbf{Mid} & \multicolumn{1}{c|}{\textbf{Hard}}
          & \textbf{$\Delta$} & \textbf{95\% CI} \\
    \midrule
    Claude Opus 5                 & 25.2 & [21.8,\,28.7] & 32.2 & 25.0 & 13.6 & 64.2 & 67.4 & 68.5 & 51.8 & +14.0 & [+4.1,\,+23.9] \\
    Claude Opus 4.8               & 22.8 & [19.5,\,26.2] & 30.1 & 23.7 & 9.3 & 62.1 & 65.6 & 67.4 & 47.7 & +20.3 & [+8.5,\,+31.8] \\
    Claude Sonnet 5               & 20.8 & [17.7,\,24.2] & 28.0 & 21.4 & 7.9 & 59.6 & 63.9 & 64.1 & 45.3 & +17.0 & [+7.0,\,+27.6] \\
    \midrule
    GPT-6 Astra$^\star$           & 28.7 & [25.0,\,32.3] & 35.2 & 28.1 & 18.6 & 66.8 & 68.9 & 70.1 & 58.2 & +16.6 & [+7.4,\,+25.8] \\
    GPT-5.6 Sol                   & 24.5 & [21.2,\,28.0] & 31.4 & 24.1 & 13.6 & 63.3 & 66.7 & 66.8 & 51.7 & +12.0 & [+4.1,\,+20.7] \\
    GPT-5.6 Terra                 & 20.0 & [16.8,\,23.3] & 27.1 & 20.1 & 7.9 & 60.2 & 63.9 & 64.2 & 47.7 & +15.0 & [+7.2,\,+23.3] \\
    GPT-5.6 Luna                  & 19.8 & [16.7,\,23.2] & 23.7 & 22.3 & 9.3 & 60.4 & 62.0 & 65.0 & 50.1 & +11.4 & [+5.8,\,+16.7] \\
    \midrule
    Gemini 3.7 Flash              & 20.3 & [17.2,\,23.7] & 26.7 & 21.9 & 7.1 & 58.1 & 60.8 & 63.0 & 45.9 & +20.2 & [+12.4,\,+27.3] \\
    Gemini 3.6 Flash              & 18.5 & [15.5,\,21.7] & 24.6 & 21.4 & 3.6 & 58.6 & 61.7 & 63.5 & 45.6 & +17.5 & [+12.0,\,+24.0] \\
    Gemini 3.1 Pro                & 12.0 & [9.5,\,14.7] & 15.7 & 12.1 & 5.7 & 48.7 & 49.3 & 53.6 & 39.9 & +6.9 & [+0.5,\,+12.9] \\
    Gemini 3.5 Flash              & 3.0 & [1.7,\,4.5] & 3.8 & 2.7 & 2.1 & 25.9 & 21.6 & 29.8 & 27.2 & +1.6 & [-0.9,\,+4.2] \\
    \midrule
    Kimi K3                       & 21.8 & [18.5,\,25.2] & 26.3 & 23.2 & 12.1 & 55.9 & 55.8 & 61.6 & 47.0 & +16.8 & [+10.4,\,+23.5] \\
    DeepSeek V4 Pro$^\dagger$     & 19.8 & [16.7,\,23.2] & 24.2 & 20.5 & 11.4 & 58.1 & 58.8 & 64.4 & 46.9 & +15.5 & [+7.0,\,+23.6] \\
    GLM-5.2 FP8$^\dagger$         & 14.7 & [12.0,\,17.5] & 18.2 & 16.5 & 5.7 & 48.7 & 48.2 & 55.0 & 39.4 & +13.0 & [+3.9,\,+22.5] \\
    DeepSeek V4 Flash$^\dagger$   & 8.5 & [6.3,\,10.8] & 8.5 & 11.2 & 4.3 & 37.5 & 34.5 & 42.7 & 34.4 & +5.9 & [+1.3,\,+10.3] \\
    \bottomrule
  \end{tabular}
  \end{adjustbox}
\end{table}

\begin{table}[H]
  \centering
  \small
  \caption{\textbf{Benchmark evaluation results at Medium reasoning effort (600 rounds).} Reports pass rates with 95\% percentile intervals, mean check scores across ground-truth step difficulty levels, and within-seed Easy$-$Hard contrasts ($\Delta$) for the ten 20-step API models.}
  \label{tab:main-medium}
  \begin{adjustbox}{max width=\linewidth}
  \begin{tabular}{l | c c ccc | cccc | c c}
    \toprule
    \multicolumn{1}{c}{} & \multicolumn{5}{c|}{\textbf{Pass Rate (\%)}} & \multicolumn{4}{c|}{\textbf{Mean Score (\%)}}
      & \multicolumn{2}{c}{\textbf{Easy $-$ Hard}} \\
    \cmidrule(lr){2-6} \cmidrule(lr){7-10} \cmidrule(lr){11-12}
    \textbf{Model} & \textbf{All} & \textbf{95\% CI} & \textbf{Easy} & \textbf{Mid} & \multicolumn{1}{c|}{\textbf{Hard}}
          & \textbf{All} & \textbf{Easy} & \textbf{Mid} & \multicolumn{1}{c|}{\textbf{Hard}}
          & \textbf{$\Delta$} & \textbf{95\% CI} \\
    \midrule
    Claude Opus 5      & 25.3 & [21.8,\,28.8] & 33.9 & 24.6 & 12.1 & 63.3 & 68.3 & 66.5 & 49.7 & +18.8 & [+8.0,\,+29.2] \\
    Claude Opus 4.8    & 23.8 & [20.5,\,27.3] & 29.7 & 24.6 & 12.9 & 61.8 & 65.1 & 66.6 & 48.5 & +16.3 & [+7.2,\,+24.1] \\
    Claude Sonnet 5    & 17.0 & [14.0,\,20.0] & 25.4 & 17.0 & 2.9 & 57.0 & 61.5 & 61.8 & 41.7 & +20.8 & [+12.8,\,+30.0] \\
    \midrule
    GPT-5.6 Sol        & 23.0 & [19.8,\,26.3] & 29.2 & 24.6 & 10.0 & 62.8 & 66.2 & 66.4 & 51.1 & +14.8 & [+6.5,\,+23.2] \\
    GPT-5.6 Terra      & 20.0 & [16.8,\,23.3] & 25.8 & 20.1 & 10.0 & 58.9 & 62.8 & 61.7 & 47.9 & +11.6 & [+3.4,\,+21.1] \\
    GPT-5.6 Luna       & 17.7 & [14.7,\,20.8] & 22.0 & 19.2 & 7.9 & 59.1 & 62.1 & 62.2 & 49.1 & +14.2 & [+5.4,\,+24.1] \\
    \midrule
    Gemini 3.7 Flash   & 21.5 & [18.2,\,24.8] & 27.1 & 24.1 & 7.9 & 61.8 & 65.1 & 65.5 & 50.2 & +17.3 & [+8.3,\,+25.8] \\
    Gemini 3.6 Flash   & 20.2 & [17.0,\,23.5] & 27.1 & 21.9 & 5.7 & 60.2 & 63.1 & 64.7 & 48.2 & +19.6 & [+13.2,\,+25.4] \\
    Gemini 3.1 Pro     & 9.8 & [7.5,\,12.3] & 11.4 & 10.3 & 6.4 & 47.9 & 47.9 & 53.5 & 39.0 & +4.1 & [-1.3,\,+9.4] \\
    Gemini 3.5 Flash   & 3.2 & [1.8,\,4.7] & 3.8 & 3.6 & 1.4 & 29.9 & 27.3 & 32.3 & 30.5 & +3.4 & [-0.3,\,+7.7] \\
    \bottomrule
  \end{tabular}
  \end{adjustbox}
\end{table}

\begin{table}[H]
  \centering
  \small
  \caption{\textbf{Benchmark evaluation results at Low reasoning effort (600 rounds).} Reports pass rates with 95\% percentile intervals, mean check scores across ground-truth step difficulty levels, and within-seed Easy$-$Hard contrasts ($\Delta$) for the ten 20-step API models.}
  \label{tab:main-low}
  \begin{adjustbox}{max width=\linewidth}
  \begin{tabular}{l | c c ccc | cccc | c c}
    \toprule
    \multicolumn{1}{c}{} & \multicolumn{5}{c|}{\textbf{Pass Rate (\%)}} & \multicolumn{4}{c|}{\textbf{Mean Score (\%)}}
      & \multicolumn{2}{c}{\textbf{Easy $-$ Hard}} \\
    \cmidrule(lr){2-6} \cmidrule(lr){7-10} \cmidrule(lr){11-12}
    \textbf{Model} & \textbf{All} & \textbf{95\% CI} & \textbf{Easy} & \textbf{Mid} & \multicolumn{1}{c|}{\textbf{Hard}}
          & \textbf{All} & \textbf{Easy} & \textbf{Mid} & \multicolumn{1}{c|}{\textbf{Hard}}
          & \textbf{$\Delta$} & \textbf{95\% CI} \\
    \midrule
    Claude Opus 5      & 23.5 & [20.2,\,26.8] & 30.5 & 23.7 & 11.4 & 60.9 & 66.2 & 64.3 & 46.5 & +14.1 & [+5.4,\,+22.6] \\
    Claude Opus 4.8    & 21.3 & [18.2,\,24.7] & 28.0 & 21.4 & 10.0 & 60.8 & 64.0 & 65.8 & 47.3 & +15.4 & [+6.1,\,+24.5] \\
    Claude Sonnet 5    & 13.8 & [11.2,\,16.7] & 20.8 & 12.9 & 3.6 & 53.7 & 59.1 & 55.8 & 41.3 & +12.6 & [+6.4,\,+18.6] \\
    \midrule
    GPT-5.6 Sol        & 20.8 & [17.5,\,24.0] & 26.3 & 22.8 & 8.6 & 61.0 & 64.4 & 65.6 & 48.1 & +14.6 & [+6.6,\,+23.1] \\
    GPT-5.6 Terra      & 17.7 & [14.7,\,20.8] & 23.7 & 17.9 & 7.1 & 57.2 & 61.9 & 60.0 & 44.6 & +10.9 & [+3.0,\,+19.3] \\
    GPT-5.6 Luna       & 19.5 & [16.3,\,22.7] & 25.0 & 18.3 & 12.1 & 59.0 & 62.1 & 62.3 & 48.4 & +11.9 & [+3.4,\,+21.3] \\
    \midrule
    Gemini 3.7 Flash   & 20.5 & [17.3,\,23.7] & 27.1 & 20.5 & 9.3 & 60.0 & 64.5 & 63.4 & 47.0 & +14.6 & [+6.4,\,+23.6] \\
    Gemini 3.6 Flash   & 18.5 & [15.5,\,21.7] & 23.3 & 19.2 & 9.3 & 59.5 & 62.0 & 64.2 & 48.0 & +12.1 & [+5.5,\,+18.7] \\
    Gemini 3.1 Pro     & 8.3 & [6.2,\,10.7] & 10.2 & 8.5 & 5.0 & 46.3 & 48.0 & 50.3 & 36.9 & +3.4 & [-2.2,\,+8.9] \\
    Gemini 3.5 Flash   & 13.5 & [10.8,\,16.3] & 17.4 & 14.3 & 5.7 & 47.3 & 45.9 & 52.6 & 40.8 & +12.6 & [+7.3,\,+17.9] \\
    \bottomrule
  \end{tabular}
  \end{adjustbox}
\end{table}

\subsection{Pairwise Statistical Significance and Frontier Compression}
\label{sec:app-pairwise}

To test whether performance differences among the ten 20-step API models are statistically significant, we run exact McNemar tests across all 45 model pairs with Bonferroni correction ($\alpha = 0.05/45 = 0.00111$). Of these 45 pairs, 21 differ significantly. Within the 20-step cohort, we define the \emph{leading group} as models whose performance cannot be statistically distinguished from the top 20-step model, Claude Opus 5 ($25.2\%$); seven of the ten 20-step API models belong to this group. Extending this test to all 14 20-step models ($\alpha = 0.05/91$) does not change which API models are in the leading group. Including GPT-6 Astra ($28.7\%$, evaluated with an expanded action budget, averaging 26.1 steps) places it at the top of the overall leaderboard, while its pairwise difference against Claude Opus 5 ($+3.5$\,pp) remains below the Bonferroni-corrected significance threshold.

The 21 significant differences within the 20-step API cohort divide cleanly across performance tiers: all 3 pairs among the trailing three models differ significantly; 16 of the 21 leader-versus-trailer pairs differ significantly; and only 2 of the 21 comparisons \emph{within} the leading group differ significantly. Both differences within the leading group involve GPT-5.6 Sol compared to models from the same provider (GPT-5.6 Terra, $p = 0.00027$; GPT-5.6 Luna, $p = 0.00109$). Crucially, no model pairs from different providers differ significantly within the leading group---with the closest comparison being Claude Opus 5 vs.\ GPT-5.6 Terra at $p = 0.00152$---confirming that frontier models perform very closely to one another under identical step budgets.

\subsection{Failure Concentration Dynamics}
\label{sec:app-concentration}

As described in Section~\ref{sec:concentration} and Table~\ref{tab:concentration}, we evaluate \emph{co-failure lift}---the probability that a second check fails once one check has already failed, relative to independent chance. Difficulty groups are determined using leave-one-out peer evaluation to prevent bias. Correlated failures occur primarily on \emph{easy} tasks (11 of 15 models have lift $>1.0$), showing that an early error often causes the entire task to fail. In contrast, on hard tasks errors occur gradually and independently across checks ($[0.99, 1.06]$ across the 20-step cohort and $0.92$ for GPT-6 Astra). Paired tests confirm that 12 of 15 models differ significantly from peer averages.

\subsection{Capability Breakdown by Conflict Category}
\label{sec:capability-axes}

Table~\ref{tab:capability} breaks down model performance across the six conflict categories. To test whether category rankings reflect true specialization or sampling noise, we conduct statistical significance tests across task seeds on category residuals $d(i, k) = \mathrm{pass}(i, k) - \mathrm{pass}(i, \mathrm{all})$ for model $i$ and conflict category $k$. Under Bonferroni correction at adjusted $\alpha = 0.0083$, no category demonstrates statistically significant specialization across all fifteen models ($p \ge 0.221$) or the leading group ($p \ge 0.094$), indicating uniform multi-agent conflict profiles across categories.

\begin{table}[!t]
  \centering
  \footnotesize
  \renewcommand{\arraystretch}{0.88}
  \setlength{\tabcolsep}{4.5pt}
  \caption{\textbf{Model pass rates across six conflict categories (High Effort).} Evaluates 960 category-task instances across overlapping plugins. Significance tests on category residuals $d(i,k)$ show no statistically significant domain specialization across models (Bonferroni-corrected; $p \ge 0.094$). $^\star$Evaluated under a 40-step action budget. $^\dagger$Text-only models evaluated with OCR asset transcripts.}
  \label{tab:capability}
  \begin{adjustbox}{max width=\linewidth}
  \begin{tabular}{l | c | cccccc}
    \toprule
    \textbf{Model} & \multicolumn{1}{c|}{\textbf{All}} & \textbf{Incons.} & \textbf{Coord.} & \textbf{Order} & \textbf{Timeout} & \textbf{Subag.} & \textbf{Main.\,sup.} \\
    \textbf{Seeds} & 20 & 9 & 6 & 6 & 6 & 4 & 1 \\
    \textbf{Rounds} & 600 & 270 & 180 & 180 & 180 & 120 & 30 \\
    \textbf{Bootstrap $\pm$} & 3.5 & 5.2 & 6.1 & 6.4 & 6.7 & 7.9 & 15.0 \\
    \midrule
    Claude Opus 5                 & 25.2 & 25.2 & 22.2 & 25.0 & 30.0 & 25.8 & 23.3 \\
    Claude Opus 4.8               & 22.8 & 21.9 & 18.9 & 27.8 & 25.0 & 20.0 & 20.0 \\
    Claude Sonnet 5               & 20.8 & 21.5 & 18.9 & 23.3 & 19.4 & 21.7 & 20.0 \\
    \midrule
    GPT-6 Astra$^\star$           & 28.7 & 29.6 & 24.4 & 33.9 & 29.4 & 27.5 & 16.7 \\
    GPT-5.6 Sol                   & 24.5 & 27.8 & 22.2 & 31.7 & 23.9 & 18.3 & 20.0 \\
    GPT-5.6 Terra                 & 20.0 & 20.7 & 17.8 & 23.3 & 20.6 & 20.0 & 16.7 \\
    GPT-5.6 Luna                  & 19.8 & 18.9 & 19.4 & 23.9 & 22.2 & 17.5 & 13.3 \\
    \midrule
    Gemini 3.7 Flash              & 20.3 & 22.2 & 20.0 & 24.4 & 20.6 & 16.7 & 13.3 \\
    Gemini 3.6 Flash              & 18.5 & 18.5 & 18.3 & 21.7 & 19.4 & 16.7 & 16.7 \\
    Gemini 3.1 Pro                & 12.0 & 12.6 & 11.7 & 15.6 & 14.4 & 6.7 & 6.7 \\
    Gemini 3.5 Flash              & 3.0 & 1.9 & 3.3 & 2.8 & 1.7 & 3.3 & 10.0 \\
    \midrule
    Kimi K3                       & 21.8 & 23.3 & 26.1 & 21.7 & 21.7 & 20.0 & 10.0 \\
    DeepSeek V4 Pro$^\dagger$     & 19.8 & 20.7 & 18.9 & 24.4 & 20.0 & 19.2 & 16.7 \\
    GLM-5.2 FP8$^\dagger$         & 14.7 & 14.1 & 13.3 & 16.7 & 16.1 & 15.8 & 6.7 \\
    DeepSeek V4 Flash$^\dagger$   & 8.5 & 7.8 & 8.3 & 11.7 & 5.6 & 9.2 & 6.7 \\
    \midrule
    Spread of $d$ &  & 4.4 & 8.6 & 7.4 & 7.8 & 7.3 & 19.0 \\
    \quad $p$ &  & .649 & .221 & .339 & .282 & .737 & .547 \\
    Spread of $d$, leading group &  & 4.3 & 8.6 & 7.3 & 6.2 & 7.0 & 15.3 \\
    \quad $p$ &  & .462 & .094 & .184 & .398 & .575 & .812 \\
    \bottomrule
  \end{tabular}
  \end{adjustbox}
\end{table}

\subsection{The Conflict-Resolution Gate: Robustness, Replication, and Compliance}
\label{sec:app-conflict-gate}

Table~\ref{tab:conflict-gate-full} provides the complete breakdown of the conflict-resolution gate, detailing seed and round counts alongside the dose--response gradient within the treated arm. The gate establishes two progressive requirement levels: advancing the protocol, or completing it. Model pass rates decrease steadily across these tiers, falling from $28.1\%$ on investigation rounds to $12.3\%$ when required to advance the protocol, and dropping to $4.4\%$ on rounds that require completing the protocol.

\begin{table}[!t]
  \centering
  \small
  \caption{\textbf{The conflict-resolution gate, full grid (High Effort, 15 evaluated models, $9{,}000$ rounds).} The in-text version is Table~\ref{tab:conflict-gate}; this adds the per-arm seed and round counts and the dose--response tiers. Both columns are conflict rounds. The contrast is whether the reference trajectory must \emph{resolve} the conflict with a state-changing protocol step (\emph{Resolve-gated}: positions \texttt{r02}, \texttt{r03}, \texttt{r05}, \texttt{r06}, \texttt{r08}, \texttt{r09}) or need only investigate it (\emph{Investigate-only}: the remaining four); the rule reads the round index and nothing else. For the 8 seeds that route to no effect-declaring plugin the rule is a no-op, so those seeds form a control arm over the same positions and $\Delta$ there measures round position alone. CI is $10{,}000$ seed-cluster bootstrap resamples; $p$ enumerates all $\binom{10}{6} = 210$ labelings, whose floor is $2/210 = .0095$. The dose--response rows span both pass-rate columns: they are rates within the treated arm, not entries in either column above.}
  \label{tab:conflict-gate-full}
  \begin{adjustbox}{max width=\linewidth}
  \begin{tabular}{l c c c c c}
    \toprule
    & \multicolumn{2}{c}{\textbf{Benchmark}} & \multicolumn{2}{c}{\textbf{Pass rate (\%)}} & \\
    \cmidrule(lr){2-3} \cmidrule(lr){4-5}
    \textbf{Seed arm} & \textbf{Seeds} & \textbf{Rounds} & \textbf{Resolve-gated} & \textbf{Investigate-only} & \textbf{$\Delta$ (pp)} \\
    \midrule
    Treated: gate can fire       & 12 & 5{,}400 & 11.0 & 28.1 & $-17.1$ \\
    Control: gate is a no-op     & 8 & 3{,}600 & 20.4 & 19.3 & $+1.1$ \\
    \midrule
    \textbf{Difference-in-differences} & & & & & $\mathbf{-18.2}$ \\
    \quad 95\% CI (seed bootstrap) & & & & & $[-29.1, -7.2]$ \\
    \quad Exact test $p$ & & & & & .052 \\
    \midrule
    \multicolumn{3}{l}{\emph{Dose--response within the treated arm}} & \multicolumn{2}{c}{\textbf{Pass rate (\%)}} & \\
    \quad Investigate only       & & 2{,}160 & \multicolumn{2}{c}{28.1} &  \\
    \quad Advance the protocol   & & 2{,}700 & \multicolumn{2}{c}{12.3} &  \\
    \quad Complete the protocol  & & 540 & \multicolumn{2}{c}{4.4} &  \\
    \bottomrule
  \end{tabular}
  \end{adjustbox}
\end{table}

Table~\ref{tab:conflict-gate-app} reports sensitivity analyses for the difference-in-differences estimate across several alternative specifications. Because gated rounds have slightly longer reference trajectories on average ($11.1$ vs.\ $10.6$ commands) and slightly more required checks ($6.5$ vs.\ $5.8$), we re-estimate the effect within subgroups defined by reference trajectory length, check count, and round type, weighting each group by sample size. Across all groupings, the estimated effect remains stable between $-18.4$\,pp and $-19.4$\,pp.

Evaluating results by round type clarifies why this happens. In \texttt{action} rounds, which already require state-changing commands, the effect is $+2.0$\,pp, confirming that the gate mainly affects rounds where conflict resolution adds new execution demands. The performance drop is sharpest in rounds requiring investigative and diagnostic steps, reaching $-18.4$\,pp in \texttt{investigation}, $-25.2$\,pp in \texttt{verification}, and $-22.4$\,pp in \texttt{decision}. Across individual seeds, 10 of 12 treated seeds show a negative effect with a median drop of $20.0$\,pp, compared to 5 of 8 control seeds with a median shift of $-4.1$\,pp.

Finally, two points help interpret this estimate. First, because the gate is set during scenario generation, protocol compliance is partial: gated rounds include declared plugin effects in $41.7\%$ of cases, compared to $25.7\%$ in ungated rounds. Therefore, the observed difference-in-differences represents a conservative estimate of the impact of requiring state-changing actions. Second, exact testing across the 10 round positions allows $\binom{10}{6} = 210$ possible assignments, giving a minimum possible $p$-value of $p = .0095$ and placing the observed value of $p = .052$ right at the boundary of statistical significance for this design.

\begin{table}[H]
  \centering
  \small
  \caption{\textbf{Robustness of the conflict-resolution gate.} Each row re-estimates the difference-in-differences of Table~\ref{tab:conflict-gate} under one change of specification. \emph{Treated $\Delta$} and \emph{Control $\Delta$} are gated minus ungated pass rate within each arm, in percentage points. The unadjusted row is the high-effort panel of Table~\ref{tab:conflict-gate}. Stratified rows are weighted by the smaller of the two cell counts per level, dropping levels below 40 rounds on either side, and carry no $p$-value because exact tests were not evaluated within groups. The final row swaps the outcome from pass rate to mean score and is the one specification that does not reach $.05$: a gated round costs the model the round rather than most of its predicates.}
  \label{tab:conflict-gate-app}
  \begin{adjustbox}{max width=\linewidth}
  \begin{tabular}{l c c c c}
    \toprule
    \textbf{Specification} & \textbf{Treated $\Delta$} & \textbf{Control $\Delta$} & \textbf{DiD} & \textbf{Exact $p$} \\
    \midrule
    \emph{Unadjusted} (high, 15 evaluated models)  & $-17.1$ & $+1.1$ & $-18.2$ & .052 \\
    Within gold-trajectory length                  & $-17.3$ & $+1.5$ & $-18.8$ &  \\
    Within required-check count                    & $-17.2$ & $+1.2$ & $-18.4$ &  \\
    Within round type                              & $-18.2$ & $+1.2$ & $-19.4$ &  \\
    \midrule
    Medium effort (10 20-step models)              & $-17.4$ & $+0.5$ & $-17.9$ & .052 \\
    Low effort (10 20-step models)                 & $-16.6$ & $+0.0$ & $-16.6$ & .071 \\
    \midrule
    Outcome: mean score                            & $-8.3$ & $-0.4$ & $-7.9$ & .252 \\
    \bottomrule
  \end{tabular}
  \end{adjustbox}
\end{table}

\subsection{Fine-Grained Diagnostic Failure Taxonomy}
\label{sec:app-taxonomy}

\begin{figure}[H]
  \centering
  \includegraphics[width=0.85\linewidth]{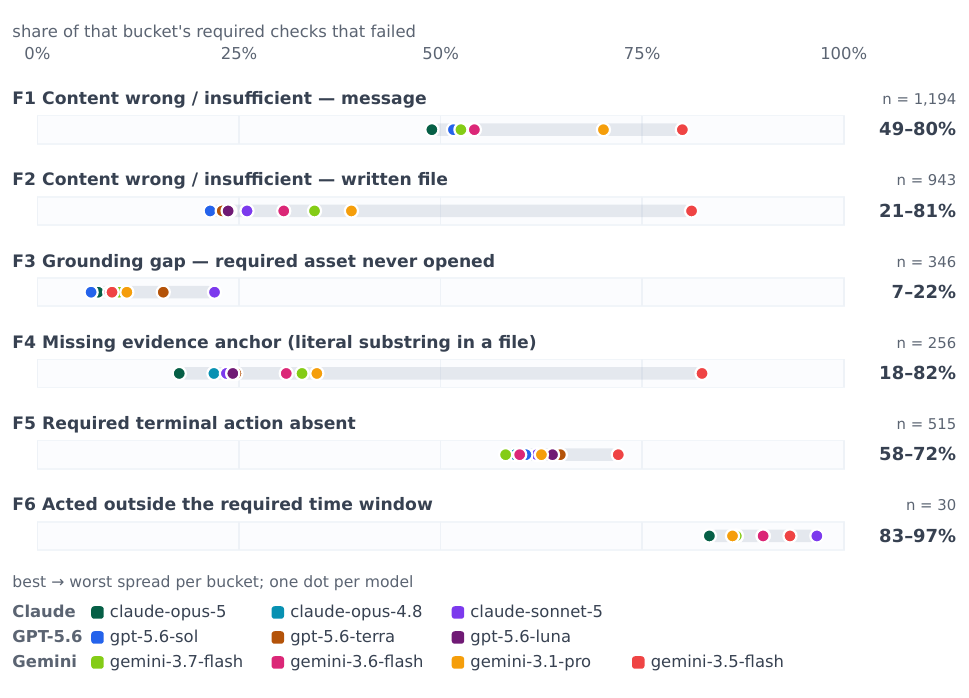}
  \caption{\textbf{Diagnostic check failure rates across failure modes (Ten 20-step API models, High Effort).} Compares failure proportions across 3,284 scored checks spanning Content (F1, F2, F4), Visual Grounding (F3), and Structural Protocol (F5, F6) failure modes.}
  \label{fig:taxonomy}
  \end{figure}

Figure~\ref{fig:taxonomy} presents check failure rates across the six diagnostic failure modes:
\begin{itemize}[leftmargin=*,itemsep=2pt]
  \item \textbf{Content Accuracy Modes (F1, F2, F4):} Separate frontier models from weaker baselines by 31 to 65 percentage points. Within the leading group, the performance spread ranges from 5.0\,pp on factual messaging (F1) to 15.2\,pp on evidence anchoring (F4).
  \item \textbf{Structural Protocol Modes (F5, F6):} Show consistently high failure rates across all evaluated API models, ranging from $58\%$--$72\%$ on terminal protocol actions (F5) and $83\%$--$97\%$ on temporal deadlines (F6).
  \item \textbf{Visual Grounding Failures (F3):} Relate directly to exploration horizon and step budget limits, where models fail to inspect required visual evidence before taking action.
\end{itemize}

\section{Compute Efficiency, Token Economics, and Planning Horizons}
\label{sec:app-efficiency}

\subsection{Economic Cost and API Breakdown}
\label{sec:app-cost}

Evaluating the ten 20-step API models over the 600-round benchmark at High effort required \$11,737 across 87,493 model API calls (or \$14,740 across 103,174 calls including GPT-6 Astra under a 40-step budget). For the four open-weight models evaluated on local infrastructure, the three with published list pricing would have incurred an estimated \$2,325 across 27,331 calls (DeepSeek V4 Flash lacks published list pricing and is excluded from dollar totals). As detailed in Table~\ref{tab:cost}, the compute cost per solved task spans a 55-fold range among 20-step models, from \$0.72 on GPT-5.6 Luna to \$39.54 on Gemini 3.5 Flash (\$17.46 for GPT-6 Astra).

\begin{table}[H]
  \centering
  \small
  \caption{\textbf{Compute cost, token expenditures, and execution dynamics across reasoning effort levels.} Reports mean dollar spend per round and per solved task (\emph{Cost (USD)}), input prompt tokens (\emph{In}), visible output tokens (\emph{Out}), internal reasoning tokens (\emph{Reas.}), and total API calls per round. \emph{Reas.\,\%}: fraction of billed output tokens allocated to reasoning. \emph{Cap-Hit}: proportion of rounds terminated by the action budget cap ($b=20$ steps for baselines; 1.0\% on GPT-6 Astra). \emph{Run Total}: cumulative dollar spend and total token volume (in millions) across all 600 rounds. DeepSeek V4 Flash lacks published pricing; open-weight dollar totals cover the remaining three models. Total rows report the ten 20-step API models alongside the all-inclusive High-effort total with GPT-6 Astra. \emph{Total} rows are computed from unrounded per-model figures.}
  \label{tab:cost}
  \begin{adjustbox}{max width=\linewidth}
  \begin{tabular}{l c | c c | c c c c | c c | c c}
    \toprule
    \multirow{2}{*}{\textbf{Model}} & \textbf{Think} & \multicolumn{2}{c|}{\textbf{Cost (USD)}}
      & \multicolumn{4}{c|}{\textbf{Per Round Tokens (k)}} & \multicolumn{2}{c|}{\textbf{Breakdown (\%)}}
      & \multicolumn{2}{c}{\textbf{Run Total}} \\
    \cmidrule(lr){3-4} \cmidrule(lr){5-8} \cmidrule(lr){9-10} \cmidrule(lr){11-12}
    & \textbf{Level} & \textbf{Round} & \multicolumn{1}{c|}{\textbf{Solved}}
          & \textbf{Calls} & \textbf{In} & \textbf{Out} & \multicolumn{1}{c|}{\textbf{Reas.}}
          & \textbf{Reas.} & \multicolumn{1}{c|}{\textbf{Cap-Hit}}
          & \textbf{\$ Total} & \textbf{Tok (M)} \\
    \midrule
    \multirow{3}{*}{Claude Opus 5}      & High & 4.38 & 17.42 & 11.9 & 845 & 5.7 & 0.6 & 8.8 & 0.5 & 2{,}631 & 511 \\
                                     & Med  & 3.74 & 14.77 & 10.4 & 728 & 3.8 & 0.2 & 4.6 & 0.0 & 2{,}245 & 439 \\
                                     & Low  & 3.08 & 13.11 & 8.8 & 604 & 2.3 & 0.0 & 1.9 & 0.0 & 1{,}848 & 364 \\
    \addlinespace[2.5pt]
    \multirow{3}{*}{Claude Opus 4.8}    & High & 4.89 & 21.41 & 12.9 & 944 & 4.3 & 2.3 & 34.5 & 6.2 & 2{,}933 & 571 \\
                                     & Med  & 4.41 & 18.50 & 11.8 & 855 & 4.0 & 1.5 & 27.3 & 2.8 & 2{,}646 & 516 \\
                                     & Low  & 3.83 & 17.94 & 10.5 & 745 & 3.3 & 0.7 & 17.6 & 0.8 & 2{,}296 & 449 \\
    \addlinespace[2.5pt]
    \multirow{3}{*}{Claude Sonnet 5}    & High & 1.72 & 8.25 & 11.4 & 820 & 5.0 & 2.8 & 36.3 & 3.3 & 1{,}031 & 497 \\
                                     & Med  & 1.39 & 8.20 & 9.6 & 670 & 4.2 & 1.1 & 20.7 & 1.0 & 836 & 405 \\
                                     & Low  & 1.18 & 8.54 & 8.2 & 572 & 3.3 & 0.4 & 10.9 & 0.3 & 708 & 345 \\
    \midrule
    GPT-6 Astra$^\star$                 & High & 5.01 & 17.46 & 26.1 & 1098 & 3.6 & 3.2 & 47.3 & 1.0 & 3{,}003 & 663 \\
    \addlinespace[2.5pt]
    \multirow{3}{*}{GPT-5.6 Sol}        & High & 3.61 & 14.73 & 14.4 & 754 & 1.7 & 1.4 & 44.2 & 1.2 & 2{,}165 & 454 \\
                                     & Med  & 3.23 & 14.05 & 13.2 & 675 & 1.5 & 0.8 & 33.5 & 2.0 & 1{,}939 & 406 \\
                                     & Low  & 2.64 & 12.65 & 10.8 & 548 & 1.2 & 0.3 & 19.9 & 0.2 & 1{,}582 & 330 \\
    \addlinespace[2.5pt]
    \multirow{3}{*}{GPT-5.6 Terra}      & High & 1.22 & 6.09 & 12.3 & 633 & 1.3 & 1.8 & 57.5 & 1.3 & 731 & 382 \\
                                     & Med  & 1.06 & 5.30 & 10.9 & 552 & 1.1 & 0.6 & 36.8 & 1.3 & 636 & 332 \\
                                     & Low  & 0.94 & 5.31 & 9.6 & 488 & 0.9 & 0.2 & 20.8 & 0.3 & 563 & 293 \\
    \addlinespace[2.5pt]
    \multirow{3}{*}{GPT-5.6 Luna}       & High & 0.14 & 0.72 & 14.2 & 746 & 1.7 & 2.0 & 54.4 & 9.2 & 86 & 450 \\
                                     & Med  & 0.11 & 0.64 & 11.7 & 593 & 1.3 & 0.6 & 32.5 & 2.7 & 68 & 357 \\
                                     & Low  & 0.10 & 0.50 & 9.9 & 508 & 1.1 & 0.3 & 21.3 & 0.5 & 59 & 305 \\
    \midrule
    \multirow{3}{*}{Gemini 3.7 Flash}   & High & 0.51 & 2.51 & 18.2 & 652 & 1.7 & 4.2 & 71.8 & 40.7 & 307 & 395 \\
                                     & Med  & 0.42 & 1.94 & 15.6 & 536 & 1.5 & 2.6 & 63.2 & 12.7 & 250 & 324 \\
                                     & Low  & 0.34 & 1.65 & 12.9 & 440 & 1.2 & 1.1 & 46.3 & 3.0 & 203 & 265 \\
    \addlinespace[2.5pt]
    \multirow{3}{*}{Gemini 3.6 Flash}   & High & 0.94 & 5.09 & 16.9 & 599 & 1.5 & 4.3 & 73.7 & 24.7 & 565 & 363 \\
                                     & Med  & 0.85 & 4.20 & 15.4 & 539 & 1.5 & 3.6 & 70.9 & 14.5 & 508 & 327 \\
                                     & Low  & 0.54 & 2.91 & 10.8 & 350 & 1.1 & 0.7 & 39.5 & 1.8 & 323 & 211 \\
    \addlinespace[2.5pt]
    \multirow{3}{*}{Gemini 3.1 Pro}     & High & 0.96 & 8.03 & 13.8 & 463 & 0.8 & 2.4 & 75.8 & 25.3 & 578 & 280 \\
                                     & Med  & 0.82 & 8.30 & 11.8 & 396 & 0.6 & 1.4 & 68.3 & 10.2 & 490 & 239 \\
                                     & Low  & 0.73 & 8.71 & 10.7 & 354 & 0.6 & 1.0 & 63.9 & 4.7 & 436 & 213 \\
    \addlinespace[2.5pt]
    \multirow{3}{*}{Gemini 3.5 Flash}   & High & 1.19 & 39.54 & 19.8 & 732 & 0.9 & 8.8 & 90.5 & 89.2 & 712 & 445 \\
                                     & Med  & 1.17 & 37.01 & 19.6 & 723 & 1.0 & 8.8 & 90.2 & 85.0 & 703 & 440 \\
                                     & Low  & 0.91 & 6.75 & 16.7 & 577 & 1.0 & 3.9 & 79.0 & 38.7 & 547 & 349 \\
    \midrule
    Kimi K3            & High & 3.62 & 16.58 & 16.1 & 1169 & 7.7 & -- & -- & 28.3 & 2{,}173 & 706 \\
    DeepSeek V4 Pro    & High & 0.05 & 0.27 & 14.1 & 112 & 5.2 & -- & -- & 13.5 & 32 & 70 \\
    GLM-5.2 FP8        & High & 0.20 & 1.37 & 15.4 & 132 & 3.5 & -- & -- & 34.5 & 120 & 82 \\
    DeepSeek V4 Flash  & High & N/A & N/A & 18.1 & 156 & 3.3 & -- & -- & 59.7 & N/A & 95 \\
    \midrule
    Total (10 API, 20-step)     & High & & & & & & & & & 11{,}737 & 4{,}347 \\
    Total (11 API, all High)    & High & & & & & & & & & 14{,}740 & 5{,}010 \\
    Total (API, billed)         & Med  & & & & & & & & & 10{,}322 & 3{,}785 \\
    Total (API, billed)         & Low  & & & & & & & & & 8{,}564 & 3{,}126 \\
    Total (open, list rate)     & High & & & & & & & & & 2{,}325 & 953 \\
    \bottomrule
  \end{tabular}
  \end{adjustbox}
\end{table} 

\paragraph{Subsequent Pricing Updates and GPT-6 Cost Estimation.} Following our benchmark execution, OpenAI announced a promotional rate reduction for GPT-5.6 Sol on August 21, 2026 (\$4.00/M input, \$20.00/M output, and \$0.40/M cached input, active through at least November 2026), which would lower its High-effort total run cost from \$2,165 to \$1,725 (\$11.73 per solved task). For GPT-6 Astra, applying OpenAI's published API rate card (\$10.00/M input, \$50.00/M output, and \$1.00/M cached input) yields an estimated benchmark run cost of \$3,003 (\$17.46 per solved task), with 62.2\% of input tokens benefiting from prompt caching across 663M total tokens.

\subsection{Reasoning Effort Inversion Dynamics}
\label{sec:effort}

Across the ten 20-step API models (open-weight models do not report reasoning tokens), reasoning tokens per round vary 16-fold (Table~\ref{tab:cost}) and correlate negatively with pass rate at $\rho = -0.697$, and at $\rho = -0.915$ against the reasoning share of output tokens. Leave-one-out sensitivity tests confirm this negative correlation holds across all ten models ($\rho \in [-0.783, -0.583]$). While executed steps vary by only a factor of $1.85$ (10.6 to 19.7 steps per round), reasoning tokens per action span a 9-fold range, from 50 to 448 tokens per step.

In contrast to cross-model comparisons, testing different effort levels within the same model shows a different picture. Raising reasoning effort from Low to High significantly improves five of the ten API models, leaves four unchanged, and harms exactly one: Gemini 3.5 Flash drops from $13.5\%$ at Low effort to $3.0\%$ at High effort, a drop of $10.5$\,pp ($95\%$ CI $[-13.5, -7.7]$). This drop occurs because repetitive reasoning loops eat up steps, causing the model to hit the 20-step budget cap on $91.9\%$ of its failed rounds.

These results indicate that spending many reasoning tokens often signals planning inefficiency in interactive computer-use tasks: models fail when internal deliberation loops crowd out necessary actions in the environment.

\subsection{Execution Step Budget Saturation}
\label{sec:app-step-budget}

Running out of steps ($\mathtt{cap\text{-}hit}$) accounts for $0.7\%$--$11.4\%$ of failed rounds across six of the seven leading 20-step models (and only $1.0\%$ on GPT-6 Astra at its 40-step budget), but rises to $51.0\%$ on Gemini 3.7 Flash, and to $30.3\%$, $28.8\%$, and $91.9\%$ on Gemini 3.6 Flash, Gemini 3.1 Pro, and Gemini 3.5 Flash respectively. Gemini 3.7 Flash shows an interesting pattern: while achieving competitive overall pass rates, it spends $71.8\%$ of its output tokens on reasoning and runs out of steps on half of its failed rounds. Running out of steps is therefore a major bottleneck for reasoning-heavy models.

\section{Extended Ablation Studies and Robustness Analyses}
\label{sec:app-corpora}

\subsection{Cross-Generator Invariance and Parallel Benchmark Design}
\label{sec:app-generator-ablation}

To verify that benchmark difficulty and model rankings do not depend on the specific generator model, we regenerated the entire benchmark from the same 20 scenario seeds using Claude Sonnet 5 as the generator (Bench B) and compared it with Gemini 3.6 Flash (Bench A). Table~\ref{tab:corpora-design} summarizes the structural parameters of both parallel benchmarks.

\begin{table}[H]
  \centering
  \small
  \caption{\textbf{Structural comparison of parallel generation benchmarks.} Contrasts Bench~A (Gemini 3.6 Flash) and Bench~B (Claude Sonnet 5) across identical 20 scenario seeds and 600 aligned conflict tasks.}
  \label{tab:corpora-design}
  \begin{tabular}{l | c | c}
    \toprule
    \multicolumn{1}{c}{} & \multicolumn{1}{c|}{\textbf{Bench A}} & \textbf{Bench B} \\
    \textbf{Generator model}                     & \multicolumn{1}{c|}{Gemini 3.6 Flash} & Claude Sonnet 5 \\
    \midrule
    Step budget                                  & 20 & 30 \\
    GT step length, mean / median / max          & 10.28 / 10 / 18 & 12.48 / 12 / 24 \\
    Budget ratio (budget $\div$ GT), mean        & $2.02\times$ & $2.51\times$ \\
    Rounds with budget ratio $< 1.5\times$       & 8.2\% (49) & 1.2\% (7) \\
    Rounds with unexposed reference strings      & 18 & 9 \\
    \midrule
    \multicolumn{3}{l}{\emph{Pairing between the two benchmarks}} \\
    GT-length correlation (Pearson $r$)          & \multicolumn{2}{c}{$0.186$} \\
    Bench B GT longer / equal / shorter          & \multicolumn{2}{c}{430 / 72 / 98} \\
    Aligned rounds per benchmark                 & \multicolumn{2}{c}{600} \\
    Ratio-matched subset ($|\Delta| \le 0.30$)   & \multicolumn{2}{c}{164} \\
    \bottomrule
  \end{tabular}
\end{table}

Evaluating seven models across 600 rounds per benchmark (Figure~\ref{fig:ablation}c, Table~\ref{tab:ablation-app}) confirms that relative model rankings remain highly stable. Rankings at high and low effort are consistent across Bench A and Bench B, with Spearman $\rho = 0.96$ at high effort, $1.00$ at low effort, and $0.97$ across all 19 comparisons, with no self-preference advantage for either generator model. Pass rate retention on Bench~B relative to Bench~A (\textbf{Ret\%} in Figure~\ref{fig:ablation}c) places both generator models side-by-side in the middle tier ($0.51$ for Gemini 3.6 Flash and $0.54$ for Claude Sonnet 5), flanked by non-generator models Gemini 3.7 Flash at $0.41$ and Claude Opus 4.8 at $0.57$.

\begin{table}[H]
  \centering
  \small
  \caption{\textbf{Comprehensive numerical results for harness parameter ablations.} \emph{(a) Step budget:} Pass rates and cap exhaustion when expanding horizon $b=20 \to 30 \to 40$ on Bench~A ($\Delta = b_{40} - b_{20}$). \emph{(b) Visual modality:} Performance of sighted vs.\ blind agents across full (600), asset-gated (346), and un-gated (254) task groups. \emph{(c) Generator difference:} Pass rates across seven models evaluated on Bench~A vs.\ Bench~B (600 rounds each; $\Delta = \text{B} - \text{A}$).}
  \label{tab:ablation-app}

  \begin{tabular*}{\linewidth}{@{\extracolsep{\fill}} l cccc cc @{}}
    \toprule
    \multicolumn{7}{@{}l}{\emph{(a) Step budget.} Same 600 rounds, high effort; only the cap moves.} \\
    \addlinespace[2pt]
    & \multicolumn{4}{c}{\textbf{Pass Rate (\%)}} & \multicolumn{2}{c}{\textbf{Ended at Cap (\%)}} \\
    \cmidrule(lr){2-5}\cmidrule(lr){6-7}
    \textbf{Model} & \textbf{20} & \textbf{30} & \textbf{40} & \textbf{$\Delta$} & \textbf{20} & \textbf{40} \\
    \midrule
    Claude Opus 5    & 25.2 & 25.7 & 27.0 & $+1.8$ & 0.5 & 0.0 \\
    Claude Sonnet 5  & 20.8 & 20.7 & 21.3 & $+0.5$ & 3.3 & 0.2 \\
    Gemini 3.7 Flash & 20.3 & 26.0 & 27.2 & $+6.8$ & 40.7 & 2.5 \\
    Gemini 3.6 Flash & 18.5 & 23.5 & 25.0 & $+6.5$ & 24.7 & 0.3 \\
    Gemini 3.5 Flash & 3.0 & 10.0 & 13.0 & $+10.0$ & 89.2 & 44.0 \\
    Gemini 3.8 Flash & 12.8 & 19.0 & 29.2 & $+16.3$ & 66.8 & 19.2 \\
    \bottomrule
  \end{tabular*}

  \vspace{6pt}

  \begin{tabular*}{\linewidth}{@{\extracolsep{\fill}} l cc cc cc @{}}
    \toprule
    \multicolumn{7}{@{}l}{\emph{(b) Visual modality.} Budget 20, high effort; assets withheld, task text unchanged.} \\
    \addlinespace[2pt]
    & \multicolumn{2}{c}{\textbf{All (600)}} & \multicolumn{2}{c}{\textbf{Asset-Gated (346)}} & \multicolumn{2}{c}{\textbf{No Asset Check (254)}} \\
    \cmidrule(lr){2-3}\cmidrule(lr){4-5}\cmidrule(lr){6-7}
    \textbf{Model} & \textbf{Sighted} & \textbf{Blind} & \textbf{Sighted} & \textbf{Blind} & \textbf{Sighted} & \textbf{Blind} \\
    \midrule
    Claude Opus 5    & 25.2 & 4.2 & 25.4 & 5.2 & 24.8 & 2.8 \\
    Claude Sonnet 5  & 20.8 & 2.8 & 18.5 & 2.6 & 24.0 & 3.1 \\
    Gemini 3.6 Flash & 18.5 & 2.0 & 17.9 & 1.7 & 19.3 & 2.4 \\
    \bottomrule
  \end{tabular*}

  \vspace{6pt}

  \begin{tabular*}{\linewidth}{@{\extracolsep{\fill}} l ccccccc @{}}
    \toprule
    \multicolumn{8}{@{}l}{\emph{(c) Generator model.} Pass rate (\%) at high effort; 600 rounds per benchmark.} \\
    \addlinespace[2pt]
    \textbf{Benchmark} & \textbf{Opus 5} & \textbf{Opus 4.8} & \textbf{Sonnet 5} & \textbf{3.7 Flash} & \textbf{3.6 Flash} & \textbf{3.1 Pro} & \textbf{3.5 Flash} \\
    \midrule
    A (Gemini 3.6 Flash) & 25.2 & 22.8 & 20.8 & 20.3 & 18.5 & 12.0 & 3.0 \\
    B (Claude Sonnet 5) & 19.2 & 13.0 & 11.2 & 8.3 & 9.5 & 4.2 & 2.3 \\
    $\Delta$ & $-6.0$ & $-9.8$ & $-9.6$ & $-12.0$ & $-9.0$ & $-7.8$ & $-0.7$ \\
    \bottomrule
  \end{tabular*}

  \vspace{3pt}
  \begin{minipage}{\linewidth}\footnotesize\raggedright
    Paired McNemar, 19 model $\times$ effort cells (A easier / n.s.\ / B): 16/3/0 full, 14/5/0 ratio-matched~($n{=}164$).
  \end{minipage}
\end{table}

\subsection{Action Budget Limits and Planning Horizon Dynamics}
\label{sec:app-budget-ablation}

Increasing the action budget from $b=20$ to $30$ and $40$ steps across identical scenarios (Figure~\ref{fig:ablation}a, Table~\ref{tab:ablation-app}) demonstrates that fixed step constraints do not penalize models uniformly. Instead, they act as an operational filter distinguishing three distinct agent planning behaviors:
\begin{enumerate}[leftmargin=1.8em,itemsep=2pt]
  \item \textbf{Concise Planners:} Models such as Claude Opus 5 and Claude Sonnet 5 rarely exhaust their budget even at $b=20$ (cap exhaustion rates of $0.5\%$ and $3.3\%$, respectively). Their trajectory lengths closely mirror human reference solutions (mean $10.28$ steps, median $10$, maximum $18$; Table~\ref{tab:corpora-design}). Expanding the budget to $b=40$ yields marginal pass rate gains ($+1.8$\,pp for Claude Opus 5 and $+0.5$\,pp for Claude Sonnet 5), establishing that a 20-step budget is generous for concise execution.
  \item \textbf{Looping and Deliberation Traps:} Weaker reasoning architectures, most notably Gemini 3.5 Flash, hit the 20-step cap on $89.2\%$ of rounds. While expanding the budget to $b=40$ raises pass rate by $+10.0$\,pp (from $3.0\%$ to $13.0\%$), $44.1\%$ of rounds still terminate at the cap. As detailed in Appendix~\S\ref{sec:app-efficiency}, these extra steps are consumed by redundant file reading, hallucinated command repetitions, and excessive reasoning tokens ($23{,}206$ tokens/task at $b=40$), inflating the cost per solved task to $\$8.69$.
  \item \textbf{Horizon-Constrained Frontier Competence:} In contrast to repetitive looping, Gemini 3.8 Flash demonstrates a capability profile characterized by deep investigative exploration and iterative hypothesis testing, discussed next.
\end{enumerate}

\subsection{Frontier Horizon Sensitivity: The Gemini 3.8 Flash Case Study}
\label{sec:app-3-8-flash}

To understand how frontier reasoning models interact with execution horizons, we conducted an extended evaluation of Gemini 3.8 Flash across four action budgets ($b=20, 30, 40, 60$) over all $600$ benchmark rounds under High reasoning effort on Vertex AI. All runs were evaluated by the corpus-standard Gemini 3.6 Flash judge. Table~\ref{tab:38-flash-ablation} summarizes empirical trajectory metrics, cap exhaustion, and token economics across the four budget arms.

\begin{table}[H]
  \centering
  \small
  \caption{\textbf{Action horizon scaling for Gemini 3.8 Flash ($n=600$ rounds, High effort, Gemini 3.6 Flash judge).} Metrics include task pass rate with $95\%$ Wilson score confidence intervals, percentage of runs terminating at the step cap, average tool calls, average reasoning tokens, cost per task at introductory pricing, and effective cost per solved task.}
  \label{tab:38-flash-ablation}
  \begin{tabular*}{\linewidth}{@{\extracolsep{\fill}} l cccccc @{}}
    \toprule
    \textbf{Budget ($b$)} & \textbf{Pass Rate (\%)} & \textbf{95\% CI} & \textbf{Cap-Hit (\%)} & \textbf{Calls/Task} & \textbf{Reason Tok} & \textbf{\$/Solved} \\
    \midrule
    $b=20$ & 12.8 (77/600)  & $[10.4, 15.7]$ & 66.8 & 19.4 & 6{,}699  & \$4.44 \\
    $b=30$ & 19.0 (114/600) & $[16.1, 22.3]$ & 46.8 & 27.3 & 8{,}976  & \$4.40 \\
    $b=40$ & 29.2 (175/600) & $[25.7, 32.9]$ & 19.2 & 32.2 & 11{,}233 & \textbf{\$3.50} \\
    $b=60$ & \textbf{30.7} (184/600) & $[27.1, 34.5]$ & \textbf{3.3} & 35.0 & 11{,}446 & \$3.71 \\
    \midrule
    \multicolumn{7}{@{}l}{\emph{Comparator: Gemini 3.7 Flash (same judge, high effort)}} \\
    $b=20$ & 20.3 (122/600) & $[17.3, 23.7]$ & 40.7 & 17.8 & 4{,}026  & \$2.51 \\
    $b=30$ & 26.0 (156/600) & $[22.7, 29.7]$ & 13.7 & 23.4 & 5{,}341  & \$2.60 \\
    $b=40$ & 27.2 (163/600) & $[23.8, 30.9]$ & 2.5  & 26.0 & 5{,}879  & \$2.80 \\
    \bottomrule
  \end{tabular*}
\end{table}

\noindent\textbf{Threshold Effects Across Step Budgets.}
The marginal return of expanding the budget for Gemini 3.8 Flash shows a clear threshold effect: increasing from $b=20$ to $b=30$ yields $+6.2$\,pp ($12.8\% \to 19.0\%$), whereas expanding from $b=30$ to $b=40$ jumps by $+10.2$\,pp ($19.0\% \to 29.2\%$). In complex multi-agent conflict scenarios, agents must first inspect contradictory logs, diagnose out-of-sync local states, and locate visual evidence before enacting coordinating fixes. High-capacity agents that conduct extensive diagnostic exploration require an operational threshold of steps before their solutions can be committed and verified. Once this threshold is surpassed, pass rates accelerate rapidly. Expanding further from $b=40$ to $b=60$ adds $+1.5$\,pp to reach $30.7\%$, while cap exhaustion drops from $19.2\%$ to $3.3\%$. The saturation of gains alongside minimal cap exhaustion indicates that $30.7\%$ represents a genuine performance ceiling for the model architecture rather than an artificial truncation floor.

\noindent\textbf{Paired Head-to-Head Comparisons vs.\ Gemini 3.7 Flash.}
Evaluating both models under identical prompts, seeds, and the corpus-standard Gemini 3.6 Flash judge allows rigorous paired McNemar tests ($n=600$):
\begin{itemize}[leftmargin=1.5em,itemsep=1.5pt]
  \item At $b=20$, Gemini 3.7 Flash leads by $+7.5$\,pp ($20.3\%$ vs.\ $12.8\%$, $p < 10^{-4}$), driven by 3.7's lower step usage ($40.7\%$ vs.\ $66.8\%$ cap-hit).
  \item At $b=30$, Gemini 3.7 Flash maintains a $+7.0$\,pp lead ($26.0\%$ vs.\ $19.0\%$, discordant pairs 55/13, exact $p = 2.75 \times 10^{-7}$).
  \item At $b=40$, Gemini 3.8 Flash overtakes Gemini 3.7 Flash by $+2.0$\,pp ($29.2\%$ vs.\ $27.2\%$, discordant pairs 40/28, exact $p = 0.182$), becoming the highest-scoring model on the benchmark.
  \item When comparing both models at their respective saturation points ($b=60$ for Gemini 3.8 Flash at $3.3\%$ cap-hit vs.\ $b=40$ for Gemini 3.7 Flash at $2.5\%$ cap-hit), Gemini 3.8 Flash establishes a statistically significant $+3.5$\,pp lead ($30.7\%$ vs.\ $27.2\%$, discordant pairs 41/20, exact $p = 0.00985$).
\end{itemize}

\noindent\textbf{Productive Execution vs.\ Repetitive Looping.}
A critical distinction between Gemini 3.8 Flash and Gemini 3.5 Flash lies in their economic efficiency curves. For Gemini 3.5 Flash, budget expansion feeds repetitive thinking loops, driving reasoning tokens from $8{,}826$ to $23{,}206$ per task and resulting in a high cost of $\$8.69$ per solved task at $b=40$. In contrast, for Gemini 3.8 Flash, the additional budget is utilized for productive state verification and execution. Consequently, its effective cost per solved task drops from $\$4.44$ at $b=20$ to $\$3.50$ at $b=40$---a $21\%$ gain in economic efficiency. This confirms that Gemini 3.8 Flash's lower score at $b=20$ is an artifact of horizon truncation rather than a lack of reasoning capability.

\subsection{Visual Necessity and Multimodal Grounding Ablation}
\label{sec:app-modality-ablation}

Withholding visual images (Figure~\ref{fig:ablation}b, Table~\ref{tab:ablation-app}) causes pass rates to drop sharply across all models. Even on the 254 rounds without a mandatory image check, blind agents keep only $11.1\%$ to $13.1\%$ of their sighted pass rates ($24.8\% \to 2.8\%$ for Claude Opus 5, $24.0\% \to 3.1\%$ for Claude Sonnet 5, and $19.3\% \to 2.4\%$ for Gemini 3.6 Flash). With Bonferroni correction ($0.05/6$), all three drops are statistically significant (largest $p = 2.6 \times 10^{-12}$), with 60, 54, and 44 rounds solved only when images are available versus just 4, 1, and 1 when blind. This confirms that visual evidence is essential for resolving conflicts, rather than decorative background.

\subsection{Judge Selection and Cross-Judge Stability}
\label{sec:app-judge-ablation}

\textbf{Offline Replay Protocol.} To test whether benchmark results depend on the specific evaluator model, we replayed all $9{,}000$ recorded high-effort rounds from the 14 20-step baseline models offline against original scenario data and reconstructed checker inputs from stored action trajectories. Deterministic checks were re-evaluated as a fidelity verification: all $24{,}331$ deterministic checks reproduced their recorded verdicts ($100.0\%$ fidelity), confirming that differences between evaluation runs stem entirely from semantic judging. The replay generated $28{,}009$ judge prompts, which were re-evaluated by three substitute judges using 3-sample majority voting: Gemini 3.6 Flash (the default judge re-run to estimate sampling variance), Claude Sonnet 5, and Claude Opus 5.

\textbf{Common Evaluation Subset.} Model endpoints occasionally decline prompts: Claude Sonnet 5 declined $161$ prompts ($0.62\%$) and Claude Opus 5 declined $3{,}089$ prompts ($11.8\%$), while the default judge completed all evaluations. To maintain fair comparisons across judges without missing-data bias, all reported metrics evaluate the common subset of $22{,}938$ checks ($87.9\%$) scored by all three judges, using recorded verdicts for un-scored prompts uniformly across columns.

\textbf{Evaluator Consistency.} Each judge ensemble demonstrates high internal consistency across its three voting samples. Disagreement rates among the three samples were $2.78\%$ for Gemini 3.6 Flash, $1.31\%$ for Claude Sonnet 5, and $0.34\%$ for Claude Opus 5, confirming stable majority verdicts across all evaluators.

\textbf{Results and Rank Invariance.} Table~\ref{tab:judge-app} summarizes evaluation results across substitute judges. Rankings on mean score show high stability across all evaluators, with rank correlations of $\rho = 0.987$ for re-run Gemini 3.6 Flash, $\rho = 0.991$ for Claude Sonnet 5, and $\rho = 0.987$ for Claude Opus 5. No model shifts by more than one rank position under either Claude judge.

Different judges affect overall score levels rather than ordinal rankings. On average, Claude Opus 5 gives slightly higher scores ($+0.021$ on mean score, two-sided sign test $p = 0.00012$) and Claude Sonnet 5 shows a modest increase ($+0.005$, $p = 0.0018$), whereas re-running the default judge produces no systematic shift ($p = 0.79$). The only rank change among top models occurs between GPT-5.6 Sol and Claude Opus 5, whose small pass rate difference of $0.67$\,pp sits comfortably within bootstrap standard errors (Table~\ref{tab:main}). Overall, the top four and trailing four models remain identical across all judges, confirming that relative benchmark standings are robust to evaluator choice.

\textbf{Analysis of the Gemini 3.7 Flash Rollouts.} Gemini 3.7 Flash includes two evaluation rollouts: the first evaluated by Gemini 3.6 Flash, and the second evaluated by Gemini 3.7 Flash itself ($+1.67$\,pp difference). Re-grading both rollouts with a common judge separates rollout variance from judge scoring differences: judge differences account for $+1.50$\,pp of the gap, while rollout variance accounts for $+0.83$\,pp (paired McNemar $p = 0.63$). We report the published run with an explicit self-scoring note for transparency.

\begin{table}[H]
  \centering
  \small
  \caption{\textbf{Cross-judge stability ablation.} Every semantic check of the high-effort panel is re-evaluated offline by three substitute three-sample ensembles with agent trajectories held fixed: the replay reproduces $24{,}331/24{,}331$ deterministic verdicts, so the columns differ only in semantic opinion. The Gemini 3.6 Flash column re-runs the default judge, so it measures sampling noise rather than a judge change. Statistics use the $22{,}938$ of $26{,}100$ panel checks ($87.9\%$) all three substitutes scored; the recorded verdict fills the rest in every column and cancels in cross-judge comparisons. Mean score leads because pass rate requires passing all checks, so one flipped check can move a model several places. The substitute judges reorder the models no more than the default judge reorders itself; they primarily shift absolute score levels. Claude Opus 5 raises all fourteen 20-step baseline models, unevenly enough to hand first place to GPT-5.6 Sol---the only change of leader in the table. The self-scored row that the two \emph{excluding} lines drop is Gemini 3.7 Flash, which issued its own recorded verdicts.}
  \label{tab:judge-app}

  \begin{tabular*}{\linewidth}{@{\extracolsep{\fill}} l ccc @{}}
    \toprule
    \textbf{Substitute judge} & Gemini 3.6 Flash & Claude Sonnet 5 & Claude Opus 5 \\
    \midrule
    \multicolumn{4}{@{}l}{\emph{Mean score (primary).}} \\
    Rank correlation $\rho$                      & 0.987 & 0.991 & 0.987 \\
    \quad excluding the self-scored row          & 0.995 & 0.995 & 0.989 \\
    Maximum rank displacement                    & 2     & 1     & 1 \\
    Models whose rank moves                      & 3/14  & 4/14  & 6/14 \\
    Mean shift                                   & $-0.001$ & $+0.005$ & $+0.021$ \\
    Largest absolute shift                       & 0.007 & 0.018 & 0.045 \\
    Models scored higher                         & 6/14  & 13/14 & 14/14 \\
    \quad sign test $p$ (two-sided)              & 0.79  & 0.0018 & 0.00012 \\
    \midrule
    \multicolumn{4}{@{}l}{\emph{Pass rate.}} \\
    Rank correlation $\rho$                      & 0.975 & 0.991 & 0.949 \\
    \quad excluding the self-scored row          & 0.982 & 0.990 & 0.974 \\
    Maximum rank displacement                    & 3     & 1     & 3 \\
    Models whose rank moves                      & 3/14  & 6/14  & 8/14 \\
    Largest move (points)                        & $-0.7$ & $+1.3$ & $+3.7$ \\
    \midrule
    \multicolumn{4}{@{}l}{\emph{Per-check agreement with the recorded verdicts.}} \\
    Agreement                                    & 98.93\% & 96.31\% & 95.46\% \\
    Cohen's $\kappa$                             & 0.977 & 0.922 & 0.902 \\
    Fail $\to$ pass flips                        & 99    & 508   & 958 \\
    Pass $\to$ fail flips                        & 147   & 339   & 83 \\
    Declined prompts (of 26{,}100)               & 0     & 161   & 3{,}089 \\
    Non-unanimous ensembles                      & 2.78\% & 1.31\% & 0.34\% \\
    \bottomrule
  \end{tabular*}
\end{table}


%
%

\section{Exhibits from the Cross-Model Behavioral Evaluation}
\label{app:report}

This appendix provides the complete set of empirical exhibits, diagnostic tables, vector charts, and behavioral panels compiled during the benchmark evaluation across all 15 evaluated models.

\noindent\smallskip\textbf{Experimental Configuration.} Unless otherwise indicated, all reported evaluations correspond to the default high reasoning effort configuration (\texttt{reasoning\_effort=high}). The reasoning effort sweep across Low, Medium, and High levels for the ten API models is detailed in \S\ref{rep:s9}. Open-weight models are evaluated at their native high-effort configuration.

\noindent\smallskip\textbf{Evaluation Runs.} All reported models are evaluated over the complete 600-round benchmark corpus. For Gemini 3.7 Flash, results correspond to run \texttt{r02} ($20.3\%$ vs.\ $18.7\%$ on \texttt{r01}, paired McNemar $p = 0.27$). All comparative claims in the main text are supported by paired statistical significance tests or bootstrap confidence intervals.

\begin{figure}[H]
\centering
\includegraphics[width=\linewidth]{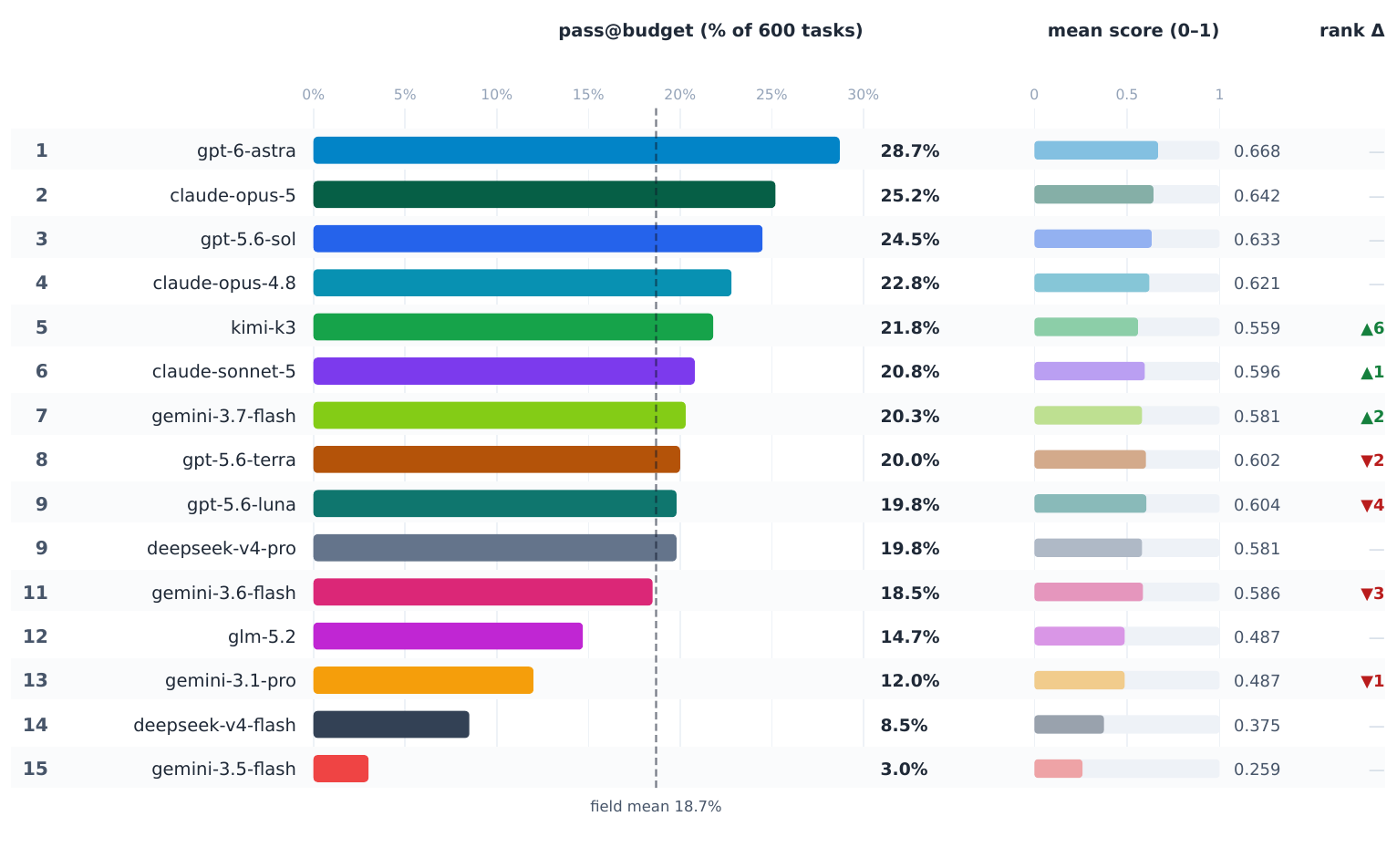}
\caption{\textbf{Overall benchmark standings across the 15 evaluated models (High Effort).} Displays task pass rate ($\mathtt{pass@budget}$, wide bar) and mean check score (narrow bar, 0--1 scale) across all 600 benchmark tasks. Models are ranked by pass rate; $\text{Rank }\Delta$ indicates the difference between mean score rank and pass rank.}
\label{fig:rep-standings}
\end{figure}

\begin{figure}[H]
\centering
\includegraphics[width=\linewidth]{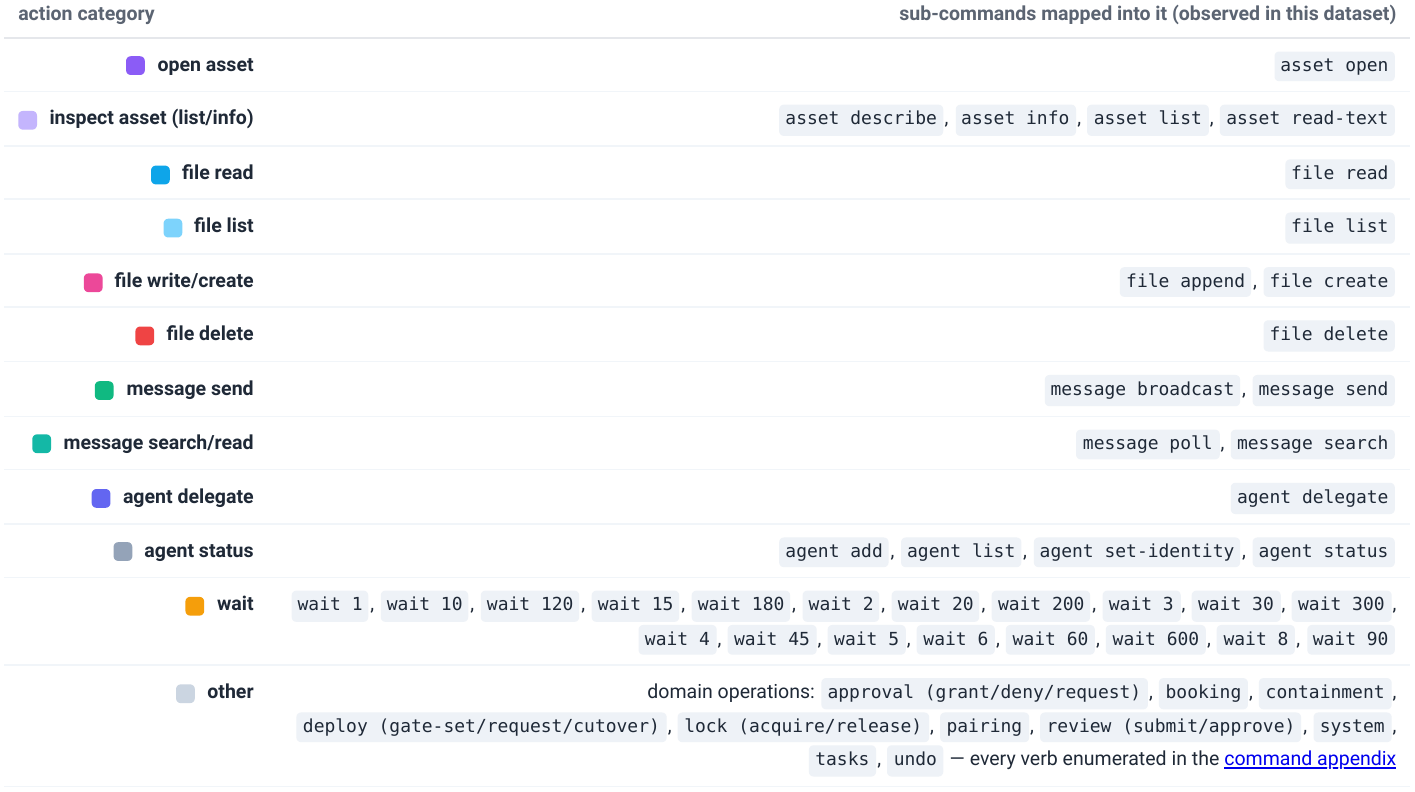}
\caption{\textbf{Action category glossary and invocation distribution.} Partitions all 88 sub-commands in the action DSL into 12 mutually exclusive functional categories (call rates normalized per task across all 600 rounds). The detailed breakdown of domain-specific operations is provided in Figure~\ref{fig:rep-other-breakdown}, and the full command inventory spans Figures~\ref{fig:rep-cmd-inventory-1}--\ref{fig:rep-cmd-inventory-3}.}
\label{fig:rep-glossary}
\end{figure}

\begin{figure}[H]
\centering
\includegraphics[width=\linewidth]{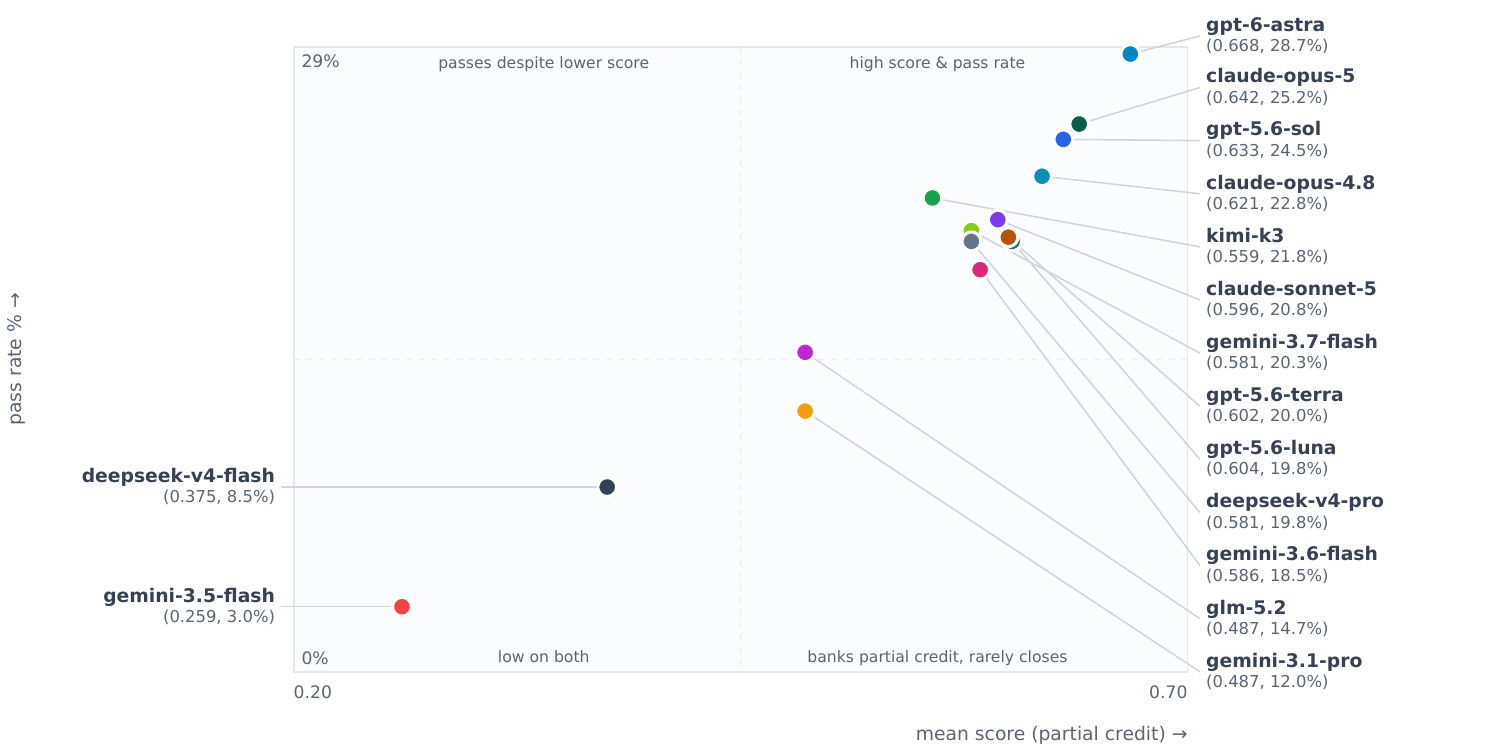}
\caption{\textbf{Pass rate versus mean check score across models (High Effort).} Shows the relationship between all-or-nothing task completion (pass rate) and mean check score across all 15 models.}
\label{fig:rep-pass-vs-score}
\end{figure}

\begin{figure}[H]
\centering
\includegraphics[width=\linewidth]{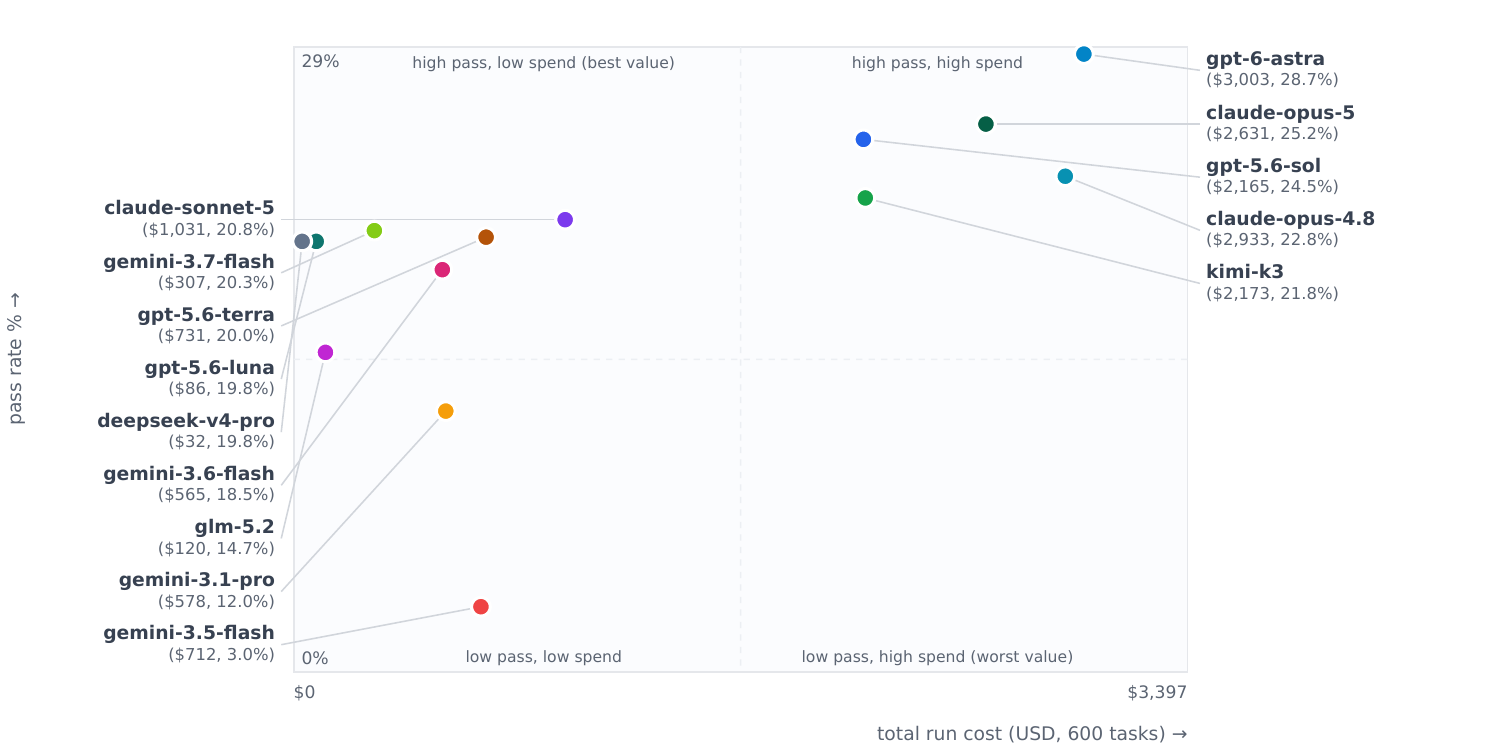}
\caption{\textbf{Pass rate versus cumulative evaluation cost (High Effort).} Compares benchmark pass rate against total evaluation cost across models.}
\label{fig:rep-pass-vs-cost}
\end{figure}

\begin{figure}[H]
\centering
\includegraphics[width=\linewidth]{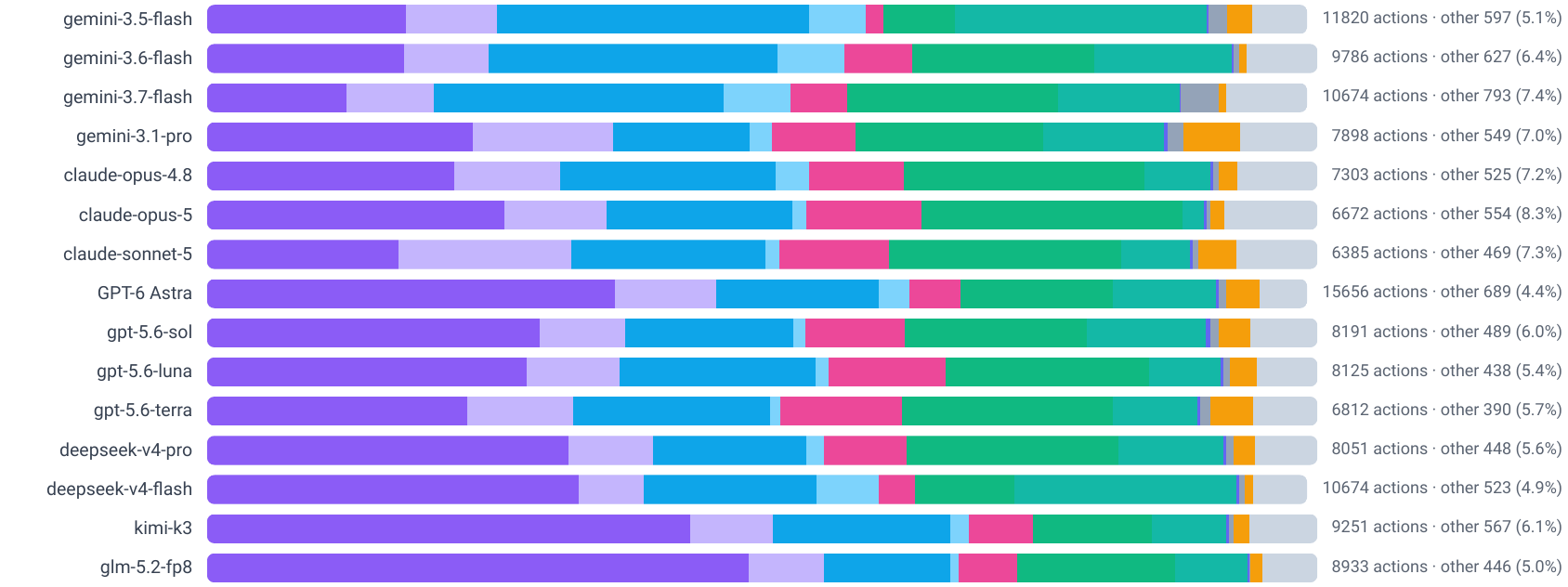}
\caption{\textbf{Empirical action-category distributions across models (High Effort).} Relative proportion of tool calls across the 12 action categories. Each bar sums to 100\% of that model's executed actions. The domain-specific operations grouped under \texttt{other} are detailed in Figure~\ref{fig:rep-other-breakdown}.}
\label{fig:rep-action-mix}
\end{figure}

\begin{figure}[H]
\centering
\includegraphics[width=\linewidth]{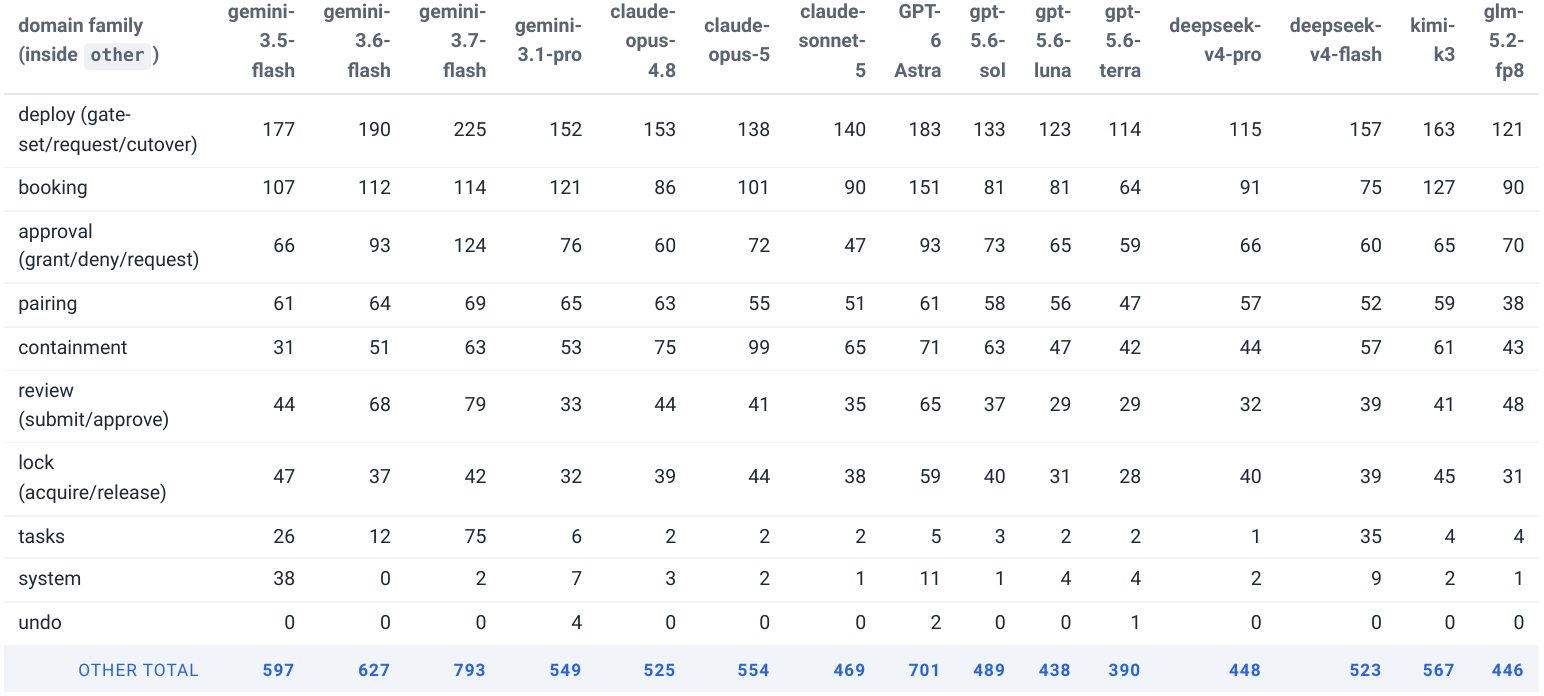}
\caption{\textbf{Itemized distribution of domain-specific operations (\texttt{other} category).} Breaks down specialized environment operations (including containment locks, deployment approvals, resource pairing, and process scheduling) across models.}
\label{fig:rep-other-breakdown}
\end{figure}

\begin{figure}[H]
\centering
\includegraphics[width=0.91\linewidth]{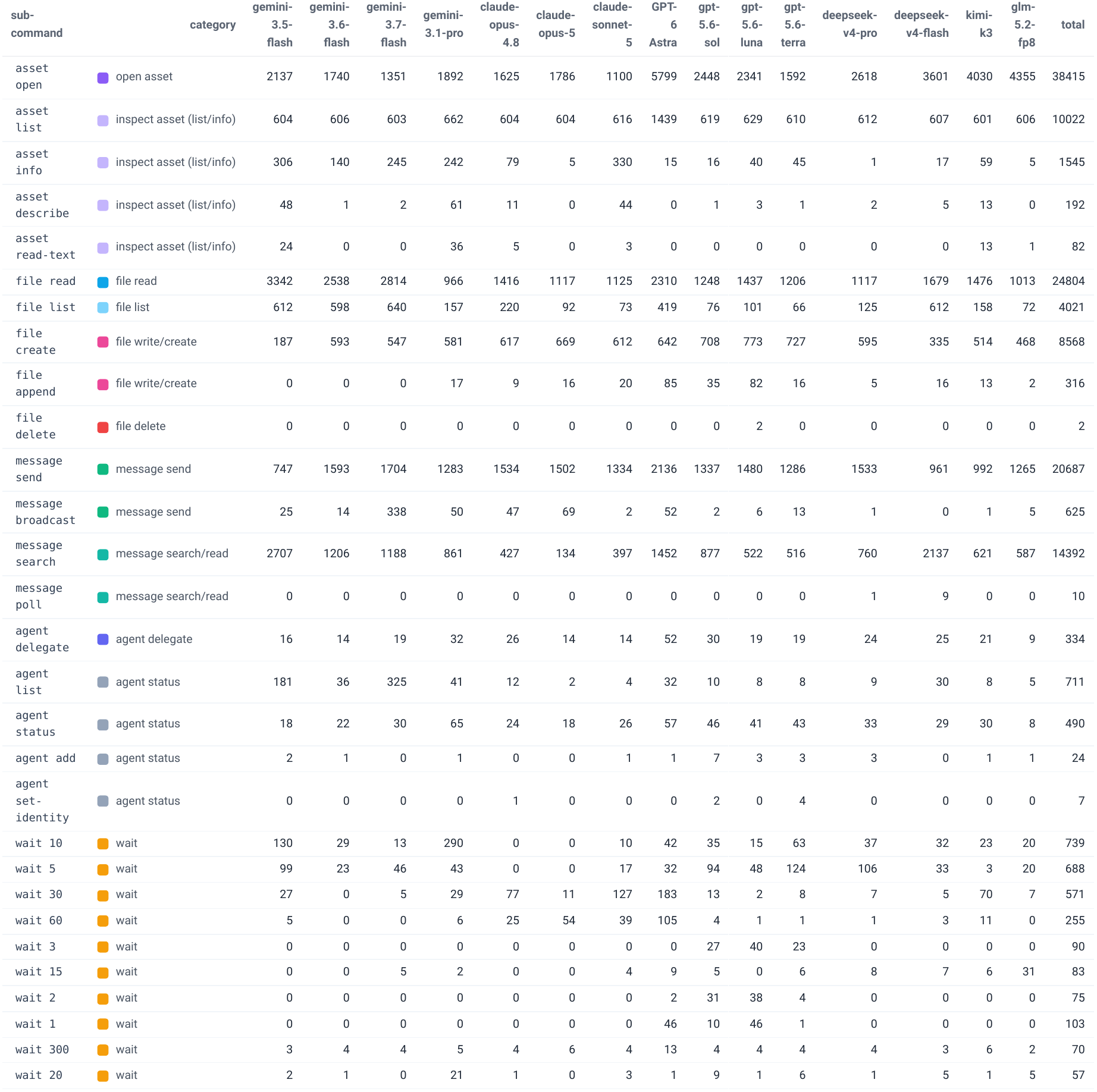}
\caption{\textbf{Complete command inventory across all evaluated models (Part 1: Rows 1--30 of 89).} Displays individual sub-command invocation counts across all 600 tasks per model, mapped to functional categories. Continued in Figure~\ref{fig:rep-cmd-inventory-2}.}
\label{fig:rep-cmd-inventory-1}
\end{figure}

\begin{figure}[H]
\centering
\includegraphics[width=\linewidth]{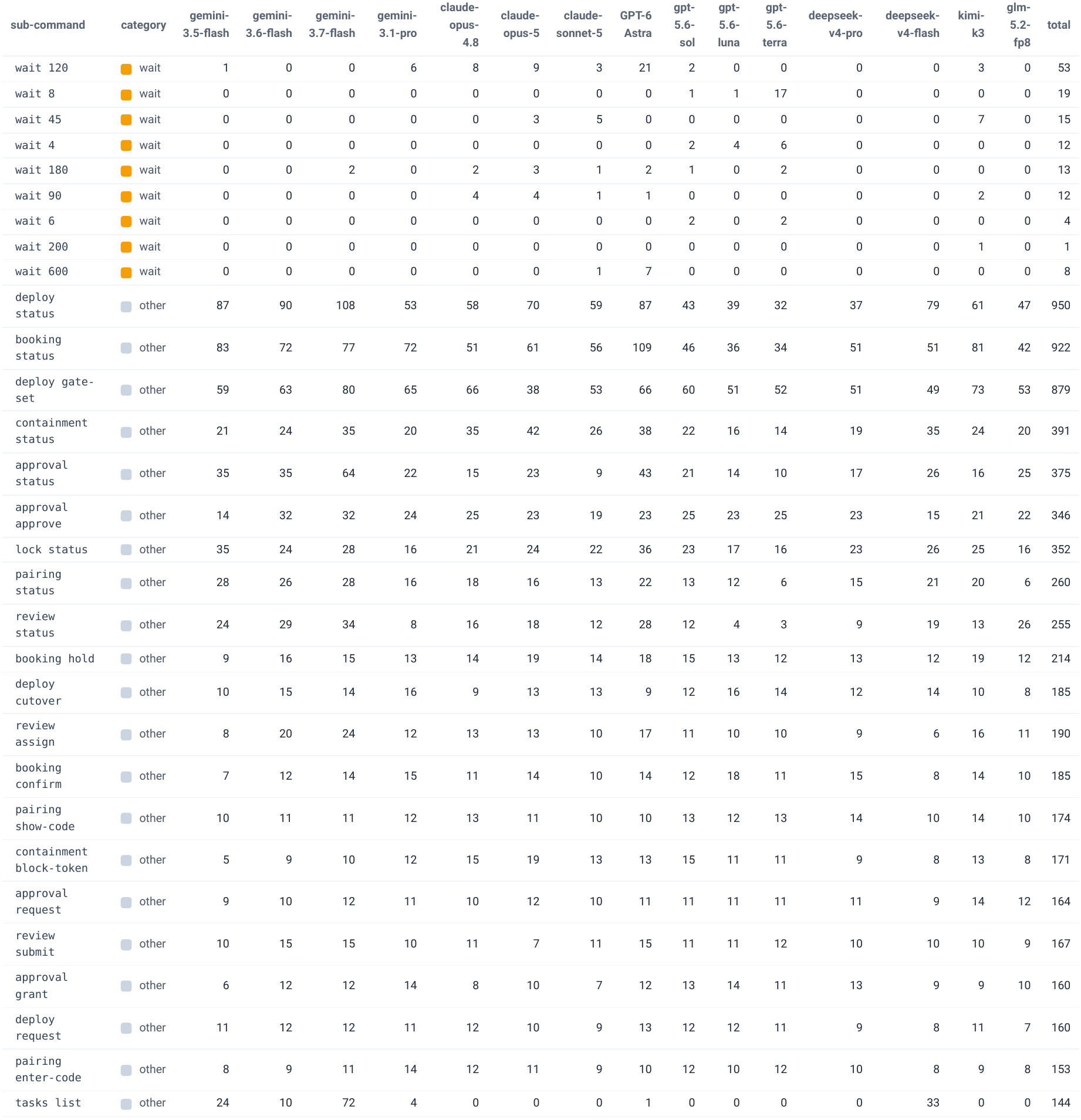}
\caption{\textbf{Complete command inventory across all evaluated models (Part 2: Rows 31--59 of 89).} Continued from Figure~\ref{fig:rep-cmd-inventory-1}; continued in Figure~\ref{fig:rep-cmd-inventory-3}.}
\label{fig:rep-cmd-inventory-2}
\end{figure}

\begin{figure}[H]
\centering
\includegraphics[width=\linewidth]{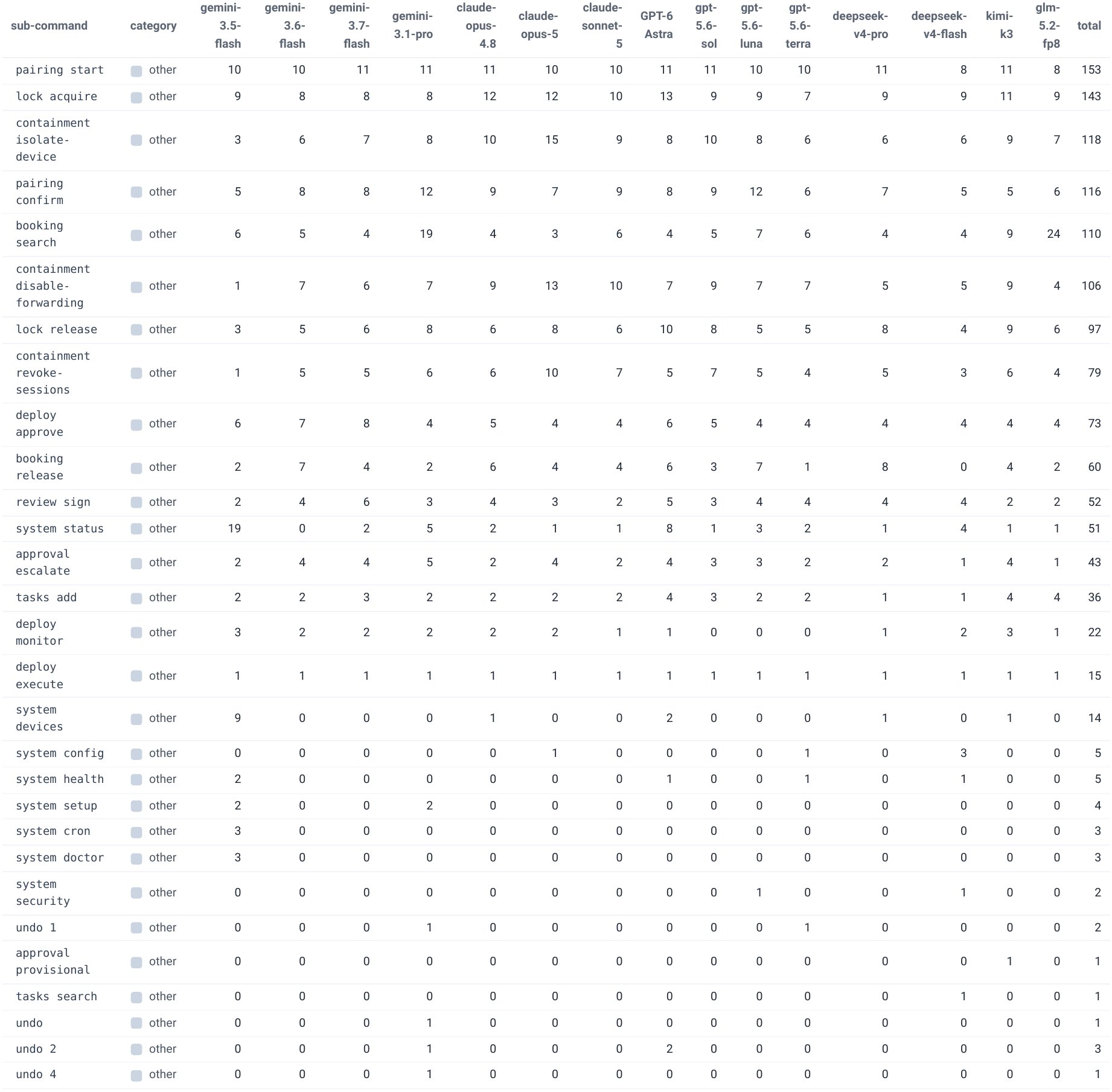}
\caption{\textbf{Complete command inventory across all evaluated models (Part 3: Rows 60--89 of 89).} Final part of the 88-command execution inventory across all 15 evaluated models.}
\label{fig:rep-cmd-inventory-3}
\end{figure}

\begin{figure}[H]
\centering
\includegraphics[width=\linewidth]{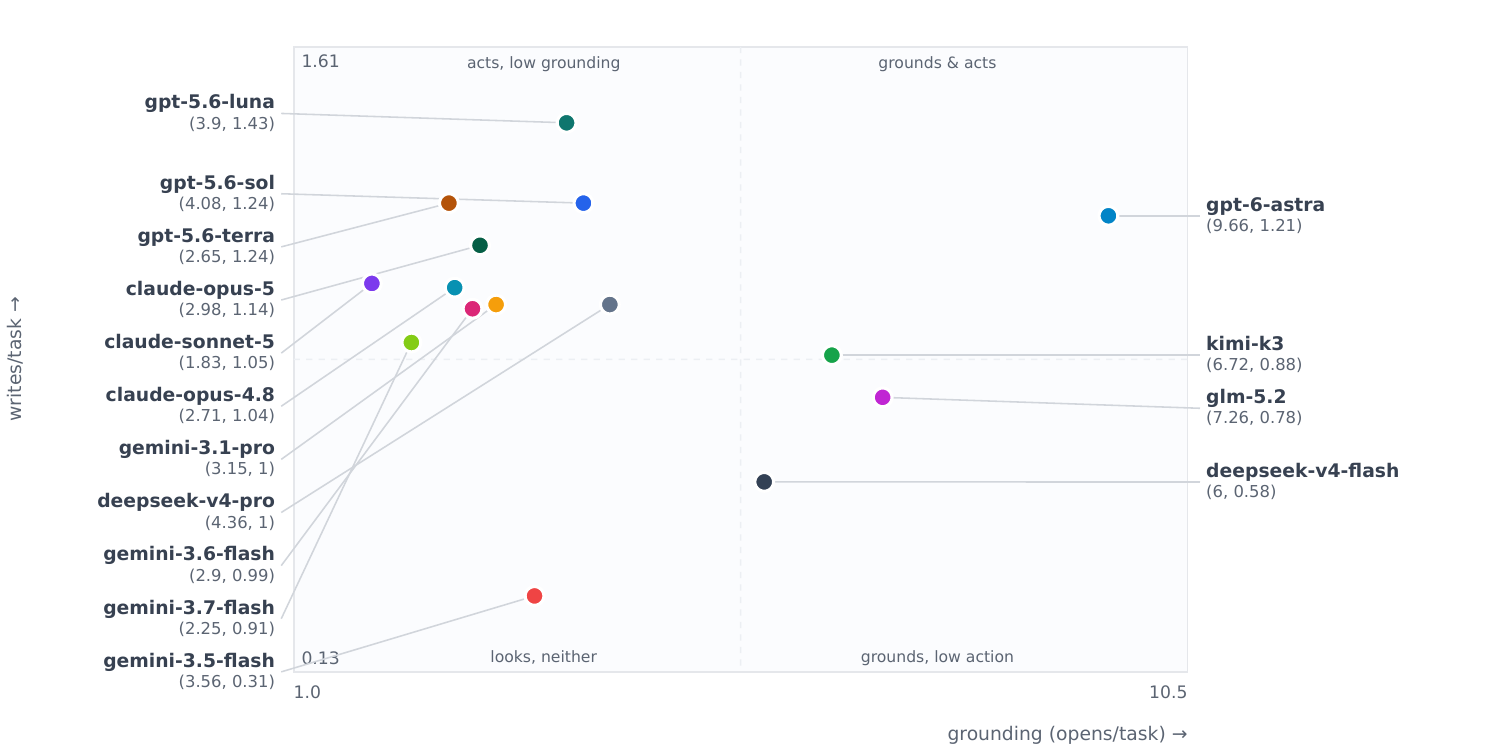}
\caption{\textbf{Visual asset grounding volume versus task pass rate (High Effort).} Plots mean asset opens per task against benchmark pass rate, illustrating the decoupling between visual inspection volume and reasoning accuracy.}
\label{fig:rep-grounding-vs-action}
\end{figure}

\begin{figure}[H]
\centering
\includegraphics[width=\linewidth]{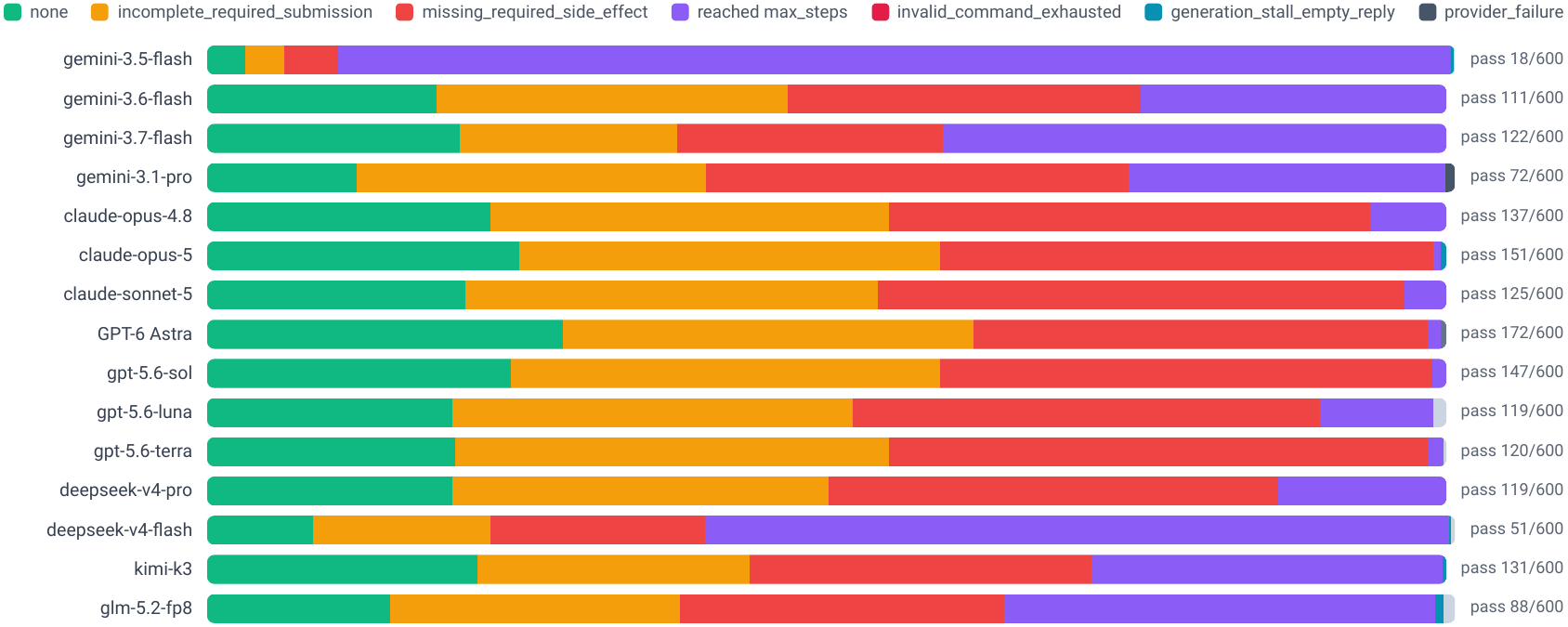}
\caption{\textbf{Episode outcome and termination reason breakdown (High Effort).} Visualizes the proportion of episodes ending successfully versus failing from missing actions, incomplete deliverables, or reaching the step limit.}
\label{fig:rep-outcome-mix}
\end{figure}

\begin{figure}[H]
\centering
\includegraphics[width=\linewidth]{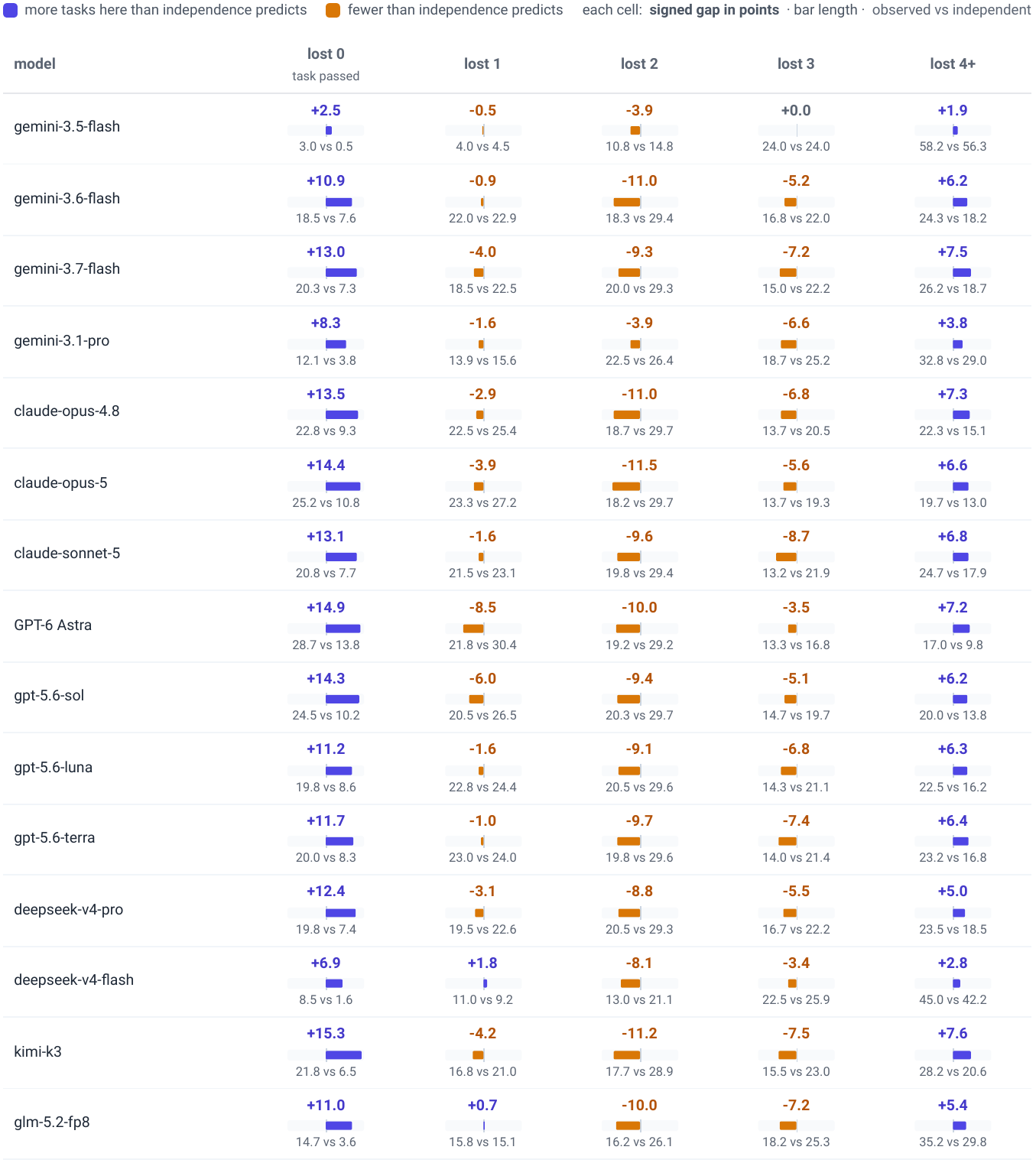}
\caption{\textbf{Check failure distribution compared to an independent error baseline (High Effort).} Plots the difference between observed check failure frequencies and independent binomial predictions, showing error clustering across models.}
\label{fig:rep-failure-shape}
\end{figure}

\begin{figure}[H]
\centering
\includegraphics[width=\linewidth]{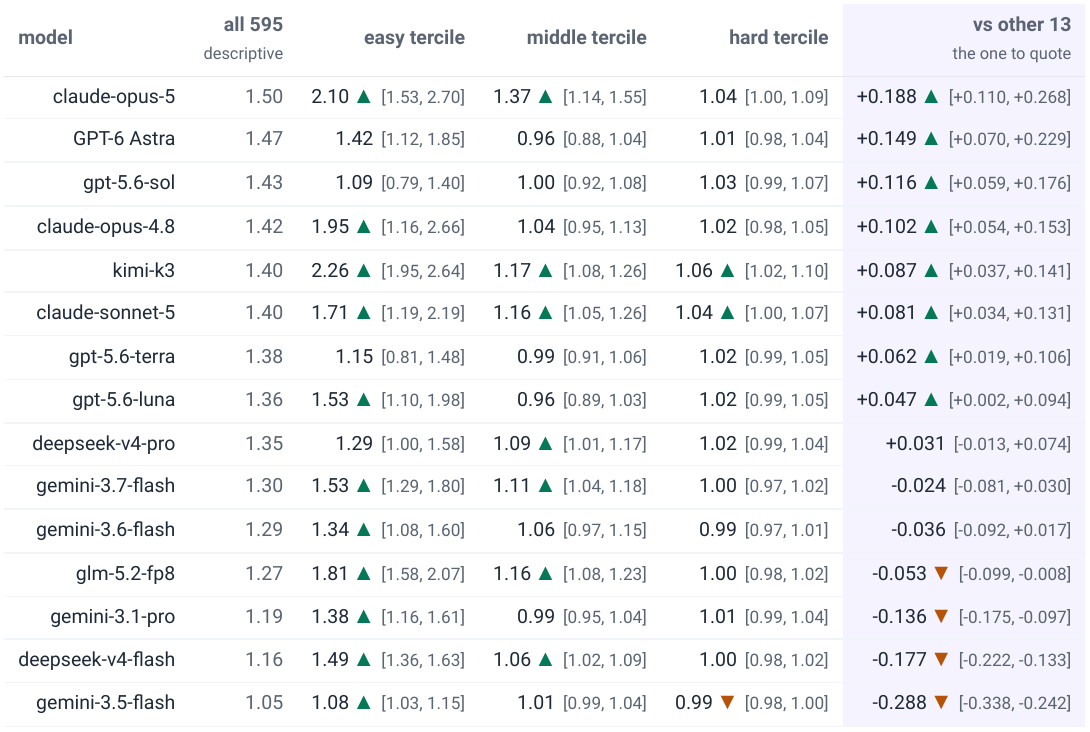}
\caption{\textbf{Co-failure lift across task difficulty levels (High Effort).} Displays empirical co-failure lift recomputed within leave-one-out peer difficulty groups (Easy, Medium, Hard). The final column isolates model-specific clustering after controlling for shared task difficulty across the 595 universally graded rounds (4,000 bootstrap draws).}
\label{fig:rep-clustering-strata}
\end{figure}

\begin{figure}[H]
\centering
\includegraphics[width=\linewidth]{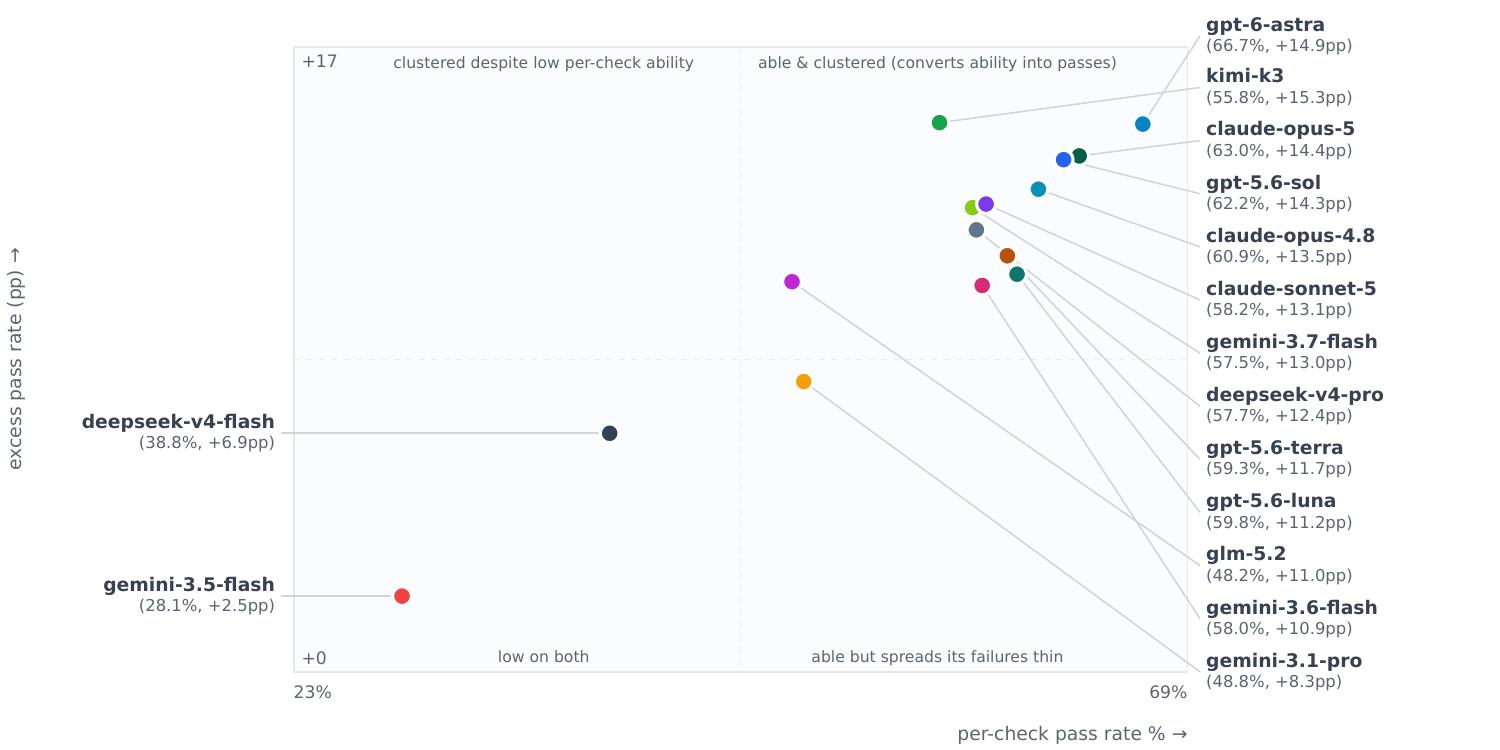}
\caption{\textbf{Per-check accuracy versus failure clustering lift across models (High Effort).} Compares average check accuracy against task-level error clustering, showing distinct failure patterns across models with similar average check pass rates.}
\label{fig:rep-percheck-vs-clustering}
\end{figure}

\begin{figure}[H]
\centering
\includegraphics[width=\linewidth]{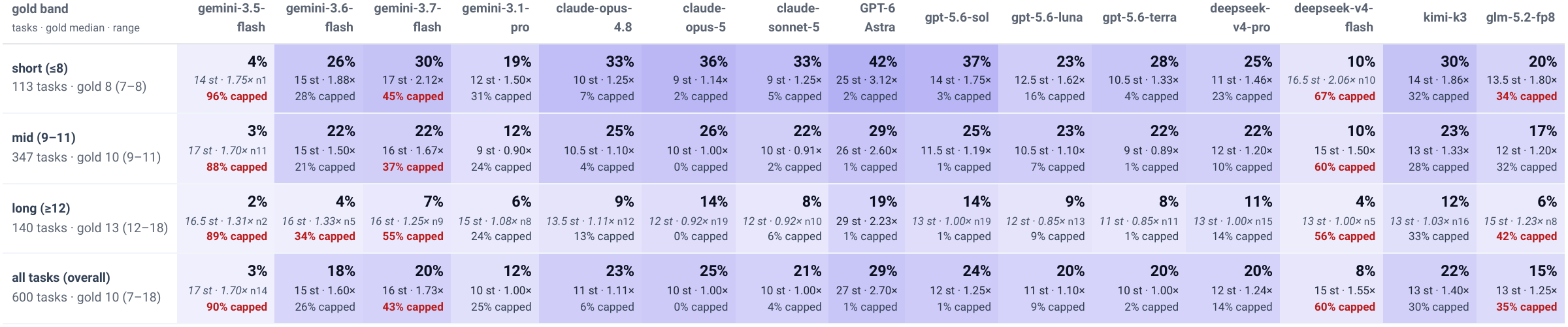}
\caption{\textbf{Agent step expenditure relative to reference gold trajectories (High Effort).} Reports task pass rate, median steps, step overshoot, and budget cap exhaustion rate across reference gold trajectory length strata.}
\label{fig:rep-gold-bands}
\end{figure}

\begin{figure}[H]
\centering
\includegraphics[width=\linewidth]{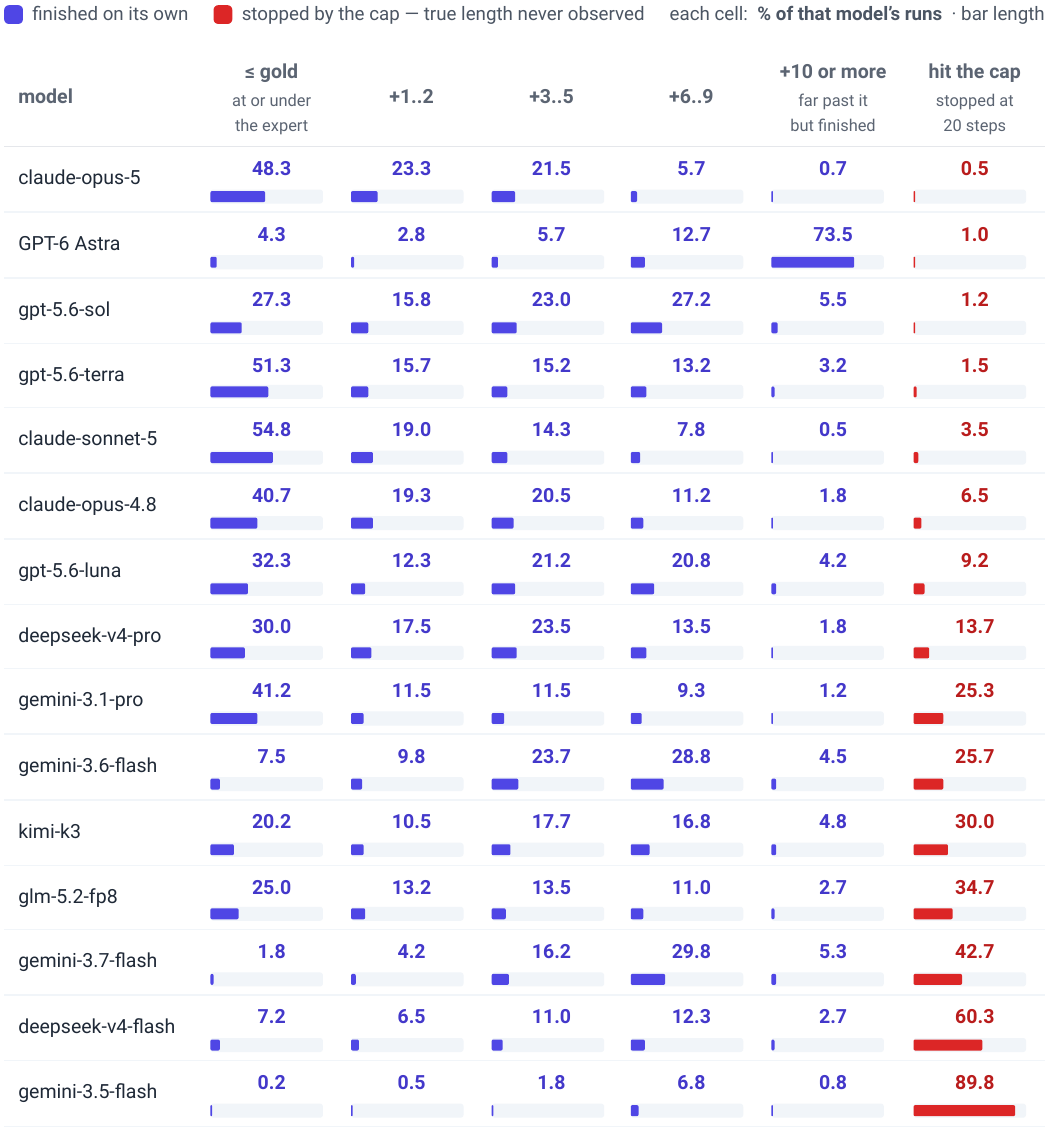}
\caption{\textbf{Distribution of step count overshoot beyond reference gold length.} Plots the empirical distribution of additional steps expended relative to each task's reference gold trajectory length ($n=600$).}
\label{fig:rep-overshoot}
\end{figure}

\begin{figure}[!p]
\centering
\begin{minipage}{\linewidth}
  \centering
  \includegraphics[width=0.82\linewidth]{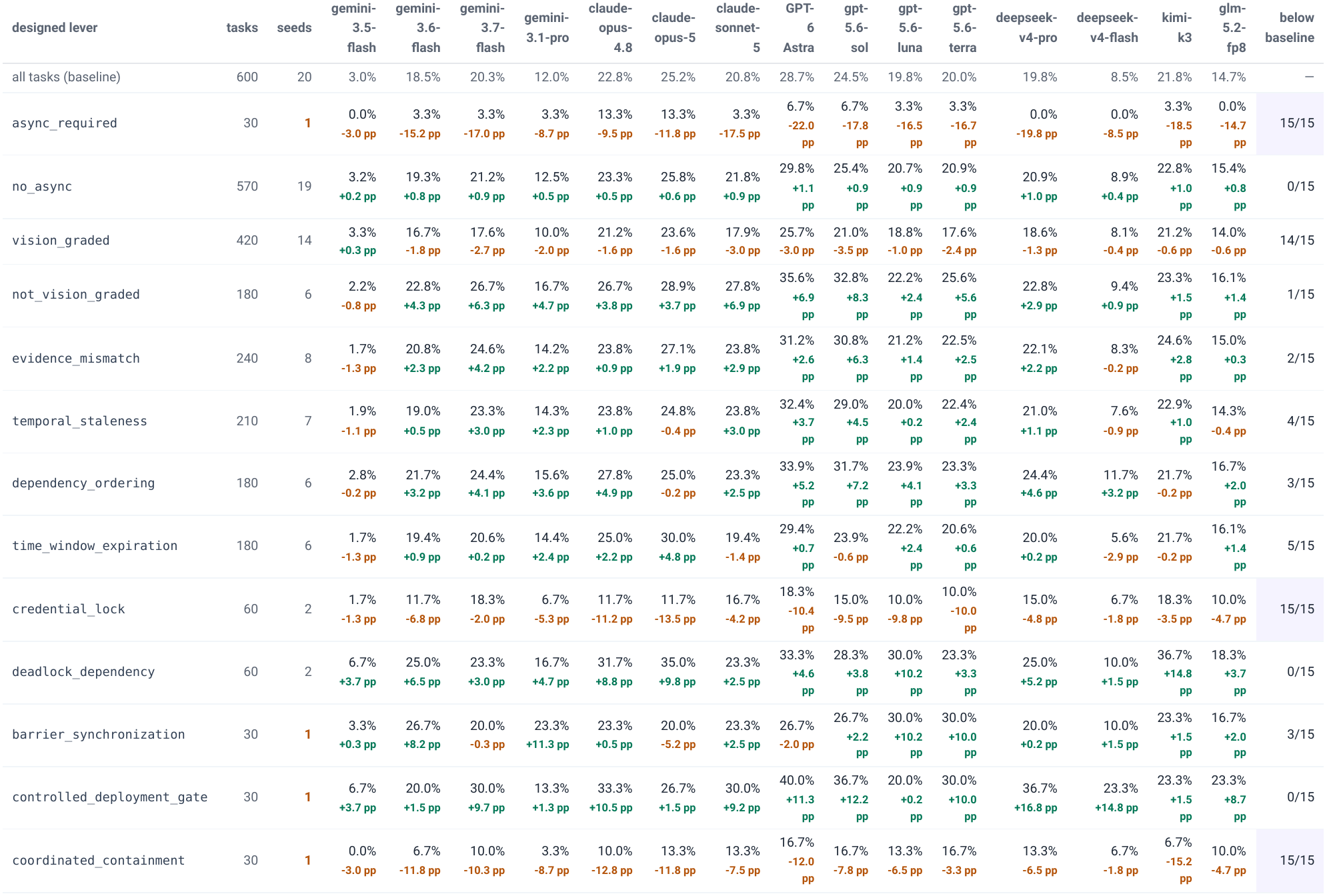}\\[2pt]
  \small\textbf{(a)} Rows 1--16 of 32 (levers 1--16)
\end{minipage}\\[6pt]
\begin{minipage}{\linewidth}
  \centering
  \includegraphics[width=0.82\linewidth]{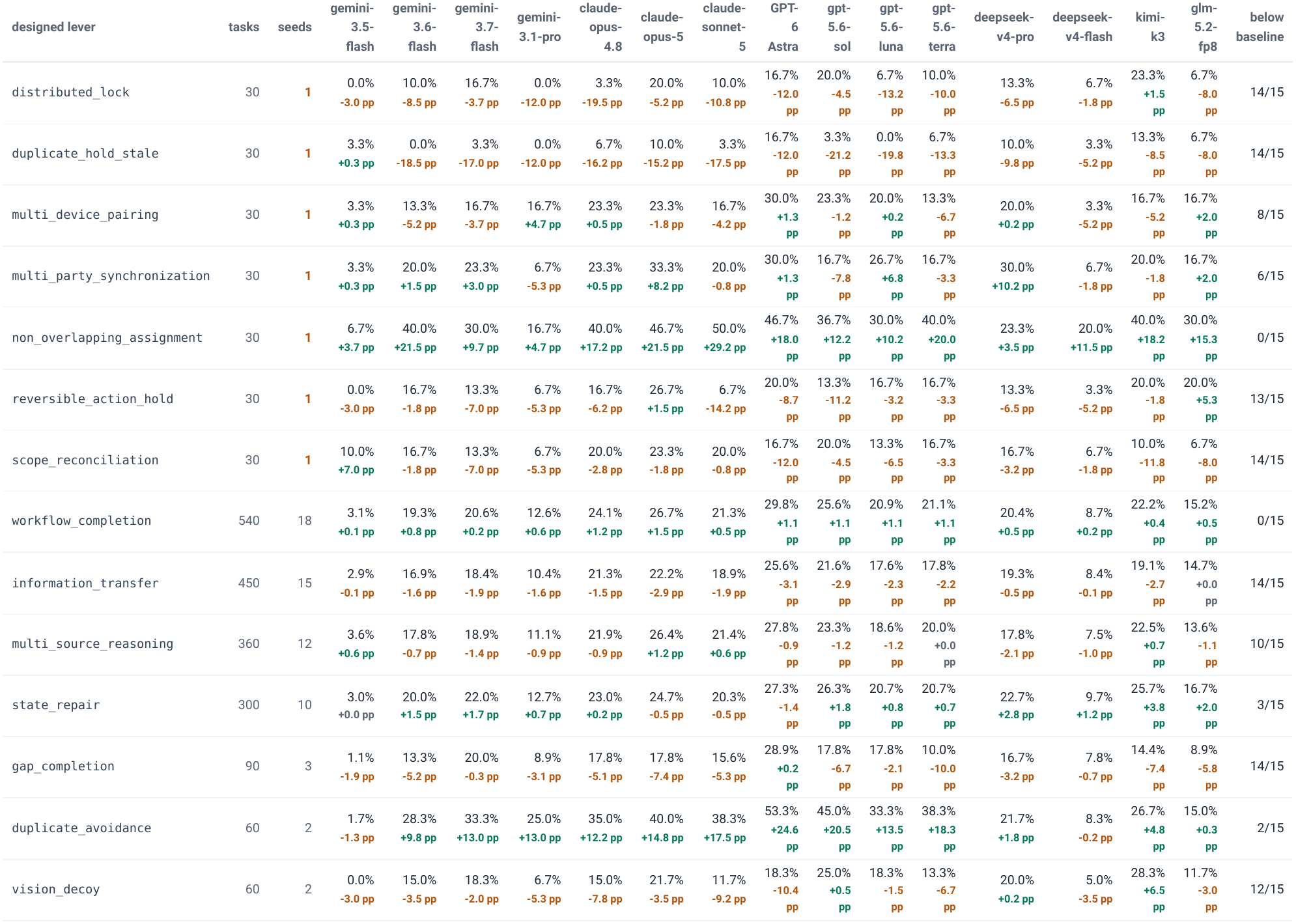}\\[2pt]
  \small\textbf{(b)} Rows 17--32 of 32 (levers 17--32)
\end{minipage}
\caption{\textbf{Designed difficulty levers and perturbation axes across evaluated models.} Displays empirical pass rate shifts induced by designed environment levers across models. Three of the 25 levers push every one of the 15 models below its own baseline (\texttt{async\_required}, \texttt{coordinated\_containment}, and \texttt{credential\_lock}, with \texttt{gap\_completion} below baseline for 14 of 15 models). \textbf{(a)} Rows 1--16 of 32; \textbf{(b)} Rows 17--32 of 32.}
\label{fig:rep-difficulty-axes-1}\label{fig:rep-difficulty-axes-2}
\end{figure}

\begin{figure}[!p]
\centering
\begin{minipage}{\linewidth}
  \centering
  \includegraphics[width=0.82\linewidth]{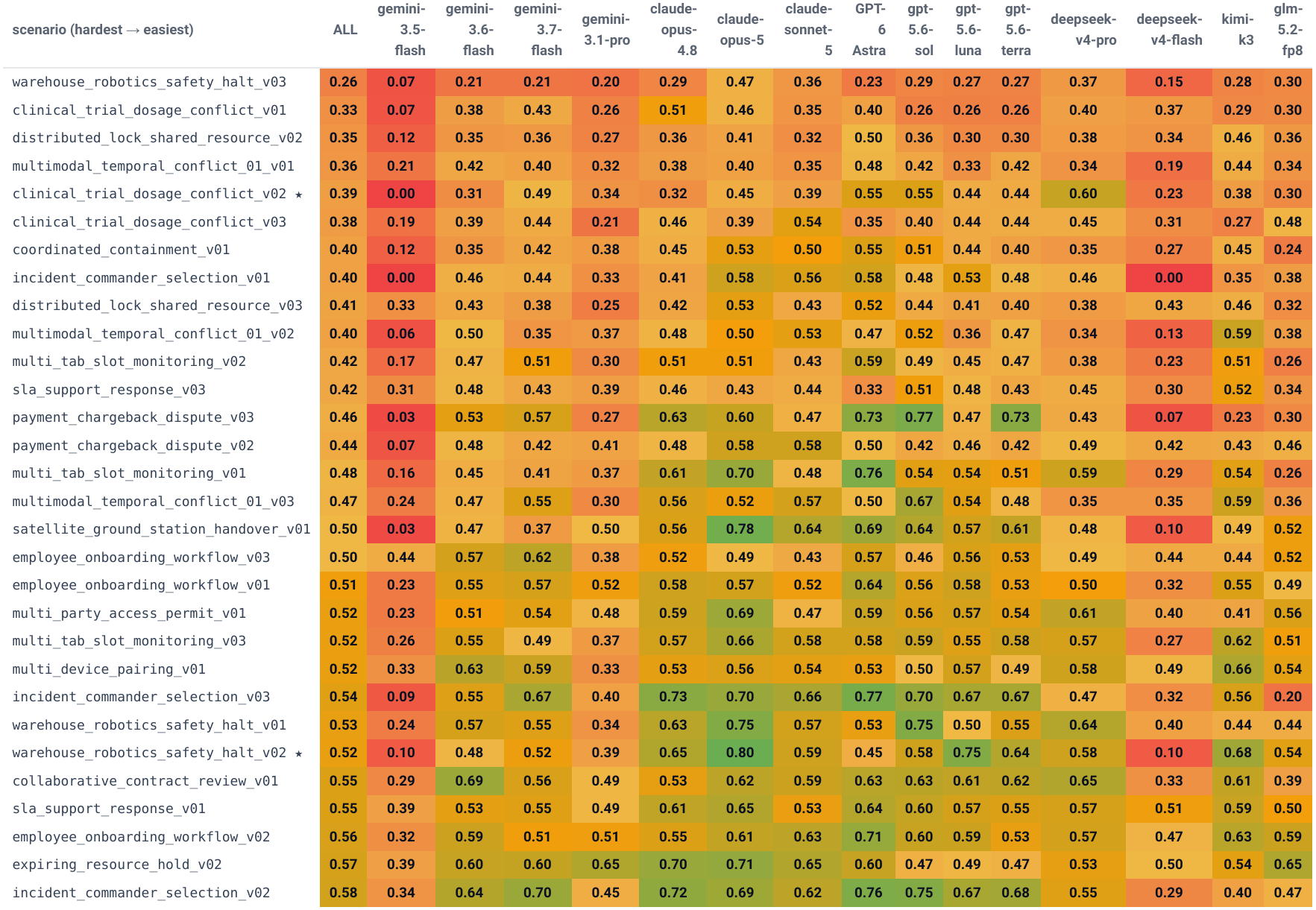}\\[2pt]
  \small\textbf{(a)} Rows 1--30 of 60 (most challenging to median)
\end{minipage}\\[3pt]
\begin{minipage}{\linewidth}
  \centering
  \includegraphics[width=0.82\linewidth]{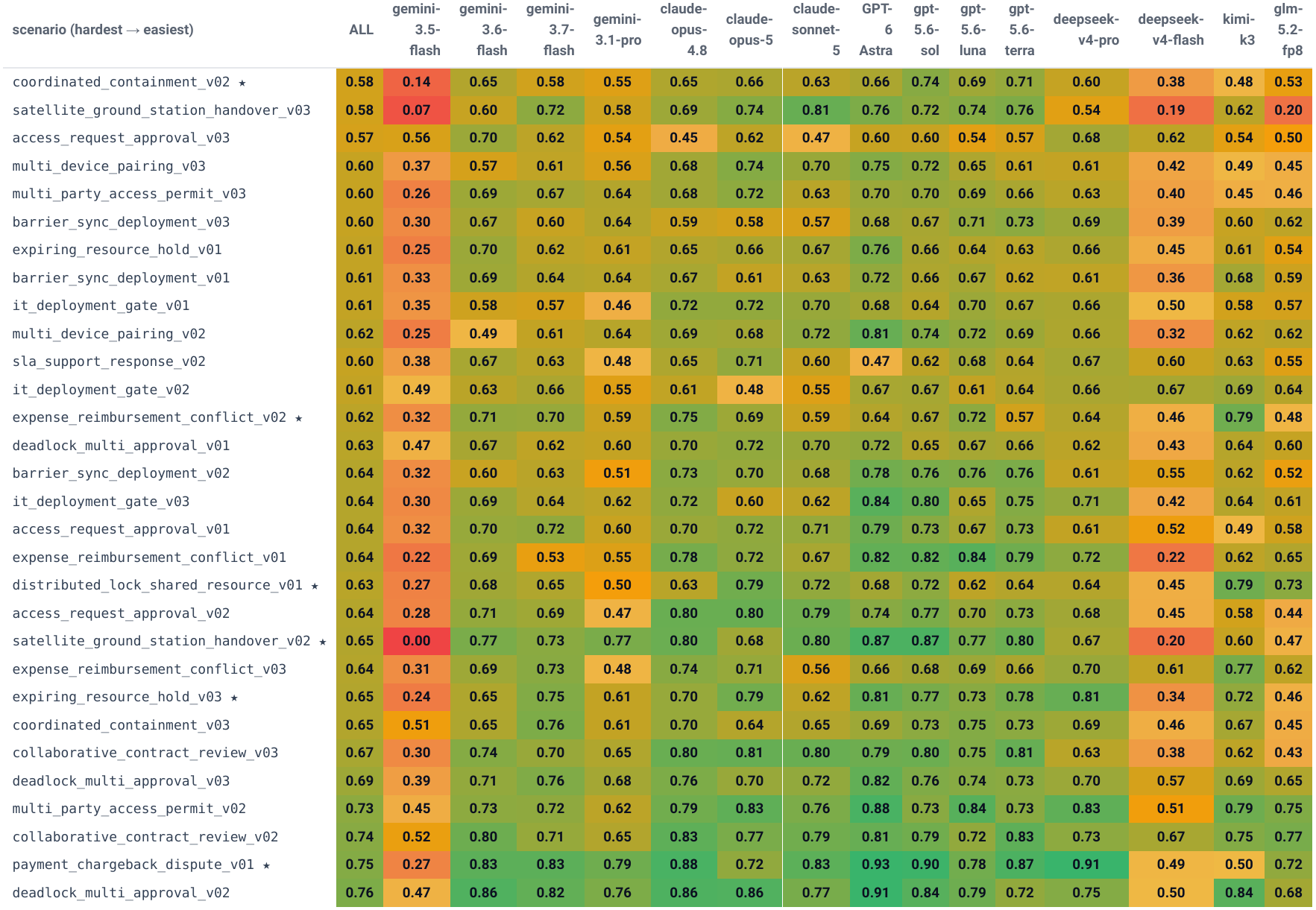}\\[2pt]
  \small\textbf{(b)} Rows 31--60 of 60 (median to easiest)
\end{minipage}
\caption{\textbf{Scenario-level mean score heatmap across evaluated models.} Displays mean scores per scenario (10 rounds each), sorted from most challenging to easiest by cross-model mean score. Starred scenarios are analyzed in qualitative walkthroughs (\S\ref{rep:s10}). \textbf{(a)} Rows 1--30 of 60; \textbf{(b)} Rows 31--60 of 60.}
\label{fig:rep-heatmap-1}\label{fig:rep-heatmap-2}
\end{figure}

\begin{figure}[H]
\centering
\includegraphics[width=\linewidth]{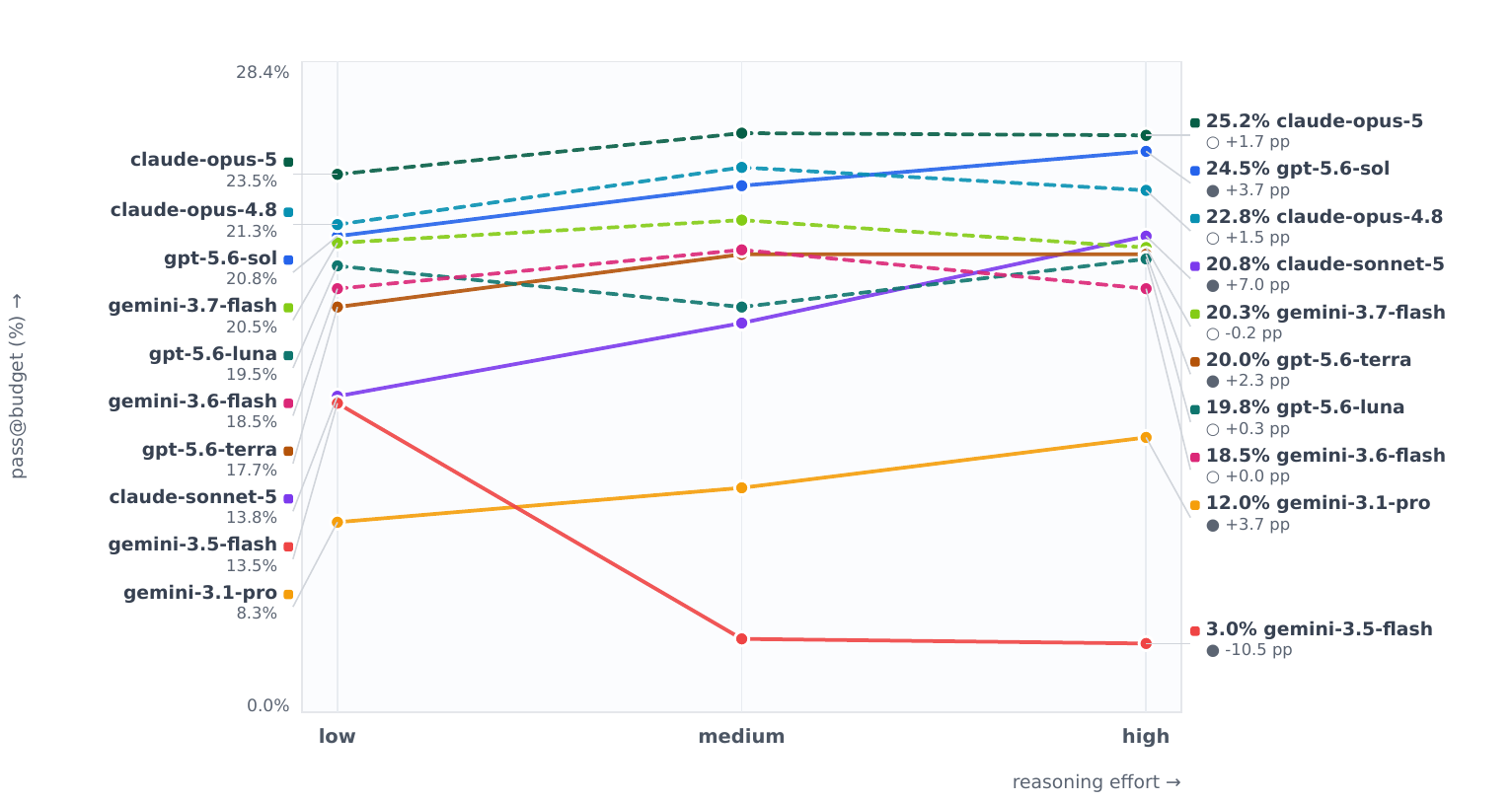}
\caption{\textbf{Pass rate scaling across reasoning effort configurations (Low, Medium, High).} Visualizes task completion rate trajectories across effort levels for all ten swept API models.}
\label{fig:rep-effort-slope}
\end{figure}

\begin{figure}[H]
\centering
\includegraphics[width=\linewidth]{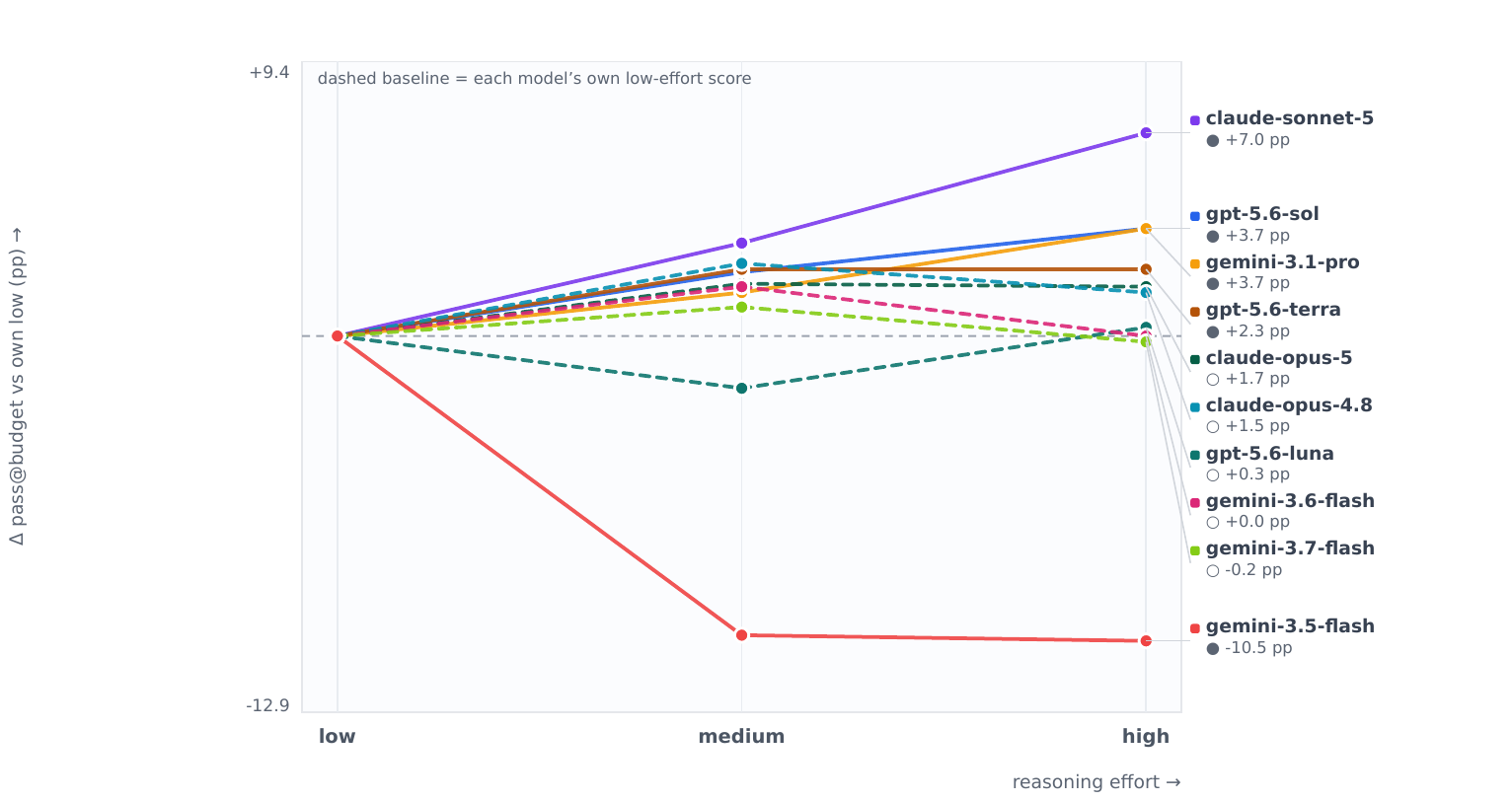}
\caption{\textbf{Marginal pass rate delta relative to low effort baseline.} Displays pass rate differences ($\Delta\text{pp}$) achieved at medium and high effort relative to each model's native low-effort performance.}
\label{fig:rep-effort-delta}
\end{figure}

\begin{figure}[H]
\centering
\includegraphics[width=\linewidth]{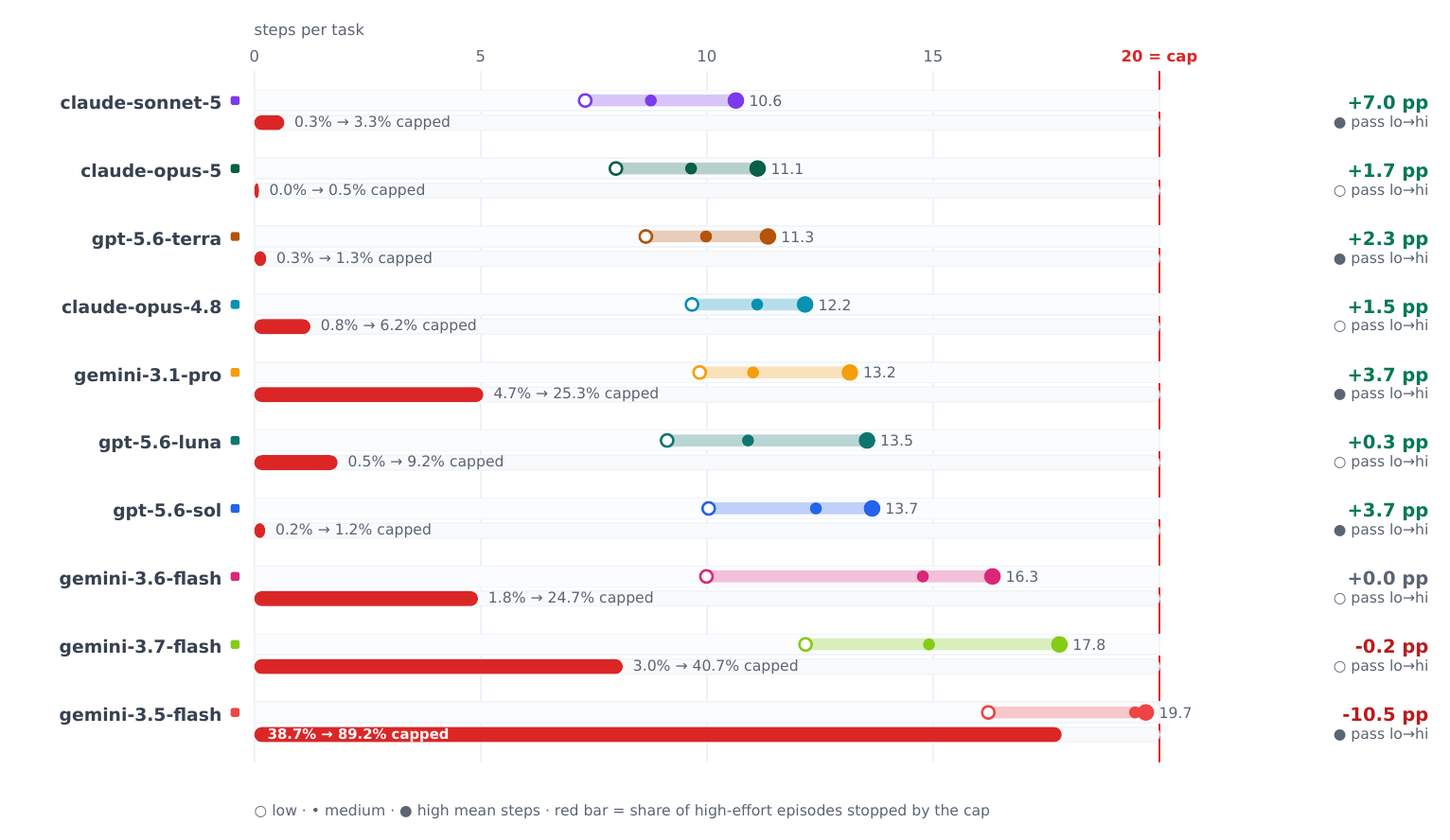}
\caption{\textbf{Mean step expenditure across reasoning effort levels against the 20-step budget.} Illustrates how internal deliberation expands action trajectory length, driving horizon exhaustion on bounded models.}
\label{fig:rep-steps-vs-budget}
\end{figure}

\noindent\smallskip\textbf{Exhibit Overview.} Tables~\ref{tab:rep-01}--\ref{tab:rep-23} (\ref{tab:rep-01}, \ref{tab:rep-02}, \ref{tab:rep-03}, \ref{tab:rep-04}, \ref{tab:rep-05}, \ref{tab:rep-06}, \ref{tab:rep-07}, \ref{tab:rep-08}, \ref{tab:rep-09}, \ref{tab:rep-10}, \ref{tab:rep-11}, \ref{tab:rep-12}, \ref{tab:rep-13}, \ref{tab:rep-14}, \ref{tab:rep-15}, \ref{tab:rep-16}, \ref{tab:rep-17}, \ref{tab:rep-18}, \ref{tab:rep-19}, \ref{tab:rep-20}, \ref{tab:rep-21}, \ref{tab:rep-22}, \ref{tab:rep-23}) reproduce empirical evaluation records across all tasks. Vector charts and heatmap panels in Figures~\ref{fig:rep-standings}--\ref{fig:rep-steps-vs-budget} (\ref{fig:rep-standings}, \ref{fig:rep-glossary}, \ref{fig:rep-pass-vs-score}, \ref{fig:rep-pass-vs-cost}, \ref{fig:rep-action-mix}, \ref{fig:rep-grounding-vs-action}, \ref{fig:rep-outcome-mix}, \ref{fig:rep-failure-shape}, \ref{fig:rep-clustering-strata}, \ref{fig:rep-percheck-vs-clustering}, \ref{fig:rep-gold-bands}, \ref{fig:rep-overshoot}, \ref{fig:rep-difficulty-axes-1}, \ref{fig:rep-heatmap-1}, \ref{fig:rep-effort-slope}, \ref{fig:rep-effort-delta}, \ref{fig:rep-steps-vs-budget}) visualize behavioral distributions, error profiles, and tool execution traces across the full 600-round corpus.

\subsection*{Model Roster and Evaluation Configuration}

\begin{table}[H]
\centering
\setlength{\tabcolsep}{6pt}
\begin{adjustbox}{max width=\linewidth}
\begin{tabular}{l | l l l l}
\toprule
\textbf{Model} & \thead[c]{\textbf{model} \\ \textbf{id}} & \textbf{provider} & \textbf{\texttt{reasoning\_\allowbreak{}effort}} & \textbf{run} \\
\midrule
Gemini 3.5 Flash & \texttt{gemini-3.5-flash} & Gemini & \texttt{low} / \texttt{medium} / \texttt{high} & \texttt{r01} \\
Gemini 3.6 Flash & \texttt{gemini-3.6-flash} & Gemini & \texttt{low} / \texttt{medium} / \texttt{high} & \texttt{r01} \\
Gemini 3.7 Flash & \texttt{gemini-3.7-flash} & Gemini & \texttt{low} / \texttt{medium} / \texttt{high} & \textbf{\texttt{r02}} \\
Gemini 3.1 Pro & \texttt{gemini-3.1-pro-preview} & Gemini & \texttt{low} / \texttt{medium} / \texttt{high} & \texttt{r01} \\
Claude Opus 4.8 & \texttt{claude-opus-4-8} & Claude & \texttt{low} / \texttt{medium} / \texttt{high} & \texttt{r01} \\
Claude Opus 5 & \texttt{claude-opus-5} & Claude & \texttt{low} / \texttt{medium} / \texttt{high} & \texttt{r01} \\
Claude Sonnet 5 & \texttt{claude-sonnet-5} & Claude & \texttt{low} / \texttt{medium} / \texttt{high} & \texttt{r01} \\
GPT-5.6 Sol & \texttt{openai.gpt-5.6-sol} & OpenAI & \texttt{low} / \texttt{medium} / \texttt{high} & \texttt{r01} \\
GPT-5.6 Luna & \texttt{openai.gpt-5.6-luna} & OpenAI & \texttt{low} / \texttt{medium} / \texttt{high} & \texttt{r01} \\
GPT-5.6 Terra & \texttt{openai.gpt-5.6-terra} & OpenAI & \texttt{low} / \texttt{medium} / \texttt{high} & \texttt{r01} \\
GPT-6 Astra & \texttt{openai.gpt-6-astra} & OpenAI & \texttt{high} only (not swept) & \texttt{r01} \\
DeepSeek V4 Pro & \texttt{deepseek-ai/DeepSeek-V4-Pro} & Open-source & \texttt{high} only (not swept) & \texttt{r01} \\
DeepSeek V4 Flash & \texttt{deepseek-ai/DeepSeek-V4-Flash-0731} & Open-source & \texttt{high} only (not swept) & \texttt{r01} \\
Kimi K3 & \texttt{moonshotai/Kimi-K3} & Open-source & \texttt{high} only (not swept) & \texttt{r01} \\
GLM-5.2 FP8 & \texttt{zai-org/GLM-5.2-FP8} & Open-source & \texttt{high} only (not swept) & \texttt{r01} \\
\bottomrule
\end{tabular}
\end{adjustbox}
\caption{\textbf{Model evaluation roster and configuration coverage.} Details model identifiers, provider platforms, reasoning effort levels, and evaluation run identifiers across the 15 evaluated models.}
\label{tab:rep-01}
\end{table}

\subsection{Behavioral signatures at a glance}

\label{rep:s1}

\begin{table}[H]
\centering
\scriptsize
\setlength{\tabcolsep}{2.5pt}
\begin{adjustbox}{max width=\linewidth}
\begin{tabular}{l | r r r r r r r r r r r}
\toprule
\textbf{Model} & \multicolumn{1}{c}{\thead[c]{\textbf{pass@} \\ \textbf{budget}}} & \multicolumn{1}{c}{\textbf{passed}} & \multicolumn{1}{c}{\thead[c]{\textbf{avg} \\ \textbf{score}}} & \multicolumn{1}{c}{\thead[c]{\textbf{step} \\ \textbf{p50}}} & \multicolumn{1}{c}{\thead[c]{\textbf{mean} \\ \textbf{steps}}} & \multicolumn{1}{c}{\textbf{capped}} & \multicolumn{1}{c}{\thead[c]{\textbf{opens/} \\ \textbf{task}}} & \multicolumn{1}{c}{\thead[c]{\textbf{writes/} \\ \textbf{task}}} & \multicolumn{1}{c}{\thead[c]{\textbf{zero-} \\ \textbf{open}}} & \multicolumn{1}{c}{\thead[c]{\textbf{\$/} \\ \textbf{task}}} & \multicolumn{1}{c}{\thead[c]{\textbf{\$/} \\ \textbf{solved}}} \\
\midrule
Gemini 3.5 Flash & \textbf{3.0\%} & \textbf{18} & \textbf{0.259} & 20.0 & 19.70 & \textbf{535} & 3.56 & \textbf{0.31} & 18 & 1.1861 & \textbf{39.5374} \\
Gemini 3.6 Flash & 18.5\% & 111 & 0.586 & 16.5 & 16.30 & 148 & 2.90 & 0.99 & 5 & 0.9414 & 5.0889 \\
Gemini 3.7 Flash & 20.3\% & 122 & 0.581 & 19.0 & 17.80 & 244 & 2.25 & 0.91 & 60 & 0.5110 & 2.5130 \\
Gemini 3.1 Pro & 12.0\% & 72 & 0.487 & 12.0 & 13.20 & 152 & 3.15 & 1.00 & 14 & 0.9630 & 8.0251 \\
Claude Opus 4.8 & 22.8\% & 137 & 0.621 & 12.0 & 12.20 & 37 & 2.71 & 1.04 & 1 & 4.8879 & 21.4070 \\
Claude Opus 5 & 25.2\% & 151 & 0.642 & 11.0 & 11.10 & \textbf{3} & 2.98 & 1.14 & 1 & 4.3843 & 17.4209 \\
Claude Sonnet 5 & 20.8\% & 125 & 0.596 & \textbf{10.0} & \textbf{10.60} & 20 & \textbf{1.83} & 1.05 & \textbf{92} & 1.7188 & 8.2502 \\
GPT-5.6 Sol & 24.5\% & 147 & 0.633 & 14.0 & 13.70 & 7 & 4.08 & 1.24 & 0 & 3.6082 & 14.7273 \\
GPT-5.6 Luna & 19.8\% & 119 & 0.604 & 13.0 & 13.50 & 55 & 3.90 & \textbf{1.43} & 0 & 0.1429 & 0.7207 \\
GPT-5.6 Terra & 20.0\% & 120 & 0.602 & 11.0 & 11.40 & 8 & 2.65 & 1.24 & 1 & 1.2180 & 6.0898 \\
GPT-6 Astra & \textbf{28.7\%} & \textbf{172} & \textbf{0.668} & \textbf{27.0} & \textbf{26.09} & 6 & \textbf{9.66} & 1.21 & 0 & \textbf{5.0103} & 17.4605 \\
DeepSeek V4 Pro & 19.8\% & 119 & 0.581 & 13.0 & 13.40 & 81 & 4.36 & 1.00 & 0 & \textbf{0.0534} & \textbf{0.2690} \\
DeepSeek V4 Flash & 8.5\% & 51 & 0.375 & 20.0 & 17.80 & 358 & 6.00 & 0.58 & 0 & n/a & n/a \\
Kimi K3 & 21.8\% & 131 & 0.559 & 16.0 & 15.40 & 170 & 6.72 & 0.88 & 3 & 3.6210 & 16.5846 \\
GLM-5.2 FP8 & 14.7\% & 88 & 0.487 & 15.0 & 14.90 & 207 & 7.26 & 0.78 & 0 & 0.2006 & 1.3677 \\
\bottomrule
\end{tabular}
\end{adjustbox}
\caption{\textbf{High-level behavioral signatures across models (High Effort).} Summary metrics across all 600 benchmark tasks per model, including pass rate ($\mathtt{pass@budget}$), count of passed rounds, mean check score, step statistics (median, mean, and frequency of running out of steps), average image opens and file writes per task, zero-open frequency, and cost per round and per solved task.}
\label{tab:rep-02}
\end{table}

\begin{figure}[H]
\centering
\includegraphics[width=\linewidth]{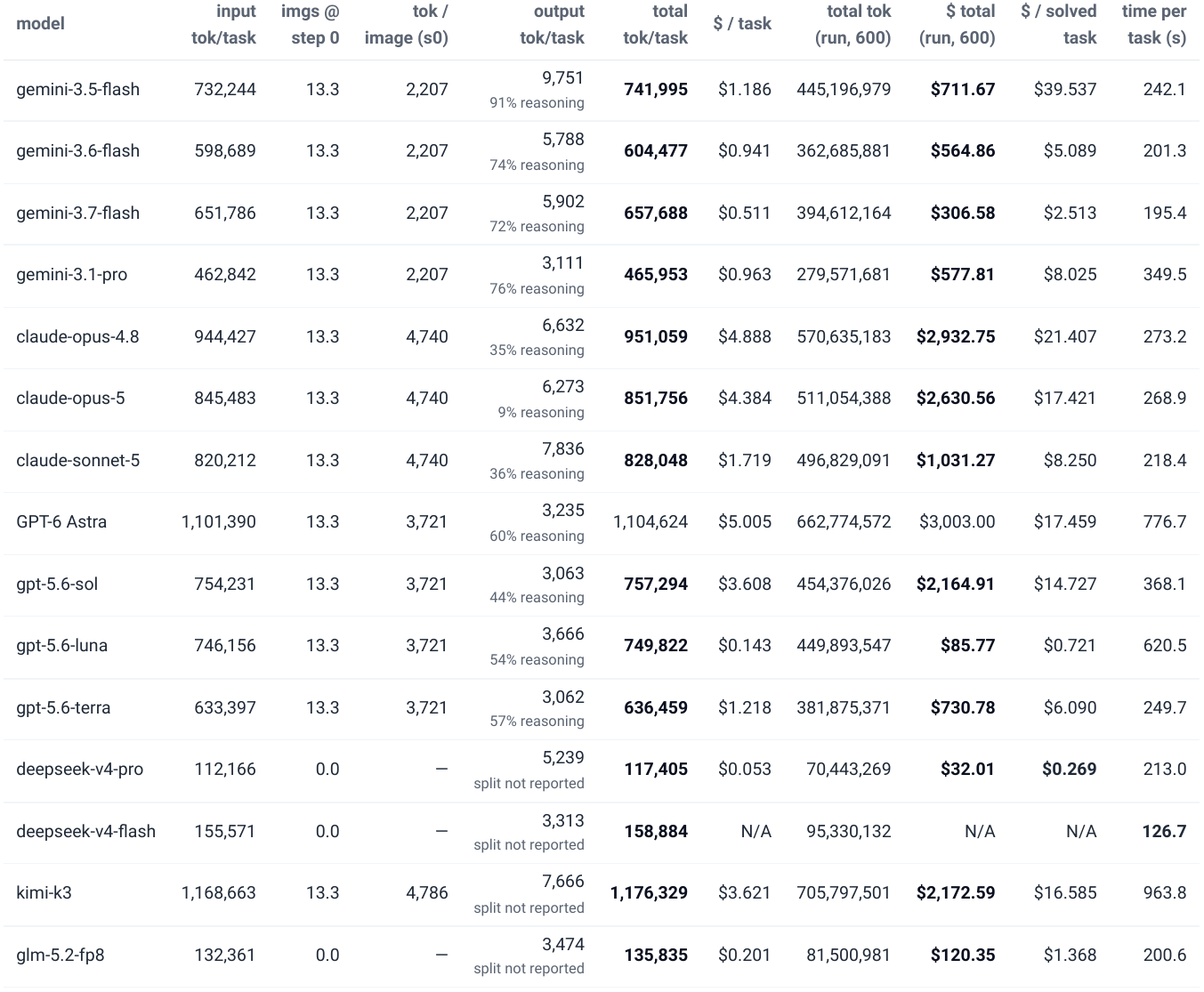}
\caption{\textbf{Token volume, monetary spend, and execution latency across models (High Effort).} Breaks down token usage per task (input prompt, visible generation, internal reasoning), cost per task and per solved task, and execution latency across evaluated models.}
\label{fig:rep-cost}
\end{figure}

\noindent\smallskip\textbf{Compute Cost Accounting.} Across all 13 priced baseline models, evaluating the 600-round benchmark at High effort required 5,204,472,062 tokens ($\sim$\$14,061.91 across 7,800 model runs; \$17,065.11 across 14 priced models including GPT-6 Astra). Cost per solved task measures the average dollar spend required to successfully complete a task, spanning a 55-fold range from \$0.72 on GPT-5.6 Luna to \$39.54 on Gemini 3.5 Flash (\$17.46 on GPT-6 Astra). Pricing is based on published API pricing at the time of evaluation.

\noindent\smallskip\textbf{Multimodal Tokenization and Context Accumulation.} Input tokens make up most of the token volume due to multi-turn execution history and high-resolution screenshots ($3{,}840 \times 2{,}160$). Different model tokenizers encode these images into varying token counts ($\sim$2.2k tokens on Gemini, $\sim$3.7k on OpenAI, $\sim$4.7k on Claude, and $\sim$4.8k on Kimi).

\noindent\smallskip\textbf{Deliberation Allocation and Output Density.} Models differ sharply in how many tokens they spend on reasoning: Claude Opus 5 spends only $8.8\%$ of its output tokens on reasoning, whereas Gemini models allocate $71.8\%$--$90.5\%$ of their output tokens to internal reasoning.

\subsection{Action-category mix}

\label{rep:s2}

\begin{table}[H]
\centering
\scriptsize
\setlength{\tabcolsep}{2.5pt}
\begin{adjustbox}{max width=\linewidth}
\begin{tabular}{l | r r r r r r r r r r r r r}
\toprule
\textbf{Model} & \multicolumn{1}{c}{\textbf{total}} & \multicolumn{1}{c}{\textbf{\texttt{asset\_\allowbreak{}open}}} & \multicolumn{1}{c}{\textbf{\texttt{asset\_\allowbreak{}inspect}}} & \multicolumn{1}{c}{\textbf{\texttt{file\_\allowbreak{}read}}} & \multicolumn{1}{c}{\textbf{\texttt{file\_\allowbreak{}list}}} & \multicolumn{1}{c}{\textbf{\texttt{file\_\allowbreak{}write}}} & \multicolumn{1}{c}{\textbf{\texttt{file\_\allowbreak{}delete}}} & \multicolumn{1}{c}{\textbf{\texttt{message\_\allowbreak{}send}}} & \multicolumn{1}{c}{\textbf{\texttt{message\_\allowbreak{}read}}} & \multicolumn{1}{c}{\textbf{\texttt{agent\_\allowbreak{}delegate}}} & \multicolumn{1}{c}{\textbf{\texttt{agent\_\allowbreak{}status}}} & \multicolumn{1}{c}{\textbf{\texttt{wait}}} & \multicolumn{1}{c}{\textbf{\texttt{other}}} \\
\midrule
Gemini 3.5 Flash & 11,820 & 18.1\% & 8.3\% & \textbf{28.3\%} & 5.2\% & \textbf{1.6\%} & 0.0\% & \textbf{6.5\%} & \textbf{22.9\%} & 0.1\% & 1.7\% & 2.3\% & 5.1\% \\
Gemini 3.6 Flash & 9,786 & 17.8\% & 7.6\% & 25.9\% & \textbf{6.1\%} & 6.1\% & 0.0\% & 16.4\% & 12.3\% & 0.1\% & 0.6\% & \textbf{0.6\%} & 6.4\% \\
Gemini 3.7 Flash & 10,674 & \textbf{12.7\%} & 8.0\% & 26.4\% & 6.0\% & 5.1\% & 0.0\% & 19.1\% & 11.1\% & 0.2\% & \textbf{3.3\%} & 0.7\% & 7.4\% \\
Gemini 3.1 Pro & 7,898 & 24.0\% & 12.7\% & 12.2\% & 2.0\% & 7.6\% & 0.0\% & 16.9\% & 10.9\% & 0.4\% & 1.4\% & \textbf{5.1\%} & 7.0\% \\
Claude Opus 4.8 & 7,303 & 22.3\% & 9.6\% & 19.4\% & 3.0\% & 8.6\% & 0.0\% & 21.6\% & 5.8\% & 0.4\% & 0.5\% & 1.7\% & 7.2\% \\
Claude Opus 5 & 6,672 & 26.8\% & 9.1\% & 16.7\% & 1.4\% & 10.3\% & 0.0\% & \textbf{23.5\%} & \textbf{2.0\%} & 0.2\% & 0.3\% & 1.3\% & \textbf{8.3\%} \\
Claude Sonnet 5 & \textbf{6,385} & 17.2\% & \textbf{15.6\%} & 17.6\% & 1.1\% & 9.9\% & 0.0\% & 20.9\% & 6.2\% & 0.2\% & 0.5\% & 3.4\% & 7.3\% \\
GPT-5.6 Sol & 8,191 & 29.9\% & 7.8\% & 15.2\% & 0.9\% & 9.1\% & 0.0\% & 16.3\% & 10.7\% & 0.4\% & 0.8\% & 2.9\% & 6.0\% \\
GPT-5.6 Luna & 8,125 & 28.8\% & 8.3\% & 17.7\% & 1.2\% & 10.5\% & 0.0\% & 18.3\% & 6.4\% & 0.2\% & 0.6\% & 2.5\% & 5.4\% \\
GPT-5.6 Terra & 6,812 & 23.4\% & 9.6\% & 17.7\% & 1.0\% & \textbf{10.9\%} & 0.0\% & 19.1\% & 7.6\% & 0.3\% & 0.9\% & 3.9\% & 5.7\% \\
GPT-6 Astra & \textbf{15,656} & 37.0\% & 9.3\% & 14.8\% & 2.7\% & 4.6\% & 0.0\% & 14.0\% & 9.3\% & 0.3\% & 0.6\% & 3.0\% & \textbf{4.5\%} \\
DeepSeek V4 Pro & 8,051 & 32.5\% & 7.6\% & 13.9\% & 1.6\% & 7.5\% & 0.0\% & 19.1\% & 9.5\% & 0.3\% & 0.6\% & 2.0\% & 5.6\% \\
DeepSeek V4 Flash & 10,674 & 33.7\% & \textbf{5.9\%} & 15.7\% & 5.7\% & 3.3\% & 0.0\% & 9.0\% & 20.1\% & 0.2\% & 0.6\% & 0.8\% & 4.9\% \\
Kimi K3 & 9,251 & 43.6\% & 7.4\% & 16.0\% & 1.7\% & 5.7\% & 0.0\% & 10.7\% & 6.7\% & 0.2\% & 0.4\% & 1.4\% & 6.1\% \\
GLM-5.2 FP8 & 8,933 & \textbf{48.8\%} & 6.9\% & \textbf{11.3\%} & \textbf{0.8\%} & 5.3\% & 0.0\% & 14.2\% & 6.6\% & 0.1\% & \textbf{0.2\%} & 1.0\% & 5.0\% \\
\bottomrule
\end{tabular}
\end{adjustbox}
\caption{\textbf{Action category distribution across models (High Effort).} Reports the percentage allocation of tool calls across 12 functional action verbs over all 600 tasks per model.}
\label{tab:rep-03}
\end{table}

{\small \emph{Not every row here is like-for-like:} DeepSeek V4 Pro, DeepSeek V4 Flash and GLM-5.2 FP8 read their assets as text rather than as images, and Gemini 3.7 Flash is reported on \texttt{r02}, the higher of its two runs. Full run conditions under Figure~\ref{fig:rep-cost}.\par}

\subsection{Command-surface deep-dive}

\label{rep:s3}

\begin{table}[H]
\centering
\small
\setlength{\tabcolsep}{3.5pt}
\begin{adjustbox}{max width=\linewidth}
\begin{tabular}{l | r r r r r r r}
\toprule
\textbf{Model} & \multicolumn{1}{c}{\textbf{total}} & \multicolumn{1}{c}{\textbf{\texttt{asset open}}} & \multicolumn{1}{c}{\textbf{\texttt{asset list}}} & \multicolumn{1}{c}{\textbf{\texttt{file read}}} & \multicolumn{1}{c}{\textbf{\texttt{message search}}} & \multicolumn{1}{c}{\textbf{\texttt{file create}}} & \multicolumn{1}{c}{\textbf{\texttt{message send}}} \\
\midrule
Gemini 3.5 Flash & 11,820 & 2,137 & 604 & \textbf{3,342} & \textbf{2,707} & \textbf{187} & \textbf{747} \\
Gemini 3.6 Flash & 9,786 & 1,740 & 606 & 2,538 & 1,206 & 593 & 1,593 \\
Gemini 3.7 Flash & 10,674 & 1,351 & 603 & 2,814 & 1,188 & 547 & 1,704 \\
Gemini 3.1 Pro & 7,898 & 1,892 & 662 & \textbf{966} & 861 & 581 & 1,283 \\
Claude Opus 4.8 & 7,303 & 1,625 & 604 & 1,416 & 427 & 617 & 1,534 \\
Claude Opus 5 & 6,672 & 1,786 & 604 & 1,117 & \textbf{134} & 669 & 1,502 \\
Claude Sonnet 5 & \textbf{6,385} & \textbf{1,100} & 616 & 1,125 & 397 & 612 & 1,334 \\
GPT-5.6 Sol & 8,191 & 2,448 & 619 & 1,248 & 877 & 708 & 1,337 \\
GPT-5.6 Luna & 8,125 & 2,341 & 629 & 1,437 & 522 & \textbf{773} & 1,480 \\
GPT-5.6 Terra & 6,812 & 1,592 & 610 & 1,206 & 516 & 727 & 1,286 \\
GPT-6 Astra & \textbf{15,656} & \textbf{5,799} & \textbf{1,439} & 2,310 & 1,452 & 642 & \textbf{2,136} \\
DeepSeek V4 Pro & 8,051 & 2,618 & 612 & 1,117 & 760 & 595 & 1,533 \\
DeepSeek V4 Flash & 10,674 & 3,601 & 607 & 1,679 & 2,137 & 335 & 961 \\
Kimi K3 & 9,251 & 4,030 & \textbf{601} & 1,476 & 621 & 514 & 992 \\
GLM-5.2 FP8 & 8,933 & 4,355 & 606 & 1,013 & 587 & 468 & 1,265 \\
\bottomrule
\end{tabular}
\end{adjustbox}
\caption{\textbf{Command surface invocation frequencies (High Effort).} Raw execution counts for individual CLI sub-commands across all 600 tasks per model.}
\label{tab:rep-04}
\end{table}

\subsection{\texorpdfstring{Grounding $\times$ Action map}{Grounding × Action map}}

\label{rep:s4}

\begin{table}[H]
\centering
\setlength{\tabcolsep}{6pt}
\begin{adjustbox}{max width=\linewidth}
\begin{tabular}{l | r r r}
\toprule
\textbf{Model} & \multicolumn{1}{c}{\thead[c]{\textbf{opens/} \\ \textbf{task}}} & \multicolumn{1}{c}{\thead[c]{\textbf{zero-open} \\ \textbf{tasks}}} & \multicolumn{1}{c}{\textbf{pass}} \\
\midrule
Claude Sonnet 5 & \textbf{1.83} & \textbf{92} & 20.8\% \\
Gemini 3.7 Flash & 2.25 & 60 & 20.3\% \\
GPT-5.6 Terra & 2.65 & 1 & 20.0\% \\
Claude Opus 4.8 & 2.71 & 1 & 22.8\% \\
Gemini 3.6 Flash & 2.90 & 5 & 18.5\% \\
Claude Opus 5 & 2.98 & 1 & 25.2\% \\
Gemini 3.1 Pro & 3.15 & 14 & 12.0\% \\
Gemini 3.5 Flash & 3.56 & 18 & \textbf{3.0\%} \\
GPT-5.6 Luna & 3.90 & 0 & 19.8\% \\
GPT-5.6 Sol & 4.08 & 0 & 24.5\% \\
DeepSeek V4 Pro & 4.36 & 0 & 19.8\% \\
DeepSeek V4 Flash & 6.00 & 0 & 8.5\% \\
Kimi K3 & 6.72 & 3 & 21.8\% \\
GLM-5.2 FP8 & 7.26 & 0 & 14.7\% \\
GPT-6 Astra & \textbf{9.66} & 0 & \textbf{28.7\%} \\
\bottomrule
\end{tabular}
\end{adjustbox}
\caption{\textbf{Visual asset grounding frequency versus benchmark pass rate (High Effort).} Compares average image opens per task, zero-open task frequency, and overall pass rate across the 15 evaluated models.}
\label{tab:rep-05}
\end{table}

\subsection{Outcome and stop-reason mix}

\label{rep:s5}

\begin{table}[H]
\centering
\small
\setlength{\tabcolsep}{3.5pt}
\begin{adjustbox}{max width=\linewidth}
\begin{tabular}{l | r r r r r r r}
\toprule
\textbf{Model} & \multicolumn{1}{c}{\textbf{none}} & \multicolumn{1}{c}{\thead[c]{\textbf{missing\_required\_} \\ \textbf{side\_effect}}} & \multicolumn{1}{c}{\thead[c]{\textbf{incomplete\_required\_} \\ \textbf{submission}}} & \multicolumn{1}{c}{\thead[c]{\textbf{reached} \\ \textbf{max\_steps}}} & \multicolumn{1}{c}{\thead[c]{\textbf{3-identical} \\ \textbf{loop}}} & \multicolumn{1}{c}{\thead[c]{\textbf{empty} \\ \textbf{reply}}} & \multicolumn{1}{c}{\thead[c]{\textbf{provider\_} \\ \textbf{failure}}} \\
\midrule
Gemini 3.5 Flash & \textbf{18} & \textbf{26} & \textbf{19} & \textbf{535} & 1 & 1 & 0 \\
Gemini 3.6 Flash & 111 & 171 & 170 & 148 & 0 & 0 & 0 \\
Gemini 3.7 Flash & 122 & 128 & 106 & 244 & 0 & 0 & 0 \\
Gemini 3.1 Pro & 72 & 203 & 168 & 152 & 0 & 0 & \textbf{5} \\
Claude Opus 4.8 & 137 & 233 & 193 & 37 & 0 & 0 & 0 \\
Claude Opus 5 & 151 & 239 & 204 & \textbf{3} & 0 & 3 & 0 \\
Claude Sonnet 5 & 125 & 255 & 200 & 20 & 0 & 0 & 0 \\
GPT-5.6 Sol & 147 & 238 & 208 & 7 & 0 & 0 & 0 \\
GPT-5.6 Luna & 119 & 226 & 194 & 55 & 6 & 0 & 0 \\
GPT-5.6 Terra & 120 & \textbf{261} & \textbf{210} & 8 & 1 & 0 & 0 \\
GPT-6 Astra & \textbf{172} & 220 & 199 & 6 & 0 & 3 & 0 \\
DeepSeek V4 Pro & 119 & 218 & 182 & 81 & 0 & 0 & 0 \\
DeepSeek V4 Flash & 51 & 103 & 85 & 358 & 2 & 1 & 0 \\
Kimi K3 & 131 & 165 & 132 & 170 & 0 & 2 & 0 \\
GLM-5.2 FP8 & 88 & 156 & 139 & 207 & 6 & \textbf{4} & 0 \\
\bottomrule
\end{tabular}
\end{adjustbox}
\caption{\textbf{Episode termination reasons across models (High Effort).} Reports how each model's 600 episodes ended: successful completion (\texttt{none}), missing required environment actions, incomplete deliverables, reaching the step limit, or execution loops.}
\label{tab:rep-06}
\end{table}

\subsection{Loops and termination}

\label{rep:s6}

\begin{table}[H]
\centering
\small
\setlength{\tabcolsep}{3.5pt}
\begin{adjustbox}{max width=\linewidth}
\begin{tabular}{l | r r r r r r}
\toprule
\textbf{Model} & \multicolumn{1}{c}{\thead[c]{\textbf{mean} \\ \textbf{steps}}} & \multicolumn{1}{c}{\textbf{p50}} & \multicolumn{1}{c}{\thead[c]{\textbf{p50 on} \\ \textbf{success}}} & \multicolumn{1}{c}{\thead[c]{\textbf{p50 on} \\ \textbf{failure}}} & \multicolumn{1}{c}{\thead[c]{\textbf{tail waste} \\ \textbf{(mean)}}} & \multicolumn{1}{c}{\thead[c]{\textbf{step} \\ \textbf{errors}}} \\
\midrule
Gemini 3.5 Flash & 19.7 & 20.0 & 18.0 & 20.0 & \textbf{6.6} & \textbf{0.3\%} \\
Gemini 3.6 Flash & 16.3 & 16.5 & 15.0 & 17.0 & 1.6 & 0.7\% \\
Gemini 3.7 Flash & 17.8 & 19.0 & 17.0 & 20.0 & 1.9 & 0.6\% \\
Gemini 3.1 Pro & 13.2 & 12.0 & 10.0 & 13.0 & 3.0 & 1.1\% \\
Claude Opus 4.8 & 12.2 & 12.0 & 11.0 & 12.0 & 1.6 & 0.7\% \\
Claude Opus 5 & 11.1 & 11.0 & 10.0 & 11.0 & 1.5 & 0.9\% \\
Claude Sonnet 5 & \textbf{10.6} & \textbf{10.0} & 10.0 & \textbf{10.0} & \textbf{1.3} & 0.9\% \\
GPT-5.6 Sol & 13.7 & 14.0 & 12.0 & 15.0 & 1.6 & 1.0\% \\
GPT-5.6 Luna & 13.5 & 13.0 & 11.0 & 14.0 & 2.5 & 1.2\% \\
GPT-5.6 Terra & 11.4 & 11.0 & 10.0 & 11.0 & 1.8 & 1.0\% \\
GPT-6 Astra & \textbf{26.1} & \textbf{27.0} & \textbf{26.0} & \textbf{27.0} & 2.3 & \textbf{1.6\%} \\
DeepSeek V4 Pro & 13.4 & 13.0 & 12.0 & 13.0 & 1.6 & 1.2\% \\
DeepSeek V4 Flash & 17.8 & 20.0 & 15.0 & 20.0 & 4.9 & 0.5\% \\
Kimi K3 & 15.4 & 16.0 & 14.0 & 17.0 & 2.8 & 0.4\% \\
GLM-5.2 FP8 & 14.9 & 15.0 & 13.0 & 16.0 & 3.3 & 0.7\% \\
\bottomrule
\end{tabular}
\end{adjustbox}
\caption{\textbf{Step execution dynamics and tail error rates (High Effort).} Breaks down step counts by outcome (successful vs.\ failed rounds), wasted steps taken after the last required environment action, and tool execution error rates.}
\label{tab:rep-07}
\end{table}

\subsection{Failure clustering: the shape of a model's losses}

\label{rep:s7}

\subsubsection{Concentration of losses}

\label{rep:s7.1}

\begin{table}[H]
\centering
\small
\setlength{\tabcolsep}{3.5pt}
\begin{adjustbox}{max width=\linewidth}
\begin{tabular}{l | r r r r r r r}
\toprule
\textbf{Model} & \multicolumn{1}{c}{\thead[c]{\textbf{per-check} \\ \textbf{pass}}} & \multicolumn{1}{c}{\thead[c]{\textbf{observed} \\ \textbf{pass}}} & \multicolumn{1}{c}{\thead[c]{\textbf{if} \\ \textbf{independent}}} & \multicolumn{1}{c}{\textbf{excess}} & \multicolumn{1}{c}{\textbf{lift}} & \multicolumn{1}{c}{\thead[c]{\textbf{95\%} \\ \textbf{CI}}} & \multicolumn{1}{c}{\textbf{overdispersion}} \\
\midrule
GPT-6 Astra & \textbf{66.7\%} & \textbf{28.7\%} & \textbf{13.8\%} & +14.9pp & 1.47 & [1.38, 1.57] & 1.91 \\
Claude Opus 5 & 63.0\% & 25.2\% & 10.8\% & +14.4pp & \textbf{1.49} & [1.39, 1.59] & \textbf{2.23} \\
GPT-5.6 Sol & 62.2\% & 24.5\% & 10.2\% & +14.3pp & 1.42 & [1.33, 1.51] & 2.05 \\
Claude Opus 4.8 & 60.9\% & 22.8\% & 9.3\% & +13.5pp & 1.41 & [1.33, 1.50] & 2.08 \\
Kimi K3 & 55.8\% & 21.8\% & 6.5\% & \textbf{+15.3pp} & 1.40 & [1.33, 1.47] & 2.11 \\
Claude Sonnet 5 & 58.2\% & 20.8\% & 7.7\% & +13.1pp & 1.39 & [1.32, 1.47] & 2.13 \\
GPT-5.6 Terra & 59.3\% & 20.0\% & 8.3\% & +11.7pp & 1.37 & [1.30, 1.45] & 2.01 \\
GPT-5.6 Luna & 59.8\% & 19.8\% & 8.6\% & +11.2pp & 1.36 & [1.28, 1.44] & 1.93 \\
DeepSeek V4 Pro & 57.7\% & 19.8\% & 7.4\% & +12.4pp & 1.35 & [1.28, 1.41] & 1.95 \\
Gemini 3.7 Flash & 57.5\% & 20.3\% & 7.3\% & +13.0pp & 1.30 & [1.24, 1.36] & 1.83 \\
Gemini 3.6 Flash & 58.0\% & 18.5\% & 7.6\% & +10.9pp & 1.28 & [1.23, 1.34] & 1.79 \\
GLM-5.2 FP8 & 48.2\% & 14.7\% & 3.6\% & +11.0pp & 1.27 & [1.22, 1.32] & 1.97 \\
Gemini 3.1 Pro & 48.8\% & 12.1\% & 3.8\% & +8.3pp & 1.19 & [1.15, 1.24] & 1.66 \\
DeepSeek V4 Flash & 38.8\% & 8.5\% & 1.6\% & +6.9pp & 1.16 & [1.12, 1.19] & 1.56 \\
Gemini 3.5 Flash & \textbf{28.1\%} & \textbf{3.0\%} & \textbf{0.5\%} & \textbf{+2.5pp} & \textbf{1.05} & [1.03, 1.08] & \textbf{1.13} \\
\bottomrule
\end{tabular}
\end{adjustbox}
\caption{\textbf{Failure clustering across models (High Effort).} Reports per-check pass rates, observed task pass rates, expected pass rates if check failures were independent, excess pass rate, empirical co-failure lift (with 95\% bootstrap CIs), and error clustering overdispersion.}
\label{tab:rep-08}
\end{table}

\begin{table}[H]
\centering
\setlength{\tabcolsep}{6pt}
\begin{adjustbox}{max width=\linewidth}
\begin{tabular}{l | r r r}
\toprule
\textbf{decomposition} & \multicolumn{1}{c}{\thead[c]{\textbf{per-check} \\ \textbf{component}}} & \multicolumn{1}{c}{\thead[c]{\textbf{clustering} \\ \textbf{component}}} & \multicolumn{1}{c}{\textbf{total}} \\
\midrule
pooled & +0.58 & $-$0.07 & 0.50 \\
stratified & $-$2.33 & +2.83 & 0.50 \\
\bottomrule
\end{tabular}
\end{adjustbox}
\caption{\textbf{Performance difference decomposition between top models (Claude Opus 5 vs.\ GPT-5.6 Sol).} Decomposes the score difference into individual per-check accuracy and joint failure clustering components ($n = 598$, 4,000 bootstrap draws).}
\label{tab:rep-09}
\end{table}

\noindent\smallskip\textbf{Error Clustering Patterns.} Twelve of the 15 models show systematic error clustering: failures are pulled away from intermediate outcomes (failing 1--3 checks) and concentrated at the extremes (passing all checks or failing $\ge 4$ checks). All 15 models show this pattern, with differences across models driven mainly by strength rather than direction.

\noindent\smallskip\textbf{Error Clustering Across Difficulty Levels.} After controlling for baseline task difficulty through peer grouping, co-failure lift remains elevated on easy and intermediate tasks for top models (GPT-6 Astra, Claude Opus 5, GPT-5.6 Sol, Kimi K3). This indicates that agents fail due to all-or-nothing misunderstandings rather than isolated slips on single checks.

\subsubsection{Check-type split}

\label{rep:s7.2}

\begin{table}[H]
\centering
\setlength{\tabcolsep}{6pt}
\begin{adjustbox}{max width=\linewidth}
\begin{tabular}{l | r r r}
\toprule
\textbf{Model} & \multicolumn{1}{c}{\textbf{\texttt{field\_\allowbreak{}semantic\_\allowbreak{}match}}} & \multicolumn{1}{c}{\textbf{\texttt{field\_\allowbreak{}contains}}} & \multicolumn{1}{c}{\textbf{\texttt{other}}} \\
\midrule
GPT-6 Astra & \textbf{65.7} & \textbf{94.7} & \textbf{43.5} \\
Kimi K3 & \textbf{65.7} & 84.9 & 42.6 \\
Claude Opus 5 & 62.7 & 87.9 & 38.2 \\
Gemini 3.7 Flash & 62.2 & 80.7 & 40.4 \\
GPT-5.6 Sol & 62.1 & 87.6 & 38.0 \\
Gemini 3.6 Flash & 61.6 & 83.4 & 38.5 \\
Claude Opus 4.8 & 61.1 & 83.2 & 39.3 \\
GPT-5.6 Luna & 60.9 & 83.8 & 34.7 \\
GPT-5.6 Terra & 60.4 & 79.6 & 33.8 \\
DeepSeek V4 Pro & 60.0 & 84.1 & 33.2 \\
Claude Sonnet 5 & 59.1 & 73.6 & 36.1 \\
GLM-5.2 FP8 & 58.9 & 83.3 & 28.3 \\
DeepSeek V4 Flash & 53.9 & 79.1 & 27.7 \\
Gemini 3.1 Pro & 46.4 & 79.9 & 36.2 \\
Gemini 3.5 Flash & \textbf{45.6} & \textbf{70.8} & \textbf{26.8} \\
\bottomrule
\end{tabular}
\end{adjustbox}
\caption{\textbf{Check-type pass rate breakdown across models (High Effort).} Reports pass rates across LLM-judged semantic matches (\texttt{field\_semantic\_match}), deterministic literal substrings (\texttt{field\_contains}), and auxiliary state assertions (\texttt{other}).}
\label{tab:rep-10}
\end{table}

\subsubsection{\texttt{field\_\allowbreak{}contains}, decomposed}

\label{rep:s7.3}

\begin{table}[H]
\centering
\setlength{\tabcolsep}{6pt}
\begin{adjustbox}{max width=\linewidth}
\begin{tabular}{l | r r r r}
\toprule
\textbf{Model} & \multicolumn{1}{c}{\thead[c]{\textbf{asset} \\ \textbf{gate}}} & \multicolumn{1}{c}{\thead[c]{\textbf{literal} \\ \textbf{content}}} & \multicolumn{1}{c}{\textbf{pooled}} & \multicolumn{1}{c}{\thead[c]{\textbf{content} \\ \textbf{$-$ asset}}} \\
\midrule
GPT-6 Astra & \textbf{98.8} & 81.6 & \textbf{94.7} & $-$17.2 \\
Claude Opus 5 & 89.7 & \textbf{82.4} & 87.9 & $-$7.3 \\
GPT-5.6 Sol & 91.2 & 75.8 & 87.6 & $-$15.4 \\
Kimi K3 & 92.5 & 60.5 & 84.9 & $-$32.0 \\
DeepSeek V4 Pro & 88.9 & 68.8 & 84.1 & $-$20.1 \\
GPT-5.6 Luna & 86.3 & 75.8 & 83.8 & $-$10.5 \\
Gemini 3.6 Flash & 87.8 & 69.1 & 83.4 & $-$18.7 \\
GLM-5.2 FP8 & 93.2 & 51.6 & 83.3 & $-$41.6 \\
Claude Opus 4.8 & 84.8 & 78.1 & 83.2 & $-$6.7 \\
Gemini 3.7 Flash & 84.9 & 67.2 & 80.7 & $-$17.7 \\
Gemini 3.1 Pro & 84.4 & 65.4 & 79.9 & $-$19.0 \\
GPT-5.6 Terra & 80.9 & 75.4 & 79.6 & $-$5.5 \\
DeepSeek V4 Flash & 91.8 & 38.3 & 79.1 & $-$53.5 \\
Claude Sonnet 5 & \textbf{72.6} & 76.6 & 73.6 & \textbf{+4.0} \\
Gemini 3.5 Flash & 87.3 & \textbf{17.6} & \textbf{70.8} & \textbf{$-$69.7} \\
\bottomrule
\end{tabular}
\end{adjustbox}
\caption{\textbf{Decomposition of \texttt{field\_contains} into visual gating and literal transcription (High Effort).} Separates substring evaluation into the visual asset access gate (\texttt{asset gate}) and faithful deliverable transcription (\texttt{literal content}). Both components are verified via deterministic checkers. Rates are over all roles, including unscored diagnostic checks (which sit entirely in the asset gate); Table~\ref{tab:validity} reports the asset gate over scored checks only and therefore prints higher rates.}
\label{tab:rep-11}
\end{table}

\clearpage
\subsection{Difficulty landscape}

\label{rep:s8}

\begingroup
\setlength{\tabcolsep}{6pt}
\begin{xltabular}{\linewidth}{>{\raggedright\arraybackslash\hsize=1.20\hsize}X | >{\raggedright\arraybackslash\hsize=1.10\hsize}X >{\raggedright\arraybackslash\hsize=0.70\hsize}X}
\caption{\textbf{Inventory and suitability of candidate task difficulty axes.} Summarizes potential difficulty dimensions across the 600 task specifications, evaluating their variety of values and usefulness for analysis.}\label{tab:rep-12} \\
\toprule
\textbf{field} & \textbf{cardinality} & \textbf{usable?} \\
\midrule
\endfirsthead
\multicolumn{3}{@{}l@{}}{\itshape Table~\thetable\ (continued)} \\
\toprule
\textbf{field} & \textbf{cardinality} & \textbf{usable?} \\
\midrule
\endhead
\midrule
\multicolumn{3}{@{}r@{}}{\itshape continued on the next page} \\
\endfoot
\bottomrule
\endlastfoot
\texttt{difficulty} & 1 (every task is \texttt{3}) & no \\
\texttt{domains} & 1 (\texttt{composite}, \texttt{conflict}) & no \\
\texttt{template\_\allowbreak{}id} & 1 & no \\
\texttt{weight} & 1 (\texttt{1.0}) & no \\
\texttt{hidden\_\allowbreak{}constraints} / \texttt{visible\_\allowbreak{}constraints} & 0 on every task & no \\
\texttt{decision\_\allowbreak{}requirements} & 1 on every task & no \\
\texttt{availability\_\allowbreak{}tier} / \texttt{online\_\allowbreak{}requirement} & null on every task & no \\
\texttt{realism\_\allowbreak{}tags} & 5, but = \texttt{conflict\_\allowbreak{}vlm\_\allowbreak{}v1} + \texttt{round\_\allowbreak{}type} & redundant \\
\texttt{conflict\_\allowbreak{}taxonomy} & \textbf{absent from v8 entirely} & no \\
\texttt{round\_\allowbreak{}type} & 5 (one with n=1) & \textbf{yes} \\
\texttt{required\_\allowbreak{}modalities} & 2 & yes, with caveats \\
\texttt{allowed\_\allowbreak{}tool\_\allowbreak{}families} & 10 & yes, confounded with scenario \\
\textbf{scored-unit count} & 2--14 per task & \textbf{yes --- the best axis} \\
\end{xltabular}
\endgroup

\subsubsection{Scored-unit count is the one clean difficulty axis}

\label{rep:s8.1}

\begin{table}[H]
\centering
\small
\setlength{\tabcolsep}{3.5pt}
\begin{adjustbox}{max width=\linewidth}
\begin{tabular}{l | r r r r r r}
\toprule
\textbf{Model} & \multicolumn{1}{c}{\thead[c]{\textbf{2--4} \\ \textbf{pass}}} & \multicolumn{1}{c}{\thead[c]{\textbf{5--6} \\ \textbf{pass}}} & \multicolumn{1}{c}{\thead[c]{\textbf{7+} \\ \textbf{pass}}} & \multicolumn{1}{c}{\thead[c]{\textbf{2--4} \\ \textbf{score}}} & \multicolumn{1}{c}{\thead[c]{\textbf{5--6} \\ \textbf{score}}} & \multicolumn{1}{c}{\thead[c]{\textbf{7+} \\ \textbf{score}}} \\
\midrule
Gemini 3.5 Flash & \textbf{3.1} & \textbf{3.5} & \textbf{2.0} & \textbf{0.160} & \textbf{0.303} & \textbf{0.311} \\
Gemini 3.6 Flash & 29.7 & 17.1 & 6.7 & 0.585 & 0.614 & 0.538 \\
Gemini 3.7 Flash & 31.8 & 19.4 & 7.3 & 0.572 & 0.619 & 0.528 \\
Gemini 3.1 Pro & 20.0 & 10.1 & 5.4 & 0.469 & 0.523 & 0.464 \\
Claude Opus 4.8 & 33.3 & 21.7 & 11.3 & 0.638 & 0.652 & 0.543 \\
Claude Opus 5 & 34.4 & 24.0 & 15.3 & 0.655 & 0.670 & 0.576 \\
Claude Sonnet 5 & 32.8 & 18.2 & 10.0 & 0.621 & 0.623 & 0.518 \\
GPT-5.6 Sol & 35.9 & 21.3 & 15.3 & 0.654 & 0.645 & 0.584 \\
GPT-5.6 Luna & 28.6 & 19.8 & 8.7 & 0.606 & 0.627 & 0.561 \\
GPT-5.6 Terra & 28.6 & 19.8 & 9.3 & 0.618 & 0.619 & 0.553 \\
GPT-6 Astra & \textbf{39.6} & \textbf{24.8} & \textbf{21.3} & \textbf{0.661} & \textbf{0.685} & \textbf{0.647} \\
DeepSeek V4 Pro & 27.6 & 18.2 & 12.7 & 0.571 & 0.608 & 0.548 \\
DeepSeek V4 Flash & 9.9 & 9.3 & 5.3 & 0.279 & 0.444 & 0.380 \\
Kimi K3 & 29.2 & 21.7 & 12.7 & 0.536 & 0.594 & 0.529 \\
GLM-5.2 FP8 & 22.4 & 12.8 & 8.0 & 0.468 & 0.518 & 0.457 \\
\textbf{pooled} & 27.1 & 17.5 & 10.1 & 0.540 & 0.583 & 0.515 \\
\bottomrule
\end{tabular}
\end{adjustbox}
\caption{\textbf{Benchmark pass rate and check score grouped by scored-unit count (High Effort).} Compares performance across low (2--4 units, $n=192$), medium (5--6 units, $n=258$), and high (7+ units, $n=150$) check count groups (3,284 total evaluated units).}
\label{tab:rep-13}
\end{table}

\subsubsection{Task difficulty varies by round type independent of check count}

\label{rep:s8.2}

\begin{table}[H]
\centering
\small
\setlength{\tabcolsep}{3.5pt}
\begin{adjustbox}{max width=\linewidth}
\begin{tabular}{l | r r r r r r}
\toprule
\thead[c]{\textbf{round\_} \\ \textbf{type}} & \multicolumn{1}{c}{\textbf{n}} & \multicolumn{1}{c}{\thead[c]{\textbf{mean} \\ \textbf{units}}} & \multicolumn{1}{c}{\thead[c]{\textbf{2--4} \\ \textbf{band}}} & \multicolumn{1}{c}{\thead[c]{\textbf{5--6} \\ \textbf{band}}} & \multicolumn{1}{c}{\thead[c]{\textbf{7+} \\ \textbf{band}}} & \multicolumn{1}{c}{\textbf{marginal}} \\
\midrule
investigation & 274 & 4.91 & 27.5 [119] & 13.4 [112] & \textbf{2.3} [43] & 17.8 \\
verification & 174 & 5.83 & \textbf{28.6} [39] & 18.1 [80] & 11.2 [55] & 18.3 \\
action & 86 & 6.55 & 21.0 [17] & \textbf{26.7} [31] & 13.7 [38] & \textbf{19.9} \\
decision & 65 & 5.49 & 17.4 [16] & 16.7 [35] & 11.2 [14] & 15.7 \\
\bottomrule
\end{tabular}
\end{adjustbox}
\caption{\textbf{Pass rate distribution across round operational types.} Compares pass rates across investigation, verification, action, and decision task types, broken down by rubric unit count.}
\label{tab:rep-14}
\end{table}

\subsubsection{Natural photographs do not increase task difficulty}

\label{rep:s8.3}

\begin{table}[H]
\centering
\setlength{\tabcolsep}{6pt}
\begin{adjustbox}{max width=\linewidth}
\begin{tabular}{l | r r r r r}
\toprule
\textbf{band} & \multicolumn{1}{c}{\thead[c]{\textbf{n (no-} \\ \textbf{img)}}} & \multicolumn{1}{c}{\thead[c]{\textbf{n} \\ \textbf{(img)}}} & \multicolumn{1}{c}{\thead[c]{\textbf{no-img} \\ \textbf{pass}}} & \multicolumn{1}{c}{\thead[c]{\textbf{img} \\ \textbf{pass}}} & \multicolumn{1}{c}{\textbf{diff}} \\
\midrule
2--4 & 120 & 72 & 24.8 & 28.6 & +3.8 \\
5--6 & 223 & 35 & 17.1 & 16.0 & $-$1.1 \\
7+ & 137 & 13 & 9.8 & 4.4 & \textbf{$-$5.4} \\
\bottomrule
\end{tabular}
\end{adjustbox}
\caption{\textbf{Stratified comparison of natural-image versus chart-only tasks.} Reports pass rates for tasks requiring natural photographic grounding versus structured charts across rubric unit bands ($n=600$).}
\label{tab:rep-15}
\end{table}

\subsubsection{Tool family (confounded, reported for completeness)}

\label{rep:s8.4}

\begin{table}[H]
\centering
\setlength{\tabcolsep}{6pt}
\begin{adjustbox}{max width=\linewidth}
\begin{tabular}{l | r r}
\toprule
\textbf{family} & \multicolumn{1}{c}{\textbf{n}} & \multicolumn{1}{c}{\thead[c]{\textbf{pooled} \\ \textbf{pass}}} \\
\midrule
lock + review & 30 & \textbf{32.2\%} \\
shell\_scripting & 150 & 22.4\% \\
approval & 90 & 20.0\% \\
deploy & 90 & 19.5\% \\
\textit{(none)} & 60 & 19.0\% \\
pairing & 30 & 16.2\% \\
lock & 30 & 10.5\% \\
containment & 30 & 10.0\% \\
booking & 60 & 9.3\% \\
tasks & 30 & \textbf{4.1\%} \\
\bottomrule
\end{tabular}
\end{adjustbox}
\caption{\textbf{Pooled pass rate across domain tool families.} Summarizes aggregate pass rates across specialized domain CLI interfaces beyond the core toolset.}
\label{tab:rep-16}
\end{table}

\subsubsection{Gold-trajectory length}

\label{rep:s8.5}

\begin{table}[H]
\centering
\setlength{\tabcolsep}{6pt}
\begin{adjustbox}{max width=\linewidth}
\begin{tabular}{l | r r r r}
\toprule
\textbf{Model} & \multicolumn{1}{c}{\textbf{short}} & \multicolumn{1}{c}{\textbf{mid}} & \multicolumn{1}{c}{\textbf{long}} & \multicolumn{1}{c}{\textbf{all}} \\
\midrule
Gemini 3.5 Flash & \textbf{3.5} & \textbf{3.2} & \textbf{2.1} & \textbf{3.0} \\
Gemini 3.6 Flash & 26.5 & 21.9 & 3.6 & 18.5 \\
Gemini 3.7 Flash & 30.1 & 22.5 & 7.1 & 20.3 \\
Gemini 3.1 Pro & 18.6 & 12.4 & 5.7 & 12.0 \\
Claude Opus 4.8 & 32.7 & 25.1 & 9.3 & 22.8 \\
Claude Opus 5 & 36.3 & 26.2 & 13.6 & 25.2 \\
Claude Sonnet 5 & 32.7 & 22.2 & 7.9 & 20.8 \\
GPT-5.6 Sol & \textbf{37.2} & 24.8 & 13.6 & 24.5 \\
GPT-5.6 Luna & 23.0 & 23.1 & 9.3 & 19.8 \\
GPT-5.6 Terra & 28.3 & 22.2 & 7.9 & 20.0 \\
GPT-6 Astra & 35.2 & \textbf{28.1} & \textbf{18.6} & \textbf{28.7} \\
DeepSeek V4 Pro & 24.8 & 21.6 & 11.4 & 19.8 \\
DeepSeek V4 Flash & 9.7 & 9.8 & 4.3 & 8.5 \\
Kimi K3 & 30.1 & 23.1 & 12.1 & 21.8 \\
GLM-5.2 FP8 & 19.5 & 16.7 & 5.7 & 14.7 \\
\bottomrule
\end{tabular}
\end{adjustbox}
\caption{\textbf{Pass rate by reference trajectory length (High Effort).} Divides tasks into short, medium, and long reference trajectory groups ($n = 113 / 347 / 140$, maximum reference length 18 steps).}
\label{tab:rep-17}
\end{table}

\noindent\smallskip\textbf{Step Usage Relative to Reference Solutions.} Comparing steps taken against expert reference solutions shows that successful episodes complete tasks close to the reference length, whereas failed runs frequently run out of steps at the 20-step limit. On longer tasks, models run out of steps much more often.

\noindent\smallskip\textbf{Step Count Overshoot Distribution.} Successful runs closely match the reference length with near-zero extra steps, whereas search loops and repeated polling cause models to hit the step limit.

\noindent\smallskip\textbf{Designed Difficulty Levers and Capability Shifts.} Three designed environment levers (\texttt{async\_required}, \texttt{coordinated\_containment}, and \texttt{credential\_lock}) reduce pass rates across all 15 evaluated models (with \texttt{gap\_completion} reducing pass rates across 14 of 15 models). Tasks requiring asynchronous polling and active security containment cause the largest performance drops across all models.

\clearpage
\subsection{The reasoning-effort sweep}

\label{rep:s9}

\subsubsection{Reasoning token consumption across effort levels}

\label{rep:s9.1}

\begin{table}[H]
\centering
\setlength{\tabcolsep}{6pt}
\begin{adjustbox}{max width=\linewidth}
\begin{tabular}{l | r r r r}
\toprule
\textbf{Model} & \multicolumn{1}{c}{\thead[c]{\textbf{reasoning tok/} \\ \textbf{call (l/m/h)}}} & \multicolumn{1}{c}{\thead[c]{\textbf{reasoning tok/} \\ \textbf{task (l/m/h)}}} & \multicolumn{1}{c}{\textbf{low$\rightarrow$high}} & \multicolumn{1}{c}{\thead[c]{\textbf{zero-reasoning} \\ \textbf{calls (l/m/h)}}} \\
\midrule
Gemini 3.5 Flash & 236.3 / 449.3 / \textbf{445.8} & 3,945 / 8,800 / 8,826 & 1.89$\times$ & 0.6\% / 0.4\% / 0.4\% \\
Gemini 3.6 Flash & 65.3 / 231.3 / 252.7 & 704 / 3,568 / 4,266 & 3.87$\times$ & \textbf{57.3\%} / 0.0\% / 0.0\% \\
Gemini 3.7 Flash & 83.4 / 165.6 / 233.0 & 1,079 / 2,578 / 4,237 & 2.80$\times$ & 39.1\% / 0.0\% / 0.0\% \\
Gemini 3.1 Pro & 93.6 / 118.0 / 171.2 & 1,002 / 1,394 / 2,359 & 1.83$\times$ & 1.0\% / 0.1\% / 0.2\% \\
Claude Opus 4.8 & 68.6 / 126.1 / 177.8 & 717 / 1,493 / 2,291 & 2.59$\times$ & 64.1\% / 57.2\% / 51.4\% \\
Claude Opus 5 & \textbf{5.1 / 17.6 / 46.5} & \textbf{45 / 184 / 553} & \textbf{9.07$\times$} & 95.3\% / 88.7\% / 76.3\% \\
Claude Sonnet 5 & 49.6 / 114.4 / 249.8 & 406 / 1,095 / 2,847 & 5.03$\times$ & 72.7\% / 64.5\% / 53.0\% \\
GPT-5.6 Sol & 27.7 / 57.8 / 94.0 & 301 / 760 / 1,354 & 3.39$\times$ & 10.6\% / 9.7\% / 10.2\% \\
GPT-5.6 Luna & 29.1 / 54.0 / 140.0 & 289 / 632 / 1,994 & 4.82$\times$ & 23.4\% / 26.6\% / 29.0\% \\
GPT-5.6 Terra & 25.3 / 57.2 / 142.5 & 243 / 625 / 1,760 & 5.63$\times$ & 0.4\% / 0.1\% / 0.1\% \\
\bottomrule
\end{tabular}
\end{adjustbox}
\caption{\textbf{Reasoning token generation across effort levels.} Reports internal reasoning tokens generated per tool call, per task, scaling factor ($\text{Low} \to \text{High}$), and frequency of zero-reasoning calls across the ten swept API models.}
\label{tab:rep-18}
\end{table}

\subsubsection{Performance impact of reasoning effort scaling}

\label{rep:s9.2}

\begin{table}[H]
\centering
\setlength{\tabcolsep}{6pt}
\begin{adjustbox}{max width=\linewidth}
\begin{tabular}{l | r r r r}
\toprule
\textbf{Model} & \multicolumn{1}{c}{\thead[c]{\textbf{pass low/} \\ \textbf{med/high}}} & \multicolumn{1}{c}{\thead[c]{\textbf{score low/} \\ \textbf{med/high}}} & \multicolumn{1}{c}{\thead[c]{\textbf{mean steps} \\ \textbf{l/m/h}}} & \multicolumn{1}{c}{\thead[c]{\textbf{capped\%} \\ \textbf{l/m/h}}} \\
\midrule
Claude Opus 5 & 23.5 / \textbf{25.3} / 25.2 & 0.609 / 0.633 / \textbf{0.642} & 7.99 / 9.65 / 11.12 & 0.0 / 0.0 / 0.5 \\
GPT-5.6 Sol & 20.8 / 23.0 / \textbf{24.5} & 0.610 / 0.628 / \textbf{0.633} & 10.04 / 12.41 / 13.65 & 0.2 / 2.0 / 1.2 \\
Claude Opus 4.8 & 21.3 / \textbf{23.8} / 22.8 & 0.608 / 0.618 / \textbf{0.621} & 9.67 / 11.11 / 12.17 & 0.8 / 2.8 / 6.2 \\
Claude Sonnet 5 & 13.8 / 17.0 / \textbf{20.8} & 0.537 / 0.570 / \textbf{0.596} & 7.31 / 8.76 / 10.64 & 0.3 / 1.0 / 3.3 \\
Gemini 3.7 Flash & 20.5 / \textbf{21.5} / 20.3 & 0.600 / \textbf{0.618} / 0.581 & 12.18 / 14.91 / 17.79 & 3.0 / 12.7 / 40.7 \\
GPT-5.6 Terra & 17.7 / 20.0 / 20.0 & 0.572 / 0.589 / \textbf{0.602} & 8.65 / 9.98 / 11.35 & 0.3 / 1.3 / 1.3 \\
GPT-5.6 Luna & 19.5 / 17.7 / \textbf{19.8} & 0.590 / 0.591 / \textbf{0.604} & 9.12 / 10.91 / 13.54 & 0.5 / 2.7 / 9.2 \\
Gemini 3.6 Flash & 18.5 / \textbf{20.2} / 18.5 & 0.595 / \textbf{0.602} / 0.586 & 9.99 / 14.77 / 16.31 & 1.8 / 14.5 / 24.7 \\
Gemini 3.1 Pro & 8.3 / 9.8 / \textbf{12.0} & 0.463 / 0.479 / \textbf{0.487} & 9.84 / 11.02 / 13.16 & 4.7 / 10.2 / 25.3 \\
Gemini 3.5 Flash & \textbf{13.5} / 3.2 / 3.0 & \textbf{0.473} / 0.299 / 0.259 & 16.22 / 19.46 / 19.70 & 38.7 / 85.0 / 89.2 \\
\bottomrule
\end{tabular}
\end{adjustbox}
\caption{\textbf{Performance metrics across reasoning effort configurations.} Reports pass rate, mean check score, average steps per episode, and budget cap exhaustion rate across Low, Medium, and High effort levels for all ten API models.}
\label{tab:rep-19}
\end{table}

\begin{table}[H]
\centering
\setlength{\tabcolsep}{6pt}
\begin{adjustbox}{max width=\linewidth}
\begin{tabular}{l | r r r r l}
\toprule
\textbf{Model} & \multicolumn{1}{c}{\textbf{$\Delta$pp}} & \multicolumn{1}{c}{\textbf{lost}} & \multicolumn{1}{c}{\textbf{gained}} & \multicolumn{1}{c}{\textbf{p}} & \textbf{verdict} \\
\midrule
Claude Sonnet 5 & \textbf{+7.0} & 11 & \textbf{53} & 1.0e-07 & \textbf{real} \\
Gemini 3.1 Pro & +3.7 & \textbf{9} & 31 & 6.8e-04 & \textbf{real} \\
GPT-5.6 Sol & +3.7 & 12 & 34 & 1.6e-03 & \textbf{real} \\
GPT-5.6 Terra & +2.3 & 15 & 29 & 0.049 & \textbf{real} \\
Claude Opus 5 & +1.7 & 23 & 33 & 0.23 & n.s. \\
Claude Opus 4.8 & +1.5 & 18 & 27 & 0.23 & n.s. \\
GPT-5.6 Luna & +0.3 & 24 & 26 & 0.89 & n.s. \\
Gemini 3.6 Flash & 0.0 & 34 & 34 & 1.00 & n.s. \\
Gemini 3.7 Flash & $-$0.2 & 34 & 33 & 1.00 & n.s. \\
Gemini 3.5 Flash & \textbf{$-$10.5} & \textbf{63} & \textbf{0} & 2.2e-19 & \textbf{real (negative)} \\
\bottomrule
\end{tabular}
\end{adjustbox}
\caption{\textbf{Paired McNemar significance tests across reasoning effort scaling ($\text{Low} \to \text{High}$).} Evaluates task-level changes over identical 600 rounds per model ($n=600$). Significance verdicts reflect uncorrected ($\alpha = 0.05$) and Bonferroni-corrected family-wise thresholds ($\alpha = 0.05 / 30 = 0.00167$).}
\label{tab:rep-20}
\end{table}

\subsection{Qualitative Analysis: Representative Conflict Scenarios and Model Signatures}

\label{rep:s10}

To examine detailed agent behavior and where models diverge, this section provides step-level walkthroughs across eight conflict scenarios with large score differences across the 14 20-step baseline models, followed by 14 case studies showing each baseline model's primary failure mode.

\subsubsection*{Paired Tool-Execution Chains across Eight Representative Scenarios}

Each exhibit shows all 14 20-step baseline models tested on the exact same task instance. Rows show sequential tool calls (colored by action type; red circles indicate tool execution errors), along with final check results, step counts, visual assets opened, files written, and termination reasons.

{\footnotesize\raggedright
\textcolor[HTML]{8B5CF6}{$\bullet$}\,\texttt{asset~open}\quad
\textcolor[HTML]{0EA5E9}{$\bullet$}\,\texttt{file~read}\quad
\textcolor[HTML]{10B981}{$\bullet$}\,\texttt{message~broadcast/send}\quad
\textcolor[HTML]{14B8A6}{$\bullet$}\,\texttt{message~poll/search}\quad
\textcolor[HTML]{C4B5FD}{$\bullet$}\,\texttt{asset~describe/info/list/read-text}\quad
\textcolor[HTML]{EC4899}{$\bullet$}\,\texttt{file~append/create}\quad
\textcolor[HTML]{CBD5E1}{$\bullet$}\,\texttt{booking~hold/status}, \texttt{containment~block-token/disable-forwarding/status}, \texttt{lock~acquire/status}, \texttt{system~cron/status}, \texttt{tasks~list}\quad
\textcolor[HTML]{F59E0B}{$\bullet$}\,\texttt{wait}\quad
\textcolor[HTML]{7DD3FC}{$\bullet$}\,\texttt{file~list}\quad
\textcolor[HTML]{94A3B8}{$\bullet$}\,\texttt{agent~list/status}\quad
\textcolor[HTML]{6366F1}{$\bullet$}\,\texttt{agent~delegate}\quad
\textcolor[HTML]{EF4444}{$\circ$}\,a~red~ring~marks~a~step~whose~tool~returned~an~error\par}

\begin{figure}[H]
\centering
\includegraphics[width=\linewidth]{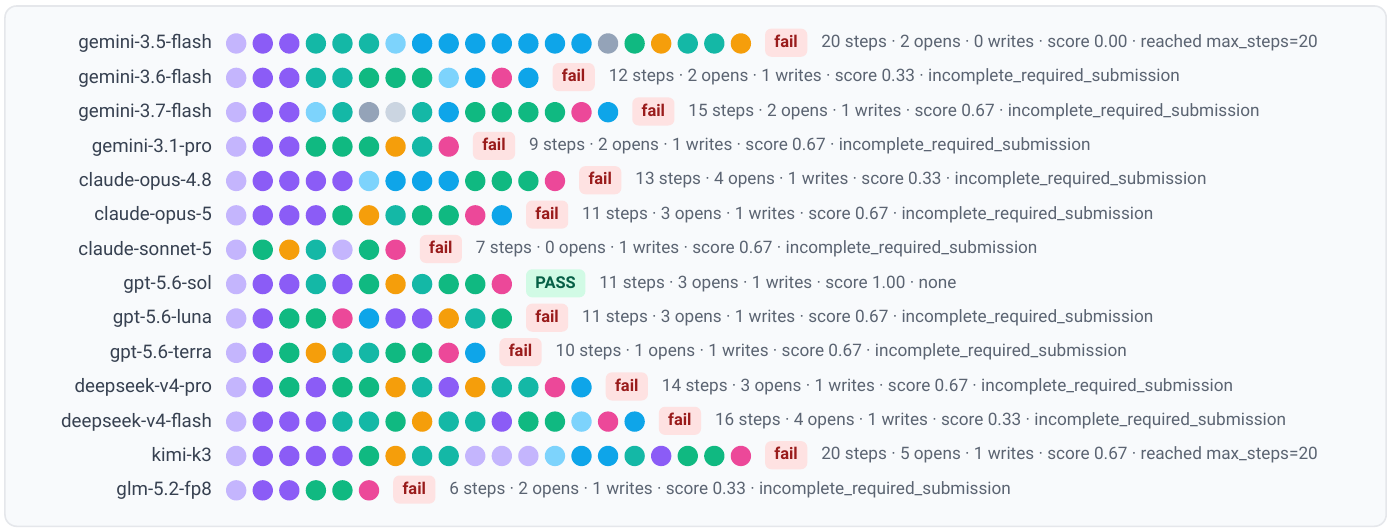}
\caption{\textbf{Representative Execution Trace: Payment Chargeback Dispute (\texttt{v01\_r04}; Figure~\ref{fig:rep-round-1-payment-chargeback-dispute}).} Compares agent action traces on credit verification, report delivery, and notifications under conflicting financial ledgers.}
\label{fig:rep-round-1-payment-chargeback-dispute}
\end{figure}

\noindent\smallskip\textbf{Case Analysis: Payment Chargeback Dispute (\texttt{v01\_r04}).}
This scenario evaluates multi-agent coordination under conflicting billing records across three scored checks: (1) extracting the validation code \texttt{PASSPORT-VERIFY-9982}, (2) reporting the chargeback dispute in \texttt{reports/support\_escrow\_key\_verification.txt}, and (3) sending the code to \texttt{Casey\_Support\_Specialist}.

All 14 baseline models send a message to \texttt{Casey\_Support\_Specialist} and write to the correct report path. However, only three models (GPT-5.6 Sol at step 5, DeepSeek V4 Pro at step 4, and GPT-5.6 Luna at step 10) include the exact required code in their message. GPT-5.6 Sol is the only model to pass completely (11 steps), correctly writing the report and sending the notification. Eight models achieve partial credit ($0.67$) by writing both required findings in a single report but missing the exact message content. Notably, Claude Sonnet 5 finishes in 7 steps without opening any images by skipping the diagnostic escrow panel, whereas Kimi K3 uses all 20 steps and opens 5 images before writing its report at step 19.

\begin{figure}[H]
\centering
\includegraphics[width=\linewidth]{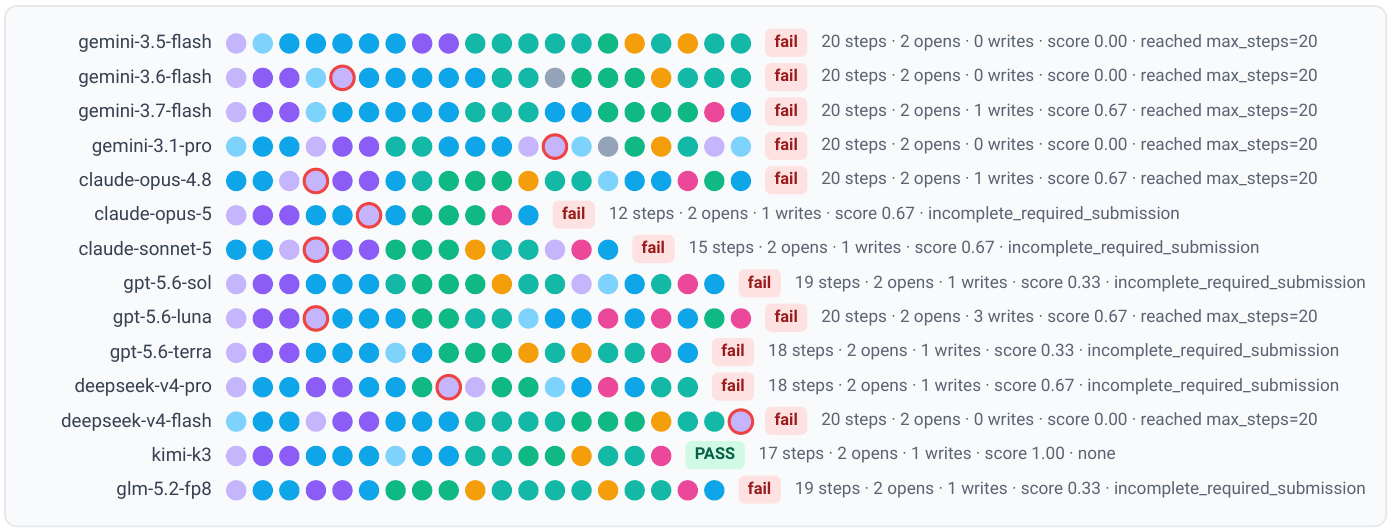}
\caption{\textbf{Representative Execution Trace: Warehouse Robotics Safety Halt (\texttt{v02\_r01}; Figure~\ref{fig:rep-round-2-warehouse-robotics-safety-halt}).} Compares emergency threshold verification, thermal sensor reading extraction (\texttt{88.5\,$^\circ$C}), and safety channel notification across models.}
\label{fig:rep-round-2-warehouse-robotics-safety-halt}
\end{figure}

\noindent\smallskip\textbf{Case Analysis: Warehouse Robotics Safety Halt (\texttt{v02\_r01}).}
This scenario requires identifying thermal sensor anomalies across three scored units: (1) extracting the temperature reading \texttt{88.5\,$^\circ$C}, (2) filing the safety halt verdict in the incident report, and (3) broadcasting the emergency alert to \texttt{Avery\_Safety\_Officer} in channel \texttt{\#ops-safety-emergency-b7}.

Kimi K3 is the sole model to pass (17 steps), issuing the alert at step 11 and committing the report at step 16. Six models (Claude Opus 5, Claude Sonnet 5, DeepSeek V4 Pro, Gemini 3.7 Flash, Claude Opus 4.8, GPT-5.6 Luna) score $0.67$ by satisfying both report checks while missing the channel notification. Four models (Gemini 3.5 Flash, Gemini 3.6 Flash, Gemini 3.1 Pro, DeepSeek V4 Flash) score $0.00$, exhausting the 20-step budget with zero file writes. Failure modes illustrate three distinct loss mechanisms: Gemini 3.1 Pro misidentifies the recipient (messaging \texttt{Blair\_Robotics\_Engineer} instead), GPT-5.6 Luna substitutes \texttt{message broadcast} for direct messaging, and the remaining models fail on message payload syntax.

\begin{figure}[H]
\centering
\includegraphics[width=\linewidth]{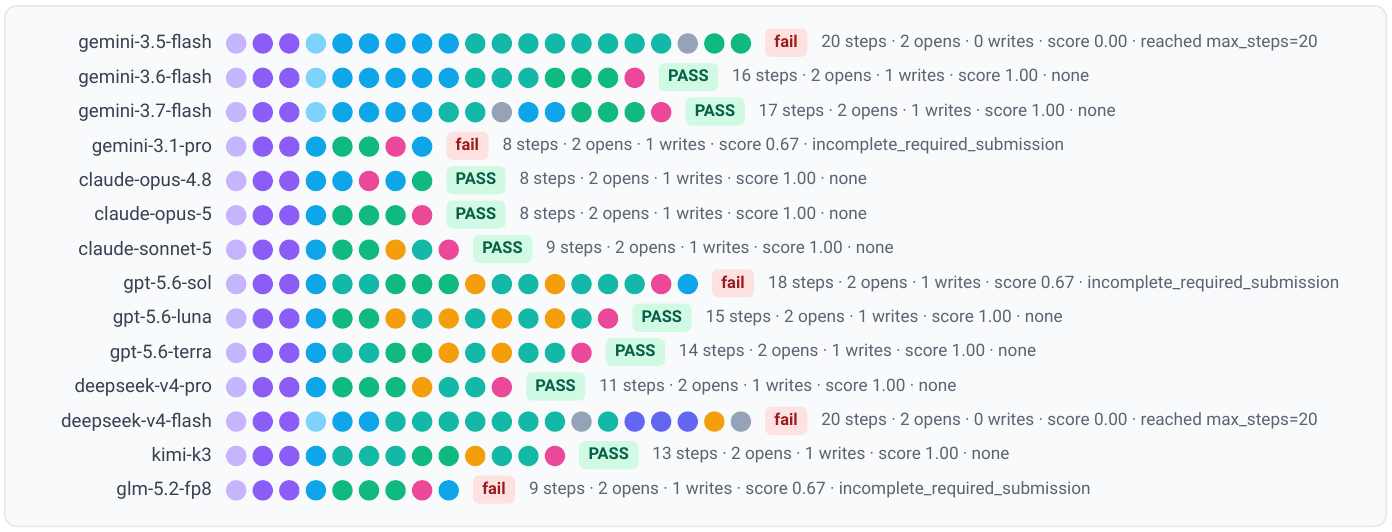}
\caption{\textbf{Representative Execution Trace: Satellite Ground Station Handover (\texttt{v02\_r01}; Figure~\ref{fig:rep-round-3-satellite-ground-station-handover}).} Illustrates telemetry alarm resolution (\texttt{ERR-LOCK-049-DRV}) and operations handover messaging under uniform visual grounding.}
\label{fig:rep-round-3-satellite-ground-station-handover}
\end{figure}

\noindent\smallskip\textbf{Case Analysis: Satellite Ground Station Handover (\texttt{v02\_r01}).}
The rubric comprises three units: (1) transcribing the diagnostic alert \texttt{ERR-LOCK-049-DRV}, (2) stating the handover status verdict, and (3) notifying \texttt{Devon\_Ground\_Ops\_Lead} in channel \texttt{\#ops-handover}.

Nine of the 14 baseline models successfully pass. Crucially, all 14 baseline models open exactly 2 visual assets, providing an empirical control where grounding volume is identical across the entire field. The primary distinguishing factor is the final verdict inference: three models (Gemini 3.1 Pro, GLM-5.2 FP8, GPT-5.6 Sol) extract the code and notify the operator but fail the conclusion check ($0.67$), while two models (Gemini 3.5 Flash, DeepSeek V4 Flash) exhaust the action budget without writing ($0.00$). Execution horizon spans 8 steps (Claude Opus 4.8, Claude Opus 5) to 16 steps (Gemini 3.6 Flash) among passing trajectories.

\begin{figure}[H]
\centering
\includegraphics[width=\linewidth]{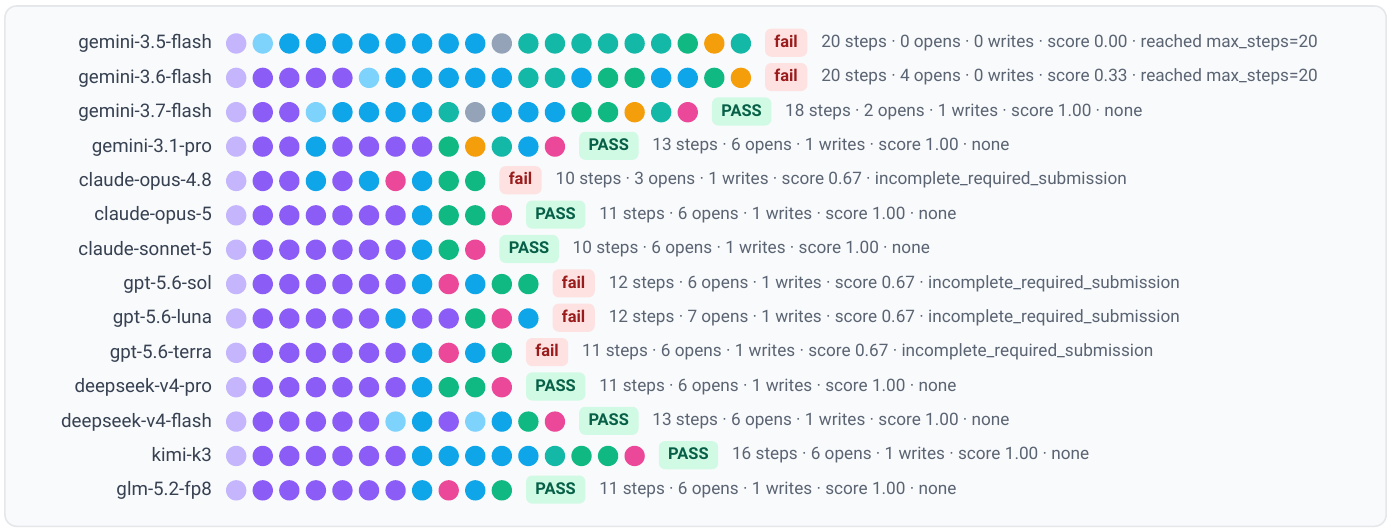}
\caption{\textbf{Representative Execution Trace: Clinical Trial Dosage Conflict (\texttt{v02\_r03}; Figure~\ref{fig:rep-round-4-clinical-trial-dosage-conflict}).} Highlights divergence in causal interpretation (control cohort mislabeling vs.\ toxicity) despite identical numerical extraction (\texttt{142\,U/L}).}
\label{fig:rep-round-4-clinical-trial-dosage-conflict}
\end{figure}

\noindent\smallskip\textbf{Case Analysis: Clinical Trial Dosage Conflict (\texttt{v02\_r03}).}
This scenario tests causal conflict resolution across three units: (1) extracting the liver enzyme reading \texttt{142\,U/L} into \texttt{reports/cohort\_audit.md}, (2) alerting \texttt{Avery\_Trial\_Safety\_Monitor} in \texttt{\#pv-safety-governance}, and (3) correctly identifying that elevated readings stem from control-cohort mislabeling rather than true drug toxicity.

Eight models successfully pass. Literal extraction and recipient routing are universal among all models that generate deliverables. However, four models (Claude Opus 4.8, GPT-5.6 Terra, GPT-5.6 Sol, GPT-5.6 Luna) score $0.67$ due to misattributing the anomaly to pharmaceutical toxicity. Grounding volume does not predict accuracy: Gemini 3.7 Flash passes with 2 asset opens, whereas Claude Opus 4.8 fails with 3 opens, and Gemini 3.5 Flash scores $0.00$ with zero opens and zero writes.

\begin{figure}[H]
\centering
\includegraphics[width=\linewidth]{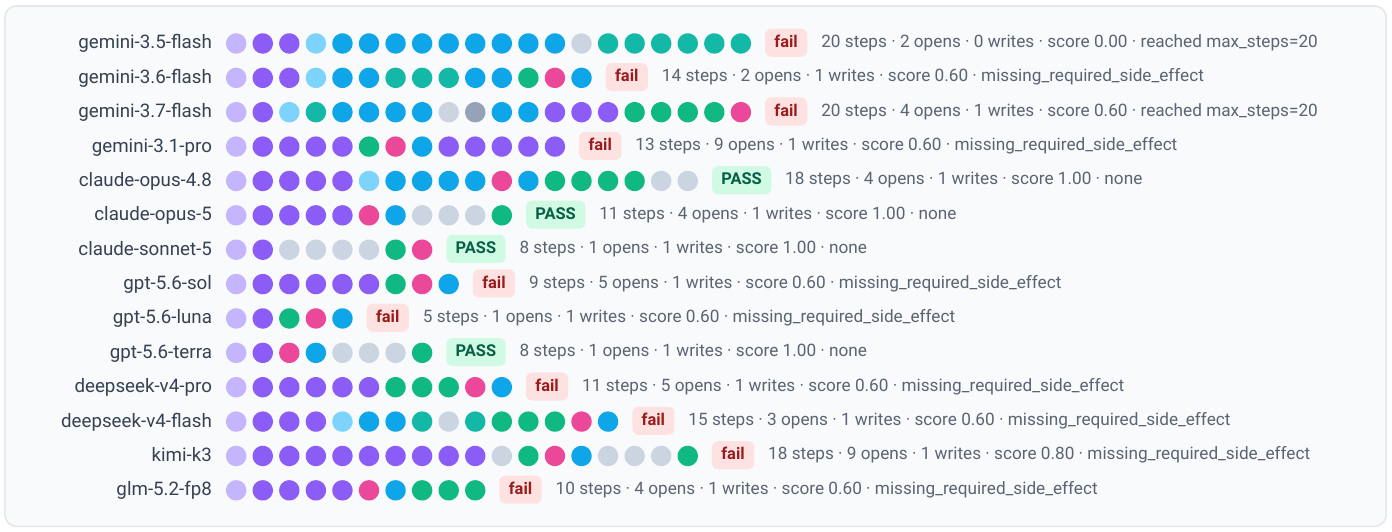}
\caption{\textbf{Representative Execution Trace: Coordinated Security Containment (\texttt{v02\_r04}; Figure~\ref{fig:rep-round-5-coordinated-containment}).} Shows the gap between writing diagnostic reports and performing required actions in the environment (revoking an API token and blocking email forwarding).}
\label{fig:rep-round-5-coordinated-containment}
\end{figure}

\noindent\smallskip\textbf{Case Analysis: Coordinated Security Containment (\texttt{v02\_r04}).}
This scenario requires coordinated incident response across five checks: two environment actions (blocking \texttt{standard\_user\_449} and revoking API token \texttt{tok\_live\_sec\_99321}), reporting the anomaly rate \texttt{142\,msgs/hr}, reporting the incident verdict, and notifying the on-call engineer.

While nearly all models write the report, taking real actions in the environment is the main challenge. Exactly four models (Claude Sonnet 5, GPT-5.6 Terra, Claude Opus 5, Claude Opus 4.8) execute both containment actions with correct parameters, accounting for all four passes. Eight models execute neither action, stopping with a \texttt{missing\_required\_side\_effect} error after writing an accurate report about a breach they failed to contain ($0.60$). Kimi K3 gets partial credit ($0.80$) due to wrong arguments (disabling forwarding for \texttt{j.doe@company.com} instead of \texttt{standard\_user\_449}).

\begin{figure}[H]
\centering
\includegraphics[width=\linewidth]{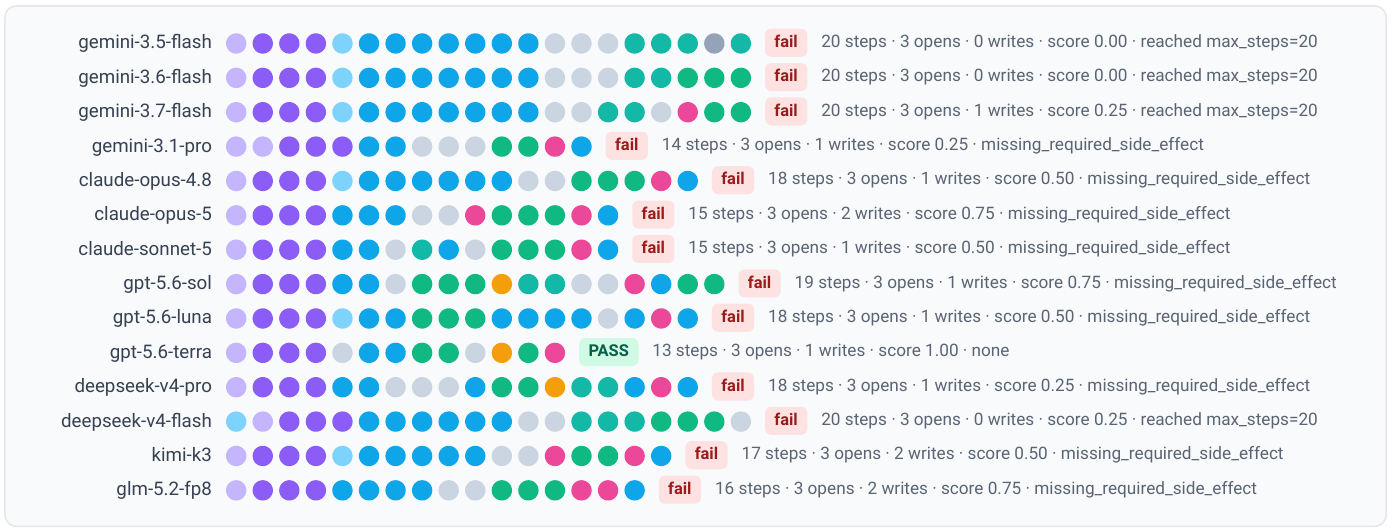}
\caption{\textbf{Representative Execution Trace: Distributed Lock Conflict Resolution (\texttt{v01\_r02}; Figure~\ref{fig:rep-round-6-distributed-lock-shared-resource}).} Examines identifying race conditions and acquiring the required lock on \texttt{inventory\_ledger}.}
\label{fig:rep-round-6-distributed-lock-shared-resource}
\end{figure}

\noindent\smallskip\textbf{Case Analysis: Distributed Lock Conflict Resolution (\texttt{v01\_r02}).}
The rubric evaluates conflict resolution across four checks: identifying the state \texttt{UNLOCKED (Un-serialized Write Blocked)}, writing the audit report, acquiring the lock on \texttt{inventory\_ledger}, and sending the confirmation notification.

GPT-5.6 Terra is the only model to pass (13 steps). All 14 baseline models open all 3 images. However, acquiring the shared ledger lock is difficult: Claude Sonnet 5 acquires the lock at step 9 but scores $0.50$ due to incomplete report and message contents, showing that acquiring locks is necessary but not sufficient without accurate deliverables.

\begin{figure}[H]
\centering
\includegraphics[width=\linewidth]{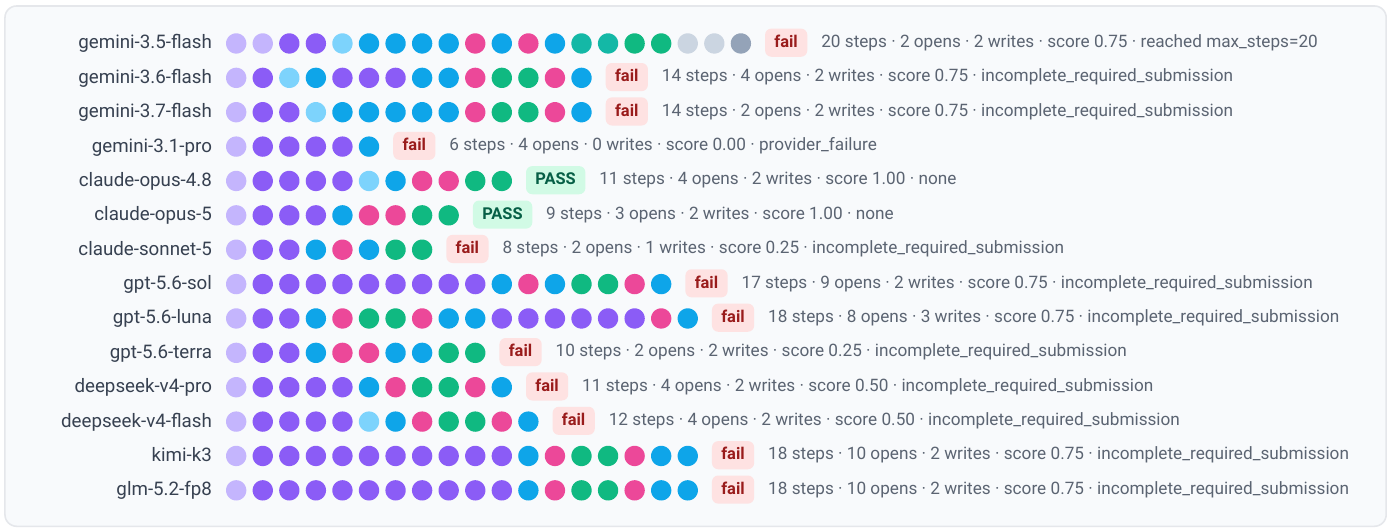}
\caption{\textbf{Representative Execution Trace: Expense Reimbursement Audit (\texttt{v02\_r05}; Figure~\ref{fig:rep-round-7-expense-reimbursement-conflict}).} Contrasts telemetry configuration changes, audit passcode delivery, and multi-recipient stakeholder notifications.}
\label{fig:rep-round-7-expense-reimbursement-conflict}
\end{figure}

\noindent\smallskip\textbf{Case Analysis: Expense Reimbursement Audit (\texttt{v02\_r05}).}
This scenario requires modifying environment configuration across four units: reading \texttt{config/telemetry\_settings.json}, rewriting it with audit passcode \texttt{AUDIT-8842-REV}, generating the report, and notifying both \texttt{Blair\_Audit\_Investigator} and \texttt{Casey\_Approver}.

Claude Opus 5 (9 steps) and Claude Opus 4.8 (11 steps) achieve clean passes. Seven models form a $0.75$ plateau by successfully rewriting the configuration file and sending notifications but missing report precision. Claude Sonnet 5 acts without modifying the configuration file (scoring $0.25$ in 8 steps), while GPT-5.6 Terra scores $0.25$ via the inverse pattern (rewriting configuration but failing on notification delivery).

\begin{figure}[H]
\centering
\includegraphics[width=\linewidth]{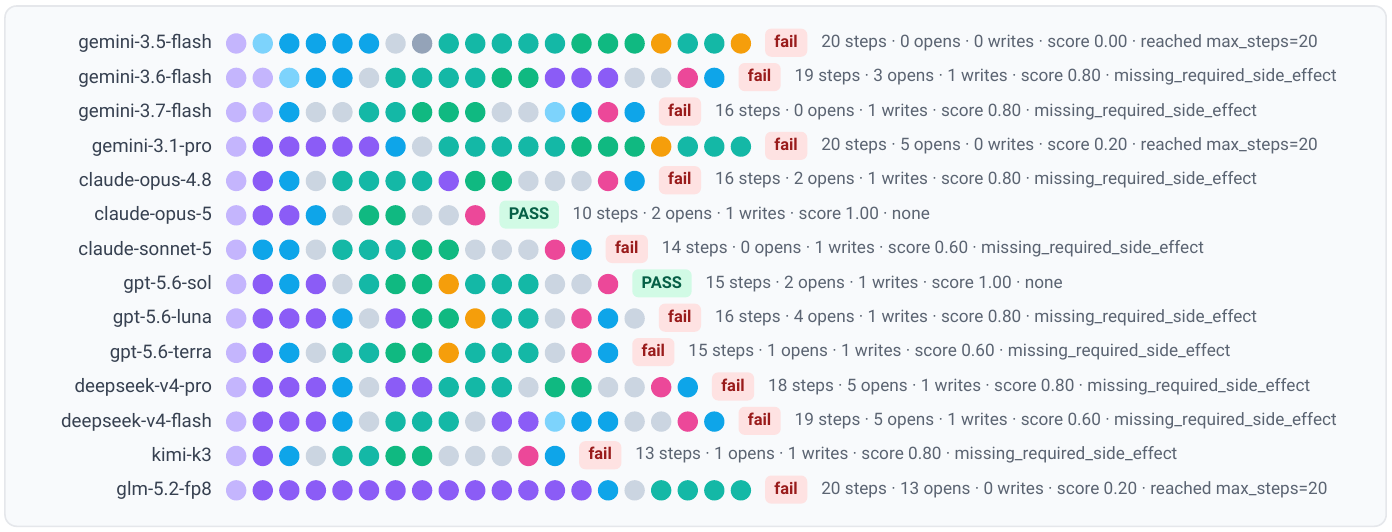}
\caption{\textbf{Representative Execution Trace: Expiring GPU Cluster Resource Hold (\texttt{v03\_r04}; Figure~\ref{fig:rep-round-8-expiring-resource-hold}).} Evaluates inspecting the reservation board, acquiring the provisional hold, and reporting the lease expiration.}
\label{fig:rep-round-8-expiring-resource-hold}
\end{figure}

\noindent\smallskip\textbf{Case Analysis: Expiring GPU Cluster Resource Hold (\texttt{v03\_r04}).}
This scenario tests visual inspection and resource management across five checks: (1) inspecting the visual dashboard \texttt{cluster\_dashboard\_reservation\_board\_active\_hold\_r04}, (2) acquiring hold token \texttt{HOLD-80G-9412-PROVISIONAL} on slot \texttt{A100-80G-02} with a 300\,s duration, (3) reporting the lease expiration, (4) stating the report verdict, and (5) notifying the stakeholder.

Claude Opus 5 (10 steps) and GPT-5.6 Sol (15 steps) pass all five checks. Six models score $0.80$ by opening the image and acquiring the hold, but failing on notification details. Claude Sonnet 5 misses the visual check (0 images opened) but still acquires the hold and writes the report ($0.60$), while GLM-5.2 FP8 opens 13 images but runs out of steps without writing any files ($0.20$).

\subsubsection*{Signature failure modes (one exemplar per model)}

\begin{figure}[H]
\centering
\includegraphics[width=\linewidth]{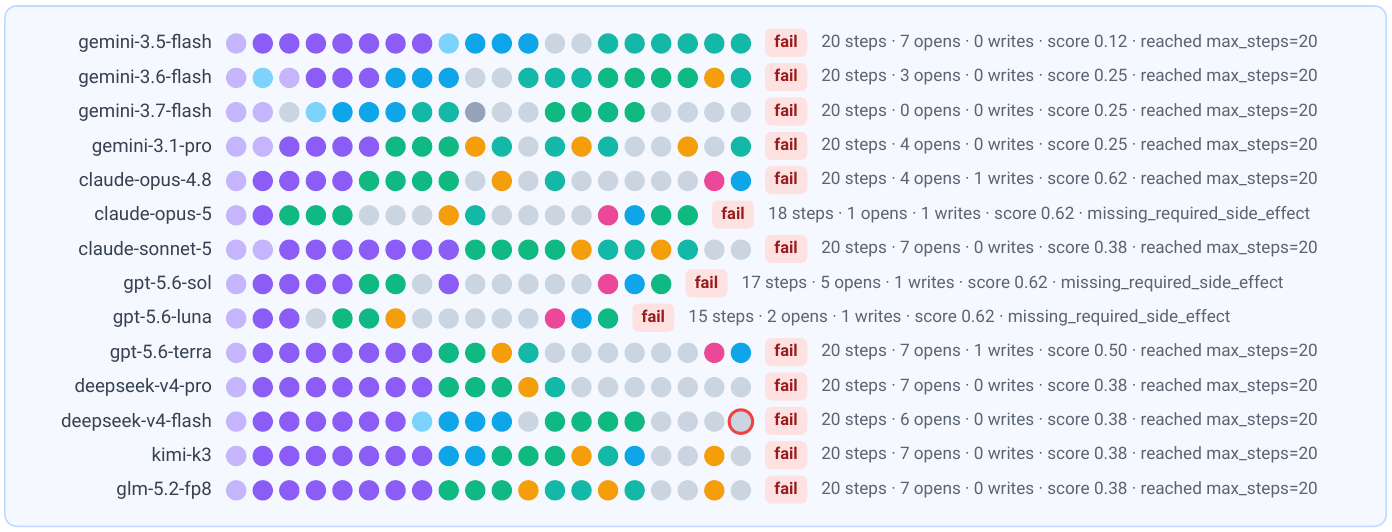}
\caption{\textbf{Gemini 3.5 Flash: Running out of steps due to search loops.} Gets trapped in repetitive search loops and reaches the 20-step limit on \textbf{89\%} of tasks (highest across all models). Task: \texttt{access\_request\_approval\_v01\_r01}.}
\label{fig:rep-sig-gemini-3-5-flash}
\end{figure}

\begin{figure}[H]
\centering
\includegraphics[width=\linewidth]{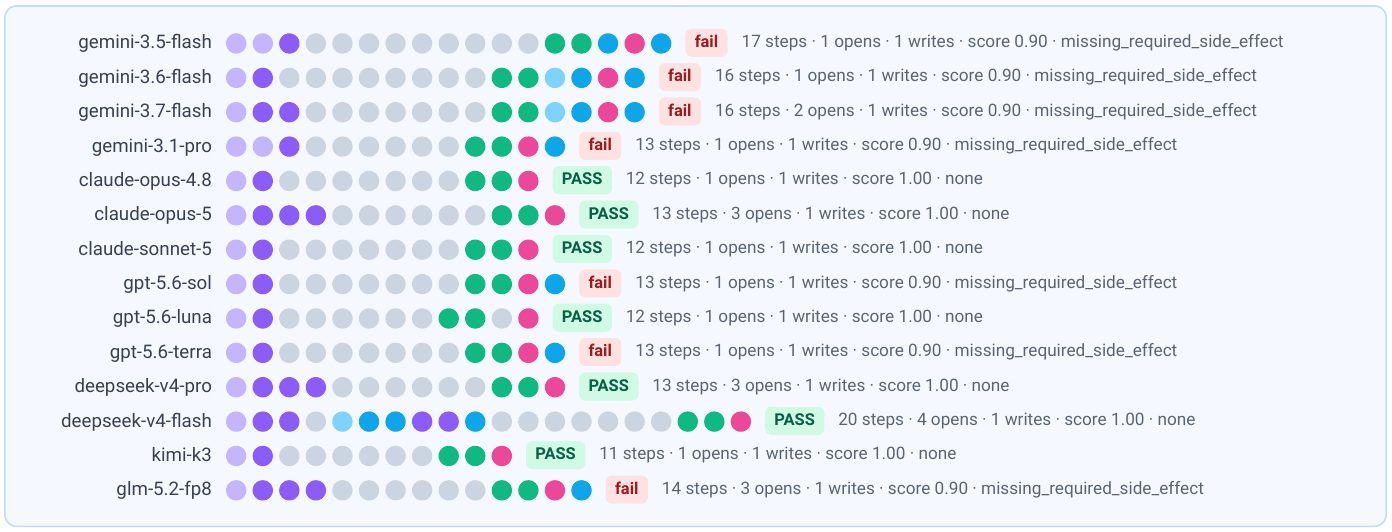}
\caption{\textbf{Gemini 3.6 Flash: Near-miss deliverable submission.} Quickly inspects images and writes reports without looping, but makes small precision errors on required text. Task: \texttt{employee\_onboarding\_workflow\_v01\_r08}.}
\label{fig:rep-sig-gemini-3-6-flash}
\end{figure}

\begin{figure}[H]
\centering
\includegraphics[width=\linewidth]{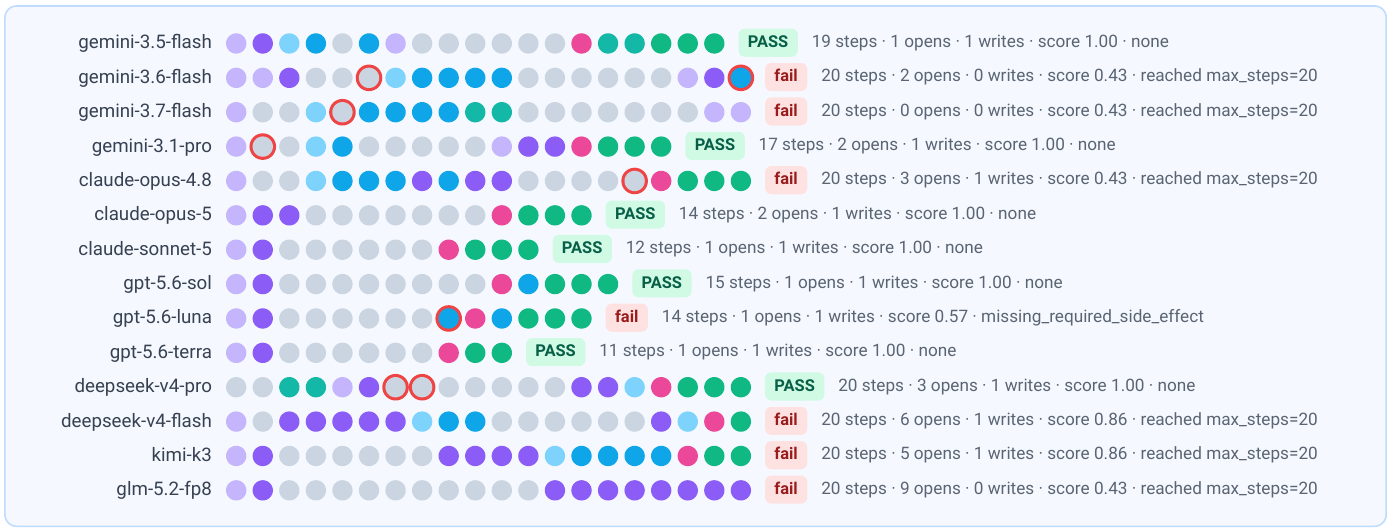}
\caption{\textbf{Gemini 3.7 Flash: Running out of steps by repeatedly checking agent status.} Inspects few images (\textbf{2.25} opens/task) while hitting the 20-step limit on \textbf{41\%} of tasks, spending \textbf{3.3\%} of all actions polling \texttt{agent status}. Task: \texttt{access\_request\_approval\_v03\_r07}.}
\label{fig:rep-sig-gemini-3-7-flash}
\end{figure}

\begin{figure}[H]
\centering
\includegraphics[width=\linewidth]{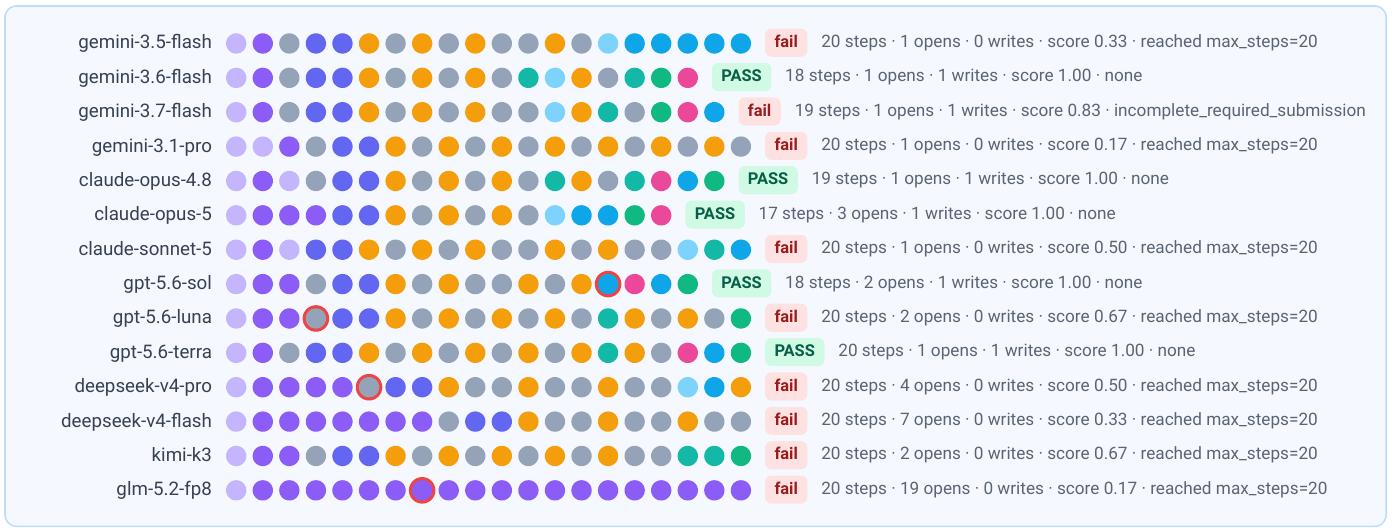}
\caption{\textbf{Gemini 3.1 Pro: Excessive waiting after sub-agent delegation.} Spends \textbf{5.1\%} of all actions on \texttt{wait} commands (highest across all models) after delegating work to sub-agents. Task: \texttt{barrier\_sync\_deployment\_v02\_r07}.}
\label{fig:rep-sig-gemini-3-1-pro}
\end{figure}

\begin{figure}[H]
\centering
\includegraphics[width=\linewidth]{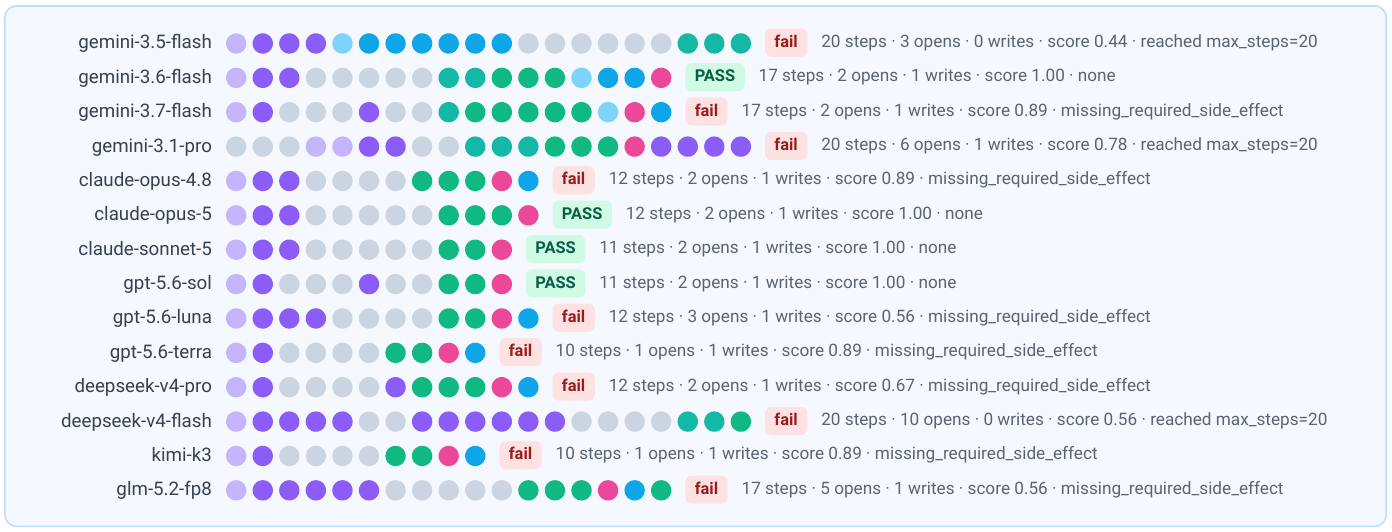}
\caption{\textbf{Claude Opus 4.8: Content precision errors under minimal image inspection.} Writes deliverables after inspecting few visual assets, missing fine-grained verification details. Task: \texttt{it\_deployment\_gate\_v03\_r05}.}
\label{fig:rep-sig-claude-opus-4-8}
\end{figure}

\begin{figure}[H]
\centering
\includegraphics[width=\linewidth]{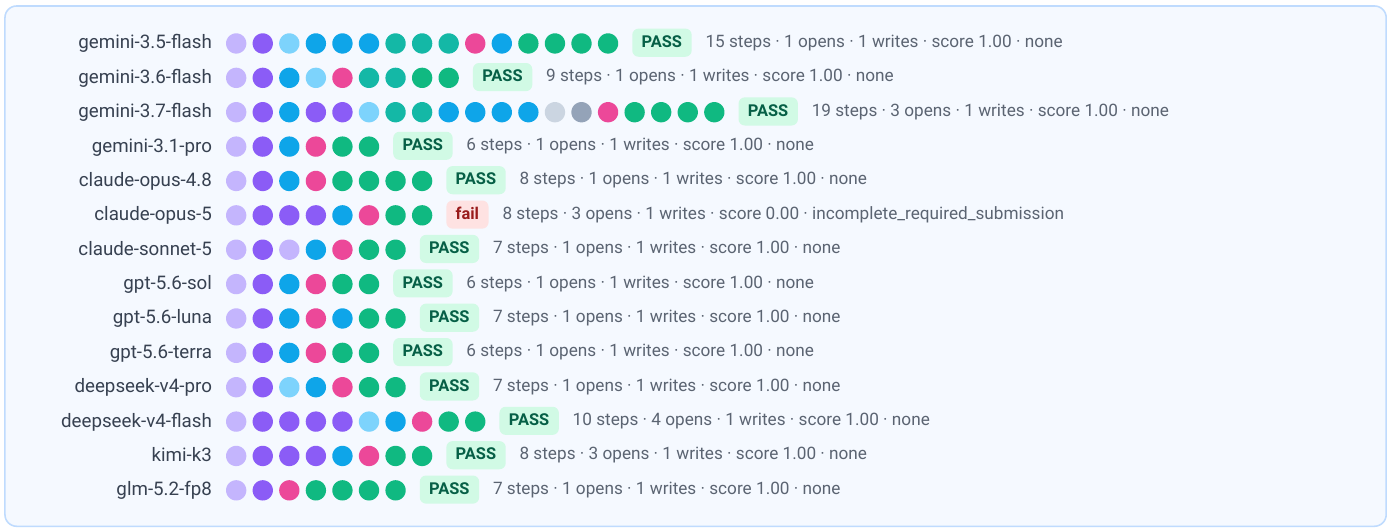}
\caption{\textbf{Claude Opus 5: High-concentration all-or-nothing task resolution.} Shows the highest failure clustering (\textbf{1.49}) and overdispersion (\textbf{2.23}), passing all checks simultaneously on \textbf{25.2\%} of tasks versus \textbf{10.8\%} expected under independence. Task: \texttt{it\_deployment\_gate\_v02\_r10}.}
\label{fig:rep-sig-claude-opus-5}
\end{figure}

\begin{figure}[H]
\centering
\includegraphics[width=\linewidth]{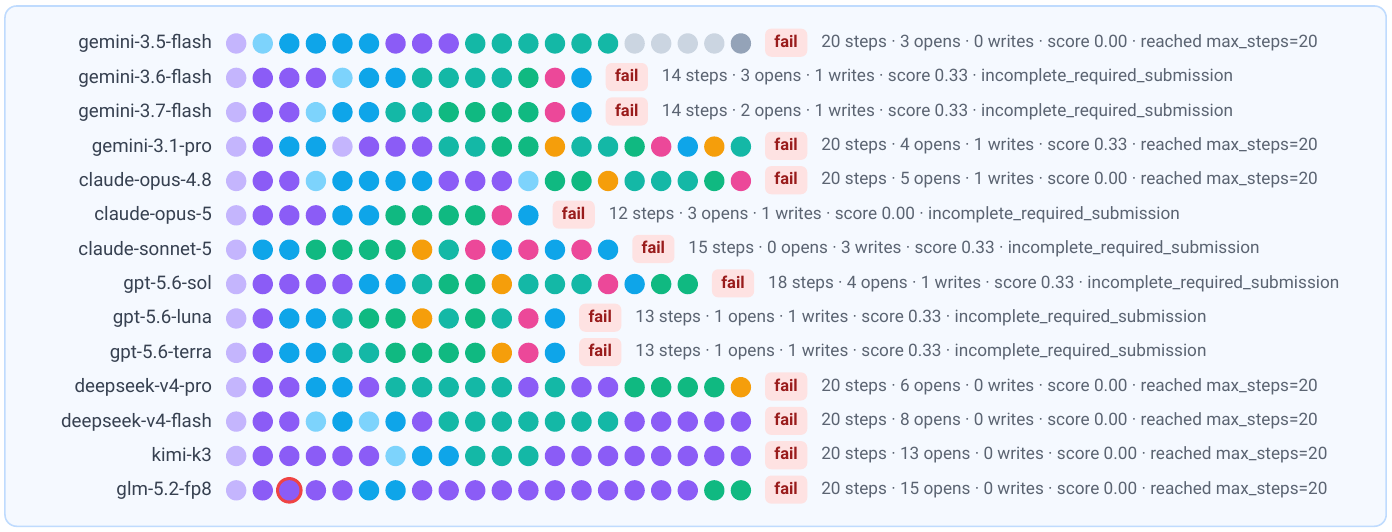}
\caption{\textbf{Claude Sonnet 5: Relying on text alone without inspecting images.} Skips image inspection entirely on \textbf{92} of 600 tasks, attempting to write reports from text context alone. Task: \texttt{satellite\_ground\_station\_handover\_v03\_r09}.}
\label{fig:rep-sig-claude-sonnet-5}
\end{figure}

\begin{figure}[H]
\centering
\includegraphics[width=\linewidth]{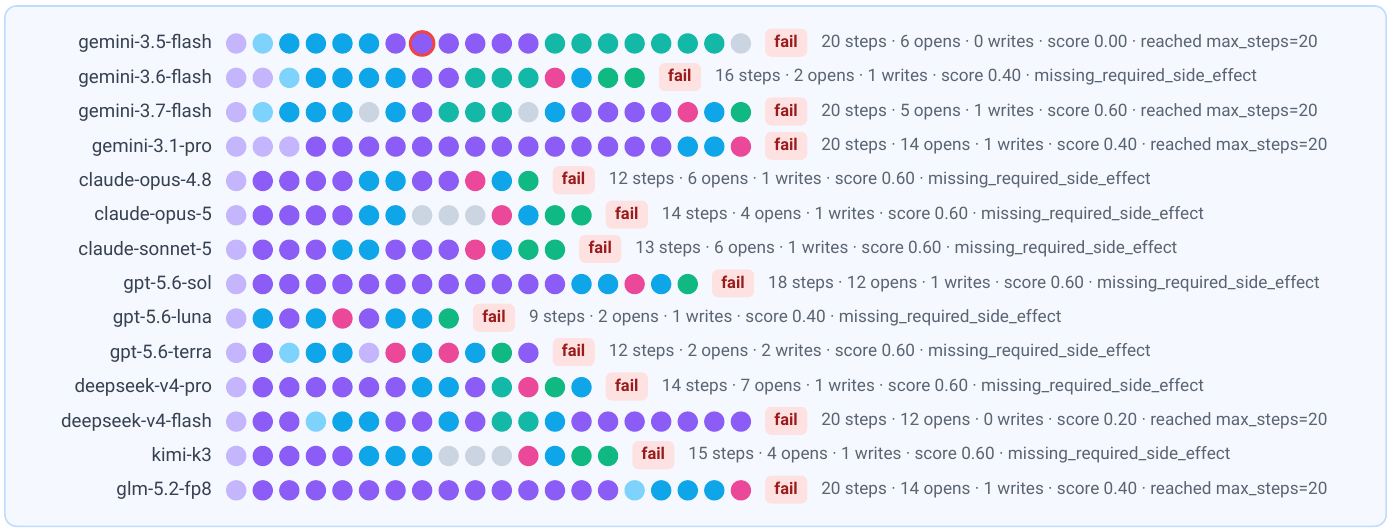}
\caption{\textbf{GPT-5.6 Sol: Thorough visual inspection with text transcription limits.} Inspects visual evidence on \textbf{100\%} of tasks and writes deliverables on 599/600 rounds, with failures driven by subtle text transcription errors. Task: \texttt{multi\_tab\_slot\_monitoring\_v02\_r05}.}
\label{fig:rep-sig-gpt-5-6-sol}
\end{figure}

\begin{figure}[H]
\centering
\includegraphics[width=\linewidth]{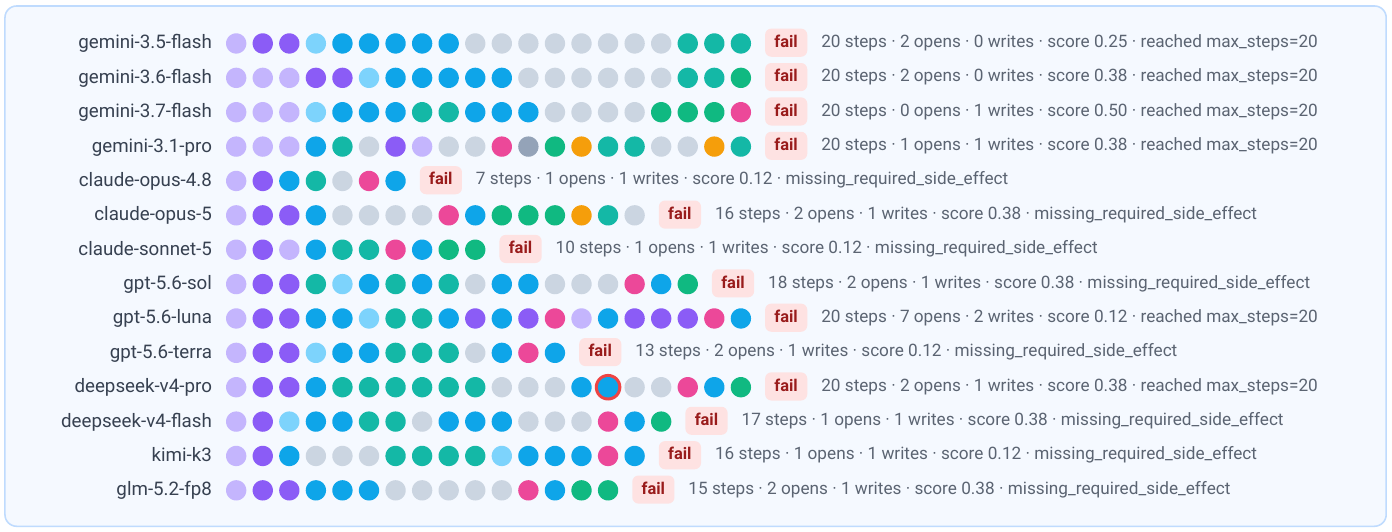}
\caption{\textbf{GPT-5.6 Luna: Re-opening images instead of finishing deliverables.} Repeatedly re-opens already-inspected images rather than writing outputs, leading to more step limit hits than other OpenAI models. Task: \texttt{multi\_tab\_slot\_monitoring\_v01\_r01}.}
\label{fig:rep-sig-gpt-5-6-luna}
\end{figure}

\begin{figure}[H]
\centering
\includegraphics[width=\linewidth]{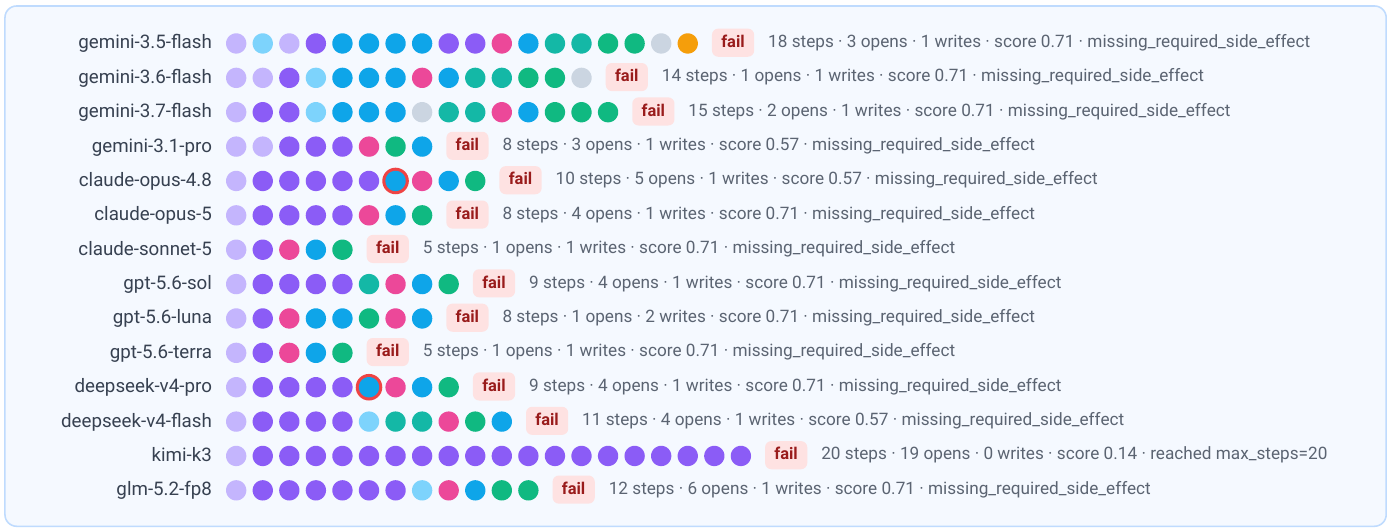}
\caption{\textbf{GPT-5.6 Terra: Stopping too early.} Concludes episodes early without looping, sometimes stopping before completing all required actions or deliverables. Task: \texttt{employee\_onboarding\_workflow\_v03\_r08}.}
\label{fig:rep-sig-gpt-5-6-terra}
\end{figure}

\begin{figure}[H]
\centering
\includegraphics[width=\linewidth]{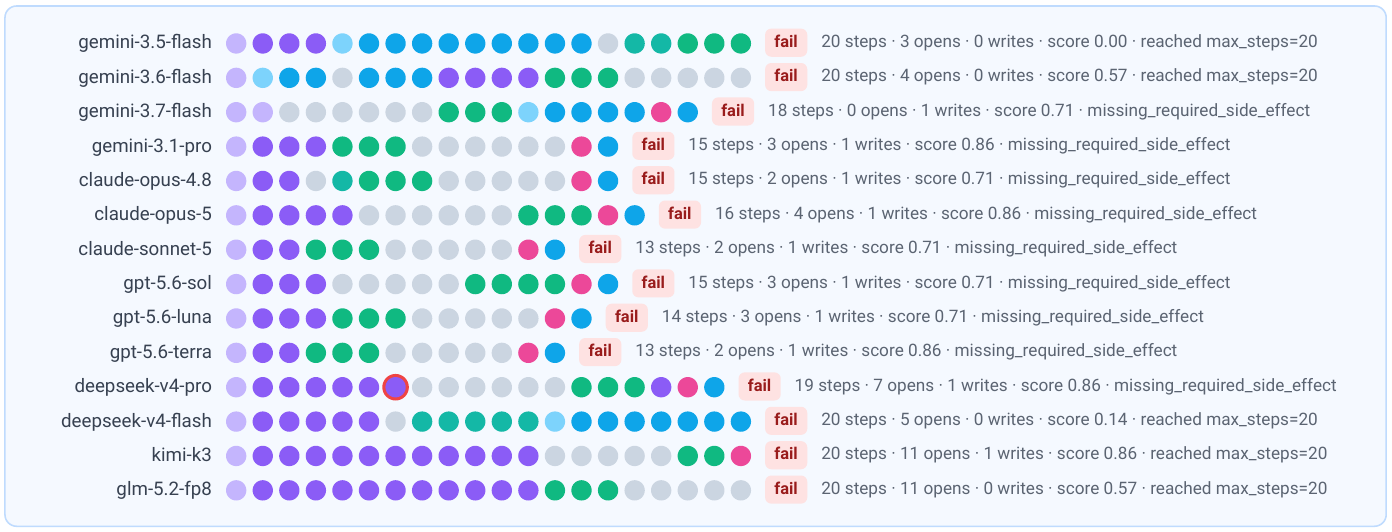}
\caption{\textbf{DeepSeek V4 Pro: Reliable completion among open-weight models.} Finishes and writes deliverables on \textbf{92.8\%} of tasks, hitting the step limit on only \textbf{13.5\%} of episodes. Task: \texttt{coordinated\_containment\_v01\_r05}.}
\label{fig:rep-sig-deepseek-v4-pro}
\end{figure}

\begin{figure}[H]
\centering
\includegraphics[width=\linewidth]{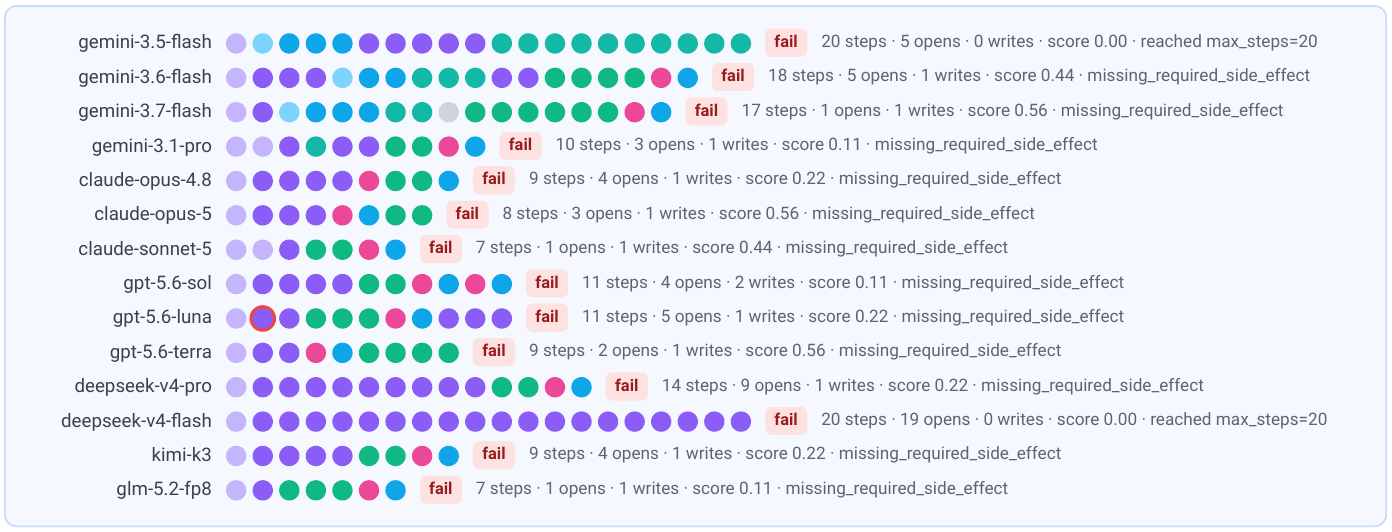}
\caption{\textbf{DeepSeek V4 Flash: Inspecting images without writing deliverables.} Opens \textbf{6.00} images per task but fails to write any deliverable on \textbf{47.3\%} of episodes. Task: \texttt{multi\_device\_pairing\_v02\_r08}.}
\label{fig:rep-sig-deepseek-v4-flash}
\end{figure}

\begin{figure}[H]
\centering
\includegraphics[width=\linewidth]{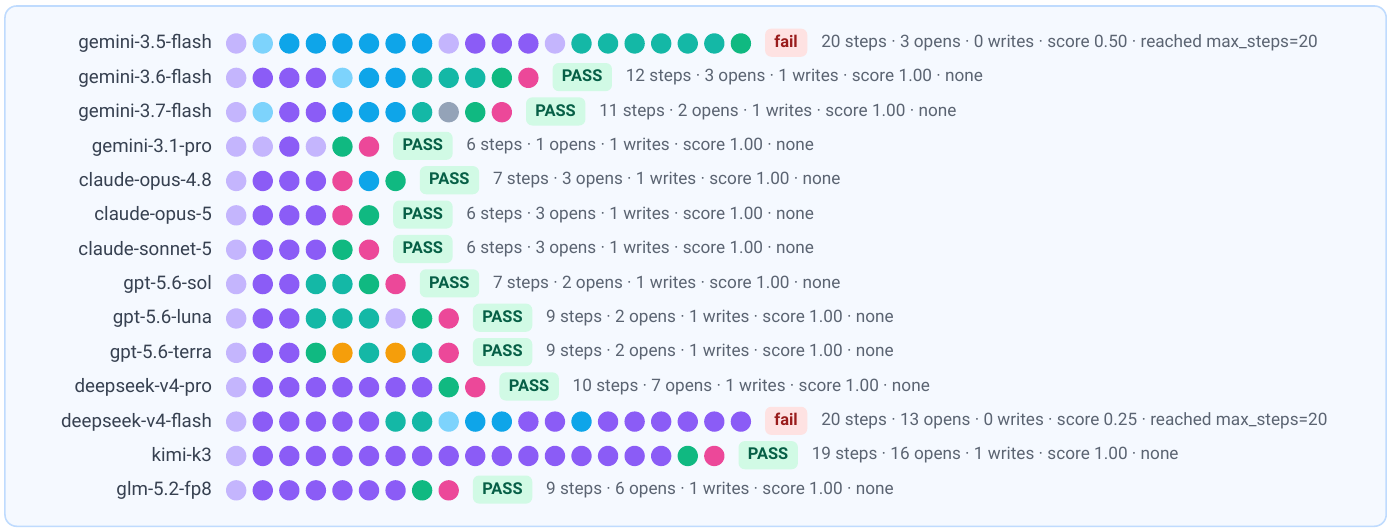}
\caption{\textbf{Kimi K3: Strong visual reasoning and image inspection.} Combines active image inspection (\textbf{6.72} opens/task) with high semantic check accuracy, achieving a \textbf{21.8\%} overall pass rate. Task: \texttt{multi\_party\_access\_permit\_v03\_r07}.}
\label{fig:rep-sig-kimi-k3}
\end{figure}

\begin{figure}[H]
\centering
\includegraphics[width=\linewidth]{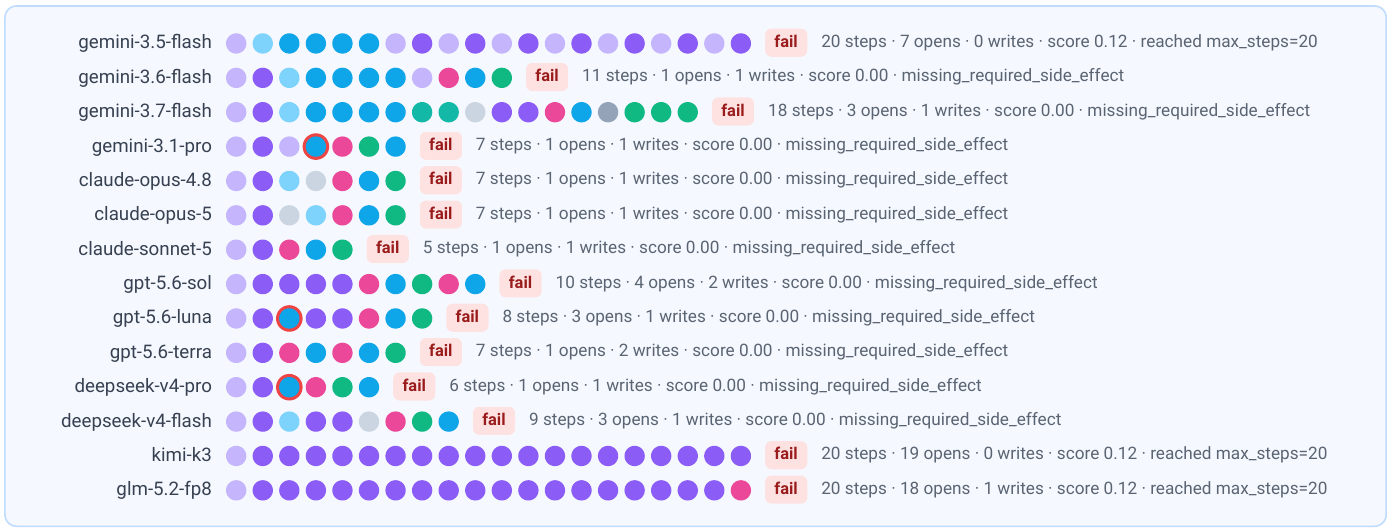}
\caption{\textbf{GLM-5.2 FP8: Heavy image inspection without writing deliverables.} Spends \textbf{48.8\%} of actions inspecting images (highest share across all models, \textbf{7.26} assets/task) but achieves a \textbf{14.7\%} pass rate because it often fails to write final deliverables. Task: \texttt{employee\_onboarding\_workflow\_v03\_r09}.}
\label{fig:rep-sig-glm-5-2-fp8}
\end{figure}

One round was chosen to clearly illustrate each model's characteristic behavior in Figures~\ref{fig:rep-sig-gemini-3-5-flash}--\ref{fig:rep-sig-glm-5-2-fp8} (\ref{fig:rep-sig-gemini-3-5-flash}, \ref{fig:rep-sig-gemini-3-6-flash}, \ref{fig:rep-sig-gemini-3-7-flash}, \ref{fig:rep-sig-gemini-3-1-pro}, \ref{fig:rep-sig-claude-opus-4-8}, \ref{fig:rep-sig-claude-opus-5}, \ref{fig:rep-sig-claude-sonnet-5}, \ref{fig:rep-sig-gpt-5-6-sol}, \ref{fig:rep-sig-gpt-5-6-luna}, \ref{fig:rep-sig-gpt-5-6-terra}, \ref{fig:rep-sig-deepseek-v4-pro}, \ref{fig:rep-sig-deepseek-v4-flash}, \ref{fig:rep-sig-kimi-k3}, \ref{fig:rep-sig-glm-5-2-fp8}). We use the same format as above---all 14 20-step baseline models tested on the same task, with action dots and the highlighted model shown first.

\subsection{Limitations}

\label{rep:s11}

\setcounter{subsubsection}{2}

\subsubsection{Notification checks depend on message body text rather than routing}

\label{rep:s11.3}

\begingroup
\setlength{\tabcolsep}{6pt}
\begin{xltabular}{\linewidth}{>{\raggedright\arraybackslash\hsize=1.15\hsize}X | >{\raggedright\arraybackslash\hsize=0.85\hsize}X r r}
\caption{\textbf{Notification check results across representative rounds.} Compares sending messages to the right person versus receiving grading credit. Across all rounds, models almost always message the right recipient (10--14 of the 14 baseline models), but scoring credit depends on matching the required body text.}\label{tab:rep-21} \\
\toprule
\textbf{round} & \textbf{named recipient} & \multicolumn{1}{c}{\thead[c]{\textbf{route} \\ \textbf{to them}}} & \multicolumn{1}{c}{\textbf{credited}} \\
\midrule
\endfirsthead
\multicolumn{4}{@{}l@{}}{\itshape Table~\thetable\ (continued)} \\
\toprule
\textbf{round} & \textbf{named recipient} & \multicolumn{1}{c}{\thead[c]{\textbf{route} \\ \textbf{to them}}} & \multicolumn{1}{c}{\textbf{credited}} \\
\midrule
\endhead
\midrule
\multicolumn{4}{@{}r@{}}{\itshape continued on the next page} \\
\endfoot
\bottomrule
\endlastfoot
\texttt{payment\_\allowbreak{}chargeback\_\allowbreak{}dispute\_\allowbreak{}v01\_\allowbreak{}r04} & \texttt{Casey\_\allowbreak{}Support\_\allowbreak{}Specialist} & 14/14 & 3/14 \\
\texttt{warehouse\_\allowbreak{}robotics\_\allowbreak{}safety\_\allowbreak{}halt\_\allowbreak{}v02\_\allowbreak{}r01} & \texttt{Avery\_\allowbreak{}Safety\_\allowbreak{}Officer} & 12/14 & \textbf{1/14} \\
\texttt{satellite\_\allowbreak{}ground\_\allowbreak{}station\_\allowbreak{}handover\_\allowbreak{}v02\_\allowbreak{}r01} & \texttt{Devon\_\allowbreak{}Ground\_\allowbreak{}Ops\_\allowbreak{}Lead} & 12/14 & 12/14 \\
\texttt{clinical\_\allowbreak{}trial\_\allowbreak{}dosage\_\allowbreak{}conflict\_\allowbreak{}v02\_\allowbreak{}r03} & \texttt{Avery\_\allowbreak{}Trial\_\allowbreak{}Safety\_\allowbreak{}Monitor} & 13/14 & 13/14 \\
\texttt{coordinated\_\allowbreak{}containment\_\allowbreak{}v02\_\allowbreak{}r04} & \texttt{Owen\_\allowbreak{}Platform\_\allowbreak{}On\_\allowbreak{}Call} & 11/14 & 13/14 \\
\texttt{distributed\_\allowbreak{}lock\_\allowbreak{}shared\_\allowbreak{}resource\_\allowbreak{}v01\_\allowbreak{}r02} & \texttt{Casey\_\allowbreak{}Inventory\_\allowbreak{}Controller} & 12/14 & 4/14 \\
\texttt{expense\_\allowbreak{}reimbursement\_\allowbreak{}conflict\_\allowbreak{}v02\_\allowbreak{}r05} & \texttt{Blair\_\allowbreak{}Audit\_\allowbreak{}Investigator} & 12/14 & 11/14 \\
\texttt{expense\_\allowbreak{}reimbursement\_\allowbreak{}conflict\_\allowbreak{}v02\_\allowbreak{}r05} & \texttt{Casey\_\allowbreak{}Approver} & 12/14 & 11/14 \\
\texttt{expiring\_\allowbreak{}resource\_\allowbreak{}hold\_\allowbreak{}v03\_\allowbreak{}r04} & \texttt{Rowan\_\allowbreak{}Research\_\allowbreak{}Lead} & 10/14 & 2/14 \\
\end{xltabular}
\endgroup

\clearpage
\subsubsection{Eighteen exact-match checks lack readable reference strings}

\label{rep:s11.4}

\begingroup
\setlength{\tabcolsep}{6pt}
\begin{xltabular}{\linewidth}{>{\raggedright\arraybackslash\hsize=1.00\hsize}X | r r r r}
\caption{\textbf{Pass rates of exact-match string checks.} Compares pass rates on checks where reference strings appear in \texttt{visible\_refs} ($n=238$, $75.8\%$ pass) versus checks where they are unexposed ($n=18$, $0.7\%$ pass).}\label{tab:rep-22} \\
\toprule
\textbf{exact-match checks} & \multicolumn{1}{c}{\textbf{n}} & \multicolumn{1}{c}{\thead[c]{\textbf{this check} \\ \textbf{passes}}} & \multicolumn{1}{c}{\thead[c]{\textbf{sibling checks,} \\ \textbf{same rounds}}} & \multicolumn{1}{c}{\thead[c]{\textbf{never} \\ \textbf{passed}}} \\
\midrule
\endfirsthead
\multicolumn{5}{@{}l@{}}{\itshape Table~\thetable\ (continued)} \\
\toprule
\textbf{exact-match checks} & \multicolumn{1}{c}{\textbf{n}} & \multicolumn{1}{c}{\thead[c]{\textbf{this check} \\ \textbf{passes}}} & \multicolumn{1}{c}{\thead[c]{\textbf{sibling checks,} \\ \textbf{same rounds}}} & \multicolumn{1}{c}{\thead[c]{\textbf{never} \\ \textbf{passed}}} \\
\midrule
\endhead
\midrule
\multicolumn{5}{@{}r@{}}{\itshape continued on the next page} \\
\endfoot
\bottomrule
\endlastfoot
value readable from \texttt{visible\_\allowbreak{}refs} & 238 & 75.8\% & 67.6\% & 17 \\
value \textbf{not} readable & 18 & \textbf{0.7\%} & 37.7\% & \textbf{15} \\
\end{xltabular}
\endgroup

\subsection{What each model would need}

\label{rep:s12}

\begingroup
\setlength{\tabcolsep}{6pt}
\begin{xltabular}{\linewidth}{l | >{\raggedright\arraybackslash\hsize=1.35\hsize}X >{\raggedright\arraybackslash\hsize=0.65\hsize}X}
\caption{\textbf{Primary bottlenecks and diagnostic evidence for each evaluated model.} Summarizes the main bottlenecks limiting task completion across models, ordered by High-effort pass rate.}\label{tab:rep-23} \\
\toprule
\textbf{Model} & \textbf{binding constraint} & \textbf{evidence} \\
\midrule
\endfirsthead
\multicolumn{3}{@{}l@{}}{\itshape Table~\thetable\ (continued)} \\
\toprule
\textbf{Model} & \textbf{binding constraint} & \textbf{evidence} \\
\midrule
\endhead
\midrule
\multicolumn{3}{@{}r@{}}{\itshape continued on the next page} \\
\endfoot
\bottomrule
\endlastfoot
GPT-6 Astra & Multi-agent closing protocols and shared state cleanup. Leads the benchmark with \textbf{28.7\%} pass rate, \textbf{66.8\%} mean score, and \textbf{98.8\%} visual gate compliance (averaging 26.1 steps/task). Main failures stem from terminating prematurely before completing closing protocols: incomplete deliverable submissions (\textbf{199} tasks) and missing environment side effects (\textbf{220} tasks), hitting its step limit on only \textbf{6} tasks. & \S\ref{rep:s1}, \S\ref{rep:s5}, \S\ref{rep:s7.1}, \S\ref{rep:s7.3} \\
Claude Opus 5 & Executing environment actions, not more reasoning. Achieves strong per-check accuracy (63.0\%) and high failure clustering (lift \textbf{1.49}), yet \textbf{239} episodes end without required environment actions while only \textbf{3} hit the step limit. \textbf{140} rounds finish exactly one check short (47 on missing environment actions). Higher reasoning effort provides negligible gain (+1.7pp, n.s.). & \S\ref{rep:s5}, \S\ref{rep:s7.1}, \S\ref{rep:s9.2} \\
GPT-5.6 Sol & Search efficiency and deliverable accuracy. Issues \textbf{877} \texttt{message search} calls and opens \textbf{4.08} images/task, reaching a \textbf{91.2\%} visual gate rate but lower text transcription accuracy (\textbf{75.8\%} vs.\ Claude Opus 5's 82.4\%). Shows steady gains from higher reasoning effort (+3.7pp, $p = 0.0016$). & \S\ref{rep:s3}, \S\ref{rep:s7.3}, \S\ref{rep:s9.2} \\
Claude Opus 4.8 & High token cost and compute efficiency. Highest cost per solved task among frontier models (\textbf{\$21.41}, and \textbf{\$4.89}/task) with a 2.4pp lower pass rate than Claude Opus 5, driven by \textbf{4.1$\times$} higher reasoning token usage (2,291 vs.\ 553 tokens/task). Higher reasoning effort yields no significant improvement (+1.5pp, n.s.). & \S\ref{rep:s1}, \S\ref{rep:s9.1}, \S\ref{rep:s9.2} \\
Kimi K3 & Writing deliverables before timing out. Achieves strong semantic accuracy (\textbf{65.7\%} on \texttt{field\_semantic\_match} and \textbf{42.6\%} on \texttt{other}), but \textbf{437} semantic checks are never tested because the model wrote no files on \textbf{128} rounds and hit the step limit on \textbf{170} rounds. & \S\ref{rep:s7.2}, \S\ref{rep:s7.3}, \S\ref{rep:s5} \\
Claude Sonnet 5 & Opening and inspecting visual assets. Opens the fewest images (\textbf{1.83}/task), skips images entirely on \textbf{92} tasks, has the lowest visual gate rate (\textbf{72.6\%}), and fails \textbf{76} image checks. When it does open images, transcription accuracy is high (\textbf{76.6\%}), and higher reasoning effort yields the largest gain in the field (\textbf{+7.0pp}, $p = 1.0\times 10^{-7}$). & \S\ref{rep:s7.2}, \S\ref{rep:s7.3}, \S\ref{rep:s9.2} \\
Gemini 3.7 Flash & Running out of steps. Hits the 20-step limit on \textbf{244} rounds (\textbf{40.7\%}, highest among competitive models) with a median of 19 steps. Higher reasoning effort increases limit exhaustion ($3.0\% \to 12.7\% \to 40.7\%$) with no accuracy gain ($-0.2$pp). & \S\ref{rep:s5}, \S\ref{rep:s3}, \S\ref{rep:s9.2} \\
GPT-5.6 Terra & Stopping too early. Has the highest rates of missing environment actions (\textbf{261}) and incomplete deliverables (\textbf{210}), while hitting the step limit on only \textbf{8} rounds (mean 11.4 steps). Higher reasoning effort yields modest gains (+2.3pp, $p = 0.049$). & \S\ref{rep:s5}, \S\ref{rep:s6} \\
GPT-5.6 Luna & High latency and slow execution. Has low financial cost (\$0.14/task, \$0.72/solved), but very slow runtimes (\textbf{620.5\,s}/task) and \textbf{6} stalls in repetitive loops. & \S\ref{rep:s1}, \S\ref{rep:s9.2}, \S\ref{rep:s5} \\
DeepSeek V4 Pro & Reasoning about final verdicts. Operates at the lowest cost (\$0.05/task, \$0.27/solved) with balanced error types (218 missing actions, 182 incomplete deliverables, 81 step-limited), but scores lower on final verdict checks (\texttt{other} at \textbf{33.2\%}). & \S\ref{rep:s1}, \S\ref{rep:s7.2} \\
Gemini 3.6 Flash & Outperformed by newer models. Gemini 3.7 Flash achieves a higher pass rate (20.3\% vs.\ \textbf{18.5\%}) at half the cost per solved task (\$2.51 vs.\ \textbf{\$5.09}). Higher reasoning effort yields no accuracy gain (+0.0pp, $p = 1.0$) while causing more step limit hits ($1.8\% \to 14.5\% \to 24.7\%$). & \S\ref{rep:s1}, \S\ref{rep:s9.2} \\
GLM-5.2 FP8 & Turning visual evidence into written deliverables. Opens \textbf{7.26} images/task with \textbf{93.2\%} visual gate compliance, but writes no files on \textbf{157} rounds and achieves only \textbf{51.6\%} text transcription accuracy (a \textbf{41.6pp} gap). & \S\ref{rep:s3}, \S\ref{rep:s7.3} \\
Gemini 3.1 Pro & Notification content precision and excessive waiting. Attains \textbf{46.4\%} semantic match accuracy and fails \textbf{831} message content checks, spending \textbf{402} actions on \texttt{wait} (highest across all models). & \S\ref{rep:s7.1}, \S\ref{rep:s7.2}, \S\ref{rep:s3} \\
DeepSeek V4 Flash & Running out of steps and omitting deliverables. Ends \textbf{358} rounds (\textbf{59.7\%}) at the 20-step limit and writes no deliverables on \textbf{284} rounds (0.58 writes/task). & \S\ref{rep:s5}, \S\ref{rep:s3}, \S\ref{rep:s7.3} \\
Gemini 3.5 Flash & Trapped in search loops. Executes \textbf{2,707} search actions (4.51/task) and writes files only \textbf{0.31} times/task with \textbf{440} zero-write episodes, causing \textbf{535} step limit hits (\textbf{89.2\%}). High reasoning effort worsens search loops ($-10.5$pp, $p = 2.2\times 10^{-19}$), whereas Low effort recovers a \textbf{13.5\%} pass rate. & \S\ref{rep:s3}, \S\ref{rep:s6}, \S\ref{rep:s7.1}, \S\ref{rep:s9.2} \\
\end{xltabular}
\endgroup

\end{document}